\documentclass[reprint,pra,aps,preprintnumbers,superscriptaddress,longbibliography,nofootinbib]{revtex4-2}
\usepackage{graphicx}
\usepackage{dcolumn}
\usepackage{bm}
\usepackage{orcidlink}
\usepackage{physics}
\usepackage{amssymb}
\usepackage{amsmath}
\usepackage{enumitem}
\usepackage{tikz}
\usepackage{slashed}
\usepackage{bm}
\usepackage{amsthm}
\usepackage{hyperref}
\usepackage{cleveref}
\usepackage{mathrsfs}
\usepackage{mathtools}
\usepackage{multirow}
\usepackage{hhline}
\usetikzlibrary{calc,arrows.meta,decorations.markings}
\usepackage{xparse}
\usepackage[table]{xcolor}
\usepackage{float}

\newlength{\twocolumnwidth}
\AtBeginDocument{
  \setlength{\twocolumnwidth}{0.45\textwidth}
}

\begin{document}

\newcommand{\Crefpanel}[2]{\Cref{#1}(#2)}

\newcommand{\SU}{\mathrm{SU}}
\newcommand{\phys}{\mathrm{phys}}
\newcommand{\hop}{\mathrm{hop}}
\newcommand{\eps}{\varepsilon}
\newcommand{\vx}{\mathcal V}
\newcommand{\lk}{\mathcal L}
\newcommand{\pq}{\mathcal P}
\newcommand{\KS}{\mathrm{KS}}
\newcommand{\ps}{\text{p.s.}}
\newcommand{\hex}{\mathrm{hex}}
\DeclarePairedDelimiter{\del}{(}{)}
\newcommand{\Tone}{{$T_1$}}
\newcommand{\Ttwo}{{$T_2$}}
\newcommand{\Tthree}{{$T_3$}}
\newcommand{\Tthreep}{{$T_3'$}}
\newcommand{\Tfour}{{$T_4$}}
\newcommand{\Tfive}{{$T_5$}}
\newcommand{\TTone}{{$TT_1$}}
\newcommand{\TTtwo}{{$TT_2$}}
\renewcommand{\figurename}{Fig.}
\renewcommand{\tablename}{Table}
\Crefname{figure}{Fig.}{Figs.}
\renewcommand{\crefrangeconjunction}{--}
\newcommand{\NM}[1]{\textcolor{red}{\textit{[NM: #1]}}}
\newcommand{\CWB}[1]{\textcolor{blue}{\textit{[CWB: #1]}}}
\newcommand{\ANC}[1]{\textcolor{green}{\textit{[ANC: #1]}}}
\newcommand{\JCH}[1]{\textcolor{purple}{\textit{[JCH: #1]}}}

\newcommand{\Square}[3][]{
  \draw[#1] #2 rectangle ++(#3,#3);
}

\newcommand{\SmallCircle}[5][]{
\pgfmathsetmacro{\GrayPct}{#4*100}
  \filldraw[
    draw=black,
    fill=black!\GrayPct,
    fill opacity=1.0,
    draw opacity=1.0,
    #1
  ] ({#2},{#3}) circle ({#5});
}

\newcommand{\EvenPlaquette}[5][]{
  \Square[draw=#5]{(#2, #3)}{#4}

  \pgfmathsetmacro{\pr}{0.075}

  \SmallCircle[draw=#5, fill=#5]{#2}{#3}{1}{\pr}
  \SmallCircle[draw=#5]{#2+#4}{#3}{0}{\pr}
  \SmallCircle[draw=#5, fill=#5]{#2+#4}{#3+#4}{1}{\pr}
  \SmallCircle[draw=#5]{#2}{#3+#4}{0}{\pr}
}

\newcommand{\EvenPlaquetteSiteUncolored}[5][]{
  \Square[draw=#5]{(#2, #3)}{#4}

  \pgfmathsetmacro{\pr}{0.075}

  \SmallCircle{#2}{#3}{1}{\pr}
  \SmallCircle{#2+#4}{#3}{0}{\pr}
  \SmallCircle{#2+#4}{#3+#4}{1}{\pr}
  \SmallCircle{#2}{#3+#4}{0}{\pr}
}

\newcommand{\OddPlaquette}[5][]{
  \Square[draw=#5]{(#2, #3)}{#4}

  \pgfmathsetmacro{\pr}{0.075}

  \SmallCircle[draw=#5]{#2}{#3}{0}{\pr}
  \SmallCircle[draw=#5, fill=#5]{#2+#4}{#3}{1}{\pr}
  \SmallCircle[draw=#5]{#2+#4}{#3+#4}{0}{\pr}
  \SmallCircle[draw=#5, fill=#5]{#2}{#3+#4}{1}{\pr}
}

\newcommand{\OddPlaquetteSiteUncolored}[5][]{
  \Square[draw=#5]{(#2, #3)}{#4}

  \pgfmathsetmacro{\pr}{0.075}

  \SmallCircle{#2}{#3}{0}{\pr}
  \SmallCircle{#2+#4}{#3}{1}{\pr}
  \SmallCircle{#2+#4}{#3+#4}{0}{\pr}
  \SmallCircle{#2}{#3+#4}{1}{\pr}
}

\newcommand{\WilsonRight}[6][]{
\draw[draw=#6] (#2, #3) -- (#4, #5)
\pgfmathsetmacro{\pr}{0.075}
\SmallCircle[draw=#5, fill=#5]{#2}{#3}{1}{\pr}
\SmallCircle[draw=#5]{#4}{#5}{0}{\pr}
}

\newcommand{\WilsonRightSiteUncolored}[6][]{
\draw[draw=#6] (#2, #3) -- (#4, #5)
\pgfmathsetmacro{\pr}{0.075}
\SmallCircle{#2}{#3}{1}{\pr}
\SmallCircle{#4}{#5}{0}{\pr}
}

\newcommand{\WilsonLeft}[6][]{
\draw[draw=#6] (#2, #3) -- (#4, #5)
\pgfmathsetmacro{\pr}{0.075}
\SmallCircle[draw=#5]{#2}{#3}{0}{\pr}
\SmallCircle[draw=#5, fill=#5]{#4}{#5}{1}{\pr}
}

\newcommand{\WilsonLeftSiteUncolored}[6][]{
\draw[draw=#6] (#2, #3) -- (#4, #5)
\pgfmathsetmacro{\pr}{0.075}
\SmallCircle{#2}{#3}{0}{\pr}
\SmallCircle{#4}{#5}{1}{\pr}
}

\newcommand{\OddHexPlaquette}[5][]{

  \pgfmathsetmacro{\Ax}{#2 + #4}
  \pgfmathsetmacro{\Ay}{#3}
  \pgfmathsetmacro{\Bx}{#2 + 1.5*#4}
  \pgfmathsetmacro{\By}{#3 - 0.5*#4}
  \pgfmathsetmacro{\Cx}{#2 + 2*#4}
  \pgfmathsetmacro{\Cy}{#3 - #4}
  \pgfmathsetmacro{\Dx}{#2 + 2*#4}
  \pgfmathsetmacro{\Dy}{#3 - 2*#4}
  \pgfmathsetmacro{\Ex}{#2 + #4}
  \pgfmathsetmacro{\Ey}{#3 - 2*#4}
  \pgfmathsetmacro{\Fx}{#2 + 0.5*#4}
  \pgfmathsetmacro{\Fy}{#3 - 1.5*#4}
  \pgfmathsetmacro{\Gx}{#2}
  \pgfmathsetmacro{\Gy}{#3 - #4}
  
  \draw[draw=#5, #1]
  ({#2}, {#3}) --
  ({\Ax}, {\Ay}) --
  ({\Bx}, {\By}) --
  ({\Cx}, {\Cy}) --
  ({\Dx}, {\Dy}) --
  ({\Ex}, {\Ey}) --
  ({\Fx}, {\Fy}) --
  ({\Gx}, {\Gy}) -- ({#2},{#3});

  \pgfmathsetmacro{\pr}{0.075}

  \pgfmathsetmacro{\Hx}{#2-0.5*#4}
  \pgfmathsetmacro{\Hy}{#3+0.5*#4}
  \pgfmathsetmacro{\Ix}{\Dx+0.5*#4}
  \pgfmathsetmacro{\Iy}{\Dy-0.5*#4}

  \draw[draw=#5] ({#2},{#3}) -- ({\Hx},{\Hy});
  \draw[draw=#5] ({\Dx},{\Dy}) -- ({\Ix},{\Iy});

  \SmallCircle[draw=#5]{\Bx}{\By}{0}{\pr}
  \SmallCircle[draw=#5]{\Fx}{\Fy}{0}{\pr}
  \SmallCircle[draw=#5, fill=#5]{\Hx}{\Hy}{1}{\pr}
  \SmallCircle[draw=#5, fill=#5]{\Ix}{\Iy}{1}{\pr}
}

\newcommand{\OddHexPlaquetteReduced}[5][]{

  \pgfmathsetmacro{\Ax}{#2 + #4}
  \pgfmathsetmacro{\Ay}{#3}
  \pgfmathsetmacro{\Bx}{#2 + 1.5*#4}
  \pgfmathsetmacro{\By}{#3 - 0.5*#4}
  \pgfmathsetmacro{\Cx}{#2 + 2*#4}
  \pgfmathsetmacro{\Cy}{#3 - #4}
  \pgfmathsetmacro{\Dx}{#2 + 2*#4}
  \pgfmathsetmacro{\Dy}{#3 - 2*#4}
  \pgfmathsetmacro{\Ex}{#2 + #4}
  \pgfmathsetmacro{\Ey}{#3 - 2*#4}
  \pgfmathsetmacro{\Fx}{#2 + 0.5*#4}
  \pgfmathsetmacro{\Fy}{#3 - 1.5*#4}
  \pgfmathsetmacro{\Gx}{#2}
  \pgfmathsetmacro{\Gy}{#3 - #4}
  
  \draw[draw=#5, #1]
  ({#2}, {#3}) --
  ({\Ax}, {\Ay}) --
  ({\Bx}, {\By}) --
  ({\Cx}, {\Cy}) --
  ({\Dx}, {\Dy}) --
  ({\Ex}, {\Ey}) --
  ({\Fx}, {\Fy}) --
  ({\Gx}, {\Gy}) -- ({#2},{#3});

  \pgfmathsetmacro{\pr}{0.075}

  \SmallCircle[draw=#5]{\Bx}{\By}{0}{\pr}
  \SmallCircle[draw=#5]{\Fx}{\Fy}{0}{\pr}
}

\newcommand{\EvenHexPlaquette}[5][]{

  \pgfmathsetmacro{\Ax}{#2 + #4}
  \pgfmathsetmacro{\Ay}{#3}
  \pgfmathsetmacro{\Bx}{#2 + 1.5*#4}
  \pgfmathsetmacro{\By}{#3 - 0.5*#4}
  \pgfmathsetmacro{\Cx}{#2 + 2*#4}
  \pgfmathsetmacro{\Cy}{#3 - #4}
  \pgfmathsetmacro{\Dx}{#2 + 2*#4}
  \pgfmathsetmacro{\Dy}{#3 - 2*#4}
  \pgfmathsetmacro{\Ex}{#2 + #4}
  \pgfmathsetmacro{\Ey}{#3 - 2*#4}
  \pgfmathsetmacro{\Fx}{#2 + 0.5*#4}
  \pgfmathsetmacro{\Fy}{#3 - 1.5*#4}
  \pgfmathsetmacro{\Gx}{#2}
  \pgfmathsetmacro{\Gy}{#3 - #4}
  
  \draw[draw=#5, #1]
  ({#2}, {#3}) --
  ({\Ax}, {\Ay}) --
  ({\Bx}, {\By}) --
  ({\Cx}, {\Cy}) --
  ({\Dx}, {\Dy}) --
  ({\Ex}, {\Ey}) --
  ({\Fx}, {\Fy}) --
  ({\Gx}, {\Gy}) -- ({#2},{#3});

  \pgfmathsetmacro{\pr}{0.075}

  \pgfmathsetmacro{\Hx}{#2-0.5*#4}
  \pgfmathsetmacro{\Hy}{#3+0.5*#4}
  \pgfmathsetmacro{\Ix}{\Dx+0.5*#4}
  \pgfmathsetmacro{\Iy}{\Dy-0.5*#4}

  \draw[draw=#5] ({#2},{#3}) -- ({\Hx},{\Hy});
  \draw[draw=#5] ({\Dx},{\Dy}) -- ({\Ix},{\Iy});

  \SmallCircle[draw=#5, fill=#5]{\Bx}{\By}{1}{\pr}
  \SmallCircle[draw=#5, fill=#5]{\Fx}{\Fy}{1}{\pr}
  \SmallCircle[draw=#5]{\Hx}{\Hy}{0}{\pr}
  \SmallCircle[draw=#5]{\Ix}{\Iy}{0}{\pr}
}

\newcommand{\EvenHexPlaquetteReduced}[5][]{

  \pgfmathsetmacro{\Ax}{#2 + #4}
  \pgfmathsetmacro{\Ay}{#3}
  \pgfmathsetmacro{\Bx}{#2 + 1.5*#4}
  \pgfmathsetmacro{\By}{#3 - 0.5*#4}
  \pgfmathsetmacro{\Cx}{#2 + 2*#4}
  \pgfmathsetmacro{\Cy}{#3 - #4}
  \pgfmathsetmacro{\Dx}{#2 + 2*#4}
  \pgfmathsetmacro{\Dy}{#3 - 2*#4}
  \pgfmathsetmacro{\Ex}{#2 + #4}
  \pgfmathsetmacro{\Ey}{#3 - 2*#4}
  \pgfmathsetmacro{\Fx}{#2 + 0.5*#4}
  \pgfmathsetmacro{\Fy}{#3 - 1.5*#4}
  \pgfmathsetmacro{\Gx}{#2}
  \pgfmathsetmacro{\Gy}{#3 - #4}
  
  \draw[draw=#5, #1]
  ({#2}, {#3}) --
  ({\Ax}, {\Ay}) --
  ({\Bx}, {\By}) --
  ({\Cx}, {\Cy}) --
  ({\Dx}, {\Dy}) --
  ({\Ex}, {\Ey}) --
  ({\Fx}, {\Fy}) --
  ({\Gx}, {\Gy}) -- ({#2},{#3});

  \pgfmathsetmacro{\pr}{0.075}

  \SmallCircle[draw=#5, fill=#5]{\Bx}{\By}{1}{\pr}
  \SmallCircle[draw=#5, fill=#5]{\Fx}{\Fy}{1}{\pr}
}

\newcommand{\Hexagon}[4][]{
  \pgfmathsetmacro{\a}{#4}
  \pgfmathsetmacro{\h}{0.866025403784*\a}

  \draw[#1]
    ({#2+0.5*\a},{#3+0})      --
    ({#2+1.5*\a},{#3+0})      --
    ({#2+2*\a},   {#3+\h})    --
    ({#2+1.5*\a},{#3+2*\h})   --
    ({#2+0.5*\a},{#3+2*\h})   --
    ({#2+0},     {#3+\h})     --
    cycle;
}

\newcommand{\OrientedEdge}[6][]{
  \pgfmathsetmacro{\mx}{0.66*(#4 - #2) + #2}
  \pgfmathsetmacro{\my}{0.66*(#5 - #3) + #3}
  \pgfmathsetmacro{\mmx}{0.48*(#4 - #2) + #2}
  \pgfmathsetmacro{\mmy}{0.48*(#5 - #3) + #3}
  \draw[#1, line width=1pt, draw=#6, -{Stealth}] ({#2},{#3}) -- ({\mx},{\my});
  \draw[#1, line width=1pt, draw=#6]             ({\mmx},{\mmy}) -- ({#4},{#5});
}
\newcommand{\ArrowHead}[6][]{
  \pgfmathsetmacro{\mx}{0.66*(#4 - #2) + #2}
  \pgfmathsetmacro{\my}{0.66*(#5 - #3) + #3}
  \draw[#1, line width=1pt, draw=#6, -{Stealth}] ({#2},{#3}) -- ({\mx},{\my});
}

\newcommand{\PlaceEq}[4][]{
  \node[#1] at ({#2},{#3}) {$#4$};
}

\NewDocumentCommand{\LatticeOneD}{ m m o O{1} }{
  \pgfmathsetmacro{\LatDX}{1.0}
  \pgfmathsetmacro{\LatY}{0.0}
  \pgfmathsetmacro{\LatR}{0.07}
  \def\LatOp{1}
  \def\WhiteOp{0}

  \pgfmathtruncatemacro{\Nsites}{#1}
  \pgfmathtruncatemacro{\FirstBlack}{#2}
  \pgfmathtruncatemacro{\DrawEnds}{#4}

  \IfNoValueTF{#3}{
    \def\EllLink{-1}
  }{
    \IfBlankTF{#3}{
      \def\EllLink{-1}
    }{
      \pgfmathtruncatemacro{\EllLink}{#3}
    }
  }

  \pgfmathtruncatemacro{\Nlinks}{\Nsites-1}
  \foreach \i in {0,...,\numexpr\Nlinks-1\relax}{
    \pgfmathtruncatemacro{\LinkIdx}{\i+1}
    \pgfmathsetmacro{\xL}{
      \i*\LatDX + ifthenelse(\EllLink>0, ifthenelse(\i>=\EllLink, \LatDX, 0), 0)
    }
    \pgfmathsetmacro{\xR}{
      (\i+1)*\LatDX + ifthenelse(\EllLink>0, ifthenelse((\i+1)>=\EllLink, \LatDX, 0), 0)
    }

    \ifnum\LinkIdx=\EllLink\relax
      \pgfmathsetmacro{\xM}{0.5*(\xL+\xR)}
      \pgfmathsetmacro{\gap}{0.18*(\xR-\xL)}
      \draw[line width=0.6pt] ({\xL},{\LatY}) -- ({\xM-\gap},{\LatY});
      \node at ({\xM},{\LatY}) {$\cdots$};
      \draw[line width=0.6pt] ({\xM+\gap},{\LatY}) -- ({\xR},{\LatY});
    \else
      \draw[line width=0.6pt] ({\xL},{\LatY}) -- ({\xR},{\LatY});
    \fi
  }

  \foreach \i in {0,...,\numexpr\Nsites-1\relax}{
    \pgfmathtruncatemacro{\IsEnd}{(\i==0) || (\i==\Nsites-1)}

    \pgfmathsetmacro{\x}{
      \i*\LatDX + ifthenelse(\EllLink>0, ifthenelse(\i>=\EllLink, \LatDX, 0), 0)
    }

    \ifnum\DrawEnds=0\relax
      \ifnum\IsEnd=1\relax
      \else
        \pgfmathtruncatemacro{\IsBlack}{mod(\i+\FirstBlack,2)}
        \ifnum\IsBlack=1\relax
          \SmallCircle[fill=black,draw=black]{\x}{\LatY}{\LatOp}{\LatR}
        \else
          \SmallCircle[fill=white,draw=black]{\x}{\LatY}{\WhiteOp}{\LatR}
        \fi
      \fi
    \else
      \pgfmathtruncatemacro{\IsBlack}{mod(\i+\FirstBlack,2)}
      \ifnum\IsBlack=1\relax
        \SmallCircle[fill=black,draw=black]{\x}{\LatY}{\LatOp}{\LatR}
      \else
        \SmallCircle[fill=white,draw=black]{\x}{\LatY}{\WhiteOp}{\LatR}
      \fi
    \fi
  }
}

\newcommand{\BubbleDownArrow}{
  \pgfmathsetmacro{\Aw}{0.2}
  \pgfmathsetmacro{\Ah}{0.2}
  \pgfmathsetmacro{\Sw}{0.05}
  \pgfmathsetmacro{\Sh}{0.8}

  \filldraw[fill=black, draw=black, line width=1.6pt, line join=round]
    (-\Sw,\Sh) -- (\Sw,\Sh) -- (\Sw,\Ah) -- (\Aw,\Ah) --
    (0,0) -- (-\Aw,\Ah) -- (-\Sw,\Ah) -- cycle;
}

\newcommand{\BubbleUpDownArrow}{
  \pgfmathsetmacro{\Aw}{0.2}
  \pgfmathsetmacro{\Ah}{0.2}
  \pgfmathsetmacro{\Sw}{0.05}
  \pgfmathsetmacro{\Sh}{0.55}

  \filldraw[fill=black, draw=black, line width=1.6pt, line join=round]
    (0,{2*\Ah+\Sh}) --
    (\Aw,{ \Ah+\Sh}) --
    (\Sw,{ \Ah+\Sh}) --
    (\Sw,{ \Ah}) --
    (\Aw,{ \Ah}) --
    (0,0) --
    (-\Aw,{ \Ah}) --
    (-\Sw,{ \Ah}) --
    (-\Sw,{ \Ah+\Sh}) --
    (-\Aw,{ \Ah+\Sh}) --
    cycle;
}

\newcommand{\BubbleRightArrow}{
  \pgfmathsetmacro{\Aw}{0.2}
  \pgfmathsetmacro{\Ah}{0.2}
  \pgfmathsetmacro{\Sw}{0.05}
  \pgfmathsetmacro{\Sh}{0.8}

  \filldraw[fill=black, draw=black, line width=1.6pt, line join=round]
    (0,\Sw) -- (0,-\Sw) -- (\Sh,-\Sw) -- (\Sh,-\Aw) --
    (\Sh+\Ah,0) -- (\Sh,\Aw) -- (\Sh,\Sw) -- cycle;
}

\newcommand{\BubbleLeftRightArrow}{
  \pgfmathsetmacro{\Aw}{0.2}
  \pgfmathsetmacro{\Ah}{0.2}
  \pgfmathsetmacro{\Sw}{0.05}
  \pgfmathsetmacro{\Sh}{0.55}

  \filldraw[fill=black, draw=black, line width=1.6pt, line join=round]
    (0,0) --
    (\Ah,\Aw) --
    (\Ah,\Sw) --
    (\Ah+\Sh,\Sw) --
    (\Ah+\Sh,\Aw) --
    (2*\Ah+\Sh,0) --
    (\Ah+\Sh,-\Aw) --
    (\Ah+\Sh,-\Sw) --
    (\Ah,-\Sw) --
    (\Ah,-\Aw) --
    cycle;
}

\newcommand{\PointSplitFirst}[3]{
\pgfmathsetmacro{\x}{#1}
\pgfmathsetmacro{\y}{#2}
\pgfmathsetmacro{\a}{#3}
\draw (\x,\y) -- ({\x+6*\a},\y);
\draw ({\a},\y) -- ({\x+\a},{\y-\a});
\draw ({\x+5*\a},\y) -- ({\x+5*\a},{\y+\a});
\SmallCircle{2*\a}{0}{1}{0.075}
\SmallCircle{4*\a}{0}{0}{0.075}
}

\newcommand{\PointSplitSecond}[3]{
\pgfmathsetmacro{\x}{#1}
\pgfmathsetmacro{\y}{#2}
\pgfmathsetmacro{\a}{#3}
\draw (\x,\y) -- ({\x+\a},{\y-\a}) -- ({\x+3*\a},{\y-\a}) -- ({\x+4*\a},{\y-2*\a});
\draw ({\x+\a},{\y-\a}) -- ({\x+\a},{\y-2*\a});
\draw ({\x+3*\a},{\y-\a}) -- ({\x+3*\a},{\y});
\SmallCircle{\x+0.5*\a}{\y-0.5*\a}{1}{0.075}
\SmallCircle{\x+3.5*\a}{\y-1.5*\a}{0}{0.075}
}

\NewDocumentCommand{\KetPic}{ o m }{
  \begingroup
  \setbox0=\hbox{
    \begin{tikzpicture}[inner sep=0pt, outer sep=0pt]
      #2
    \end{tikzpicture}
  }

  \pgfmathsetlengthmacro{\StateW}{\wd0}
  \pgfmathsetlengthmacro{\StateH}{\ht0+\dp0}
  \dimen2=\dimexpr(\ht0-\dp0)/2\relax

  \pgfmathsetlengthmacro{\HPad}{4pt}
  \pgfmathsetlengthmacro{\VPad}{2pt}
  \pgfmathsetlengthmacro{\Tip}{7pt}
  \pgfmathsetlengthmacro{\LW}{1.2pt}

  \pgfmathsetlengthmacro{\HalfW}{0.5*\StateW}
  \pgfmathsetlengthmacro{\HalfH}{0.5*\StateH}
  \pgfmathsetlengthmacro{\YTop}{\HalfH+\VPad}
  \pgfmathsetlengthmacro{\YBot}{-\HalfH-\VPad}
  \pgfmathsetlengthmacro{\XL}{-\HalfW-\HPad}
  \pgfmathsetlengthmacro{\XR}{\HalfW+\HPad}
  \pgfmathsetlengthmacro{\XTip}{\XR+\Tip}

  \dimen4=\dimexpr(\ht0+\dp0)+2\VPad\relax

  \begin{tikzpicture}[baseline={(STATE.center)}, x=1pt, y=1pt]
    \node[inner sep=0pt, outer sep=0pt] (STATE) at (0,0)
      {\raisebox{-\dimen2}{\box0}};

    \draw[line width=\LW, line cap=round] (\XL,\YBot) -- (\XL,\YTop);

    \draw[line width=\LW, line cap=round, line join=round]
      (\XR,\YTop) -- (\XTip,0pt) -- (\XR,\YBot);
  \end{tikzpicture}

  \IfNoValueTF{#1}{}{
    \IfBlankTF{#1}{}{
      \kern0.4em
      \raisebox{0pt}[0pt][0pt]{\resizebox{!}{\dimen4}{$#1$}}
    }
  }
  \endgroup
}

\NewDocumentCommand{\BraPic}{ m }{
  \begingroup
  \setbox0=\hbox{
    \begin{tikzpicture}[inner sep=0pt, outer sep=0pt]
      #1
    \end{tikzpicture}
  }

  \pgfmathsetlengthmacro{\StateW}{\wd0}
  \pgfmathsetlengthmacro{\StateH}{\ht0+\dp0}
  \dimen2=\dimexpr(\ht0-\dp0)/2\relax
  
  \pgfmathsetlengthmacro{\HPad}{4pt}
  \pgfmathsetlengthmacro{\VPad}{2pt}
  \pgfmathsetlengthmacro{\Tip}{7pt}
  \pgfmathsetlengthmacro{\LW}{1.2pt}

  \pgfmathsetlengthmacro{\HalfW}{0.5*\StateW}
  \pgfmathsetlengthmacro{\HalfH}{0.5*\StateH}
  \pgfmathsetlengthmacro{\YTop}{\HalfH+\VPad}
  \pgfmathsetlengthmacro{\YBot}{-\HalfH-\VPad}
  \pgfmathsetlengthmacro{\XL}{-\HalfW-\HPad}
  \pgfmathsetlengthmacro{\XR}{\HalfW+\HPad}
  \pgfmathsetlengthmacro{\XTip}{\XL-\Tip}

  \begin{tikzpicture}[baseline={(STATE.center)}, x=1pt, y=1pt]
    \node[inner sep=0pt, outer sep=0pt] (STATE) at (0,0)
      {\raisebox{-\dimen2}{\box0}};

    \draw[line width=\LW, line cap=round, line join=round]
      (\XL,\YTop) -- (\XTip,0pt) -- (\XL,\YBot);

    \draw[line width=\LW, line cap=round] (\XR,\YBot) -- (\XR,\YTop);
  \end{tikzpicture}
  \endgroup
}

\title{Large \texorpdfstring{$N_c$}{N} Truncations for \texorpdfstring{$\SU(N_c)$}{SU(N)} Lattice Yang-Mills Theory with Fermions}

\author{Neel S.~Modi\,\orcidlink{0009-0002-8308-9882}}
\email{neel\_modi@berkeley.edu}
\affiliation{Leinweber Institute for Theoretical Physics and Department of Physics,\\ University of California, Berkeley, Berkeley, CA 94720, USA}
\affiliation{Physics Division, Lawrence Berkeley National Laboratory, Berkeley, CA 94720, USA}

\author{Anthony N.~Ciavarella\,\orcidlink{0000-0003-3918-4110}}
\email{anciavarella@lbl.gov}
\affiliation{Physics Division, Lawrence Berkeley National Laboratory, Berkeley, CA 94720, USA}

\author{Jad C.~Halimeh\,\orcidlink{0000-0002-0659-7990}}
\email{jad.halimeh@physik.lmu.de}
\affiliation{Department of Physics and Arnold Sommerfeld Center for Theoretical Physics (ASC),\\ Ludwig Maximilian University of Munich, 80333 Munich, Germany}
\affiliation{Max Planck Institute of Quantum Optics, 85748 Garching, Germany}
\affiliation{Munich Center for Quantum Science and Technology (MCQST), 80799 Munich, Germany}
\affiliation{Department of Physics, College of Science, Kyung Hee University, Seoul 02447, Republic of Korea}

\author{Christian W.~Bauer\,\orcidlink{0000-0001-9820-5810}}
\email{cwbauer@lbl.gov}
\affiliation{Physics Division, Lawrence Berkeley National Laboratory, Berkeley, CA 94720, USA}

\begin{abstract}

Quantum simulations of quantum chromodynamics (QCD) require a representation of gauge fields and fermions on the finitely many degrees of freedom available on a quantum computer. We introduce a truncation of lattice QCD coupled to staggered fermions that includes (i) a local Krylov truncation that generates allowed basis states; (ii) a maximum allowed electric energy per link; (iii) a limit on the number of fermions per site; and (iv) a truncation in the large $N_c$ scaling of Hamiltonian matrix elements. Explicit truncated Hamiltonians for $1+1$D and $2+1$D lattices are given, and numerical simulations of string-breaking dynamics are performed.
\end{abstract}

\maketitle

\tableofcontents

\section{Introduction}\label{sec:intro}

Gauge theories, a fundamental framework of modern physics describing interactions between elementary particles as mediated through gauge bosons, underpin the Standard Model of Particle Physics \cite{Weinberg1995QuantumTheoryFields,Gattringer2009QuantumChromodynamicsLattice,Zee2003QuantumFieldTheory}. Their lattice formulations, lattice gauge theories (LGTs) \cite{Kogut1975HamiltonianFormulationWilsons,Kogut1979AnIntroductionToLatticeGaugeTheory,Rothe2012LatticeGaugeTheories}, were conceived to study quark confinement \cite{Wilson1974ConfinementQuarks,Wilson1977QuarksStringsLattice}, with progress made on this problem traditionally through Monte Carlo (MC) techniques \cite{Creutz1979MonteCarloStudy,Creutz1980MonteCarloStudy,Creutz1983MonteCarloComputations,Creutz1988LatticeGaugeTheory,Creutz1989LatticeGaugeTheories,montvay1994quantum,Kieu1994MonteCarloSimulations,Philipsen1998StringBreakingNonAbelianGaugeTheories,Knechtli1998StringBreakingSU2GaugeTheory,Bali2005ObservationStringBreakingQCD,Pepe2009FromDecayToCompleteBreaking,Bulava2019StringBreakingLightStrangeQuarksQCD,Hackett2019DigitizingGaugeFields}.

Non-perturbative predictions of quantum chromodynamics (QCD), the gauge theory of the strong nuclear force, are challenging to compute due to a number of difficulties. These include sign problems \cite{Troyer2005ComputationalComplexityFundamental,deforcrand2010simulatingqcdfinitedensity} in MC calculations. The traditional approach is to work in the imaginary-time setting and to perform MC simulations that build ensemble-averaged Euclidean correlation functions from which physical observables can be inferred. However, this approach can rarely provide information about the non-equilibrium dynamics or real-time evolution of the theory. One important example of real-time evolution is the fragmentation of quarks and gluons into hadrons \cite{Ellis2003QCDColliderPhysics}, a process which affects almost all experimental observables used to probe the Standard Model of particle physics and its extensions. There are alternative classical frameworks that can handle dynamics, with tensor networks \cite{Schollwock2011DensitymatrixRenormalizationGroup,Orus2014PracticalIntroductionTensorNetworks,Montangero2018IntroductionTensorNetwork,Orus2019TensorNetworksComplex,Paeckel2019TimeevolutionMethodsMatrixproduct} being a prominent example. They have been used extensively to study LGTs \cite{Dalmonte2016LatticeGaugeTheory,Magnifico2024TensorNetworksLattice} particularly when it comes to confinement and flux string dynamics \cite{Kuhn2015NonAbelianStringBreaking,Pichler2016RealTimeDynamicsU1,Kasper2016SchwingerPairProduction,Buyens2017RealTimeSimulationSchwingerEffect,Kuno2017QuantumSimulation$1+1$dimensional,Sala2018VariationalStudyU1,Spitz2019SchwingerPairProduction,Park2019GlassyDynamicsFromQuarkConfinement,Surace2020LatticeGaugeTheories,Magnifico2020RealTimeDynamics,Notarnicola2020RealtimedynamicsQuantumSimulation,Chanda2020ConfinementLackThermalization,Borla2025StringBreaking$2+1$D,Xu2025StringBreakingDynamics,Tian2025RolePlaquetteTerm,Barata2025RealTimeSimulationJetEnergyLoss,Marcantonio2025RougheningDynamicsElectric,Xu2025TensorNetworkStudyRoughening,artiaco2025outofequilibriumdynamicsu1lattice,cataldi2025realtimestringdynamics21d,cao2026stringbreakingglueballdynamics}. However, the large Hilbert spaces involved limit the efficacy of tensor networks to toy model LGTs in one or two spatial dimensions because of the rapid growth of entanglement entropy with evolution time \cite{Magnifico2025TensorNetworksLattice}. 

Quantum simulation of LGTs \cite{Byrnes2006SimulatingLatticeGauge, Zohar2015QuantumSimulationsLattice, Aidelsburger:2021mia, Zohar2021QuantumSimulationLattice, 
Barata2022MediumInducedJetBroadening,Klco2022StandardModelPhysics,Barata2023QuantumSimulationInMediumQCDJets,Barata2023RealTimeDynamicsofHyperonSpin, Bauer2023QuantumSimulationHighEnergy, Bauer2023QuantumSimulationFundamental,
DiMeglio2024QuantumComputingHighEnergy,Hariprakash:2023tla,Cheng2024EmergentGaugeTheory, Halimeh2022StabilizingGaugeTheories, Cohen2021QuantumAlgorithmsTransport,Barata2025ProbingCelestialEnergy, Lee2025QuantumComputingEnergy, Turro2024ClassicalQuantumComputing,Halimeh2023ColdatomQuantumSimulators,Bauer2025EfficientUseQuantum,Halimeh2025QuantumSimulationOutofequilibrium,Bauer:2021gup,hanada2026gaugesymmetryquantumsimulation} is the strongest available candidate for an approach that can overcome these issues.
The idea is to directly compile the time-evolution operator in the Hamiltonian formulation down to a sequence of quantum gates that can efficiently be implemented on quantum hardware. Recent years have witnessed many successful quantum simulation experiments probing various features of toy model LGTs \cite{Martinez2016RealtimeDynamicsLattice, Klco2018QuantumclassicalComputationSchwinger,Gorg2019RealizationDensitydependentPeierls, Schweizer2019FloquetApproachZ2, Mil2020ScalableRealizationLocal, Yang2020ObservationGaugeInvariance, Wang2022ObservationEmergent$mathbbZ_2$, Su2023ObservationManybodyScarring, Zhou2022ThermalizationDynamicsGauge, Wang2023InterrelatedThermalizationQuantum, Zhang2025ObservationMicroscopicConfinement, Zhu2024ProbingFalseVacuum, Ciavarella2021TrailheadQuantumSimulation, Ciavarella2022PreparationSU3Lattice, Ciavarella2023QuantumSimulationLattice-1, Ciavarella2024QuantumSimulationSU3, 
Gustafson2024PrimitiveQuantumGates, Gustafson2024PrimitiveQuantumGates-1, Lamm2024BlockEncodingsDiscrete, Farrell2023PreparationsQuantumSimulations-1, Farrell2023PreparationsQuantumSimulations, 
Farrell2024ScalableCircuitsPreparing,
Farrell2024QuantumSimulationsHadron, Li2024SequencyHierarchyTruncation, Zemlevskiy2025ScalableQuantumSimulations, Lewis2019QubitModelU1, Atas2021SU2HadronsQuantum, ARahman:2022tkr, Atas2023SimulatingOnedimensionalQuantum, Mendicelli2023RealTimeEvolution, Kavaki2024SquarePlaquettesTriamond, Than2024PhaseDiagramQuantum, Angelides:2023noe, Gyawali2025ObservationDisorderfreeLocalization,  
Mildenberger2025Confinement$$mathbbZ_2$$Lattice, Schuhmacher2025ObservationHadronScattering, Davoudi2025QuantumComputationHadron, Saner2025RealTimeObservationAharonovBohm, Xiang2025RealtimeScatteringFreezeout, Wang2025ObservationInelasticMeson,li2025frameworkquantumsimulationsenergyloss,mark2025observationballisticplasmamemory,froland2025simulatingfullygaugefixedsu2,Hudomal2025ErgodicityBreakingMeetsCriticality,hayata2026onsetthermalizationqdeformedsu2,Cochran2025VisualizingDynamicsCharges, Gonzalez-Cuadra2025ObservationStringBreaking, Crippa2024AnalysisConfinementString, De2024ObservationStringbreakingDynamics, Liu2024StringBreakingMechanism, Alexandrou:2025vaj,Cobos2025RealTimeDynamics2+1D}.

An important precursor step that must be taken in such quantum simulations is to truncate the infinitely-many degrees of freedom in the field theory down to a finite-dimensional Hamiltonian matrix that can be represented in finite quantum memory.
Numerous methods have been developed for simulating Hamiltonians of lattice field theories, both from the point of view of constructing a truncated Hamiltonian and for compiling its time-evolution operator.

However, a significant difficulty arises for lattice gauge theories, in that some errors can take the simulated system outside of the gauge-invariant Hilbert space, leading to unphysical outputs.
For this reason, an important paradigm is to keep gauge invariance manifest, both when truncating the Hamiltonian, and when approximating the time-evolution operator.
When gauge invariance is part of the simulation scheme in this way, it can also enable a reduction in the overhead necessary for quantum error correction on actual hardware \cite{Rajput:2021trn}. 
An alternative to restricting the simulation to the physical Hilbert space is to use the gauge redundancy for error correction~\cite{Stryker:2018efp,Spagnoli:2024mib, Yao:2025cxs}. 

Various proposals exist for the quantum simulation of non-Abelian LGTs on various analog and digital quantum-hardware platforms \cite{kadam2025loopstringhadronapproachsu3latticei,Fontana2025EfficientFiniteResourceFormulation,Halimeh2024SpinExchangeEnabledQuantumSimulator,Calajo2024digital,Illa2025ImprovedHoneycomb,halimeh2025universalframeworkexponentialspeedup,depaciani2025quantumsimulationfermionicnonabelian,kadam2025loopstringhadronapproachsu3latticeii,ilcic2025physicalityoraclesu3loopstringhadron,Balaji:2025yua}. Recent developments \cite{Ciavarella:2025bsg, Ciavarella:2024fzw} have shown that one practical approach for efficient, gauge-invariant simulations is to truncate the $\SU(N_c)$ Kogut-Susskind Hamiltonian in the $1/N_c$ expansion.
The motivation for the $1/N_c$ expansion is that $\SU(N_c)$ gauge theories typically exhibit classical behavior at large $N_c$, with the $1/N_c$ contributions corresponding to quantum corrections~\cite{yaffe1982large}.
This was initially noticed by 't Hooft who showed that Feynman diagrams for $\SU(N_c)$ gauge theories naturally reorganize themselves into a genus expansion in powers of $1/N_c$, with planar diagrams dominating as $N_c \to \infty$~\cite{tHooft:1973alw}.
Since then, this concept has found utility in a number of fields, such as random matrix theory, quantum gravity, and parton shower algorithms~\cite{maldacena1999large,LUCINI201393,PICH_2002,Bahr:2008pv,Sjostrand:2006za}.
The aspect of the expansion most important to quantum simulation is that it gives rise to a dramatic reduction in the complexity of the required Hamiltonian.

For the truncated Kogut-Susskind Hamiltonian, a consistent large $N_c$ truncation requires two considerations:
\begin{enumerate}
    \item Introducing a consistent energy density cutoff $\Lambda$.
    \item Truncating interaction matrix elements based on their order of appearance in the $1/N_c$ expansion.
\end{enumerate}
Prior work \cite{Ciavarella:2025bsg, Ciavarella:2024fzw} solved these conditions at leading order and next-to-leading order in $1/N_c$ for \textit{pure gauge theory} with the following truncation scheme:
\begin{enumerate}
    \item Truncate to a Krylov-type subspace generated from the electric vacuum by all possible plaquettes and Wilson lines.
    \item Pick an electric energy cutoff per link, and delete all basis states in non-compliance with this cutoff.
\end{enumerate}

In this work, we extend this formalism to $\SU(N_c)$ lattice gauge theory coupled to fermions, i.e. lattice QCD.
Our extension takes the following truncation steps:
\begin{enumerate}
    \item Truncate to a Krylov-type subspace generated from an arbitrary initial state by all possible plaquettes, Wilson lines, fermion hopping operators, and fermion creation / annihilation operators.
    \item Pick an electric energy cutoff per link and a fermion energy cutoff per lattice site, and delete all basis states in non-compliance with these cutoffs.
    \item Pick an order in the $1/N_c$ expansion, and keep only interactions in the Hamiltonian up to that order (and optionally refine those interactions).
\end{enumerate}
In other words, our truncation scheme performs a $1/N_c$ expansion \textit{on top} of a Krylov subspace truncation, where the latter is known to keep the dominant contributions to real-time evolution according to recent work \cite{Ciavarella:2025tdl}.
Note that these truncations only lead to a controlled approximation of the underlying theory in certain regions of parameter space. For example, in the continuum limit, where $g \to 0$, the truncation based on representations breaks down.

An important point arises here that is not present in pure gauge theory: there are two types of interactions (pure gauge interactions on plaquettes and fermion interactions that require hopping from one lattice site to a neighboring one). 
The matrix elements of plaquette operators scale with integer powers of $1/N_c$, while the matrix elements of the hopping term scale with $N_c^{-1/2}$. Therefore, the large $N_c$ expansion in this work is an expansion in powers of $N_c^{-1/2}$, not $N_c^{-1}$.

Additionally, it is possible that one type of interaction (e.g. fermion hopping) generates gauge-invariant states at some order in $1/N_c$, but those same gauge-invariant states are only reached at higher orders in $1/N_c$ by the other type of interaction (e.g. plaquette term).
Due to these complications, and other reasons to be discussed in the main text, we leave open the option of refining any truncation in the $1/N_c$ expansion.
This is the reason for the third step in the above list.

After defining our truncation scheme, we will explicitly construct the Hamiltonian for the first few low-lying truncations and discuss their relationship with the large $N_c$ expansion.
This will allow us to form analogies with the pure gauge truncations \cite{Ciavarella:2025bsg, Ciavarella:2024fzw}.
We will also discuss how to formulate Hamiltonians containing external charges at a given level of truncation, which requires modifying our prescription.

Finally, we will use exact diagonalization to simulate the truncated Hamiltonians in $1+1$D and $2+1$D, focusing on the real-time evolution of the free vacuum and meson excitations, as well as string-breaking dynamics.
This leads to several notable observations.
For instance, although the quantitative convergence of our truncations is best seen in the limit of strong coupling, the qualitative features at weak coupling appear to be maintained across all truncations.
An important example is the appearance of fidelity revivals that are reminiscent of many-body quantum scarring (discussed in more detail in \Cref{sec:simuls}).
Another interesting observation from our results is that when a string is pinned by external charges, the resonance condition to break the string can be preserved in the 't Hooft limit.
This enables us to probe the dynamical reason for suppression of string breaking in the large $N_c$ limit.
In summary, we perform practical real-time simulations in a QCD-like theory, and the results can be connected to both prior knowledge and hypothesized properties of non-abelian gauge theories.

This paper is organized as follows:
in \Cref{sec:KS}, we define the Kogut-Susskind Hamiltonian coupled to staggered fermions on an arbitrary bipartite lattice;
in \Cref{sec:trunc}, we define our truncation scheme that incorporates lattice fermions, and provide the general formalism that can be used to deduce the gauge-invariant basis and Hamiltonian at any level of truncation;
in \Cref{sec:hams}, we explicitly derive several different truncations of Kogut-Susskind in $1+1$D and $2+1$D, show how the formalism can be extended to higher spatial dimensions, and also discuss the prescription for adding external charges;
in \Cref{sec:simuls}, we discuss the results from our real-time simulations, and provide a comparison between different truncations;
and in \Cref{sec:summary}, we summarize all results and discuss future avenues of work.
\Cref{app:electric,app:gauge,app:calc,app:tab,app:calc2} contain further details on the definitions and calculations assumed in the main text.

\section{Kogut-Susskind Hamiltonian}\label{sec:KS}

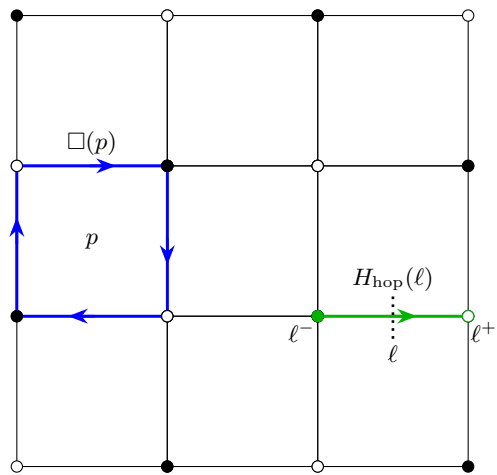
\begin{figure}[htp]
    \centering
    \begin{minipage}{\linewidth}
    \centering
    \begin{tikzpicture}
    \pgfmathsetmacro{\side}{2.0}
    
    \OrientedEdge{1*\side}{2*\side}{2*\side}{2*\side}{blue}
    \OrientedEdge{2*\side}{2*\side}{2*\side}{1*\side}{blue}
    \OrientedEdge{2*\side}{1*\side}{1*\side}{1*\side}{blue}
    \OrientedEdge{1*\side}{1*\side}{1*\side}{2*\side}{blue}
    \OrientedEdge{3*\side}{1*\side}{4*\side}{1*\side}{black!30!green}
    
    \OddPlaquette{1*\side}{0*\side}{1.0*\side}{black}
    \EvenPlaquette{2*\side}{0*\side}{1.0*\side}{black}
    \OddPlaquette{3*\side}{0*\side}{1.0*\side}{black}
    \EvenPlaquette{1*\side}{1*\side}{1.0*\side}{black}
    \OddPlaquette{2*\side}{1*\side}{1.0*\side}{black}
    \EvenPlaquette{3*\side}{1*\side}{1.0*\side}{black}
    \OddPlaquette{1*\side}{2*\side}{1.0*\side}{black}
    \EvenPlaquette{2*\side}{2*\side}{1.0*\side}{black}
    \OddPlaquette{3*\side}{2*\side}{1.0*\side}{black}

    \draw[draw=black!30!green, line width=1pt] (4*\side-0.08, 1*\side) -- (3*\side+0.08, 1*\side);
    \SmallCircle[draw=black!30!green, fill=black!30!green]{3*\side}{\side}{1}{0.075}
    \SmallCircle[draw=black!30!green]{4*\side}{\side}{0}{0.075}
    \draw[draw=blue, line width=1pt] (1*\side+0.08, 2*\side) -- (2*\side-0.08, 2*\side);
    \draw[draw=blue, line width=1pt] (2*\side, 2*\side-0.08) -- (2*\side, 1*\side+0.08);
    \draw[draw=blue, line width=1pt] (2*\side-0.08, 1*\side) -- (1*\side+0.08, 1*\side);
    \draw[draw=blue, line width=1pt] (1*\side, 1*\side+0.08) -- (1*\side, 2*\side-0.08);

    \pgfmathsetmacro{\norm}{0.707}
    \PlaceEq{1.5*\side}{2*\side+0.3}{\square(p)}
    \PlaceEq{1.5*\side}{1.5*\side}{p}
    \PlaceEq{3.5*\side}{\side+0.5}{H_\hop(\ell)}
    \PlaceEq{3.5*\side}{\side-0.5}{\ell}
    \PlaceEq{3*\side-0.3*\norm}{\side-0.3*\norm}{\ell^-}
    \PlaceEq{4*\side+0.3*\norm}{\side-0.3*\norm}{\ell^+}

    \draw[dotted, line width=1pt] (3.5*\side,\side-0.3) -- (3.5*\side,\side+0.3);
    \end{tikzpicture}
    \end{minipage}
    \caption{Graphical representation of interaction terms in the Kogut-Susskind Hamiltonian on a square lattice. Blue indicates a Wilson loop for plaquette $p$. Green indicates a Wilson line and lattice sites for hopping operator at link $\ell$. The odd and even endpoints of $\ell$ are indicated as $\ell^-$ and $\ell^+$, respectively. The notation $\ell^+$, $\ell^-$ can also be used for the half-links of $\ell$ attached to these respective endpoints (residing on each side of the dotted line that cuts link $\ell$).}\label{fig:KS}
\end{figure}

The seminal work of Kogut and Susskind provides the prototypical Hamiltonian formulation of lattice gauge theory ~\cite{Kogut1975HamiltonianFormulationWilsons}.
Our work focuses on the Kogut-Susskind Hamiltonian with staggered fermions, which is defined on a bipartite lattice where $\vx$, $\lk$, and $\pq$ are the lattice sites, links, and plaquettes respectively. Note that we may sometimes refer to lattice sites, links, and plaquettes as \textit{vertices}, \textit{edges}, and \textit{facets}, respectively.
The bipartite structure consists of even (fermion-hosting) sites and odd (anti-fermion-hosting) sites.
For any site $\mathbf n \in \vx$, we write $(-1)^{\mathbf n}$ to mean $+1$ if $\mathbf{n}$ is even, and $-1$ if $\mathbf{n}$ is odd. For a $d$-dimensional hypercubic lattice, $\mathbf{n} = (n_1,\dots,n_d)\in \mathbb Z^d$ and $(-1)^{\mathbf n} = (-1)^{n_1+\dots+n_d}$.
For any link $\ell \in \lk$, we use the notations $\ell^+$ and $\ell^-$ for its even and odd endpoints, respectively.
We also use $\eta_{\ell} \in \{+1,-1\}$ to denote the staggered phases corresponding to link $\ell$. \Cref{fig:KS} provides an example lattice with the relevant data.

For Yang-Mills coupling $g$, fermion mass $m$, unit lattice spacing, and gauge group $\SU(N_c)$, the Hamiltonian is given by
\begin{widetext}
\begin{equation}\label{eq:KS}
    H_{\KS} = \frac{g^2}{2}
     \sum_{\ell\in\lk} E^2(\ell) + m \sum_{\mathbf{n}\in\vx} (-1)^{\mathbf n} \chi^\dagger(\mathbf n)_a \chi(\mathbf n)_a
     - \frac 1{2g^2} \sum_{p\in\pq} \left[\square(p)^\dagger + \square(p)\right]
     + \frac 1{2} \sum_{\ell\in\lk} \eta_{\ell} \left[\chi^\dagger(\ell^+)_a U(\ell)_{ab} \chi(\ell^-)_b + \mathrm{h.c.}\right],
\end{equation}
\end{widetext}
where, for $1 \le a, b \le N_c$,
\begin{itemize}
    \item $E^2(\ell) \equiv E(\ell)_a E(\ell)_a$, where $E(\ell)_a$ denotes the component of the chromo-electric field operator at $\ell \in \lk$ with color index $a$;
    \item $\chi^\dagger(\mathbf n)_a$ and $\chi(\mathbf n)_a$ respectively denote the fermion creation and annihilation operators at $\mathbf n \in \vx$ with color index $a$;
    \item $U(\ell)_{ab}$ denotes the component of the Wilson line operator at $\ell \in \lk$ (oriented from $\ell^-$ to $\ell^+$) with color indices $(a, b)$; and
    \item $\square(p)$ denotes the traced Wilson loop operator around plaquette $p$ (with a consistently chosen orientation, e.g. clockwise in $2+1$D).
\end{itemize}
Note that we employ the summation convention for repeated color indices and/or representation indices.
We will also sometimes use the notation
\begin{equation}
    H_\hop(\ell) \equiv \chi^\dagger(\ell^+)_a U(\ell)_{ab} \chi(\ell^-)_b
\end{equation}
for the \textit{raising} hopping interaction at link $\ell$, which adds a fermion to site $\ell^+$ and adds an anti-fermion to site $\ell^-$.

The Hilbert space for this theory is naturally expressed in terms of finite-dimensional irreducible representations (irreps) of $\SU(N_c)$.
Every link $\ell \in \lk$ carries an irrep that matches the transformation law for the quantum state of the Wilson line across that link.
Additionally, every lattice site $\textbf{n} \in \vx$ carries an irrep that matches the transformation law for the fermion / anti-fermion color states, at that site.
The notation we will use for irreps is shown in \Cref{tab:irrepnotation}. Other choices for representing the Hilbert space are possible as well~\cite{haase2021resource,Grabowska:2024emw,DAndrea:2023qnr,Kane:2022ejm,Bauer:2021gek}.

\begin{table}[htp]
\centering
\begin{tabular}{|c|c|}

\hline

\begin{minipage}[c][0.75cm][c]{0.75\linewidth}
\textbf{Irreducible Representation of} $\SU(N_c)$
\end{minipage}
&
\begin{minipage}[c][0.75cm][c]{0.2\linewidth}
\textbf{Symbol}
\end{minipage}
\\
\hline

\begin{minipage}[c][0.75cm][c]{0.75\linewidth}
Singlet
\end{minipage}
&
\begin{minipage}[c][0.75cm][c]{0.2\linewidth}
$\mathbf{1}$
\end{minipage}
\\
\hline

\begin{minipage}[c][0.75cm][c]{0.75\linewidth}
Fundamental
\end{minipage}
&
\begin{minipage}[c][0.75cm][c]{0.2\linewidth}
$\mathbf{N}$
\end{minipage}
\\
\hline

\begin{minipage}[c][0.75cm][c]{0.75\linewidth}
Anti-fundamental
\end{minipage}
&
\begin{minipage}[c][0.75cm][c]{0.2\linewidth}
$\overline{\mathbf{N}}$
\end{minipage}
\\
\hline

\begin{minipage}[c][0.75cm][c]{0.75\linewidth}
Adjoint
\end{minipage}
&
\begin{minipage}[c][0.75cm][c]{0.2\linewidth}
$\mathbf{Ad}$
\end{minipage}
\\
\hline

\begin{minipage}[c][0.75cm][c]{0.75\linewidth}
Antisymmetric $p$-fold product of $\mathbf N$
\end{minipage}
&
\begin{minipage}[c][0.75cm][c]{0.2\linewidth}
$\mathbf{A^p}$
\end{minipage}
\\
\hline

\begin{minipage}[c][0.75cm][c]{0.75\linewidth}
Antisymmetric $p$-fold product of $\overline{\mathbf{N}}$
\end{minipage}
&
\begin{minipage}[c][0.75cm][c]{0.2\linewidth}
$\overline{\mathbf{A^p}}$
\end{minipage}
\\
\hline

\begin{minipage}[c][0.75cm][c]{0.75\linewidth}
Symmetric $p$-fold product of $\mathbf{N}$
\end{minipage}
&
\begin{minipage}[c][0.75cm][c]{0.2\linewidth}
$\mathbf{S^p}$
\end{minipage}
\\
\hline

\begin{minipage}[c][0.75cm][c]{0.75\linewidth}
Symmetric $p$-fold product of $\overline{\mathbf{N}}$
\end{minipage}
&
\begin{minipage}[c][0.75cm][c]{0.2\linewidth}
$\overline{\mathbf{S^p}}$
\end{minipage}
\\
\hline

\begin{minipage}[c][0.75cm][c]{0.75\linewidth}
Conjugate of representation $\mathbf{R}$
\end{minipage}
&
\begin{minipage}[c][0.75cm][c]{0.2\linewidth}
$\overline{\mathbf{R}}$
\end{minipage}
\\
\hline

\end{tabular}
\caption{Notation used extensively throughout this paper to denote irreps of $\SU(N_c)$.}\label{tab:irrepnotation}
\end{table}

For a link in irrep $\textbf{R}$, there are a total of $(\dim \textbf{R})^2$ orthonormal states that each carry electric energy $g^2 C_2(\textbf{R})/2$, where $C_2(\textbf{R})$ denotes the quadratic Casimir of irrep $\textbf{R}$.
At a link $\ell \in \lk$, we can denote these states as $\ket{\textbf{R}, a, b}_{\ell}$ where $1\le a,b\le \dim \textbf{R}$, or more informatively using the notation $\ket{\overline{\textbf{R}}, a}_{\ell^+}\ket{\textbf{R}, b}_{\ell^-}$, as elaborated in \Cref{app:electric}.
When using the latter notation, the subscripts $\ell^+$ and $\ell^-$ emphasize that the tensor factors live in \textit{half-links} corresponding to the even and odd endpoints of $\ell$, respectively.
It is important to note that in our convention, the irrep $\textbf{R}$ that labels a link is \textit{always} the irrep under which the \textit{odd half-link} $\ell^-$ transforms.
This has important implications when writing down truncated Hamiltonians, as we discuss in \Cref{sec:hams}.

A lattice site cannot live in an arbitrary irrep, due to the Pauli exclusion principle.
Instead, the admissible irreps are those capable of supporting totally anti-symmetric quantum states.
Due to the $(-1)^{\textbf{n}}$ phase factors in the Dirac mass term within the Hamiltonian \eqref{eq:KS}, it is prudent to separate our notation for even and odd sites.
We take the convention that an even site lives in an anti-symmetrized product of ``fundamentals'', and an odd site lives in an anti-symmetrized product of ``anti-fundamentals''.
The plural terminology in quotation marks refers to the idea that fermions (resp. anti-fermions) transform under fusions of multiple copies of the fundamental (resp. anti-fundamental) representation of $\SU(N_c)$.
Note that anti-symmetrized products of fundamentals and anti-symmetrized products of anti-fundamentals generate the same collection of irreps for $\SU(N_c)$.
This means every lattice site has the same Fock space, but the interpretation of how that Fock space is built from irreps is dependent on whether the lattice site is even or odd.
For this reason, we say that excitations at an even site are \textit{fermion-like}, while excitations at an odd site are \textit{anti-fermion-like}.

For a moment, let us interpret the Fock space at a lattice site (whether odd or even) in terms of its decomposition into anti-symmetrized products of fundamentals.
Then a single integer $0\le p \le N_c$ is sufficient to specify which irrep sector the site inhabits---it is the sector transforming under the $p$-fold antisymmetric product of fundamentals, i.e. $\mathbf{A^p}$ (in the case of the singlet sector, we can also determine \textit{which} singlet is inhabited, based on whether $p=0$ or $p=N_c$).
The basis states for a single site in an $\SU(N_c)$ lattice gauge theory can therefore be labeled by this integer $p \in \{0, \dots, N_c\}$, which we refer to as the \textit{fermion occupancy} of that site, along with representation indices within the irrep $\mathbf{A^p}$.
The \textit{anti-fermion occupancy} of a site is defined to be $N_c - p$, where $p$ is its fermion occupancy.
Although both terminologies are well-defined for every site, we will generally refer to states on fermion-like (resp. anti-fermion-like) sites with the corresponding fermion occupancy (resp. anti-fermion occupancy), to ensure that Dirac energies don't pick up sign factors.

For example, if an even site $\textbf{n}_1$ contains $p_1$ fermions, and an odd site $\textbf{n}_2$ contains $p_2$ anti-fermions, then site $\textbf{n}_1$ transforms under the irrep $\mathbf{A^{p_1}}$, and site $\textbf{n}_2$ transforms under the irrep $\overline{\mathbf{A^{p_2}}}$.
In this case, there are a total of $\dim \mathbf{A^{p_1}} = \binom{N_c}{p_1}$ orthonormal states at site $\textbf{n}_1$ that each carry Dirac energy $m \cdot p_1$, and there are a total of $\dim \overline{\mathbf{A^{p_2}}} = \binom{N_c}{p_2}$ orthonormal states at site $\textbf{n}_2$ that each carry Dirac energy $m \cdot p_2$, neglecting the overall constant term $-m\cdot N_c$ in the Hamiltonian.
We can denote these states with the notation $\ket{\mathbf{A^{p_1}}, a_1}_{\textbf{n}_1}$ for $1\le a_1 \le \binom{N_c}{p_1}$, and $\ket{\overline{\mathbf{A^{p_2}}}, a_2}_{\textbf{n}_2}$ for $1 \le a_2 \le \binom{N_c}{p_2}$, respectively.
Note that the subscripts $\textbf{n}_1$ and $\textbf{n}_2$ emphasize the lattice sites at which these states reside.

The Hamiltonian \eqref{eq:KS} also preserves gauge invariance.
This means the Hilbert space that can be explored by time evolution consists of states which satisfy Gauss's law at every vertex.
By taking advantage of this fact, and ensuring that our truncations continue to be consistent with gauge invariance, we can greatly reduce the number of qubits necessary for simulation on quantum hardware.
\Cref{app:gauge} provides more precise definitions of gauge transformations and gauge invariance on the lattice Hilbert space.

A key point (discussed more in \Cref{sec:trunc} and \Cref{sec:hams}) is that the singlets contained in a fusion of three finite-dimensional irreps of $\SU(N_c)$ are uniquely characterized by all possible invariant tensor contractions, but distinct tensor contractions can produce linearly dependent singlets in a fusion of four or more irreps.
In $1+1$D (without the presence of external charges), we will only require three irreps to be combined at each vertex, and it will turn out that the singlets are uniquely defined from the irrep data, provided they exist.\footnote{This is not always the case when fusing three irreps of $\SU(N_c)$---you can get more than one singlet. But if one of the factors is guaranteed to be an antisymmetric irrep, then the singlet is unique if it exists.}
For the $2+1$D lattice, or indeed $1+1$D simulations with external charges, four or more irreps will need to be combined at each vertex.
The latter case requires a technique known as \textit{virtual point-splitting}~\cite{Anishetty:2018vod,Raychowdhury:2018tfj,Raychowdhury:2019iki,Burbano:2024uvn} to construct a local basis that only contains gauge invariant states.

\begin{figure}[htp]
\usetikzlibrary{calc,arrows.meta}
\begin{tikzpicture}[scale=1.2]

\newcommand{\hexagon}[4][]{

  \pgfmathsetmacro{\h}{0.8660254}
  \pgfmathsetmacro{\shade}{#4*100}

  \pgfmathsetmacro{\xtop}{#2}
  \pgfmathsetmacro{\ytop}{#3 + \h}
  \pgfmathsetmacro{\xbottom}{#2}
  \pgfmathsetmacro{\ybottom}{#3 - \h}

  \draw[#1]
    (#2-1,   #3)       --
    (#2-0.5, #3+\h)    --
    (#2+0.5, #3+\h)    --
    (#2+1,   #3)       --
    (#2+0.5, #3-\h)    --
    (#2-0.5, #3-\h)    --
    cycle;

  \filldraw[draw=black, fill=black!\shade]
      (\xtop, \ytop) circle (0.075);
  \filldraw[draw=black, fill=black!\shade]
      (\xbottom, \ybottom) circle (0.075);
}

\hexagon{0}{0}{0}
\hexagon{1.5}{0.866}{1}
\hexagon{0}{-1.732}{0}
\hexagon{1.5}{-0.866}{1}

\end{tikzpicture}
\caption{Example polyhex tile that can be used to generate the full hexagonal lattice structure obtained by point-splitting a square lattice. White sites are fermion-like and black sites are anti-fermion-like.}\label{fig:hextile}
\end{figure}
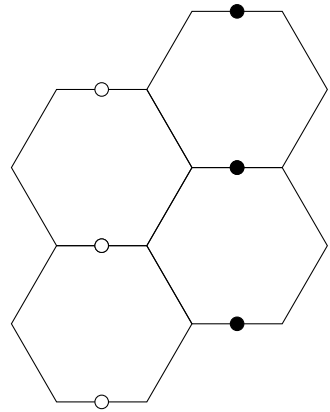

To simplify our analysis, we will arrange for lattices with irrep networks where the point-splitting is not strictly necessary to write down the truncated gauge-invariant basis.
For example, our $2+1$D discussion will take place on a hexagonal lattice (see \Cref{fig:hextile}).
It is important to note that the placement of fermions in \Cref{fig:hextile} maps directly onto the vertices of the square lattice (as discussed in \Cref{sec:2+1d}), so there are no new fermion doublers in the truncated theory we will write down, beyond those already present for the square lattice.

\section{Truncation Formalism}\label{sec:trunc}

As introduced in \Cref{sec:intro}, we have a three-step approach to defining our truncations.
This section is divided into three subsections, each dedicated to one step of truncation.
By the end, we will have simplified the constraints imposed by the full truncation scheme, so that one only needs to input Clebsch-Gordan coefficients or other known data associated with the Lie group $\SU(N_c)$.

Additionally, we will tabulate all relevant data and matrix elements for the truncations considered in this work.
For more detail on tabulation and derivations, see \Cref{app:tab} and \Cref{app:calc2}.

\subsection{Krylov Constraints}

A \textit{Krylov subspace} of a Hilbert space is the subspace that can be generated by repeated applications of a set of operators on an initial state, where each operator can be applied a prescribed maximum number of times.
Krylov-based methods can improve the efficiency of simulating quantum systems~\cite{doi:10.1021/j100319a003, doi:10.1137/S0036142995280572, Paeckel_2019, nandy2025quantumdynamicskrylovspace, Shen_2023, takahashi2025krylovsubspacemethodsquantum, Park_1986, JoséGarcía-Ripoll_2006, PhysRevB.85.205119, Wall_2012, Hubig2015StrictlySinglesiteDMRG, PhysRevB.72.020404, doi:10.1021/acs.jctc.7b00682, 10.1063/1.2080353, Rodriguez_2006}.

Prior large $N_c$ truncations for pure Yang-Mills LGTs \cite{Ciavarella:2025bsg, Ciavarella:2024fzw} used a Krylov subspace constructed through the application of plaquette operators. This requires specifying an integer $n_p$ which limits how many times an individual plaquette can be applied, and results in the subspace $\mathcal H(n_p)$ defined by
\begin{equation}
    \mathcal H(n_p) \equiv \mathrm{span}
    \left\{
        P\ket{0} \; \middle| \;
        \begin{aligned}
        \centering
        & P \text{: product of up to } n_p\\
        & \text{plaquettes per facet, and } \\
        & \hspace{-0.25em} \ket{0} \text{is the electric vacuum.}
        \end{aligned}
    \right\}\,.
\end{equation}
Note that the restriction of $n_p$ plaquette operators on a given facet refers to the total number of operators of \textit{any orientation} at that facet.
For example, if $n_p = 1$, then a state that can only be reached by $\square(p)^\dagger \square(p) \ket{0}$ for $p \in \mathcal P$ is disallowed, because it requires a total of two plaquette operations at $p$.

Our starting point will instead allow an \textit{arbitrary} but fixed initial state $\ket{\psi}$, and in addition to $n_p$, we will also require an integer $n_h$ that strictly limits the number of \textit{hopping} transitions per link.
In other words, we begin by truncating down to the Krylov subspace
\begin{equation}
    \mathcal H_\psi(n_p,n_h) \equiv \mathrm{span}
    \left\{
        P\ket{\psi} \; \middle| \;
        \begin{aligned}
        \centering
        & P \text{: product of up to } n_p\\
        & \text{plaquettes per facet, and} \\
        & \text{up to } n_h \text{ hops per link.}
        \end{aligned}
    \right\}
\end{equation}
In $1+1$D, the plaquette condition is not applicable, and the starting point is instead $\mathcal H(n_h) \equiv \mathcal H_0(0, n_h)$.
Most commonly, we will end up choosing the initial state $\ket{\psi}$ to be the electric vacuum $\ket{0}$, but this will be modified when dealing with external charges. Note that we will mostly drop the subscript $\psi$ on $ H_\psi(n_p,n_h)$ from now on, where the initial state $\ket{\psi}$ can be inferred depending on whether or not we have external charges in the system.

Next, we extend the approach taken by \cite{Ciavarella:2025bsg} and introduce two more integers $n_{\ell}$ and $n_\chi$ which further limit the allowed states as follows: we keep only the states $\ket{\phi} \in \mathcal H(n_p,n_h)$ (or $\mathcal H(n_h)$ as applicable) whose generating sequence (i.e., the sequence of operators that acts successively on $\ket{\psi}$ to produce $P\ket{\psi} \equiv \ket{\phi}$) of plaquette and/or hopping operators \textit{contains at most an} $n_\ell$-\textit{fold product of Wilson line operators per link, and at most an} $n_\chi$-\textit{fold product of fermion operators per site}.\footnote{If multiple generating sequences exist for $\ket{\phi}$, at least one of those sequences needs to satisfy this constraint to keep the state $\ket{\phi}$.}
Similarly to the case of $n_p$, the constraint $n_h$ limits the total number of hopping operators of either orientation at each link.
In $1+1$D, $n_{\ell} = n_h$ identically, so we only need to consider $n_h$ and $n_{\chi}$.

One benefit of the choice of parameters $(n_p, n_h, n_\ell, n_\chi)$ is that the lowest-lying truncation $(1,1,1,1)$ keeps interactions only to order $\mathcal O(1)$, whereas $(1,1,2,2)$ adds in all corrections at order $\mathcal O(1/N_c)$.
This is true in any number of dimensions, but we have to consider an additional technicality since our expansion is really in powers of $N_c^{-1/2}$ due to the scaling of hopping interactions---in particular, we consider interactions at order $\mathcal O(\sqrt{N_c})$ to be part of the ``de facto" order $\mathcal O(1)$ contribution, while transitions at order $\mathcal O(1/\sqrt{N_c})$ are grouped into the de facto $\mathcal O(1/N_c)$ corrections.
In order to properly view $(1,1,2,2)$ as a $1/N_c$ refinement of $(1,1,1,1)$ in this manner, these truncations must be equipped with compatible energy density cutoffs, which brings us to the next step of our truncation scheme.

\subsection{Electric Energy Density and Quark Mass Cutoff}

Independently of Krylov truncation, any quantum field theory needs to clarify its regulation by an energy cutoff.
In our case, we first set an upper limit on the electric energy (quadratic Casimir) that is allowed on any individual link, and we separately set an upper limit on the largest possible Dirac energy (mass contribution) that can be supported on any lattice site.

An arbitrary irrep for $\SU(N_c)$ can be characterized by a set of $N_c - 1$ integers called \textit{Dynkin labels}, where the $j^{\mathrm{th}}$ label counts the number of rows with $j$ boxes in the Young tableau for that irrep.
For an irrep $\textbf{R}$ with Dynkin labels $p_1$, $p_2$, $\dots$, $p_{N_c-1}$, the quadratic Casimir takes the asymptotic form
\begin{equation}
    C_2(\textbf{R}) \sim \frac{N_c}{2} \sum_{j=1}^{N_c-1} \min(j, N_c-j)\cdot p_j,
\end{equation}
valid for large $N_c$.
We refer to the sum in this expression as the \textit{number of electric field lines} carried by representation $\textbf{R}$.

This motivates the following energy density cutoff scheme for the gauge field: choose an integer $n_e$, and keep only the gauge-invariant basis states that have at most $n_e$ electric field lines per link.
This means that the maximum allowed electric energy per link is set to
\begin{equation}
    E_{\mathrm{max}}(\ell) \sim \frac{g^2N_c}{4} \cdot n_e\,.
\end{equation}

In addition to truncating the gauge field energy density, truncations of the fermion number per site can be justified when working in the heavy quark limit $m a \gg 1$.
To keep the discussion as general as possible, we impose a cutoff on the number of fermions (or anti-fermions, as appropriate) allowed per lattice site \textit{independently} of the electric field cutoff.
In other words, we choose an integer $n_f \ge 0$ and restrict to gauge-invariant basis states that have at most $n_f$ fermions on every even site, and at most $n_f$ anti-fermions on every odd site.
This means that, up to an additive constant, the maximum allowed mass energy per site is set to
\begin{equation}
    E_{\mathrm{max}}(\textbf{n}) = m \cdot n_f\,.
\end{equation}
Whether this truncation leads to a controlled approximation depends on the parameters used for a given simulation.

In this work, we often take the lowest energy cutoff $(n_e, n_f) = (1,1)$ (particularly in $2+1$D), but we will also derive results in $1+1$D at cutoffs $(1,2)$ and $(2,1)$, to illustrate how to systematically raise the allowed electric and mass energy.
Note that all cutoffs are most accurate when both $m$ and $g$ are large, which we graphically illustrate in our results in \Cref{sec:freevac}.

\subsection{Interactions at Large \texorpdfstring{$N_c$}{N}}\label{sec:largeN}

We now turn to the question of the large $N_c$ asymptotics of interactions in the Kogut-Susskind theory with fermions.
Our goal here is to provide a systematic approach that can be used to derive arbitrary matrix elements for interactions in the physical Hilbert space.
We will conclude this subsection by tabulating the matrix elements relevant for all low-lying truncations that will be used in subsequent sections.

\begin{figure*}[t]
\begin{minipage}{0.49\linewidth}
    \begin{tikzpicture}[x=2cm,y=2cm]
        \node[fill=black, circle, inner sep=0pt, minimum size=0.15cm] (v1) at (0,0) {};
        \node[fill=white, draw=black, circle, inner sep=0pt, minimum size=0.15cm] (v2) at (1.5,0) {};
        \node[above] at (v1.north) {$x$};
        \node[above] at (v2.north) {$y$};
        \draw (v1) -- node[below] {$\ell$} (v2);
        \draw (v1) -- ($ (v1) + (-0.6,0.6) $) node[left] {$i_1$};
        \draw (v1) -- ($ (v1) + (-0.6,0.3) $) node[left] {$i_2$};
        \draw (v1) -- ($ (v1) + (-0.6,-0.4) $) node[left] {$i_m$};
        \node at ($ (v1) + (-0.4,0.02) $) {$\vdots$};
        \draw (v2) -- ($ (v2) + (0.6,0.6) $) node[right] {$j_1$};
        \draw (v2) -- ($ (v2) + (0.6,0.3) $) node[right] {$j_2$};
        \draw (v2) -- ($ (v2) + (0.6,-0.4) $) node[right] {$j_n$};
        \node at ($ (v2) + (0.4,0.02) $) {$\vdots$};
    \end{tikzpicture}
    
    \vspace{0.4\baselineskip}
    {(a) Before local point-splitting.}
\end{minipage}
\hfill
\begin{minipage}{0.49\linewidth}
    \begin{tikzpicture}[x=2cm,y=2cm]
        \node[fill=black, circle, inner sep=0pt, minimum size=0.15cm] (v1) at (0,0) {};
        \node[fill=white, draw=black, circle, inner sep=0pt, minimum size=0.15cm] (v2) at (1.5,0) {};
        \node[above] at (v1.north) {$x$};
        \node[above] at (v2.north) {$y$};
        \draw (v1) -- node[below] {$\ell$} (v2);
        \draw (v1) -- ($ (v1) + (-0.6,0) $) node[left] {$i$};
        \draw (v2) -- ($ (v2) + (0.6,0) $) node[right] {$j$};
    \end{tikzpicture}

    \vspace{0.4\baselineskip}
    {(b) After local point-splitting.}
\end{minipage}
\caption{Illustration of double-vertex cuts associated with a generic hopping transition at arbitrary link $\ell \in \mathcal E$. The odd and even endpoints of $\ell$ are labeled $x$ and $y$, respectively. Odd sites are indicated with a black vertex, and even sites are indicated with a white vertex. Hopping acts on sites $x$ and $y$, as well as link $\ell$, but subsequently intertwines the irreps on all external legs into new singlets. In (a), two neighboring sites and their half-links are shown as a subgraph in a non-point-split lattice. In (b), point-splitting is performed \textit{only around link $\ell$} to yield the simplified subgraph that participates in hopping transitions at link $\ell$.}
\label{fig:cut}
\end{figure*}
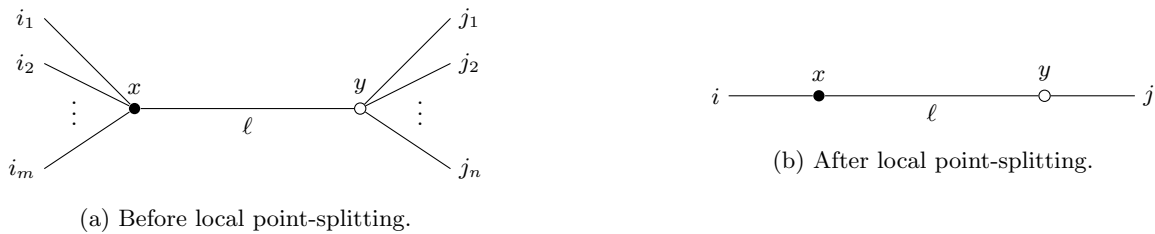

We restrict our attention here exclusively to the matrix elements of hopping operators, since the large $N_c$ asymptotics of plaquette matrix elements have already been discussed in Ref.~\cite{Ciavarella:2025bsg}. Additional methods for plaquette matrix elements at the lowest energy cutoff $(n_e, n_f) = (1, 1)$ are discussed in \Cref{sec:2+1d}.
Therefore, the most general transition we consider occurs on a subgraph in our lattice containing two neighboring lattice sites, each of which can be connected to several links.
We refer to these diagrams (see~\Cref{fig:cut}), as \textit{double-vertex cuts} on the lattice.

By applying point-splitting, it suffices to compute the interaction amplitude for the case of just two external legs $i$ and $j$, attached to the odd and even endpoints, respectively, of an arbitrary link $\ell \in \lk$.\footnote{Note that point-splitting cannot simultaneously arrange for \textit{all} pairs of adjacent lattice sites to appear in the form of \Crefpanel{fig:cut}{b}. This means that a Gram-Schmidt orthogonalization must be performed between different local point-splittings to extract the transitions between global lattice basis states in $2+1$D or higher. For the truncations we perform in this work, this step can be avoided.} 
Using the terminology of \Cref{app:gauge}, the most generic gauge-invariant state supported on this data can be expressed as the product of two \textit{vertex singlets}:
\begin{align}\label{eq:V}
    \ket{V}
    \equiv &\Big[V^1_{abc} \ket{\mathbf{R^i},a}_{i^-}\ket{\overline{\mathbf{A^p}},b}_x\ket{\mathbf{R},c}_{\ell^-}\Big]\nonumber\\
    &\,
    \otimes \Big[V^2_{def} \ket{\overline{\mathbf{R}},d}_{\ell^+}\ket{\mathbf{A^q},e}_y\ket{\overline{\mathbf{R^j}},f}_{j^+}\Big].
\end{align}
The written order of tensor factors here corresponds to their left-to-right appearance in \Crefpanel{fig:cut}{b}.

For any pair of integers $p$ and $q$,  representing the number of charged excitations on sites $x$ and $y$, respectively, and an arbitrary collection of irreps $\mathbf{R^i}$, $\mathbf{R^j}$, and $\mathbf{R}$, the \textit{vertex tensors} $V^1_{abc}$ and $V^2_{def}$ are proportional to Clebsch-Gordan coefficients for combining the irreps at each vertex into the singlet representation.
Note that there can be at most one singlet in the fusion of any three irreps of $\SU(N_c)$, assuming one of those three irreps is an anti-symmetric irrep. 
This means the entire state $\ket{V}$ supported on this cut-out portion of the lattice can be uniquely labeled up to an overall phase factor by $p$, $q$, $\mathbf{R^i}$, $\mathbf{R^j}$, and $\mathbf{R}$, if it exists. The phase conventions for the basis states considered in this paper are discussed in \Cref{app:calc2}.
This implies we can write the state $\ket{V}$ in the verbose notation
\begin{align}
    \ket{V} = \ket{\mathbf{R^i},p,\mathbf{R},q,\mathbf{R^j}}
    \,.
\end{align}
This notation can be simplified even more by an observation from the fusion rules for $\SU(N_c)$: any double-vertex singlet defined by $\ket{\mathbf{R^i},p,\mathbf{R},q,\mathbf{R^j}}$ must satisfy the relations
\begin{subequations}
\begin{align}
    p &\equiv B(\mathbf{R^i}) + B(\mathbf{R}) \pmod{N_c}\,,\\
    q &\equiv B(\mathbf{R^j}) + B(\mathbf{R}) \pmod{N_c}\,,
\end{align}
\end{subequations}
where $B(\mathbf{R})$ denotes the number of boxes in the Young tableau for irrep $\mathbf{R}$.
These relations have a unique solution with occupancies $p, q \in \{0, 1, \dots, N_c-1\}$, which implies that the occupancies can be inferred directly from the \textit{triple-irrep notation}, which we write as
\begin{equation}
    \ket{V} = \ket{\mathbf{R^i},\mathbf{R},\mathbf{R^j}}\,.
\end{equation}

We now have all the ingredients necessary to compute an \textit{arbitrary} transition amplitude of the hopping operator between two the lattice states at vertices $x$ and $y$, which depends entirely on the local data associated to the diagram shown in \Crefpanel{fig:cut}{b}.
The most general transition we need to consider is between two states $\ket{V}$ and $\ket{W}$, where $\ket{V}$ is defined exactly as in \eqref{eq:V}, and $\ket{W}$ is defined similarly in the condensed notation
\begin{equation}
\label{eq:W}
    \ket{W} = \ket{\mathbf{R^i},\mathbf{R'},\mathbf{R^j}}\,.
\end{equation}
for some new irrep $\mathbf{R'}$ on link $\ell$, but same irreps $\mathbf{R^i}$ and $\mathbf{R^j}$.

Since $H_\hop(\ell)$ always increments both anti-fermion number (on odd sites) and fermion number (on even sites) by exactly one, it suffices to consider only transitions that satisfy $B(\mathbf{R'}) = B(\mathbf{R})+1$.
We call these hopping transitions \textit{raising operations}.
For a \textit{lowering operation}, i.e., a hop that decrements anti-fermion number and fermion number, the appropriate non-zero matrix element will be contained in $H_\hop(\ell)^\dagger$, and its value will be the complex conjugate of the corresponding matrix element in $H_\hop(\ell)$ for the reversed transition.

We therefore wish to compute
\begin{align}
    \bra{W} H_\hop(\ell) \ket{V} &\equiv \bra{W} \chi^\dagger(y)_a U(\ell)_{ab} \chi(x)_b\ket{V}\,.\label{eq:hop}
\end{align}
To write down the final result of this computation, we begin by defining a set of symbols $T^r_{abc}$ implicitly via the expansion
\begin{equation}\label{eq:T}
    \chi^\dagger_b \ket{\mathbf{A^r},c} \equiv T^r_{abc} \ket{\mathbf{A^{r+1}},a}\,.
\end{equation}
These symbols will serve as known quantities that can be computed by recognizing that $\ket{\mathbf{A^r},c}$ is an $r$-fold wedge product of single-fermion basis states, where $c$ labels all $\binom{N_c}{r}$ ways of selecting the colors to appear in the wedge product.
The action of $\chi^\dagger_b$ on $\ket{\mathbf{A^r},c}$ is then simply to append a fermion with color $c$ followed by anti-symmetrization and normalization.
Since the wedge product can be defined purely with real coefficients in the tensor product basis, we can assume without loss of generality that all coefficients $T^r_{abc}$ are real. The exact values of $T^r_{abc}$ for special cases of interest are provided in \Cref{app:calc2}.

With this definition in hand, we can write down a ``hopping master equation" that expresses the value of the hopping matrix element in closed form, entirely in terms of known quantities:
\begin{align}\label{eq:master}
    \bra{W} H_\hop(\ell) \ket{V}
    = &\sqrt{\dim \mathbf R'\cdot\dim\mathbf R} \, \times\nonumber\\
    & \times V^1_{abc} \left(W^1_{a\beta\gamma}\right)^* T^{p}_{\beta \rho b} \times\nonumber\\
    & \times V^2_{cef} \left(W^2_{\gamma\eps f}\right)^* T^{q}_{\eps\rho e}.
\end{align}
The full derivation of this result is provided in \Cref{app:calc}.

Remarkably, this final expression \textit{does not} explicitly contain Clebsch-Gordan coefficients for intertwining $\mathbf{R}\otimes\mathbf{N} \to \mathbf{R'}$ where $\mathbf{N}$ is the fundamental representation of SU(N), as one would expect for such a transition.
Instead, all Clebsch-Gordan coefficients appear implicitly in the vertex tensors and $T$-symbols.
In other words, this master equation computes the fully general hopping matrix element directly from the definitions of the initial states, independent of all phase conventions (apart from the $T$-symbols being real).

For the remainder of this section, we will characterize \eqref{eq:master} for special cases that we will need later.
The full characterization for a generic transition requires three steps:
\begin{enumerate}[label=(\roman*)]
    \item Specify the selection rules (or equivalently, determine when \eqref{eq:master} is guaranteed to vanish).
    \item Systematically tabulate all possible gauge-invariant states that can occupy a double-vertex cut.
    \item Specify the matrix elements for the non-zero transitions between those allowed states.
\end{enumerate}

In our triple-irrep notation, the selection rules immediately reduce to fusion rules for $\SU(N_c)$.
In particular, a transition
\begin{equation}\label{eq:raise}
    \ket{\mathbf{R^i},\mathbf{R},\mathbf{R^j}} \to \ket{\mathbf{R^i},\mathbf{R'},\mathbf{R^j}}
\end{equation}
can only have non-zero hopping matrix element for a raising operation if the following conditions are satisfied:
\begin{enumerate}
    \item The fusions $\mathbf{R^i} \otimes \mathbf{R}$, $\mathbf{R} \otimes \mathbf{R^j}$, $\mathbf{R^i} \otimes \mathbf{R'}$, and $\mathbf{R'} \otimes \mathbf{R^j}$ all contain an anti-symmetric irrep.
    \item The fusion $\mathbf{R} \otimes \mathbf{N}$ contains $\mathbf{R'}$.
\end{enumerate}
The first condition is just the requirement that the states $\ket{\mathbf{R^i},\mathbf{R},\mathbf{R^j}}$ and $\ket{\mathbf{R^i},\mathbf{R'},\mathbf{R^j}}$ actually exist, while the second condition picks the transitions that are in principle allowed.

Next, we need to systematically tabulate all allowed gauge-invariant double-vertex cut states.
The energy cutoffs we consider in this paper are
\begin{equation}\label{eq:cutoffs}
    (n_e, n_f) \in \left\{(1, 1), (1, 2), (2, 1)\right\}.
\end{equation}
A detailed systematic tabulation is performed in \Cref{app:tab}, but the procedure can be summarized as follows:
\begin{itemize}
    \item There are $17$ vertex singlets, consistent with the cutoffs \eqref{eq:cutoffs}, that can be written down at an odd lattice site (assuming the local point-splitting from \Crefpanel{fig:cut}{b}, i.e., with vertices of degree two).
    The vertex singlets at an even lattice site are in one-to-one correspondence with the singlets at an odd site, by charge conjugation.
    \item Out of the $17^2 = 289$ ordered pairs of an odd vertex singlet with an even vertex singlet, there are  $43$ ordered pairs where the two middle half-links $\ell^-$ and $\ell^+$ carry compatible irreps to form a valid link wavefunction on $\ell$. Among these $43$ pairs, $41$ of them continue to be consistent with the energy cutoffs \eqref{eq:cutoffs}, and therefore constitute all allowed double-vertex cut states.
    \item Between those $41$ valid states, there are $12$ raising operations satisfying the selection rules.
    Many of the $41$ states do not participate in any hopping transition on their central link $\ell$.
    \item Of the $12$ transitions admitted by selection rules, $11$ continue to be consistent with the energy cutoffs \eqref{eq:cutoffs}.
    Those $11$ allowed hopping transitions involve $19$ distinct double-vertex cut states. The one case that remains inconsistent is discussed further in \Cref{app:tab}.
\end{itemize}

The $19$ participating states are tabulated in \Cref{tab:cuts}.
\Cref{tab:cuts} also provides a graphical notation showing the number and orientation (if any) of electric field lines occupying the double-vertex cut illustrated in \Crefpanel{fig:cut}{b}.
This will make it easier to qualitatively discuss the relationship between allowed transitions at any truncation and their relationship with the large $N_c$ expansion.

\input{double-vertex-table.tex}

Finally, we need to specify the matrix elements of the $11$ allowed hopping transitions ($22$ if one counts both raising operations and their conjugate lowering operations).
Recall that matrix elements computed by \eqref{eq:master} are specifically for raising operations, i.e., transitions that increment fermion and anti-fermion occupancies.
Any transition that increments these occupancies must either strictly increase or strictly decrease the number of electric field lines on the shared link $\ell$ in the double-vertex cut.
To better understand the states that can be accessed in our truncations, it is useful to re-orient some of the raising operations in our presentation, so that the number of electric field lines on the shared link $\ell$ strictly increases through every transition.
The $11$ allowed hopping transitions are provided with this reformatting in \Cref{tab:hops}, where as before phase conventions and derivations for \Cref{tab:hops} are provided in \Cref{app:calc2}.

\begin{table*}[t]
\centering
\begin{tabular}{|c|c|c|c|}

\hline
\textbf{Hopping Transition} & \textbf{Matrix Element} & \textbf{Energy $(n_e,n_f)$} & \textbf{Order in $1/N_c$} \\
\hline\hline

\begin{minipage}[c][1cm][c]{0.30\linewidth}
$\bra{\mathbf 1, \mathbf N, \mathbf 1} H_{\hop}(\ell)\ket{\mathbf 1, \mathbf 1, \mathbf 1}$
\end{minipage}
&
\begin{minipage}[c][1cm][c]{0.15\linewidth}
$\sqrt{N_c}$
\end{minipage}
&
\begin{minipage}[c][1cm][c]{0.10\linewidth}
$(1,1)$
\end{minipage}
&
\begin{minipage}[c][1cm][c]{0.15\linewidth}
$\mathcal O(1)$
\end{minipage}
\\
\hline

\begin{minipage}[c][1cm][c]{0.30\linewidth}
$\bra{\mathbf{N}, \overline{\mathbf{N}}, \mathbf{N}}H_{\hop}(\ell)^\dagger\ket{\mathbf{N}, \mathbf{1}, \mathbf{N}}$
\end{minipage}
&
\begin{minipage}[c][1cm][c]{0.15\linewidth}
$\dfrac{1}{\sqrt{N_c}}$
\end{minipage}
&
\begin{minipage}[c][1cm][c]{0.10\linewidth}
$(1,1)$
\end{minipage}
&
\begin{minipage}[c][1cm][c]{0.15\linewidth}
$\mathcal O(1/N_c)$
\end{minipage}
\\
\hline

\begin{minipage}[c][1cm][c]{0.30\linewidth}
$\bra{\mathbf N, \mathbf N, \mathbf 1} H_{\hop}(\ell)\ket{\mathbf N, \mathbf 1, \mathbf 1}$
\end{minipage}
&
\begin{minipage}[c][1cm][c]{0.15\linewidth}
$\sqrt{N_c-1}$
\end{minipage}
&
\begin{minipage}[c][1cm][c]{0.10\linewidth}
$(1,2)$
\end{minipage}
&
\begin{minipage}[c][1cm][c]{0.15\linewidth}
$\mathcal O(1)$
\end{minipage}
\\
\hline

\begin{minipage}[c][1cm][c]{0.30\linewidth}
$\bra{\mathbf 1, \mathbf N, \mathbf N} H_{\hop}(\ell)\ket{\mathbf 1, \mathbf 1, \mathbf N}$
\end{minipage}
&
\begin{minipage}[c][1cm][c]{0.15\linewidth}
$\sqrt{N_c-1}$
\end{minipage}
&
\begin{minipage}[c][1cm][c]{0.10\linewidth}
$(1,2)$
\end{minipage}
&
\begin{minipage}[c][1cm][c]{0.15\linewidth}
$\mathcal O(1)$
\end{minipage}
\\
\hline

\begin{minipage}[c][1cm][c]{0.30\linewidth}
$\bra{\mathbf N, \mathbf N, \mathbf N} H_{\hop}(\ell)\ket{\mathbf N, \mathbf 1, \mathbf N}$
\end{minipage}
&
\begin{minipage}[c][1cm][c]{0.15\linewidth}
$\dfrac{N_c-1}{\sqrt{N_c}}$
\end{minipage}
&
\begin{minipage}[c][1cm][c]{0.10\linewidth}
$(1,2)$
\end{minipage}
&
\begin{minipage}[c][1cm][c]{0.15\linewidth}
$\mathcal O(1)$
\end{minipage}
\\
\hline

\begin{minipage}[c][1cm][c]{0.30\linewidth}
$\bra{\mathbf{N}, \mathbf{Ad}, \mathbf{N}} H_{\hop}(\ell) \ket{\mathbf{N}, \overline{\mathbf{N}}, \mathbf{N}}$
\end{minipage}
&
\begin{minipage}[c][1cm][c]{0.15\linewidth}
$\sqrt{\dfrac{N_c^2-1}{N_c}}$
\end{minipage}
&
\begin{minipage}[c][1cm][c]{0.10\linewidth}
$(2,1)$
\end{minipage}
&
\begin{minipage}[c][1cm][c]{0.15\linewidth}
$\mathcal O(1)$
\end{minipage}
\\
\hline

\begin{minipage}[c][1cm][c]{0.30\linewidth}
$\bra{\mathbf{Ad}, \mathbf{Ad}, \mathbf{Ad}} H_{\hop}(\ell)^\dagger \ket{\mathbf{Ad}, \mathbf{N}, \mathbf{Ad}}$
\end{minipage}
&
\begin{minipage}[c][1cm][c]{0.15\linewidth}
$\sqrt{\dfrac{N_c}{N_c^2-1}}$
\end{minipage}
&
\begin{minipage}[c][1cm][c]{0.10\linewidth}
$(2,1)$
\end{minipage}
&
\begin{minipage}[c][1cm][c]{0.15\linewidth}
$\mathcal O(1/N_c)$
\end{minipage}
\\
\hline

\begin{minipage}[c][1cm][c]{0.30\linewidth}
$\bra{\overline{\mathbf{N}}, \mathbf{A^2}, \overline{\mathbf{N}}} H_{\hop}(\ell) \ket{\overline{\mathbf{N}}, \mathbf{N}, \overline{\mathbf{N}}}$
\end{minipage}
&
\begin{minipage}[c][1cm][c]{0.15\linewidth}
$\sqrt{\dfrac{N_c-1}{2}}$
\end{minipage}
&
\begin{minipage}[c][1cm][c]{0.10\linewidth}
$(2,1)$
\end{minipage}
&
\begin{minipage}[c][1cm][c]{0.15\linewidth}
$\mathcal O(1)$
\end{minipage}
\\
\hline

\begin{minipage}[c][1cm][c]{0.30\linewidth}
$\bra{\mathbf{A^2}, \overline{\mathbf{A^2}}, \mathbf{A^2}} H_{\hop}(\ell)^\dagger \ket{\mathbf{A^2}, \overline{\mathbf{N}}, \mathbf{A^2}}$
\end{minipage}
&
\begin{minipage}[c][1cm][c]{0.15\linewidth}
$\sqrt{\dfrac{2}{N_c-1}}$
\end{minipage}
&
\begin{minipage}[c][1cm][c]{0.10\linewidth}
$(2,1)$
\end{minipage}
&
\begin{minipage}[c][1cm][c]{0.15\linewidth}
$\mathcal O(1/N_c)$
\end{minipage}
\\
\hline

\begin{minipage}[c][1cm][c]{0.30\linewidth}
$\bra{\overline{\mathbf{N}}, \mathbf{S^2}, \overline{\mathbf{N}}} H_{\hop}(\ell) \ket{\overline{\mathbf{N}}, \mathbf{N}, \overline{\mathbf{N}}}$
\end{minipage}
&
\begin{minipage}[c][1cm][c]{0.15\linewidth}
$\sqrt{\dfrac{N_c+1}{2}}$
\end{minipage}
&
\begin{minipage}[c][1cm][c]{0.10\linewidth}
$(2,1)$
\end{minipage}
&
\begin{minipage}[c][1cm][c]{0.15\linewidth}
$\mathcal O(1)$
\end{minipage}
\\
\hline

\begin{minipage}[c][1cm][c]{0.30\linewidth}
$\bra{\mathbf{S^2}, \overline{\mathbf{S^2}}, \mathbf{S^2}} H_{\hop}(\ell)^\dagger \ket{\mathbf{S^2}, \overline{\mathbf{N}}, \mathbf{S^2}}$
\end{minipage}
&
\begin{minipage}[c][1cm][c]{0.15\linewidth}
$\sqrt{\dfrac{2}{N_c+1}}$
\end{minipage}
&
\begin{minipage}[c][1cm][c]{0.10\linewidth}
$(2,1)$
\end{minipage}
&
\begin{minipage}[c][1cm][c]{0.15\linewidth}
$\mathcal O(1/N_c)$
\end{minipage}
\\
\hline

\end{tabular}
\caption{Matrix elements for all $11$ distinct (i.e., up to conjugation) hopping transitions. The matrix element is shown for $H_{\hop}(\ell)$ if the raising operation increases the number of electric fluxes, or for $H_{\hop}(\ell)^\dagger$ if the \textit{lowering} operation increases the number of electric fluxes. The minimum energy cutoff needed for the transition and the de facto order in $1/N_c$ (relevant for truncations by powers of $1/N_c$ in $2+1$D and higher) are provided in the third and fourth columns, respectively.}\label{tab:hops}
\end{table*}

The key pattern to extract from \Cref{tab:hops} is that the large $N_c$ scaling of a hopping matrix element is $\propto \sqrt{N_c}$ if and only if the raising operation \textit{also} increases the number of electric fluxes.
Similarly, the large $N_c$ scaling is $\propto 1/\sqrt{N_c}$ if and only if the \textit{lowering} operation increases the number of electric fluxes. 

Assuming we work with the free vacuum as the initial state for our Krylov constraints, this immediately implies that the $(1, 1, 1, 1)$ truncation lives at de facto order $\mathcal O(1)$. This follows from the fact that the only allowed hopping transition is creating or annihilating an isolated quark-antiquark pair joined by a single electric field line, which has matrix element $\sqrt{N_c}$. The relevant plaquette matrix elements were already calculated in Ref.~\cite{Ciavarella:2025bsg}, and point-splitting adds no complications here.
It also follows that the $(1, 1, 2, 2)$ truncation at energy cutoff $(n_e, n_f) = (1, 1)$ simply adds in the remaining states consistent with the flux and occupancy limits, which is now automatically the $\mathcal O(1/N_c)$ refinement of the $(1, 1, 1, 1)$ truncation, according to the matrix elements in \Cref{tab:hops}.\footnote{Again, the plaquette matrix elements were previously discussed in Ref.~\cite{Ciavarella:2025bsg}, but this time point-splitting does introduce complications. For instance, Gram-Schmidt orthogonalization is required to write down a global lattice basis for a hypercubic lattice. Nevertheless, the fact that \Cref{tab:hops} contains all local point-split transitions implies that all transitions between global lattice states will just be linear combinations of these local transitions; and therefore the full expansion to order $\mathcal O(1/N_c)$ is recovered at the chosen energy cutoff.}

\section{Truncated Hamiltonians}\label{sec:hams}

With the results in hand from \Cref{sec:trunc}, we can now write down the Kogut-Susskind Hamiltonian at various levels of truncation.
We will focus on writing down Hamiltonians that we intend to simulate (simulations are performed in \Cref{sec:simuls}), but our strategy extends to arbitrary bipartite lattice geometries, and to arbitrary truncations.

In \Cref{sec:1+1d} and \Cref{sec:2+1d}, we will implement Krylov constraints by assuming the initial vacuum state $\ket{0}$, which is consistent with our later simulations.
In \Cref{sec:external}, we will discuss how to add external charges by modifying Gauss's law.
When truncating this modified theory, we will focus on Hamiltonians relevant to simulating string dynamics.
In that specific case, it will be useful to implement the Krylov constraints by replacing the initial free vacuum state with a meson-like string state.

An important consideration for $1+1$D lattice gauge theory is that the gauge field can be exactly integrated out.
In principle, this means the $1+1$D Hamiltonian derivations can be discussed in the context of a dual pure-fermion description.
To stay on track with our main focus, we will restrict ourselves to the description that contains both fermions and gauge fields, since that will also naturally translate to higher dimensions.
In this case, it will prove useful to attribute quantum memory to each \textit{link}, rather than each lattice site.

Note that assigning an irrep to each link cannot always provide a complete description for the quantum state of the global lattice, even if that lattice has already been point-split, because it is possible for three $\SU(N_c)$ irreps to fuse into more than one singlet.
However, the examples we consider in this paper will be simple enough to avoid this issue, so it will suffice for our purposes to work purely with link irreps.
How to deal with more complicated cases has been explored in~\cite{Kadam:2024ifg,Hayata:2023bgh,Hayata:2026xeo}.

To this end, we introduce some notation here that will be useful throughout this section.
For any generic truncation (henceforth denoted T), the set of allowed link irreps is denoted $S_{\text{T}}$, and we allocate one computational basis state per irrep.
For clarity, the allowed link states are denoted directly by the irrep they represent, i.e.,
\begin{equation}
    \ket{\mathbf{R}},\quad\mathbf{R}\in S_{\text{T}}.
\end{equation}

We also provide a new notation for Pauli matrices (including their generalizations to qudits) and projection operators, since we will be using these regularly in our discussion.
For $\mathbf{u}, \mathbf{v} \in S_{\text{T}}$ with $\mathbf{u} \neq \mathbf{v}$, we define
\begin{align}
\label{eq:op_def}
    P^{\mathbf{u}} &\equiv \ket{\mathbf{u}}\bra{\mathbf{u}}, \nonumber \\
    \mathcal{X}^{\mathbf{u}\mathbf{v}} &\equiv \ket{\mathbf{u}}\bra{\mathbf{v}} + \ket{\mathbf{v}}\bra{\mathbf{u}}, \nonumber \\
    \mathcal{Z}^{\mathbf{u}} &\equiv I - 2 P^{\mathbf{u}}, \nonumber \\
    b^{\mathbf{u}\mathbf{v}} & \equiv \ket{\mathbf{u}}\bra{\mathbf{v}},
\end{align}
where $I$ denotes the $|S_{\text{T}}|$-dimensional identity matrix,
\begin{equation}
    I \equiv \sum_{\mathbf{w} \in S_{\text{T}}} \ket{\mathbf{w}}\bra{\mathbf{w}}.
\end{equation}
This notation reduces to the standard projectors and Pauli operators on qubits when $|S_{\text{T}}| = 2$, and gives useful generalizations for arbitrary truncations T.

When writing down Hamiltonians, it is useful to keep in mind the following interpretations for these operators: $P^{\mathbf{u}}$ enforces $\mathbf{u}$ as a constraint; $X^{\mathbf{u}\mathbf{v}}$ swaps between $\mathbf{u}$ and $\mathbf{v}$; and $Z^{\mathbf{u}}$ is a phase oracle for a one-hot encoding of $\mathbf{u}$.
This will allow for improved readability of the (often obtrusively long) Hamiltonians we will find.

In order to avoid repeating the same discussion for each truncation, let us briefly remark here on the general presentation and methodology we will use throughout this section.
We will decompose our Hamiltonians into three terms that respectively correspond to the electric, mass, and interaction terms from lattice gauge theory.
Specifically, we write
\begin{equation}
    H_{\text{T}} = H^{\mathrm{elec}}_{\text{T}} + H^{\mathrm{mass}}_{\text{T}} + H^{\mathrm{int}}_{\text{T}},
\end{equation}
where T denotes an arbitrary truncation.
In dimensions $2+1$D and higher, we further decompose $H^{\mathrm{int}}_{\text{T}}$ into
\begin{equation}
    H^{\mathrm{int}}_{\text{T}} = H^{\mathrm{plaq}}_{\text{T}} + H^{\mathrm{hop}}_{\text{T}},
\end{equation}
to emphasize the distinction between plaquette and hopping interactions.

The simplest way to write the electric term at any truncation is to use a sum of one-body Pauli-$Z$ operators that separately encode the electric energy of each link, conditioned on its irrep.
This is performed essentially identically at every truncation, and in every dimension.

It turns out that the mass term can also be written as a sum of one-body Pauli-$Z$'s.
This might initially seem surprising, for instance, because a local mass term needs to detect \textit{pairs} of links to deduce the endpoints of strings on the lattice.
However, our notation convention leads to a convenient simplification, based on the following observation regarding the matter and antimatter content for any gauge-invariant, global lattice state:
\begin{widetext}
\begin{equation}\label{eq:flow}
    \#\,\text{quarks} = \#\,\text{antiquarks} = \sum_{\text{irreps}\,\mathbf{R}} \left(\#\,\text{occurrences of }\textbf{R}\text{ on links}\right) \times \left(\text{net electric flux of }\textbf{R} \right).
\end{equation}
\end{widetext}
In other words, by simply weighting each link irrep that appears on the lattice by its \textit{net electric flux} (defined in \Cref{tab:flow}), we can deduce the exact number of quarks and antiquarks present on the lattice (in the zero-charge sector of the theory).
This is true unless the total number of fermions at a single lattice site is $N_c$ in which case it can form a color singlet baryon. 

An intuitive understanding of this fact is as follows: the number of excitations at an odd (resp. even) lattice site is determined by counting the net charge flow leaving (resp. entering) that site.
In our notation, this value is purely determined by knowing the irrep, and agnostic to whether the site carries fermions or anti-fermions.
For instance, when working modulo $N_c$, this is just the total number of boxes in the Young tableaux over all irreps incident to the lattice site (cf. \Cref{sec:largeN}).
More generally, adding the net electric flux on all links adjacent to a lattice site gives its exact occupancy number (e.g., this holds in \Cref{tab:cuts}).
Therefore, on the full lattice, the total count of all lattice site excitations (including both quarks and antiquarks) is exactly \textit{twice} the sum of the net charge flow carried by each link irrep (it's doubled because every link has two endpoints), which leads to \eqref{eq:flow}.

\begin{table}[htp]
\centering
\begin{tabular}{|c|c|c|c|c|c|c|c|c|}

\hline

\begin{minipage}[c][1cm][c]{0.35\linewidth}
\textbf{Representation:}
\end{minipage}
&
\begin{minipage}[c][1cm][c]{0.06\linewidth}
$\mathbf{1}$
\end{minipage}
&
\begin{minipage}[c][1cm][c]{0.06\linewidth}
$\mathbf{N}$
\end{minipage}
&
\begin{minipage}[c][1cm][c]{0.06\linewidth}
$\overline{\mathbf{N}}$
\end{minipage}
&
\begin{minipage}[c][1cm][c]{0.06\linewidth}
$\mathbf{Ad}$
\end{minipage}
&
\begin{minipage}[c][1cm][c]{0.06\linewidth}
$\mathbf{A^2}$
\end{minipage}
&
\begin{minipage}[c][1cm][c]{0.06\linewidth}
$\overline{\mathbf{A^2}}$
\end{minipage}
&
\begin{minipage}[c][1cm][c]{0.06\linewidth}
$\mathbf{S^2}$
\end{minipage}
&
\begin{minipage}[c][1cm][c]{0.06\linewidth}
$\overline{\mathbf{S^2}}$
\end{minipage}
\\
\hline

\begin{minipage}[c][1cm][c]{0.35\linewidth}
\textbf{Net Electric Flux:}
\end{minipage}
&
\begin{minipage}[c][1cm][c]{0.06\linewidth}
$0$
\end{minipage}
&
\begin{minipage}[c][1cm][c]{0.06\linewidth}
$1$
\end{minipage}
&
\begin{minipage}[c][1cm][c]{0.06\linewidth}
$-1$
\end{minipage}
&
\begin{minipage}[c][1cm][c]{0.06\linewidth}
$0$
\end{minipage}
&
\begin{minipage}[c][1cm][c]{0.06\linewidth}
$2$
\end{minipage}
&
\begin{minipage}[c][1cm][c]{0.06\linewidth}
$-2$
\end{minipage}
&
\begin{minipage}[c][1cm][c]{0.06\linewidth}
$2$
\end{minipage}
&
\begin{minipage}[c][1cm][c]{0.06\linewidth}
$-2$
\end{minipage}
\\
\hline

\end{tabular}
\caption{Net electric flux for irreps considered in this work. Mathematically, this is the net fermion charge carried by the irrep from an anti-fermion endpoint to a fermion endpoint (with a negative sign if fermion charge is carried in the opposite direction).}\label{tab:flow}
\end{table}

Plaquette and hopping interactions serve as the off-diagonal entries in the Hamiltonian matrix, and their exact form is more truncation-specific than the electric and mass terms.
In general, we will use projections of Pauli-$X$ operators to isolate specific matrix elements for a conjugate pair of raising and lowering operations, followed by weighting them with the desired matrix element.
In $1+1$D this leads to simple PXP terms, but in $2+1$D and higher, the structure of the projected subspace is more complicated and will require superpositions of products of the single-irrep projectors $P^{\mathbf{u}}$.

\subsection{$1+1$D Truncations}\label{sec:1+1d}

Recall that in $1+1$D, we only require two Krylov parameters, $n_h$ and $n_{\chi}$, in addition to the energy cutoffs $n_e$ and $n_f$.
Making use of this fact, we notate several examples of low-lying $1+1$D truncations below:
\begin{itemize}
    \item \Tone: $(n_h, n_\chi) = (1, 1)$ at $(n_e, n_f) = (1, 1)$;
    \item \Ttwo: $(n_h, n_\chi) = (1, 2)$ at $(n_e, n_f) = (1, 1)$;
    \item \Tthree: $(n_h, n_\chi) = (1, 2)$ at $(n_e, n_f) = (1, 2)$;
    \item \Tthreep: \Tthree\ at order $\mathcal O(\sqrt{N_c})$;
    \item \Tfour: $(n_h, n_\chi) = (2, 3)$ at $(n_e, n_f) = (2, 1)$;
    \item \Tfive: $(n_h, n_\chi) = (2, 4)$ at $(n_e, n_f) = (2, 1)$.
\end{itemize}
We will derive Hamiltonians for all of these truncations in terms of qubits, qutrits, and qudits, as necessary.
Simulation results are provided in \Cref{sec:simuls}.

For some truncations (such as \Tone \ and \Ttwo \ above), the Krylov constraints exactly correspond to truncation by powers of $1/N_c$ at fixed energy cutoff.
For other truncations (such as \Tthreep \ and \Tthree \ above), it is useful to view the $\mathcal O(\sqrt{N_c})$ piece of the truncation separately from its correction at order $\mathcal O(1/\sqrt{N_c})$, because one cannot choose Krylov constraints at the same energy cutoff that capture only the $\mathcal O(\sqrt{N_c})$ interactions.
For yet other truncations (such as \Tfour \ and \Tfive), the Krylov constraints keep \textit{some baseline} of interactions to all orders in $1/N_c$ (for instance, all interactions from \Ttwo are kept in \Tfour and \Tfive), while limiting only very specific string-breaking or string-joining operations.

\subsubsection{Truncation \Tone}

This is the lowest nontrivial level of truncation for the Kogut-Susskind Hamiltonian.
The Krylov constraints allow at most one hopping operation per link, and at most one fermion operation per site.
This implies that the truncated Hilbert space can be generated purely by raising operations, because a lowering operation annihilates any state that does not already have excitations at both its participating lattice sites.

\begin{figure}[htp]
\centering

\begin{tikzpicture}
\LatticeOneD{4}{0}[][0]
\end{tikzpicture}

\vspace{4mm}

\begin{tikzpicture}
\bigskip
\BubbleDownArrow
\end{tikzpicture}

\vspace{4mm}

\begin{tikzpicture}
\OrientedEdge{1}{0}{2}{0}{black}
\LatticeOneD{4}{0}[][0]
\PlaceEq{1}{0.3}{\overline{q}}
\PlaceEq{2}{0.25}{q}
\end{tikzpicture}

\caption{Transition allowed at \Tone: $\ket{\mathbf{1},\mathbf{1},\mathbf{1}}\to\ket{\mathbf{1},\mathbf{N},\mathbf{1}}$.}\label{fig:T1new}
\end{figure}
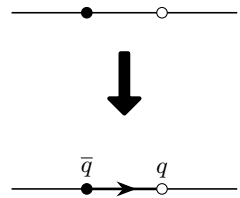

At a given link $\ell \in \lk$, the raising operation hops a fermion from the odd site $x$ of the link to its even site $y$, thereby leaving an anti-fermion excitation at $x$ and a fermion excitation at $y$.\footnote{The reverse, an \textit{annihilation} of a quark-antiquark pair, is also allowed because it reverts to a state already included in the Hilbert space.}
This also leaves the link $\ell$ in the fundamental representation, with an electric field line oriented from $x \to y$.
Moreover, this hopping is only allowed when both $x$ and $y$ are initially in the Dirac vacuum, which in turn implies that neighboring links cannot be simultaneously excited.
This transition is shown diagrammatically in \Cref{fig:T1new}

Reading off from \Cref{tab:hops}, this interaction has matrix element $\sqrt{N_c}$.
In other words, truncation \Tone provides the leading order contribution to the energy cutoff $(n_e, n_f) = (1, 1)$ in the large $N_c$ expansion.
Based on this information, any gauge-invariant basis state on a lattice with $L$ links is summarized by a sequence of $L$ irreps, each chosen to be $\mathbf{1}$ or $\mathbf{N}$, such that no two consecutive terms are both $\mathbf{N}$.
In other words, the state of each link can be summarized by a single qubit, with slight redundancy (since any instance of $\mathbf{N}$ instantly guarantees that its neighbors are both $\mathbf{1}$).
In this case, we utilize the notation
\begin{equation}
    S_{\text{\Tone}} \equiv \{\mathbf{1}, \mathbf{N}\}
\end{equation}
for the allowed link qubit states.
The translation to standard qubit notation can be performed by relabeling $\ket{0} \equiv \ket{\mathbf{1}}$ and $\ket{1} \equiv \ket{\mathbf{N}}$.

Physically, the truncated Hilbert space contains isolated quark-antiquark pairs, each occupying the space of a single link.
The energy carried by one of these pairs is the sum of its electric energy and mass energy, namely
\begin{equation}
    E = 2m + \frac{g^2}{2} C_2(\mathbf{N}) = 2m + \frac{g^2}{2} \frac{N_c^2-1}{2N_c}\,.
\end{equation}
The diagonal part of the Hamiltonian therefore involves summing this contribution from every quark-antiquark pair that is physically present on the lattice.

The electric term can be implemented with Pauli-$Z$'s by raising the energy level for every link excited to the fundamental representation (see~\eqref{eq:op_def} for the definition of the local operators):
\begin{equation}
    H^{\mathrm{elec}}_{\text{\Tone}} = -\frac{g^2}{4}  \frac{N_c^2-1}{2N_c} \sum_{j=1}^L \mathcal{Z}^{\mathbf{N}}_j\,,
\end{equation}
where the subscript $j$ indicates the Pauli-$Z$ on the $j^{\mathrm{th}}$ qubit.
The mass term is similarly given by
\begin{equation}
    H^{\mathrm{mass}}_{\text{\Tone}} = -m \sum_{j=1}^L \mathcal{Z}^{\mathbf{N}}_j\,.
\end{equation}

The off-diagonal part of the Hamiltonian needs to introduce a $\sqrt{N_c}/2$ interaction strength to switch a link between $\ket{\mathbf{1}}$ and $\ket{\mathbf{N}}$, conditioned on the requirement that its neighbors are both in the singlet representation.
This condition can be implemented by sandwiching a Pauli-$X$ between two projectors, giving rise to the interaction term
\begin{equation}
    H^{\mathrm{int}}_{\text{\Tone}} = \frac{\sqrt{N_c}}{2} \sum_{j=1}^L P^{\mathbf{1}}_{j-1} \mathcal{X}^{\mathbf{1}\mathbf{N}}_j P^{\mathbf{1}}_{j+1},
\end{equation}
where we again use the subscript $j$ on operators to indicate their action on the $j^{\mathrm{th}}$ qubit, and we also identify indices periodically so that $0 \sim L$ and $(L+1) \sim 1$.

The Hamiltonian at this truncation is therefore precisely the PXP model with an external field.
This has been studied extensively both experimentally~\cite{Bernien2017ProbingManyBodyDynamics,Su2023ObservationManybodyScarring} and theoretically~\cite{Surace2020LatticeGaugeTheories,Hudomal2022DrivingQuantumManyBodyScars}, and its thermalization is known to feature weak ergodicity breaking in the form of many-body scarring~\cite{Turner2018WeakErgodicityBreaking}.

For our purposes, it is useful to note that this truncation does not permit string breaking, because states containing strings spanning more than one link are simply not part of the Hilbert space.
This can be interpreted as a consequence of the fact that interactions permitting string breaking are $1/N_c$ suppressed relative to the interactions kept at this truncation.
In other words, we expect that string breaking is explored very weakly at large $N_c$.
This fact is already understood from traditional studies of $\SU(N_c)$ Yang-Mills theory, where the 't Hooft expansion keeps only planar Feynman diagrams in the perturbative large $N_c$ limit~\cite{tHooft:1973alw}.

\subsubsection{Truncation \Ttwo}

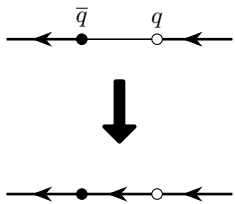
\begin{figure}[htp]
\centering

\begin{tikzpicture}
\OrientedEdge{1}{0}{0}{0}{black}
\OrientedEdge{3}{0}{2}{0}{black}
\LatticeOneD{4}{0}[][0]
\PlaceEq{1}{0.3}{\overline{q}}
\PlaceEq{2}{0.25}{q}
\end{tikzpicture}

\vspace{4mm}

\begin{tikzpicture}
\bigskip
\BubbleDownArrow
\end{tikzpicture}

\vspace{4mm}

\begin{tikzpicture}
\OrientedEdge{1}{0}{0}{0}{black}
\OrientedEdge{2}{0}{1}{0}{black}
\OrientedEdge{3}{0}{2}{0}{black}
\LatticeOneD{4}{0}[][0]
\end{tikzpicture}

\caption{New transition at \Ttwo: $\ket{\mathbf{N},\mathbf{1},\mathbf{N}}\to\ket{\mathbf{N},\overline{\mathbf{N}},\mathbf{N}}$.}\label{fig:T2new}
\end{figure}

The situation becomes even more interesting once we allow a second fermion operation per site.
Now it is possible to apply a lowering operation to a link $\ell$ in the singlet representation, if it is sandwiched between two links that are both in the fundamental.
This is achieved by hopping the quark at $x$ into the hole provided by the antiquark at $y$ (see \Cref{fig:T2new}).
Both quarks are annihilated as a result, while releasing energy by exciting link $\ell$ into the anti-fundamental representation.
This lowering operation increases electric flux count, so according to \Cref{sec:largeN}, it will contribute at order $\mathcal O(1/\sqrt{N_c})$.

In graphical terms, this operation joins the endpoints of two electric field lines carrying the same orientation---initially separated by a gap of one link---into a single elongated electric field line, i.e. a string.
The matrix element for this lowering operation (and its reverse, breaking a string) can be read off as $1/\sqrt{N_c}$ from \Cref{tab:hops}, and allowing this interaction completes the entire hopping term at energy cutoff $(n_e, n_f) = (1, 1)$.
For this reason, truncation \Ttwo\ is the $\mathcal O(1/\sqrt{N_c})$ correction to truncation \Tone (with a relative scaling of $1/N_c$).

Unlike the case with \Tone, each link can now occupy three different irreps, and we have
\begin{equation}
    S_{\text{\Ttwo}} \equiv \left\{\mathbf{1}, \mathbf{N}, \overline{\mathbf{N}}\right\}.
\end{equation}
We are thus dealing with qutrits, instead of qubits.

A generic gauge-invariant basis state at this truncation can be thought of as a collection of \textit{isolated} strings, meaning that every string maintains a gap of at least one link from every other string.
In our notation, the quantum state of a string is represented by an alternating sequence of $\ket{\mathbf{N}}$ and $\ket{\overline{\mathbf{N}}}$ along its links, with the first and last links both being in the fundamental.

Since the quadratic Casimir is invariant under conjugation, a string of length $r$ must carry a total energy
\begin{equation}
    E(r) = 2m + \frac{g^2}{2} \sum_{\ell \in \mathrm{string}} C_2(\mathbf{R}_{\ell}) = 2m + r \frac{g^2}{2}\frac{N_c^2-1}{2N_c}\,,
\end{equation}
where the sum runs over all links $\ell$ composing the string.
(In particular, the irrep labels $\mathbf{R}_\ell$ at each link $\ell$ simply alternate between $\mathbf{N}$ and $\overline{\mathbf{N}}$.)
The total string energy is obtained by summing this over all strings of every possible length that may be physically present in the lattice.

The electric part of the energy can be implemented by using Pauli-$Z$'s to count all links excited to either $\ket{\mathbf{N}}$ or $\ket{\overline{\mathbf{N}}}$.
We can thus write
\begin{equation}
    H_{\text{\Ttwo}}^{\mathrm{elec}} = -\frac{g^2}{4}\frac{N_c^2-1}{2N_c} \sum_{j=1}^L \left(\mathcal{Z}^{\mathbf{N}}_j + \mathcal{Z}^{\overline{\mathbf{N}}}_j\right)\,.
\end{equation}

The quark mass term can be implemented by recognizing that for any string, the number of links in the fundamental is exactly one more than the number of links in the anti-fundamental.
This means the difference between the counts of fundamental and anti-fundamental irreps across all links in the lattice gives the total number of strings present.\footnote{Note that this is a special case of \eqref{eq:flow}.}
We can therefore write down
\begin{equation}
    H^{\mathrm{mass}}_{\text{\Ttwo}} = -m \sum_{j=1}^L \left(\mathcal{Z}^{\mathbf{N}}_j - \mathcal{Z}^{\overline{\mathbf{N}}}_j\right)\,.
\end{equation}

To write down the off-diagonal piece of the qutrit Hamiltonian, we need to borrow the PXP term from truncation \Tone, and add in all string-joining and string-breaking operations.
This can be achieved by adding interactions with strength $1/(2\sqrt{N_c})$ to switch a link between $\ket{\mathbf{1}}$ and $\ket{\overline{\mathbf{N}}}$, conditioned on the requirement that its neighbors are both in the fundamental representation.
The result is
\begin{equation}
    H^{\mathrm{int}}_{\text{\Ttwo}} = \frac{\sqrt{N_c}}{2} \sum_{j=1}^L P^{\mathbf{1}}_{j-1} \mathcal{X}^{\mathbf{1}\mathbf{N}}_j P^{\mathbf{1}}_{j+1} + \frac{1}{2\sqrt{N_c}} \sum_{j=1}^L P^{\mathbf{N}}_{j-1} \mathcal{X}^{\mathbf{1}\overline{\mathbf{N}}}_j P^{\mathbf{N}}_{j+1}.
\end{equation}

In the resulting qutrit Hamiltonian, string breaking \textit{is} permitted at finite $N_c$, and will be investigated in \Cref{sec:simuls}.

\subsubsection{Truncations \Tthreep\ and \Tthree}

Truncation \Tthree\ can be obtained by starting with truncation \Ttwo, and adding the interactions that spawn quark-antiquark pairs adjacent to existing quark-antiquark pairs (i.e., sharing a lattice site so that two excitations are created on that site). 
In particular, it already includes interactions at both $\mathcal O(\sqrt{N_c})$ and $\mathcal O(1/\sqrt{N_c})$.
For this reason, we exercise here the optional step to separate our discussion for the $\mathcal O(\sqrt{N_c})$ piece (truncation \Tthreep), before including the $\mathcal O(1/\sqrt{N_c})$ corrections.

\begin{figure*}[t]
\centering

\begin{minipage}{0.3\linewidth}
\begin{tikzpicture}
\OrientedEdge{1}{0}{0}{0}{black}
\LatticeOneD{4}{0}[][0]
\PlaceEq{1}{0.3}{\overline{q}}
\end{tikzpicture}

\vspace{4mm}

\begin{tikzpicture}
\bigskip
\BubbleDownArrow
\end{tikzpicture}

\vspace{4mm}

\begin{tikzpicture}
\OrientedEdge{1}{0}{0}{0}{black}
\OrientedEdge{1}{0}{2}{0}{black}
\LatticeOneD{4}{0}[][0]
\PlaceEq{1}{0.3}{\overline{q}\,\,\overline{q}}
\PlaceEq{2}{0.25}{q}
\end{tikzpicture}
\end{minipage}%
\begin{minipage}{0.3\linewidth}
\begin{tikzpicture}
\OrientedEdge{3}{0}{2}{0}{black}
\LatticeOneD{4}{0}[][0]
\PlaceEq{2}{0.25}{q}
\end{tikzpicture}

\vspace{4mm}

\begin{tikzpicture}
\bigskip
\BubbleDownArrow
\end{tikzpicture}

\vspace{4mm}

\begin{tikzpicture}
\OrientedEdge{1}{0}{2}{0}{black}
\OrientedEdge{3}{0}{2}{0}{black}
\LatticeOneD{4}{0}[][0]
\PlaceEq{1}{0.3}{\overline{q}}
\PlaceEq{2}{0.25}{q\,\,q}
\end{tikzpicture}
\end{minipage}%
\begin{minipage}{0.3\linewidth}
\begin{tikzpicture}
\OrientedEdge{1}{0}{0}{0}{black}
\OrientedEdge{3}{0}{2}{0}{black}
\LatticeOneD{4}{0}[][0]
\PlaceEq{1}{0.3}{\overline{q}}
\PlaceEq{2}{0.25}{q}
\end{tikzpicture}

\vspace{4mm}

\begin{tikzpicture}
\bigskip
\BubbleDownArrow
\end{tikzpicture}

\vspace{4mm}

\begin{tikzpicture}
\OrientedEdge{1}{0}{0}{0}{black}
\OrientedEdge{1}{0}{2}{0}{black}
\OrientedEdge{3}{0}{2}{0}{black}
\LatticeOneD{4}{0}[][0]
\PlaceEq{1}{0.3}{\overline{q}\,\,\overline{q}}
\PlaceEq{2}{0.25}{q\,\,q}
\end{tikzpicture}
\end{minipage}

\caption{New transitions at truncation \Tthree'. \textit{Left}: $\ket{\mathbf{N},\mathbf{1},\mathbf{1}}\to\ket{\mathbf{N},\mathbf{N},\mathbf{1}}$. \textit{Middle}: $\ket{\mathbf{1},\mathbf{1},\mathbf{N}}\to\ket{\mathbf{1},\mathbf{N},\mathbf{N}}$. \textit{Right}: $\ket{\mathbf{N},\mathbf{1},\mathbf{N}}\to\ket{\mathbf{N},\mathbf{N},\mathbf{N}}$.}\label{fig:T3pnew}
\end{figure*}
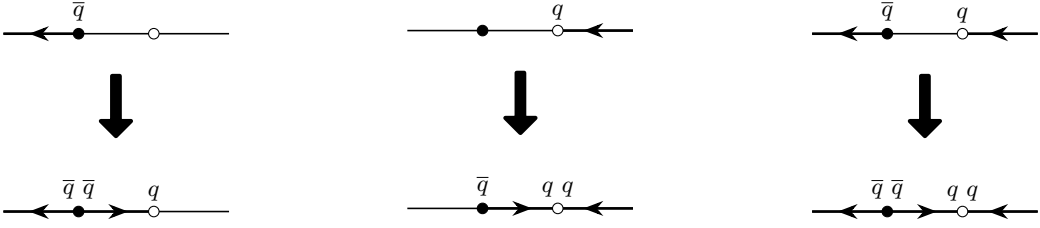

At $\mathcal O(\sqrt{N_c})$, the only allowed interactions are creation or annihilation of quark-antiquark pairs, without joining any strings.
This means the allowed link irreps are
\begin{equation}
    S_{\text{\Tthreep}} = \left\{\mathbf{1},\mathbf{N}\right\}\,,
\end{equation}
just as for truncation \Tone.
The energy for all quark-antiquark pairs follows exactly the same logic as it did for truncation \Tone, so we can directly borrow the exact same diagonal terms from that discussion:
\begin{align}
    H^{\mathrm{elec}}_{\text{\Tthreep}} &= -\frac{g^2}{4} \frac{N_c^2-1}{2N_c} \sum_{j=1}^L \mathcal{Z}^{\mathbf{N}}_j\,, \nonumber \\
    H^{\mathrm{mass}}_{\text{\Tthreep}} &= -m \sum_{j=1}^L \mathcal{Z}^{\mathbf{N}}_j\,.
\end{align}
There are also two new types of transitions allowed (see \Cref{fig:T3pnew}), based on \Cref{tab:hops}: (i) creating a quark-antiquark excitation directly left-adjacent or right-adjacent from an existing quark-antiquark excitation, with matrix element $\sqrt{N_c-1}$) and (ii) creating an excitation \textit{in between} two existing excitations, with matrix element $(N_c-1)N_c^{-1/2}$.
By putting these together with the PXP terms from truncation \Tone, we find the full off-diagonal contribution to the Hamiltonian
\begin{align}
    H^{\mathrm{int}}_{\text{\Tthreep}} = &\frac{\sqrt{N_c}}{2} \sum_{j=1}^L P^{\mathbf{1}}_{j-1} \mathcal{X}^{\mathbf{1}\mathbf{N}}_j P^{\mathbf{1}}_{j+1}\nonumber\\
    &+\frac{\sqrt{N_c-1}}{2}\sum_{j=1}^L \left(P^{\mathbf{1}}_{j-1} \mathcal{X}^{\mathbf{1}\mathbf{N}}_j P^{\mathbf{N}}_{j+1} + P^{\mathbf{N}}_{j-1} \mathcal{X}^{\mathbf{1}\mathbf{N}}_j P^{\mathbf{1}}_{j+1}\right)\nonumber\\& + \frac{N_c-1}{2\sqrt{N_c}}\sum_{j=1}^L P^{\mathbf{N}}_{j-1} \mathcal{X}^{\mathbf{1}\mathbf{N}}_j P^{\mathbf{N}}_{j+1}\,.
\end{align}

The full Hamiltonian,
\begin{equation}
    H_{\text{\Tthreep}} = H^{\mathrm{elec}}_{\text{\Tthreep}} + H^{\mathrm{mass}}_{\text{\Tthreep}} + H^{\mathrm{int}}_{\text{\Tthreep}}
\end{equation}
turns out to be an almost entirely non-interacting gas of spins in a tilted external field, with small PXP-type interactions that vanish as $N_c \to \infty$.
To see this interpretation, it is useful to rewrite the interaction term as
\begin{align}
    H^{\mathrm{int}}_{\text{\Tthreep}} = &\frac{\sqrt{N_c}}{2} \sum_{j=1}^L \mathcal{X}^{\mathbf{1}\mathbf{N}}_j\nonumber\\
     &+\frac 12\left(\sqrt{N_c-1} - \sqrt{N_c}\right)\sum_{j=1}^L P^{\mathbf{1}}_{j-1} \mathcal{X}^{\mathbf{1}\mathbf{N}}_j P^{\mathbf{N}}_{j+1}\nonumber\\
     &+\frac 12\left(\sqrt{N_c-1} - \sqrt{N_c}\right)\sum_{j=1}^L P^{\mathbf{N}}_{j-1} \mathcal{X}^{\mathbf{1}\mathbf{N}}_j P^{\mathbf{1}}_{j+1}\nonumber\\
     &+\frac 12\left(\frac{N_c-1}{\sqrt{N_c}} - \sqrt{N_c}\right)\sum_{j=1}^L P^{\mathbf{N}}_{j-1} \mathcal{X}^{\mathbf{1}\mathbf{N}}_j P^{\mathbf{N}}_{j+1}\,,
\end{align}
where we used the fact that $P^{\mathbf{1}}_j + P^{\mathbf{N}}_j$ is the identity for every qubit index $j$.
The new coefficients appearing on the remaining PXP terms scale as
\begin{align}
    \frac 12\left(\sqrt{N_c - 1} - \sqrt{N_c}\right) &\sim - \frac 1{4\sqrt{N_c}}\,, \nonumber \\
    \frac 12\left(\frac{N_c-1}{\sqrt{N_c}} - \sqrt{N_c}\right) &\sim - \frac 1{2\sqrt{N_c}}\,,
\end{align}
which implies that keeping hopping interactions that appear only with scaling $\sim \sqrt{N_c}$ in the Kogut-Susskind Hamiltonian can actually lead to higher order interactions in the corresponding spin model.

Switching gears to the full truncation \Tthree, we must include all remaining terms from \Cref{tab:hops} that are consistent with our energy cutoff.
This means we need to add back the string-joining and string-breaking interactions that were present in truncation \Ttwo\ (see \Cref{fig:T2new}).
The Hilbert space for truncation \Tthree, therefore, consists of lattice states that represent collections of strings, \textit{without} requiring gaps between strings.

The allowed link irreps are
\begin{equation}
    S_{\text{\Tthree}} = \left\{\mathbf{1},\mathbf{N},\overline{\mathbf{N}}\right\},
\end{equation}
which implies that the electric and mass terms can be borrowed from truncation \Ttwo,
\begin{align}
    H^{\mathrm{elec}}_{\text{\Tthree}} &= -\frac{g^2}{4} \frac{N_c^2-1}{2N_c} \sum_{j=1}^L \left(\mathcal{Z}^{\mathbf{N}}_j + \mathcal{Z}^{\overline{\mathbf{N}}}_j\right)\nonumber\\
    H^{\mathrm{mass}}_{\text{\Tthree}} &= -m\sum_{j=1}^L \left(\mathcal{Z}^{\mathbf{N}}_j - \mathcal{Z}^{\overline{\mathbf{N}}}_j\right)\,.
\end{align}

The off-diagonal part of the Hamiltonian is obtained by simply taking all interactions from $H^{\mathrm{int}}_{\text{\Tthreep}}$ above, and adding the sub-leading interactions from truncation \Ttwo.
We have to be careful when simplifying the result, however, because $P^{\mathbf{1}} + P^{\mathbf{N}}$ evaluates to $I - P^{\overline{\mathbf{N}}}$ at truncation \Tthree, instead of simply $I$ as in truncation \Tthreep.
The simplest way to make this adjustment is by writing
\begin{widetext}
\begin{align}
    H^{\mathrm{int}}_{\text{\Tthree}} = \frac{\sqrt{N_c}}{2} &\sum_{j=1}^L \left(P^{\mathbf{1}}_{j-1} + P^{\mathbf{N}}_{j-1}\right) \mathcal{X}^{\mathbf{1}\mathbf{N}}_j \left(P^{\mathbf{1}}_{j+1} + P^{\mathbf{N}}_{j+1}\right) + \frac{1}{2\sqrt{N_c}} \sum_{j=1}^L P^{\mathbf{N}}_{j-1} \mathcal{X}^{\mathbf{1}\overline{\mathbf{N}}}_j P^{\mathbf{N}}_{j+1}\\
     &+\frac 12\left(\sqrt{N_c-1} - \sqrt{N_c}\right)\sum_{j=1}^L \left(P^{\mathbf{1}}_{j-1} \mathcal{X}^{\mathbf{1}\mathbf{N}}_j P^{\mathbf{N}}_{j+1} + P^{\mathbf{N}}_{j-1} \mathcal{X}^{\mathbf{1}\mathbf{N}}_j P^{\mathbf{1}}_{j+1}\right)
     +\frac 12\left(\frac{N_c-1}{\sqrt{N_c}} - \sqrt{N_c}\right)\sum_{j=1}^L P^{\mathbf{N}}_{j-1} \mathcal{X}^{\mathbf{1}\mathbf{N}}_j P^{\mathbf{N}}_{j+1}\,.\nonumber
\end{align}
\end{widetext}

\subsubsection{Truncations \Tfour\ and \Tfive}

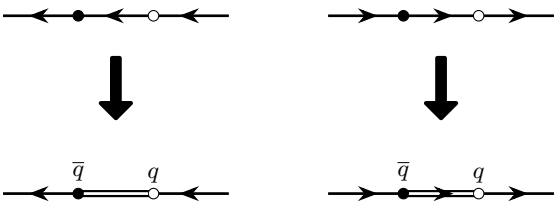
\begin{figure}[htp]
\centering

\begin{minipage}{0.5\linewidth}
\centering
\begin{tikzpicture}
\OrientedEdge{1}{0}{0}{0}{black}
\OrientedEdge{2}{0}{1}{0}{black}
\OrientedEdge{3}{0}{2}{0}{black}
\LatticeOneD{4}{0}[][0]
\end{tikzpicture}

\vspace{4mm}

\begin{tikzpicture}
\bigskip
\BubbleDownArrow
\end{tikzpicture}

\vspace{4mm}

\begin{tikzpicture}
\OrientedEdge{1}{0}{0}{0}{black}
\OrientedEdge{3}{0}{2}{0}{black}
\draw[line width=1pt] (1,0.03) -- (2,0.03);
\draw[line width=1pt] (1,-0.03) -- (2,-0.03);
\LatticeOneD{4}{0}[][0]
\PlaceEq{1}{0.3}{\overline{q}}
\PlaceEq{2}{0.25}{q}
\draw[draw=white, fill=white, opacity=1.0] (1.08,-0.01) rectangle ++(0.84,0.02);
\end{tikzpicture}
\end{minipage}%
\begin{minipage}{0.5\linewidth}
\centering
\begin{tikzpicture}
\OrientedEdge{0}{0}{1}{0}{black}
\OrientedEdge{1}{0}{2}{0}{black}
\OrientedEdge{2}{0}{3}{0}{black}
\LatticeOneD{4}{0}[][0]
\end{tikzpicture}

\vspace{4mm}

\begin{tikzpicture}
\bigskip
\BubbleDownArrow
\end{tikzpicture}

\vspace{4mm}

\begin{tikzpicture}
\OrientedEdge{0}{0}{1}{0}{black}
\OrientedEdge{2}{0}{3}{0}{black}
\draw[line width=1pt] (1,0.03) -- (2,0.03);
\draw[line width=1pt] (1,-0.03) -- (2,-0.03);
\LatticeOneD{4}{0}[][0]
\PlaceEq{1}{0.3}{\overline{q}}
\PlaceEq{2}{0.25}{q}
\draw[draw=white, fill=white, opacity=1.0] (1.08,-0.01) rectangle ++(0.84,0.02);
\ArrowHead{1.6}{0}{1.7}{0}{black}
\end{tikzpicture}
\end{minipage}%

\caption{New transitions available at both \Tfour\ and \Tfive. \textit{Left}: $\ket{\mathbf{N},\overline{\mathbf{N}},\mathbf{N}}\to\ket{\mathbf{N},\mathbf{Ad},\mathbf{N}}$. \textit{Right}: $\ket{\overline{\mathbf{N}},\mathbf{N},\overline{\mathbf{N}}}\to\ket{\overline{\mathbf{N}},\mathbf{A^2},\overline{\mathbf{N}}}$ or $\ket{\overline{\mathbf{N}},\mathbf{N},\overline{\mathbf{N}}}\to\ket{\overline{\mathbf{N}},\mathbf{S^2},\overline{\mathbf{N}}}$}.\label{fig:T4new}
\end{figure}

Truncations \Tfour\ and \Tfive\ are very similar to truncation \Ttwo, with the additional ability to carry up to two electric field lines per link.
For instance, starting from the free vacuum, we can use at most one hop per link and at most two fermion operators per site to build a meson-like string.
On top of this string, we can then apply a \textit{raising} operation, which effectively creates a new quark-antiquark pair \textit{overlaid} on the existing string.
This transition either produces the adjoint irrep $\mathbf{Ad}$ on the link, or a superposition of $\mathbf{A^2}$ and $\mathbf{S^2}$, depending on the orientation of the new quark-antiquark pair relative to the orientation of the original string (see \Cref{fig:T4new}).

\begin{figure}[!th]
\centering

\begin{minipage}{0.5\linewidth}
\begin{tikzpicture}
\draw[line width=1pt] (0,0.03) -- (1,0.03);
\draw[line width=1pt] (0,-0.03) -- (1,-0.03);
\draw[line width=1pt] (2,0.03) -- (3,0.03);
\draw[line width=1pt] (2,-0.03) -- (3,-0.03);
\OrientedEdge{1}{0}{2}{0}{black}
\LatticeOneD{4}{0}[][0]
\PlaceEq{1}{0.3}{\overline{q}}
\PlaceEq{2}{0.25}{q}
\draw[draw=white, fill=white, opacity=1.0] (-0.08,-0.01) rectangle ++(1.00,0.02);
\draw[draw=white, fill=white, opacity=1.0] (2.08,-0.01) rectangle ++(1.00,0.02);
\end{tikzpicture}

\vspace{4mm}

\begin{tikzpicture}
\bigskip
\BubbleDownArrow
\end{tikzpicture}

\vspace{4mm}

\begin{tikzpicture}
\draw[line width=1pt] (0,0.03) -- (1,0.03);
\draw[line width=1pt] (0,-0.03) -- (1,-0.03);
\draw[line width=1pt] (1,0.03) -- (2,0.03);
\draw[line width=1pt] (1,-0.03) -- (2,-0.03);
\draw[line width=1pt] (2,0.03) -- (3,0.03);
\draw[line width=1pt] (2,-0.03) -- (3,-0.03);
\LatticeOneD{4}{0}[][0]
\draw[draw=white, fill=white, opacity=1.0] (-0.08,-0.01) rectangle ++(1.0,0.02);
\draw[draw=white, fill=white, opacity=1.0] (1.08,-0.01) rectangle ++(0.84,0.02);
\draw[draw=white, fill=white, opacity=1.0] (2.08,-0.01) rectangle ++(1.00,0.02);
\end{tikzpicture}
\end{minipage}%
\begin{minipage}{0.5\linewidth}
\begin{tikzpicture}
\draw[line width=1pt] (0,0.03) -- (1,0.03);
\draw[line width=1pt] (0,-0.03) -- (1,-0.03);
\draw[line width=1pt] (2,0.03) -- (3,0.03);
\draw[line width=1pt] (2,-0.03) -- (3,-0.03);
\OrientedEdge{2}{0}{1}{0}{black}
\LatticeOneD{4}{0}[][0]
\PlaceEq{1}{0.3}{\overline{q}}
\PlaceEq{2}{0.25}{q}
\draw[draw=white, fill=white, opacity=1.0] (-0.08,-0.01) rectangle ++(1.0,0.02);
\draw[draw=white, fill=white, opacity=1.0] (2.08,-0.01) rectangle ++(1.0,0.02);
\ArrowHead{0.4}{0}{0.3}{0}{black}
\ArrowHead{2.4}{0}{2.3}{0}{black}
\end{tikzpicture}

\vspace{4mm}

\begin{tikzpicture}
\bigskip
\BubbleDownArrow
\end{tikzpicture}

\vspace{4mm}

\begin{tikzpicture}
\draw[line width=1pt] (0,0.03) -- (1,0.03);
\draw[line width=1pt] (0,-0.03) -- (1,-0.03);
\draw[line width=1pt] (1,0.03) -- (2,0.03);
\draw[line width=1pt] (1,-0.03) -- (2,-0.03);
\draw[line width=1pt] (2,0.03) -- (3,0.03);
\draw[line width=1pt] (2,-0.03) -- (3,-0.03);
\LatticeOneD{4}{0}[][0]
\draw[draw=white, fill=white, opacity=1.0] (-0.08,-0.01) rectangle ++(1.0,0.02);
\draw[draw=white, fill=white, opacity=1.0] (1.08,-0.01) rectangle ++(0.84,0.02);
\draw[draw=white, fill=white, opacity=1.0] (2.08,-0.01) rectangle ++(1.0,0.02);
\ArrowHead{0.4}{0}{0.3}{0}{black}
\ArrowHead{1.4}{0}{1.3}{0}{black}
\ArrowHead{2.4}{0}{2.3}{0}{black}
\end{tikzpicture}
\end{minipage}%

\caption{New transitions available only at \Tfive\ (and higher truncations). \textit{Left}: $\ket{\mathbf{Ad},\mathbf{N},\mathbf{Ad}}\to\ket{\mathbf{Ad},\mathbf{Ad},\mathbf{Ad}}$. \textit{Right}: $\ket{\mathbf{A^2},\overline{\mathbf{N}},\mathbf{A^2}}\to\ket{\mathbf{A^2},\overline{\mathbf{A^2}},\mathbf{A^2}}$ or $\ket{\mathbf{S^2},\overline{\mathbf{N}},\mathbf{S^2}}\to\ket{\mathbf{S^2},\overline{\mathbf{S^2}},\mathbf{S^2}}$.}\label{fig:T5new}
\end{figure}
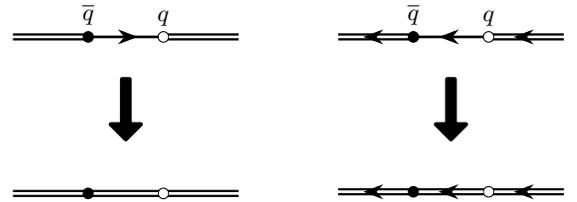

All such transitions require a second hop per link and a third fermion operator per lattice site, and are therefore allowed by both \Tfour\ and \Tfive.
The process of joining together quark-antiquark pairs (or longer strings) with compatible orientations \textit{on top} of a pre-existing string is allowed only by \Tfive, since it requires a fourth fermion operation on the lattice sites involved (see \Cref{fig:T5new}).

Truncations \Tfour\ and \Tfive\ have the irrep requirements
\begin{align}
    S_{\text{\Tfour}} &= \left\{\mathbf{1},\mathbf{N},\overline{\mathbf{N}},\mathbf{Ad},\mathbf{A^2},\mathbf{S^2}\right\},\nonumber\\
    S_{\text{\Tfive}} &= \left\{\mathbf{1},\mathbf{N},\overline{\mathbf{N}},\mathbf{Ad},\mathbf{A^2},\mathbf{S^2},\overline{\mathbf{A^2}},\overline{\mathbf{S^2}}\right\},
\end{align}
respectively.
This requires qudits with $d=6$ dimensions for \Tfour, and qudits with $d=8$ dimensions for \Tfive. On hardware, one might prefer to use a qubit or qutrit encoding for these high-dimensional qudits.

At truncation \Tfour, the electric term needs to provide energies proportional to the Casimir for each irrep.
By using projectors, we can force this conditioning as follows:
\begin{widetext}
\begin{align}
    H^{\mathrm{elec}}_{\text{\Tfour}} &= \frac{g^2}{2}\left(C_2(\mathbf{N})\sum_{j=1}^L \left(P^{\mathbf{N}}_j + P^{\overline{\mathbf{N}}}_j\right) + C_2(\mathbf{Ad}) \sum_{j=1}^L P^{\mathbf{Ad}}_j + C_2(\mathbf{A^2}) \sum_{j=1}^L P^{\mathbf{A^2}}_j + C_2(\mathbf{S^2}) \sum_{j=1}^L P^{\mathbf{S^2}}_j\right)\nonumber\\
    & \cong -\frac{g^2}{4} \left(\frac{N_c^2-1}{2N_c}\sum_{j=1}^L \left(\mathcal{Z}^{\mathbf{N}}_j + \mathcal{Z}^{\overline{\mathbf{N}}}_j\right) + N_c \sum_{j=1}^L \mathcal{Z}^{\mathbf{Ad}}_j + \frac{N_c^2-N_c-2}{N_c}\sum_{j=1}^L \mathcal{Z}^{\mathbf{A^2}}_j + \frac{N_c^2+N_c-2}{N_c}\sum_{j=1}^L \mathcal{Z}^{\mathbf{S^2}}_j\right),
\end{align}
\end{widetext}
where we dropped constant terms in going from the first to the second line.

To write down the mass term for truncation \Tfour, we can directly use \eqref{eq:flow}.
By expressing this condition as a mass condition with Pauli-Z's, we obtain
\begin{equation}
    H^{\mathrm{mass}}_{\text{\Tfour}} = -m \sum_{j=1}^L \left(\mathcal{Z}^{\mathbf{N}}_j - \mathcal{Z}^{\overline{\mathbf{N}}}_j + 2\mathcal{Z}^{\mathbf{A^2}}_j + 2\mathcal{Z}^{\mathbf{S^2}}_j\right).
\end{equation}

The types of interactions allowed at truncation \Tfour\ can be summarized as: (i) creating an isolated meson-like excitation in the free vacuum; (ii) joining two excitations into a longer meson string; and (iii) creating an isolated meson-like excitation on top of a string.
The interactions (i) and (ii) were treated in truncation \Ttwo.
Interaction (iii) is applicable for raising $\ket{\overline{\mathbf{N}}}$ into $\ket{\mathbf{Ad}}$ in between two fundamentals, raising $\ket{\mathbf{N}}$ into $\ket{\mathbf{A^2}}$ in between two anti-fundamentals, and raising $\ket{\mathbf{N}}$ into $\ket{\mathbf{S^2}}$ in between two anti-fundamentals.
Using the results from \Cref{tab:hops}, this leads to
\begin{widetext}
\begin{align}
    H^{\mathrm{int}}_{\text{\Tfour}} = &\frac{\sqrt{N_c}}{2}\sum_{j=1}^L P^{\mathbf{1}}_{j-1} \mathcal{X}^{\mathbf{1}\mathbf{N}}_j P^{\mathbf{1}}_{j+1}
    + \frac{1}{2\sqrt{N_c}}\sum_{j=1}^L P^{\mathbf{N}}_{j-1} \mathcal{X}^{\mathbf{1}\overline{\mathbf{N}}}_j P^{\mathbf{N}}_{j+1}
    + \frac12 \sqrt{\frac{N_c^2-1}{N_c}}\sum_{j=1}^L P^{\mathbf{N}}_{j-1} \mathcal{X}^{\overline{\mathbf{N}}\mathbf{Ad}}_j P^{\mathbf{N}}_{j+1}
    \nonumber\\
    &\, 
    + \frac12 \sqrt{\frac{N_c-1}2}\sum_{j=1}^L P^{\overline{\mathbf{N}}}_{j-1} \mathcal{X}^{\mathbf{N}\mathbf{A^2}}_j P^{\overline{\mathbf{N}}}_{j+1}
    + \frac12 \sqrt{\frac{N_c+1}2}\sum_{j=1}^L P^{\overline{\mathbf{N}}}_{j-1} \mathcal{X}^{\mathbf{N}\mathbf{S^2}}_j P^{\overline{\mathbf{N}}}_{j+1}\,.
\end{align}
\end{widetext}
Note that this Hamiltonian contains terms at both $\mathcal O(\sqrt{N_c})$ as well as $\mathcal O(1/\sqrt{N_c})$, but if we were to restrict to just the leading order piece, the entire truncation collapses down to truncation \Tone. 

Let us now repeat the same process to write down the Hamiltonian for truncation \Tfive.
The electric term at \Tfive is almost the same as we had for \Tfour, but it now contains the conjugates of the higher irreps, so we can write
\begin{align}
    H^{\mathrm{elec}}_{\text{\Tfive}} = &-\frac{g^2}{4} \bigg(\frac{N_c^2-1}{2N_c}\sum_{j=1}^L \left(\mathcal{Z}^{\mathbf{N}}_j + \mathcal{Z}^{\overline{\mathbf{N}}}_j\right)+ N_c \sum_{j=1}^L \mathcal{Z}^{\mathbf{Ad}}_j \nonumber\\
    &+\frac{N_c^2-N_c-2}{N_c}\sum_{j=1}^L \left(\mathcal{Z}^{\mathbf{A^2}}_j + \mathcal{Z}^{\overline{\mathbf{A^2}}}_j\right)
    \nonumber\\
    &
    + \frac{N_c^2+N_c-2}{N_c}\sum_{j=1}^L \left(\mathcal{Z}^{\mathbf{S^2}}_j + \mathcal{Z}^{\overline{\mathbf{S^2}}}_j\right)\bigg)\,.
\end{align}
The mass term is similarly modified by using \eqref{eq:flow}:
\begin{align}
    H^{\mathrm{mass}}_{\text{\Tfive}} =& -m \sum_{j=1}^L\left(\mathcal{Z}^{\mathbf{N}}_j - \mathcal{Z}^{\overline{\mathbf{N}}}_j\right)\nonumber\\
    &- 2m \sum_{j=1}^L\left(\mathcal{Z}^{\mathbf{A^2}}_j - \mathcal{Z}^{\overline{\mathbf{A^2}}}_j\right)\nonumber\\ 
    &-2m\sum_{j=1}^L\left(\mathcal{Z}^{\mathbf{S^2}}_j - \mathcal{Z}^{\overline{\mathbf{S^2}}}_j\right).
\end{align}

Finally, by using \Cref{tab:hops}, we can write down the off-diagonal piece of the Hamiltonian at this truncation.
This requires reusing the same interactions from truncation \Tfour, and adding in the three different types of string-joining / string-breaking operations.
This yields
\begin{widetext}
\begin{align}
    H^{\mathrm{int}}_{\text{\Tfive}} = &\frac{\sqrt{N_c}}{2}\sum_{j=1}^L P^{\mathbf{1}}_{j-1} \mathcal{X}^{\mathbf{1}\mathbf{N}}_j P^{\mathbf{1}}_{j+1} + \frac{1}{2\sqrt{N_c}}\sum_{j=1}^L P^{\mathbf{N}}_{j-1} \mathcal{X}^{\mathbf{1}\overline{\mathbf{N}}}_j P^{\mathbf{N}}_{j+1}\nonumber\\
    & \, + \frac12 \sqrt{\frac{N_c^2-1}{N_c}}\sum_{j=1}^L P^{\mathbf{N}}_{j-1} \mathcal{X}^{\overline{\mathbf{N}}\mathbf{Ad}}_j P^{\mathbf{N}}_{j+1} + \frac12 \sqrt{\frac{N_c-1}2}\sum_{j=1}^L P^{\overline{\mathbf{N}}}_{j-1} \mathcal{X}^{\mathbf{N}\mathbf{A^2}}_j P^{\overline{\mathbf{N}}}_{j+1} + \frac12 \sqrt{\frac{N_c+1}2}\sum_{j=1}^L P^{\overline{\mathbf{N}}}_{j-1} \mathcal{X}^{\mathbf{N}\mathbf{S^2}}_j P^{\overline{\mathbf{N}}}_{j+1}\nonumber\\
    & \,+ \frac12 \sqrt{\frac{N_c}{N_c^2-1}}\sum_{j=1}^L P^{\mathbf{Ad}}_{j-1} \mathcal{X}^{\mathbf{N}\mathbf{Ad}}_j P^{\mathbf{Ad}}_{j+1} + \frac12 \sqrt{\frac2{N_c-1}}\sum_{j=1}^L P^{\mathbf{A^2}}_{j-1} \mathcal{X}^{\overline{\mathbf{N}}\overline{\mathbf{A^2}}}_j P^{\mathbf{A^2}}_{j+1} + \frac12 \sqrt{\frac2{N_c+1}}\sum_{j=1}^L P^{\mathbf{S^2}}_{j-1} \mathcal{X}^{\overline{\mathbf{N}}\overline{\mathbf{S^2}}}_j P^{\mathbf{S^2}}_{j+1}.
\end{align}
\end{widetext}

To conclude this section, we remark that for the case of $N_c = 3$ (i.e. QCD in the real world), the anti-symmetric representation $\mathbf{A^2}$ is actually equivalent to the anti-fundamental.
This means the truncated Hilbert space in our formalism here does not exactly match the truncated Hilbert space for QCD, if we define Krylov constraints in the analogous way (QCD is slightly simpler).
Nevertheless, in order to stay faithful to the large $N_c$ expansion of QCD, it is important to keep expanding the Hilbert space in the manner we have shown.

\subsection{$2+1$D and Higher Dimensions}\label{sec:2+1d}

We are now in a position to generalize our previous discussion to higher dimensions.
For simplicity, we will focus on truncations at the energy cutoff $(n_e, n_f) = (1, 1)$, and we will restrict attention to $2+1$D, but our constructions can be extended to higher truncations and to higher dimensions, using the same process as in \Cref{sec:1+1d}.

In $2+1$D, we require all four Krylov parameters $(n_p, n_h, n_\ell, n_\chi)$, and the truncations we consider will be denoted
\begin{itemize}
    \item \TTone: $(n_p, n_h, n_\ell, n_\chi, n_e, n_f) = (1, 1, 1, 1, 1, 1)$;
    \item \TTtwo: $(n_p, n_h, n_\ell, n_\chi, n_e, n_f) = (1, 1, 2, 2, 1, 1)$ at $\mathcal O(1/N_c)$.
\end{itemize}
In fact, there are several different orders in $1/N_c$ at which we could restrict truncation \TTtwo, but to keep things simple, we will only discuss the $\mathcal O(1/N_c)$ truncation.

\begin{figure*}[t]
    \centering
    \begin{minipage}[][5cm]{0.5\linewidth}
    \centering
    \begin{tikzpicture}
    \pgfmathsetmacro{\side}{1.4}
    
    \OrientedEdge{1*\side}{2*\side}{2*\side}{2*\side}{blue}
    \OrientedEdge{2*\side}{2*\side}{2*\side}{1*\side}{blue}
    \OrientedEdge{2*\side}{1*\side}{1*\side}{1*\side}{blue}
    \OrientedEdge{1*\side}{1*\side}{1*\side}{2*\side}{blue}
    \OrientedEdge{3*\side}{1*\side}{2*\side}{1*\side}{black!30!green}
    
    \EvenPlaquette{0*\side}{0*\side}{1.0*\side}{black}
    \OddPlaquette{1*\side}{0*\side}{1.0*\side}{black}
    \EvenPlaquette{2*\side}{0*\side}{1.0*\side}{black}
    \OddPlaquette{3*\side}{0*\side}{1.0*\side}{black}
    \OddPlaquette{0*\side}{1*\side}{1.0*\side}{black}
    \EvenPlaquette{1*\side}{1*\side}{1.0*\side}{black}
    \OddPlaquette{2*\side}{1*\side}{1.0*\side}{black}
    \EvenPlaquette{3*\side}{1*\side}{1.0*\side}{black}
    \EvenPlaquette{0*\side}{2*\side}{1.0*\side}{black}
    \OddPlaquette{1*\side}{2*\side}{1.0*\side}{black}
    \EvenPlaquette{2*\side}{2*\side}{1.0*\side}{black}
    \OddPlaquette{3*\side}{2*\side}{1.0*\side}{black}

    \draw[draw=black!30!green, line width=1pt] (3*\side-0.08, 1*\side) -- (2*\side+0.08, 1*\side);
    \draw[draw=blue, line width=1pt] (1*\side+0.08, 2*\side) -- (2*\side-0.08, 2*\side);
    \draw[draw=blue, line width=1pt] (2*\side, 2*\side-0.08) -- (2*\side, 1*\side+0.08);
    \draw[draw=blue, line width=1pt] (2*\side-0.08, 1*\side) -- (1*\side+0.08, 1*\side);
    \draw[draw=blue, line width=1pt] (1*\side, 1*\side+0.08) -- (1*\side, 2*\side-0.08);
    \end{tikzpicture}
    \end{minipage}%
    \begin{minipage}[][5cm]{0.5\linewidth}
    \begin{tikzpicture}
    \SmallCircle{0}{0}{1}{0.075}
    \pgfmathsetmacro{\side}{1.4}
    \draw (0,0) -- (\side,0);
    \draw (0,0) -- (-\side,0);
    \draw (0,0) -- (0,\side);
    \draw (0,0) -- (0,-\side);
    \PlaceEq{0}{1*\side+0.3}{\mathbf{R_1}}
    \PlaceEq{1*\side+0.3}{0}{\mathbf{R_2}}
    \PlaceEq{0}{-1*\side-0.3}{\mathbf{R_3}}
    \PlaceEq{-1*\side-0.3}{0}{\mathbf{R_4}}
    \PlaceEq{0.3}{0.3}{\mathbf{R_5}}
    \end{tikzpicture}
    \end{minipage}
    
    \vspace{1cm}
    
    \begin{minipage}{0.5\linewidth}
    \begin{tikzpicture}
    \BubbleDownArrow
    \end{tikzpicture}
    \end{minipage}%
    \begin{minipage}{0.5\linewidth}
    \begin{tikzpicture}
    \BubbleDownArrow
    \end{tikzpicture}
    \end{minipage}%
    
    \begin{minipage}{0.5\linewidth}
    \begin{tikzpicture}
    \pgfmathsetmacro{\hexside}{0.8}

    \OrientedEdge{3*\hexside}{-3*\hexside}{4*\hexside}{-3*\hexside}{blue}
    \OrientedEdge{4*\hexside}{-3*\hexside}{4.5*\hexside}{-3.5*\hexside}{blue}
    \OrientedEdge{4.5*\hexside}{-3.5*\hexside}{5*\hexside}{-4*\hexside}{blue}
    \OrientedEdge{5*\hexside}{-4*\hexside}{5*\hexside}{-5*\hexside}{blue}
    \OrientedEdge{5*\hexside}{-5*\hexside}{4*\hexside}{-5*\hexside}{blue}
    \OrientedEdge{4*\hexside}{-5*\hexside}{3.5*\hexside}{-4.5*\hexside}{blue}
    \OrientedEdge{3.5*\hexside}{-4.5*\hexside}{3*\hexside}{-4*\hexside}{blue}
    \OrientedEdge{3*\hexside}{-4*\hexside}{3*\hexside}{-3*\hexside}{blue}
    
    \OrientedEdge{7.5*\hexside}{-6.5*\hexside}{7*\hexside}{-6*\hexside}{black!30!green}    
    \OrientedEdge{7*\hexside}{-6*\hexside}{6*\hexside}{-6*\hexside}{black!30!green}    
    \OrientedEdge{6*\hexside}{-6*\hexside}{5.5*\hexside}{-5.5*\hexside}{black!30!green}
    
    \EvenHexPlaquette{0*\hexside}{0*\hexside}{\hexside}{black}
    \OddHexPlaquette{2*\hexside}{-1*\hexside}{\hexside}{black}
    \EvenHexPlaquette{4*\hexside}{-2*\hexside}{\hexside}{black}
    \OddHexPlaquette{6*\hexside}{-3*\hexside}{\hexside}{black}
    \OddHexPlaquette{1*\hexside}{-2*\hexside}{\hexside}{black}
    
    \OddHexPlaquette{5*\hexside}{-4*\hexside}{\hexside}{black}
    \EvenHexPlaquette{7*\hexside}{-5*\hexside}{\hexside}{black}
    \EvenHexPlaquette{2*\hexside}{-4*\hexside}{\hexside}{black}
    \OddHexPlaquette{4*\hexside}{-5*\hexside}{\hexside}{black}
    \EvenHexPlaquette{6*\hexside}{-6*\hexside}{\hexside}{black}
    \OddHexPlaquette{8*\hexside}{-7*\hexside}{\hexside}{black}

    \EvenHexPlaquette[draw=blue, line width=1pt]{3*\hexside}{-3*\hexside}{\hexside}{black}
    \draw[draw=black!30!green, line width=1pt] (7.5*\hexside-0.057, -6.5*\hexside+0.057) -- (7*\hexside, -6*\hexside);
    \draw[draw=black!30!green, line width=1pt] (7*\hexside, -6*\hexside) -- (6*\hexside, -6*\hexside);
    \draw[draw=black!30!green, line width=1pt] (6*\hexside, -6*\hexside) -- (5.5*\hexside+0.057, -5.5*\hexside-0.057);
    \end{tikzpicture}
    \end{minipage}%
    \begin{minipage}{0.5\linewidth}
    \begin{tikzpicture}
    \SmallCircle{0}{0}{1}{0.075}
    \pgfmathsetmacro{\side}{1.4}
    \pgfmathsetmacro{\norm}{0.707}
    \draw (0,0) -- (-\norm*\side, \norm*\side);
    \draw (0,0) -- (\norm*\side, -\norm*\side);
    \draw (-\norm*\side, \norm*\side) -- (-\norm*\side, \side+\norm*\side);
    \draw (-\norm*\side, \norm*\side) -- (-\side-\norm*\side, \norm*\side);
    \draw (\norm*\side, -\norm*\side) -- (\side+\norm*\side, -\norm*\side);
    \draw (\norm*\side, -\norm*\side) -- (\norm*\side, -\side-\norm*\side);
    \PlaceEq{-\norm*\side}{(1+\norm)*\side+0.3}{\mathbf{R_1}}
    \PlaceEq{(1+\norm)*\side+0.3}{-\norm*\side}{\mathbf{R_2}}
    \PlaceEq{\norm*\side}{-(1+\norm)*\side-0.3}{\mathbf{R_3}}
    \PlaceEq{-(1+\norm)*\side-0.3}{\norm*\side}{\mathbf{R_4}}
    \PlaceEq{0.3}{0.3}{\mathbf{R_5}}
    \PlaceEq{-0.3*\norm-0.5*\norm*\side}{-0.3*\norm+0.5*\norm*\side}{\mathbf{V_1}}
    \PlaceEq{-0.3*\norm+0.5*\norm*\side}{-0.3*\norm-0.5*\norm*\side}{\mathbf{V_2}}
    \end{tikzpicture}
    \end{minipage}
    \caption{\textit{Left}: Point-splitting a square lattice by adding virtual (diagonal) links. Blue Wilson loops and green Wilson lines illustrate equivalent operators on the two lattices. In particular, four-body plaquettes and three-body hopping operators on a square lattice become eight-body plaquettes and five-body hopping operators on the hexagonal lattice. \textit{Right}: Point-splitting decomposes a five-point vertex singlet on the square lattice into three trivalent vertex singlets. This keeps the original five physical irreps, and adds two virtual irreps.}\label{fig:split}
\end{figure*}
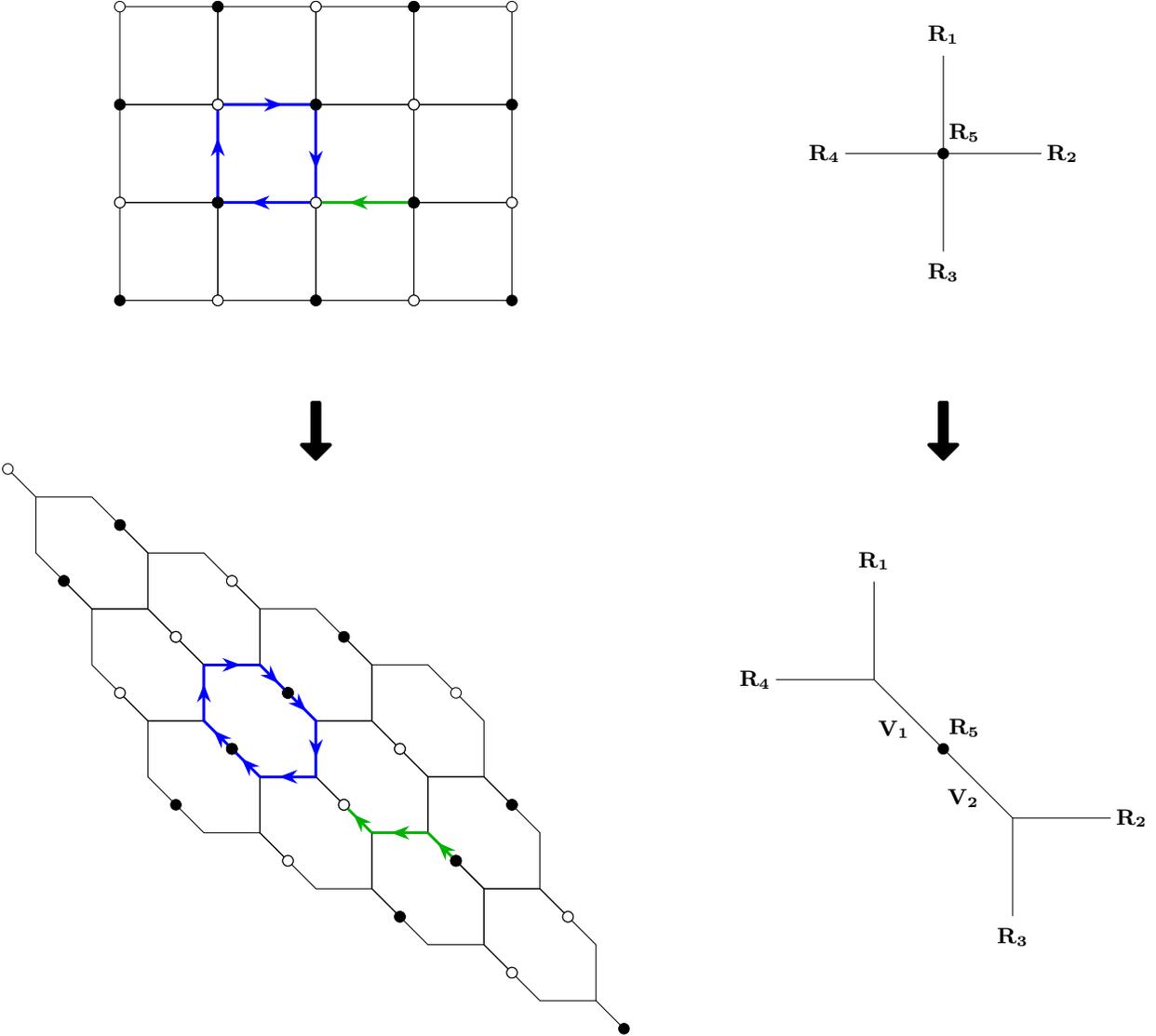

Traditional lattice QCD is often performed on a square lattice, such as the one shown in the top left of~\Cref{fig:split}.
To achieve a more practical Hamiltonian for simulation, the $2+1$D theory that we will truncate in this paper lives instead on the \textit{point-split} (hexagonal) lattice shown in the bottom left of~\Cref{fig:split}.
Before we can write down the truncations we will simulate, we need to discuss the relationship between these two lattices and the theories defined on each of them.

As shown in~\Cref{fig:split}, the point-split lattice has \textit{physical} degrees of freedom (horizontal and vertical links, and fermion fields), as well as \textit{virtual} degrees of freedom (diagonal links).
Despite the addition of these extra degrees of freedom, there exists a Hamiltonian $H_{\KS}^{\ps}$ defined on the point-split lattice that leads to \textit{numerical equivalence} with the Kogut-Susskind Hamiltonian $H_{\KS}$ on the corresponding square lattice, even when the lattice is finite.
At the Hilbert space level, this is enabled by the fact that an orthonormal basis of five-point vertex singlets on the square lattice can be decomposed uniquely into an orthonormal basis on three trivalent vertex singlets in the point-split lattice, as illustrated on the right side in~\Cref{fig:split}. 

\begin{figure}[htp]
    \centering

    \begin{tikzpicture}
    \pgfmathsetmacro{\side}{0.8}
    \pgfmathsetmacro{\norm}{0.707}
    \EvenHexPlaquetteReduced{2*\side}{-\side}{\side}{black}
    \EvenHexPlaquetteReduced{\side}{-2*\side}{\side}{black}
    \OddHexPlaquetteReduced{3*\side}{-3*\side}{\side}{black}
    \OddHexPlaquetteReduced[draw=blue, line width=1pt]{0}{0}{\side}{black}

    \draw[draw=black!30!green, line width=1pt] (3.5*\side+0.057,-1.5*\side-0.057) -- (4*\side,-2*\side);
    \draw[draw=black!30!green, line width=1pt] (4*\side,-2*\side) -- (4*\side,-3*\side);
    \draw[draw=black!30!green, line width=1pt] (4*\side,-3*\side) -- (4.5*\side-0.057,-3.5*\side+0.057);

    \PlaceEq{0.5*\side}{0.3}{p_1}
    \PlaceEq{1.25*\side+0.3*\norm}{-0.25*\side+0.3*\norm}{p_2}
    \PlaceEq{1.75*\side+0.3*\norm}{-0.75*\side+0.3*\norm}{p_3}
    \PlaceEq{2*\side+0.3}{-1.5*\side}{p_4}
    \PlaceEq{1.5*\side}{-2*\side-0.3}{p_5}
    \PlaceEq{0.75*\side-0.3*\norm}{-1.75*\side-0.3*\norm}{p_6}
    \PlaceEq{0.25*\side-0.3*\norm}{-1.25*\side-0.3*\norm}{p_7}
    \PlaceEq{-0.3}{-0.5*\side}{p_8}

    \PlaceEq{4*\side+0.14}{-2.5*\side}{\ell}
    \PlaceEq{3.75*\side+0.3*\norm}{-1.75*\side+0.3*\norm}{\ell_{\bullet}}
    \PlaceEq{3.25*\side+0.3*\norm}{-1.25*\side+0.3*\norm}{\ell_{\bullet}^*}
    \PlaceEq{4.25*\side-0.3*\norm}{-3.25*\side-0.3*\norm}{\ell_{\circ}}
    \PlaceEq{4.75*\side-0.3*\norm}{-3.75*\side-0.3*\norm}{\ell_{\circ}^*}
    \end{tikzpicture}
    
    \caption{The data associated with plaquette and hopping operators in the Kogut-Susskind Hamiltonian for the point-split lattice. For any hexagonal plaquette $p$, the links are labeled clockwise as $(p_1, \dots, p_8)$, beginning at the top-most link on the facet. For any cardinal link $\ell$, the links on the Wilson line for the hopping operator are $(\ell_{\bullet}, \ell, \ell_{\circ})$, in order of appearance on the path from $\ell^-$ to $\ell^+$. The links $\ell_{\bullet}^*$ and $\ell_{\circ}^*$ (adjacent to the Wilson line) that will be needed to classify hopping transitions are also labeled.}\label{fig:plaqhop}
\end{figure}
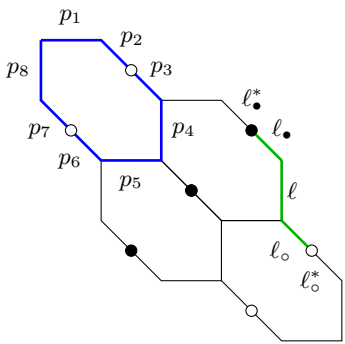

At the level of dynamics, numerical equivalence is achieved by choosing the Hamiltonian
\begin{align}\label{eq:KSps}
    H^{\ps}_{\KS} = &\frac{g^2}{2} \sum_{\ell \in \lk_c^{\mathrm{hex}}} E^2(\ell) + m\sum_{\mathbf{n} \in \vx^{\mathrm{hex}}} (-1)^{\mathbf{n}} \chi^\dagger(\mathbf{n})_a \chi(\mathbf{n})_a 
    \nonumber\\
    &\, - \frac1{2g^2} \sum_{p\in\pq^{\mathrm{hex}}} \left[\square(p)^\dagger + \square(p)\right]\nonumber\\
    &\, + \frac 12\sum_{\ell \in \lk_c^{\mathrm{hex}}} \eta_{\ell} \left[\chi^\dagger(\ell^-)_a U(\ell^-,\ell^+)_{ab} \chi(\ell^+)_b + \mathrm{h.c.}\right]\,,
\end{align}
where the electric and fermion operators are defined analogously to the square lattice, and we introduce the following additional notation for the hexagonal lattice:
\begin{itemize}
    \item $\vx^{\mathrm{hex}}$, $\lk^{\mathrm{hex}}$, and $\pq^{\mathrm{hex}}$ respectively denote the lattice sites, links, and plaquettes on the hexagonal lattice. Note that $\vx^{\mathrm{hex}}$ and $\pq^{\mathrm{hex}}$ are in one-to-one correspondence with the lattice sites $\vx$ and plaquettes $\pq$ on the square lattice, but we keep the notation separate to avoid confusion.
    \item $\lk_c^{\mathrm{hex}}$ and $\lk_d^{\mathrm{hex}}$ respectively denote the links oriented in the \textit{cardinal} (i.e. horizontal and vertical) and \textit{diagonal} directions. Note that $\lk_c^{\mathrm{hex}}$ contains the same set of links as $\lk$ for the square lattice, but we again keep the notation separate to avoid confusion.
    \item At $p \in \pq^{\mathrm{hex}}$, the plaquette operator $\square(p)$ denotes the traced holonomy around the loop passing through all \textit{eight} links surrounding plaquette $p$, oriented clockwise. We will sometimes write $p \equiv (p_1, p_2, p_3, p_4, p_5, p_6, p_7, p_8)$ to make the eight links explicit, as shown in \Cref{fig:plaqhop}.
    \item Each $\ell \in \lk_c^{\mathrm{hex}}$ is uniquely associated to a pair of \textit{nearest} lattice sites. On the contracted square lattice, these would be the odd endpoint $\ell^-$ and even endpoint $\ell^+$. We borrow this same notation for the hexagonal lattice, and write $U(\ell^-, \ell^+)$ for the Wilson line operator from $\ell^-$ to $\ell^+$.
    We will sometimes reference the unique set of five links $(\ell^*_{\bullet}, \ell_{\bullet}, \ell, \ell_{\circ}, \ell^*_{\circ})$ that surround link $\ell$, as illustrated in \Cref{fig:plaqhop}.\footnote{Strictly speaking, this notation is only guaranteed to exist at every $\ell \in \lk_c^{\mathrm{hex}}$ if the lattice has periodic boundaries.}
\end{itemize}

Note that \eqref{eq:KSps} does not include any electric energy contribution from the diagonally-oriented links.
That is precisely the sense in which these links are \textit{virtual} after the point-splitting process.

In this paper, we consider a slight modification to the lattice gauge theory defined by \eqref{eq:KSps}.
Specifically, we make the simple adjustment of replacing the sum of electric energies over \textit{cardinal links} with a sum over \textit{all links}.
To be explicit, we write
\begin{equation}\label{eq:KShex}
    H^{\mathrm{hex}}_{\KS} \equiv H^{\ps}_{\KS} + \frac{g^2}{2} \sum_{\ell \in \lk_d^{\mathrm{hex}}} E^2(\ell)\,.
\end{equation}
This change promotes all virtual links to physical links, which breaks \textit{exact} equivalence to the finite square lattice theory.
Nevertheless, the continuum limit is still preserved---for instance, this change introduces no new fermion doublers relative to those already inherited from the square lattice.

The point is now that all truncations we consider are restricted to the irreps
\begin{equation}
S_{\text{\TTone}} = S_{\text{\TTtwo'}} = S_{\text{\TTtwo}} = \left\{\mathbf{1},\mathbf{N},\overline{\mathbf{N}}\right\},
\end{equation}
which implies, by Gauss's law, that every trivalent vertex on the hexagonal lattice must have a singlet on at least one of its legs.
This has two benefits: first, it allows us to provide Hamiltonians that require only qutrits on links (as opposed to higher dimensional qudits); and second, it simplifies the process of computing matrix elements for the interaction terms.

Additionally, in \Cref{sec:simuls}, we will consider a hexagonal plaquette chain with open boundary conditions, which is not \textit{exactly} equivalent to any point-split square lattice.
This means that the notion of ``cardinal links" $\lk_c^{\mathrm{hex}}$ for the lattice we will simulate must be restricted to \textit{only} those links that reside at the center of a path between two nearest-neighbor lattice sites of opposite parity.
This slight redefinition ensures that the entirety of our qutrit Hamiltonian construction works both for periodic boundaries on a general hexagonal lattice, as well as the hexagonal plaquette chain which we will simulate.

\subsubsection{Interaction Matrix Elements}

To clarify the coefficients used to write down our truncated Hamiltonians, we sketch here two methods to compute interaction matrix elements in $2+1$D and higher dimensions.
The first method compares two different local point-splittings of the square lattice to form a relationship between the hopping matrix elements computed in \Cref{tab:hops} and matrix elements between globally point-split lattice states (the plaquette case has been considered in~\cite{Burbano:2024uvn}).
This approach directly generalizes to all truncations and all dimensions, but requires a (somewhat tedious) computation to relate the two point-split bases by linear transformation.
The second method we will provide is a faster approach that works specifically for truncations \TTone, \TTtwo, and \TTtwo' (but generalizes to any number of dimensions), which computes matrix elements for any gauge-invariant operator built from a single, non-intersecting Wilson line.
The general formula for interaction matrix elements we will ultimately provide in this section will be most easily derivable from the second method.

\begin{figure}[htp]
    \centering
    \begin{minipage}{\linewidth}
    \begin{tikzpicture}
    \SmallCircle{0}{0}{1}{0.075}
    \pgfmathsetmacro{\side}{0.7}
    \pgfmathsetmacro{\norm}{0.707}
    \draw (0,0) -- (2*\side,0);
    \draw (0,0) -- (-\side,0);
    \draw (0,0) -- (0,\side);
    \draw (0,0) -- (0,-\side);
    \draw (2*\side,0) -- (2*\side,\side);
    \draw (2*\side,0) -- (2*\side,-\side);
    \draw (2*\side,0) -- (3*\side,0);
    \SmallCircle{2*\side}{0}{0}{0.075}
    \PlaceEq{-0.3*\norm}{-0.3*\norm}{x}
    \PlaceEq{\side}{-0.3*\norm}{\ell}
    \PlaceEq{2*\side+0.3*\norm}{-0.3*\norm}{y}
    \end{tikzpicture}
    \end{minipage}
    
    \vspace{1cm}
    
    \begin{minipage}{0.5\linewidth}
    \begin{tikzpicture}
    \pgfmathsetmacro{\side}{0.7}
    \pgfmathsetmacro{\norm}{0.8}
    \PointSplitFirst{0}{0}{\side}
    \PlaceEq{2*\side}{-0.3*\norm}{x}
    \PlaceEq{3*\side}{-0.3*\norm}{\ell}
    \PlaceEq{4*\side}{-0.3*\norm}{y}
    \end{tikzpicture}
    \end{minipage}%
    \begin{minipage}{0.5\linewidth}
    \begin{tikzpicture}
    \pgfmathsetmacro{\side}{0.9}
    \pgfmathsetmacro{\norm}{0.707}
    \PointSplitSecond{0}{0}{\side}
    \PlaceEq{0.5*\side-0.3*\norm}{-0.5*\side-0.3*\norm}{x}
    \PlaceEq{2*\side}{-\side-0.3*0.8}{\ell}
    \PlaceEq{3.5*\side+0.3*\norm}{-1.5*\side+0.3*\norm}{y}
    \end{tikzpicture}
    \end{minipage}
    \caption{\textit{Top}: Double-vertex cut on a square lattice before point-splitting. \textit{Bottom}: Point-splitting to match \Crefpanel{fig:cut}{b} (left) and the hexagonal lattice (right).}\label{fig:hopsplit}
\end{figure}
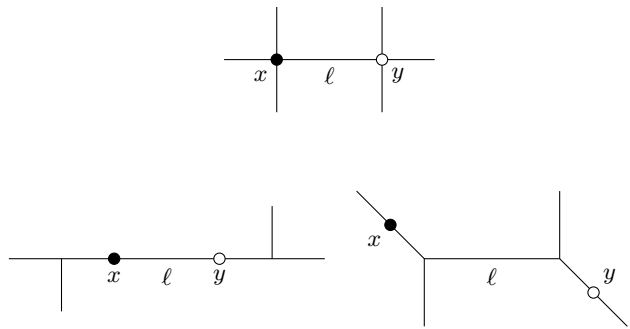

\begin{figure*}[t]
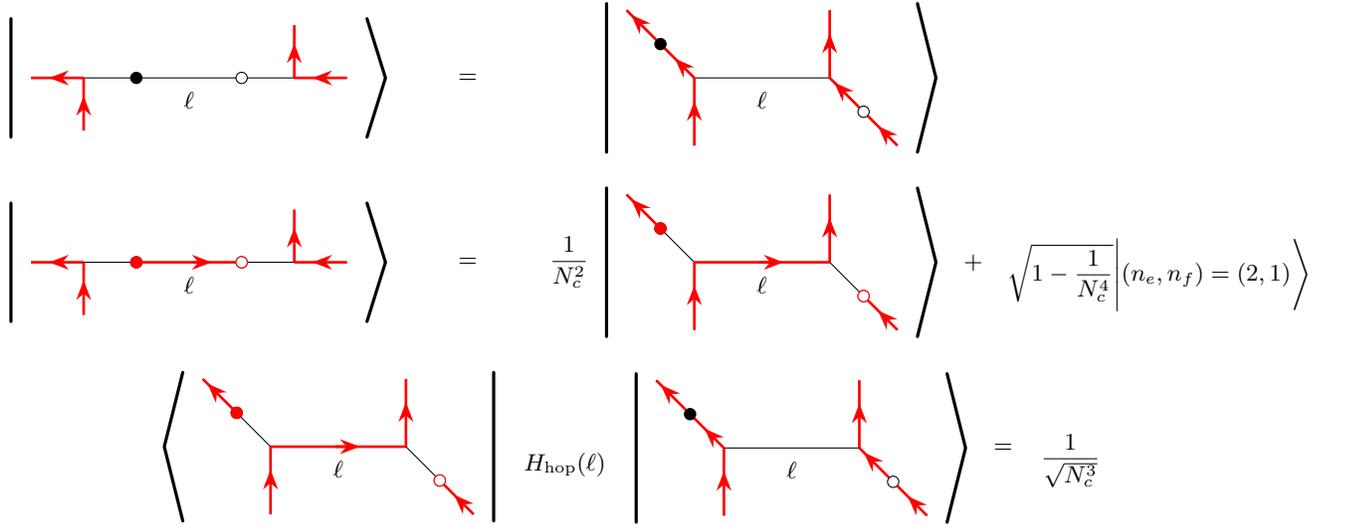

    \centering
    \begin{minipage}{0.3\linewidth}
    \KetPic{
    \pgfmathsetmacro{\side}{0.7}
    \PointSplitFirst{0}{0}{\side}
    \OrientedEdge{\side}{-\side}{\side}{0}{red}
    \OrientedEdge{\side}{0}{0}{0}{red}
    \OrientedEdge{6*\side}{0}{5*\side}{0}{red}
    \OrientedEdge{5*\side}{0}{5*\side}{\side}{red}
    \PlaceEq{3*\side}{-0.3}{\ell}
    }
    \end{minipage}%
    \begin{minipage}{0.1\linewidth}
    \centering
    $=$
    \end{minipage}%
    \begin{minipage}{0.05\linewidth}
    $\quad$
    \end{minipage}%
    \begin{minipage}{0.25\linewidth}
    \centering
    \KetPic{
    \pgfmathsetmacro{\side}{0.9}
    \pgfmathsetmacro{\norm}{0.707}
    \PointSplitSecond{0}{0}{0.9}
    \OrientedEdge{\side}{-2*\side}{\side}{-1*\side}{red}
    \OrientedEdge{\side}{-1*\side}{0.5*\side+0.08*\norm}{-0.5*\side-0.08*\norm}{red}
    \OrientedEdge{0.5*\side-0.08*\norm}{-0.5*\side+0.08*\norm}{0}{0}{red}
    \OrientedEdge{4*\side}{-2*\side}{3.5*\side+0.08*\norm}{-1.5*\side-0.08*\norm}{red}
    \OrientedEdge{3.5*\side-0.08*\norm}{-1.5*\side+0.08*\norm}{3*\side}{-1*\side}{red}
    \OrientedEdge{3*\side}{-1*\side}{3*\side}{0}{red}
    \PlaceEq{2*\side}{-\side-0.3}{\ell}
    }
    \end{minipage}%
    \begin{minipage}{0.25\linewidth}
    $\quad$
    \end{minipage}

    \vspace{4mm}

    \begin{minipage}{0.3\linewidth}
    \KetPic{
    \pgfmathsetmacro{\side}{0.7}
    \PointSplitFirst{0}{0}{\side}
    \OrientedEdge{\side}{-\side}{\side}{0}{red}
    \OrientedEdge{\side}{0}{0}{0}{red}
    \OrientedEdge{6*\side}{0}{5*\side}{0}{red}
    \OrientedEdge{5*\side}{0}{5*\side}{\side}{red}
    \OrientedEdge{2*\side+0.08}{0}{4*\side-0.08}{0}{red}
    \SmallCircle[draw=red, fill=red]{2*\side}{0}{1}{0.075}
    \SmallCircle[draw=red]{4*\side}{0}{0}{0.075}
    \PlaceEq{3*\side}{-0.3}{\ell}
    }
    \end{minipage}%
    \begin{minipage}{0.1\linewidth}
    \centering
    $=$
    \end{minipage}%
    \begin{minipage}{0.05\linewidth}
    $\dfrac{1}{N_c^2}$
    \end{minipage}%
    \begin{minipage}{0.25\linewidth}
    \centering
    \KetPic{
    \pgfmathsetmacro{\side}{0.9}
    \pgfmathsetmacro{\norm}{0.707}
    \PointSplitSecond{0}{0}{0.9}
    \OrientedEdge{\side}{-2*\side}{\side}{-1*\side}{red}
    \OrientedEdge{\side}{-1*\side}{3*\side}{-1*\side}{red}
    \OrientedEdge{0.5*\side-0.08*\norm}{-0.5*\side+0.08*\norm}{0}{0}{red}
    \OrientedEdge{4*\side}{-2*\side}{3.5*\side+0.08*\norm}{-1.5*\side-0.08*\norm}{red}
    \OrientedEdge{3*\side}{-1*\side}{3*\side}{0}{red}
    \SmallCircle[draw=red, fill=red]{0.5*\side}{-0.5*\side}{1}{0.075}
    \SmallCircle[draw=red]{3.5*\side}{-1.5*\side}{0}{0.075}
    \PlaceEq{2*\side}{-\side-0.3}{\ell}
    }
    \end{minipage}%
    \begin{minipage}{0.05\linewidth}
    $+$
    \end{minipage}%
    \begin{minipage}{0.2\linewidth}
    $$\sqrt{1-\frac{1}{N_c^4}} \Bigg|(n_e, n_f) = (2, 1)\Bigg\rangle$$
    \end{minipage}

    \vspace{4mm}

    \begin{minipage}{0.12\linewidth}
    $\quad$
    \end{minipage}%
    \begin{minipage}{0.25\linewidth}
    \centering
    \BraPic{
    \pgfmathsetmacro{\side}{0.9}
    \pgfmathsetmacro{\norm}{0.707}
    \PointSplitSecond{0}{0}{0.9}
    \OrientedEdge{\side}{-2*\side}{\side}{-1*\side}{red}
    \OrientedEdge{\side}{-1*\side}{3*\side}{-1*\side}{red}
    \OrientedEdge{0.5*\side-0.08*\norm}{-0.5*\side+0.08*\norm}{0}{0}{red}
    \OrientedEdge{4*\side}{-2*\side}{3.5*\side+0.08*\norm}{-1.5*\side-0.08*\norm}{red}
    \OrientedEdge{3*\side}{-1*\side}{3*\side}{0}{red}
    \SmallCircle[draw=red, fill=red]{0.5*\side}{-0.5*\side}{1}{0.075}
    \SmallCircle[draw=red]{3.5*\side}{-1.5*\side}{0}{0.075}
    \PlaceEq{2*\side}{-\side-0.3}{\ell}
    }
    \end{minipage}%
    \begin{minipage}{0.1\linewidth}
    $$H_{\hop}(\ell)$$
    \end{minipage}%
    \begin{minipage}{0.25\linewidth}
    \centering
    \KetPic{
    \pgfmathsetmacro{\side}{0.9}
    \pgfmathsetmacro{\norm}{0.707}
    \PointSplitSecond{0}{0}{0.9}
    \OrientedEdge{\side}{-2*\side}{\side}{-1*\side}{red}
    \OrientedEdge{\side}{-1*\side}{0.5*\side+0.08*\norm}{-0.5*\side-0.08*\norm}{red}
    \OrientedEdge{0.5*\side-0.08*\norm}{-0.5*\side+0.08*\norm}{0}{0}{red}
    \OrientedEdge{4*\side}{-2*\side}{3.5*\side+0.08*\norm}{-1.5*\side-0.08*\norm}{red}
    \OrientedEdge{3.5*\side-0.08*\norm}{-1.5*\side+0.08*\norm}{3*\side}{-1*\side}{red}
    \OrientedEdge{3*\side}{-1*\side}{3*\side}{0}{red}
    \PlaceEq{2*\side}{-\side-0.3}{\ell}
    }
    \end{minipage}%
    \begin{minipage}{0.05\linewidth}
    $=$
    \end{minipage}%
    \begin{minipage}{0.05\linewidth}
    $$\frac{1}{\sqrt{N_c^3}}$$
    \end{minipage}
    \begin{minipage}{0.12\linewidth}
    $\quad$
    \end{minipage}
    
    \caption{Local depictions of two basis states related by hopping transition (flux strings and their endpoints are shown in red for emphasis). In the top two equations, the LHS shows a normalized, point-split lattice state compatible with \Crefpanel{fig:cut}{b}, and the RHS shows how to re-express the state as a linear combination of point-split states in the hexagonal lattice. The coefficients in the linear combination follow from fusing irreps of $\SU(N_c)$ in the different ways shown diagrammatically. Since we take all links to be physical in the hexagonal lattice, our truncation projects out the higher-energy states that appear in the second configuration. The bottom equation shows the resulting hopping matrix element at $(n_e, n_f) = (1, 1)$.}\label{fig:combo}
\end{figure*}

The first method begins with the contracted (square) lattice, which can be locally point-split around an arbitrary link $\ell$ in two different ways: (i) by arranging for vertices $x \equiv \ell^-$ and $y \equiv \ell^+$ to match \Crefpanel{fig:cut}{b}; and (ii) by performing the same point-splitting used in \Cref{fig:split} to match the hexagonal lattice.
These two local point-splittings around link $\ell$ are shown in \Cref{fig:hopsplit}.
The Hilbert space of double-vertex cut states does not depend on how the point-splitting is performed---any basis state in one point-splitting is a linear combination of basis states in the other point-splitting.
By writing this linear combination explicitly, we can thus use the matrix elements from \Cref{tab:hops} to derive the hopping matrix element between two states on the hexagonal lattice.
An example of this procedure is shown in~\Cref{fig:combo}.

The second method focuses on an arbitrary gauge-invariant operator $O_{\gamma}$ constructed from a Wilson line along some path $\gamma$ (for simplicity, we will assume that the path $\gamma$ does not traverse any link more than once).
Such an operator is either a Wilson loop (if $\gamma$ is a closed loop), or a meson-like string operator (if $O_{\gamma}$ is contracted against fermion fields at its endpoints).
In both cases, the action of $O_{\gamma}$ follows from the action of the Wilson line operator $U(\gamma)_{ab}$ along the path $\gamma$.

To this end, note that at energy cutoff $(n_e, n_f) = (1, 1)$, we can view $U(\gamma)_{ab}$ as performing three types of operations on the links that comprise the path $\gamma$: (i) if the link is in the singlet representation, it gets excited into an electric field line oriented along $\gamma$; (ii) if the link contains an electric field line oriented \textit{against} $\gamma$, then the link gets de-excited to the singlet representation; and (iii) if the link contains an electric field line oriented \textit{along} $\gamma$, then the entire quantum state gets annihilated.
We can therefore assume without loss of generality that the path $\gamma$ will always pass through links that either live in the singlet representation, or contain a single electric field line oriented opposite to the path $\gamma$.

The relevant matrix elements for a Wilson line $U(\ell)_{ab}$ along an arbitrary link $\ell \in \lk^{\mathrm{hex}}$ (oriented from the odd endpoint to the even endpoint) are given by
\begin{align}\label{eq:<U>}
    \bra{\mathbf{N},c,d}_{\ell} U(\ell)_{ab}\ket{\mathbf{1}}_{\ell} &= \frac1{\sqrt{N_c}} \delta_{ac}\delta_{bc}\,, \nonumber\\
    \bra{\mathbf{1}}_{\ell} U(\ell)_{ab}\ket{\overline{\mathbf{N}},c,d}_{\ell} &= \frac1{\sqrt{N_c}} \delta_{ac}\delta_{bc}\,.
\end{align}
The analogous matrix elements for the Wilson line oriented in the opposite direction are simply obtained by conjugation.
These relations imply that the action of $O_{\gamma}$ on a gauge-invariant basis state must pick up a factor $1/\sqrt{N_c}$ for every link along the path $\gamma$.
If $\gamma$ passes through $r$ such links, this gives a factor $N_c^{-r/2}$.

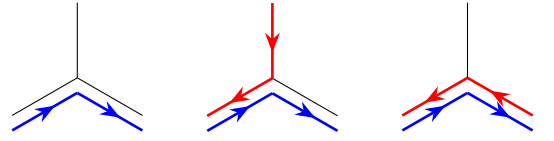
\begin{figure}[htp]
    \centering

    \begin{minipage}{0.3\linewidth}
    \begin{tikzpicture}
    \pgfmathsetmacro{\norm}{0.866}
    \pgfmathsetmacro{\side}{1.0}
    \pgfmathsetmacro{\offset}{0.2}
    \pgfmathsetmacro{\temp}{-0.5*\side-\offset}
    \draw (0,0) -- (0,\side);
    \draw (0,0) -- (-\norm*\side,-0.5*\side);
    \draw (0,0) -- (\norm*\side,-0.5*\side);
    \OrientedEdge{-\norm*\side}{\temp}{0}{-\offset}{blue}
    \OrientedEdge{0}{-\offset}{\norm*\side}{-0.5*\side-\offset}{blue}
    \end{tikzpicture}
    \end{minipage}%
    \begin{minipage}{0.3\linewidth}
    \begin{tikzpicture}
    \pgfmathsetmacro{\norm}{0.866}
    \pgfmathsetmacro{\side}{1.0}
    \pgfmathsetmacro{\offset}{0.2}
    \pgfmathsetmacro{\temp}{-0.5*\side-\offset}
    \draw (0,0) -- (0,\side);
    \draw (0,0) -- (-\norm*\side,-0.5*\side);
    \draw (0,0) -- (\norm*\side,-0.5*\side);
    \OrientedEdge{-\norm*\side}{\temp}{0}{-\offset}{blue}
    \OrientedEdge{0}{-\offset}{\norm*\side}{-0.5*\side-\offset}{blue}
    \OrientedEdge{0}{1}{0}{0}{red}
    \OrientedEdge{0}{0}{-\norm*\side}{-0.5*\side}{red}
    \end{tikzpicture}
    \end{minipage}%
    \begin{minipage}{0.3\linewidth}
    \begin{tikzpicture}
    \pgfmathsetmacro{\norm}{0.866}
    \pgfmathsetmacro{\side}{1.0}
    \pgfmathsetmacro{\offset}{0.2}
    \pgfmathsetmacro{\temp}{-0.5*\side-\offset}
    \draw (0,0) -- (0,\side);
    \draw (0,0) -- (-\norm*\side,-0.5*\side);
    \draw (0,0) -- (\norm*\side,-0.5*\side);
    \OrientedEdge{-\norm*\side}{\temp}{0}{-\offset}{blue}
    \OrientedEdge{0}{-\offset}{\norm*\side}{-0.5*\side-\offset}{blue}
    \OrientedEdge{\norm*\side}{-0.5*\side}{0}{0}{red}
    \OrientedEdge{0}{0}{-\norm*\side}{-0.5*\side}{red}
    \end{tikzpicture}
    \end{minipage}%
    
    \caption{The three possible placements of a gauge-invariant operator $O_{\gamma}$ (shown in blue) relative to the electric field content at a trivalent vertex (red), up to rotations, reflections, and charge conjugation. Note that this also applies to fermion and anti-fermion sites on the lattice, which count as trivalent vertices. \textit{Left}: The case of no excited links and no overlap (contributes unit factor). \textit{Middle}: The case of two excited links and one overlap (contributes unit factor). \textit{Right}: The case of two excited links and two overlaps (contributes a factor of $N_c$).}\label{fig:trivalent}
\end{figure}

Additionally, the color indices in the Kronecker delta symbols must be contracted after using \eqref{eq:<U>} and/or applying the fermion algebra at the endpoints of $\gamma$.
At every trivalent vertex, this either gives a factor $N_c$ (if $O_{\gamma}$ acts on two excited legs), or a factor $1$ (if $O_{\gamma}$ acts on at most a single excited leg).
See \Cref{fig:trivalent} for an illustration of these different cases.
If there are $s$ occurrences where $O_{\gamma}$ acts on two excited legs, this gives a factor $N_c^s$.

Now suppose that $\ket{i}$ and $\ket{f}$ are the two normalized ``initial" and ``final" states for a transition connected by $O_{\gamma}$, so that $O_{\gamma} \ket{i} \propto \ket{f}$.
Both $\ket{i}$ and $\ket{f}$ can be expanded in the basis of color-indexed states.
For instance, each vertex singlet in this expansion must either be of the form 
$\ket{\mathbf{1}}\ket{\mathbf{1}}\ket{\mathbf{1}}$ (if no legs are excited), or of the form
\begin{equation}
    \left(\frac1{\sqrt{N_c}} \ket{\mathbf{N},a}\ket{\overline{\mathbf{N}},a}\right) \otimes \ket{\mathbf{1}},
\end{equation}
if it contains exactly two excited legs, up to charge conjugation and permutations of tensor factors.
The latter type of vertex singlet contributes a factor $1/\sqrt{N_c}$ to the normalization coefficients for $\ket{i}$ and $\ket{f}$ in this color-indexed expansion, while the former contributes only a unit factor.
If we use $v_i$ and $v_f$ to denote the number of vertex singlets with two excited legs in $\ket{i}$ and $\ket{f}$, then the normalization coefficients are precisely $N_c^{-v_i/2}$ and $N_c^{-v_f/2}$, respectively.

\begin{figure}[htp]
    \centering

    \begin{minipage}{\linewidth}
    \begin{tikzpicture}
    \pgfmathsetmacro{\side}{0.9}
    \pgfmathsetmacro{\norm}{0.707}
    \PointSplitSecond{0}{0}{0.9}
    \OrientedEdge{\side}{-2*\side}{\side}{-1*\side}{red}
    \OrientedEdge{\side}{-1*\side}{0.5*\side+0.08*\norm}{-0.5*\side-0.08*\norm}{red}
    \OrientedEdge{0.5*\side-0.08*\norm}{-0.5*\side+0.08*\norm}{0}{0}{red}
    \OrientedEdge{4*\side}{-2*\side}{3.5*\side+0.08*\norm}{-1.5*\side-0.08*\norm}{red}
    \OrientedEdge{3.5*\side-0.08*\norm}{-1.5*\side+0.08*\norm}{3*\side}{-1*\side}{red}
    \OrientedEdge{3*\side}{-1*\side}{3*\side}{0}{red}

    \pgfmathsetmacro{\offset}{0.1}
    \pgfmathsetmacro{\y}{-0.5*\side + \offset*\norm}
    \pgfmathsetmacro{\yy}{-\side+\offset*\norm}
    \pgfmathsetmacro{\x}{0.5*\side+\offset*\norm}
    \pgfmathsetmacro{\xx}{\side + \offset*\norm}
    \pgfmathsetmacro{\xxx}{\xx+2*\side-4*\offset*\norm}
    \pgfmathsetmacro{\xxxx}{\xx+2.5*\side-2*\offset*\norm}
    \pgfmathsetmacro{\yyy}{-1.5*\side-\offset*\norm}
    \OrientedEdge{\x}{\y}{\xx}{\yy}{blue}
    \OrientedEdge{\xx}{\yy}{\xxx}{\yy}{blue}
    \OrientedEdge{\xxx}{\yy}{\xxxx}{\yyy}{blue}
    \SmallCircle[draw=blue, fill=blue]{\x}{\y}{1}{0.075}
    \SmallCircle[draw=blue]{\xxxx}{\yyy}{0}{0.075}
    \end{tikzpicture}
    \end{minipage}

    \vspace{4mm}

    \begin{minipage}{\linewidth}
    \begin{tikzpicture}
    \BubbleDownArrow
    \end{tikzpicture}
    \end{minipage}

    \vspace{4mm}
    
    \begin{minipage}{\linewidth}
    \begin{tikzpicture}
    \pgfmathsetmacro{\side}{0.9}
    \pgfmathsetmacro{\norm}{0.707}
    \PointSplitSecond{0}{0}{0.9}
    \OrientedEdge{\side}{-2*\side}{\side}{-1*\side}{red}
    \OrientedEdge{\side}{-1*\side}{3*\side}{-1*\side}{red}
    \OrientedEdge{0.5*\side-0.08*\norm}{-0.5*\side+0.08*\norm}{0}{0}{red}
    \OrientedEdge{4*\side}{-2*\side}{3.5*\side+0.08*\norm}{-1.5*\side-0.08*\norm}{red}
    \OrientedEdge{3*\side}{-1*\side}{3*\side}{0}{red}
    \SmallCircle[draw=red, fill=red]{0.5*\side}{-0.5*\side}{1}{0.075}
    \SmallCircle[draw=red]{3.5*\side}{-1.5*\side}{0}{0.075}
    \end{tikzpicture}
    \end{minipage}
    
    \caption{Example showing how to compute the matrix element for the same hopping transition shown in \Cref{fig:combo}. The parameters $v_i=4$, $v_f=4$, $r=3$, and $s=0$ can be read off from the diagram. Since the transition amplitude \eqref{eq:Omaster} depends only on the difference $v_f - v_i$, any lattice with this local transition will yield the same matrix element. By plugging into \eqref{eq:Omaster}, we find $\bra{f} O_{\gamma} \ket{i} = 1/\sqrt{N_c^3}$ in this case.}\label{fig:hopformula}
\end{figure}
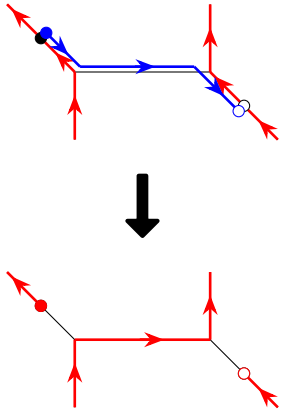

Putting it all together, the action of $O_{\gamma}$ on $\ket{i}$ produces an overall coefficient $N_c^{-v_i/2} N_c^{-r/2} N_c^s$ in the expansion of the final state in the color-indexed basis.
The desired matrix element $\bra{f} O_{\gamma} \ket{i}$ is the factor by which this overshoots the normalization coefficient $N_c^{-v_f/2}$ in the final state, i.e.,
\begin{equation}\label{eq:Omaster}
    \bra{f} O_{\gamma} \ket{i} = \sqrt{N_c^{v_f - v_i - r + 2s}}.
\end{equation}
Since this holds for any gauge-invariant operator $O_{\gamma}$, it also immediately provides the matrix elements for the plaquette and hopping operators by appropriate diagrammatic interpretation. Note that this only holds for states at the truncation we are working in.
An example of applying this approach is shown in \Cref{fig:hopformula}.

Note that the full state at a trivalent vertex requires knowing the status of at least two of its links.
This implies that plaquette operators require knowledge only of the eight links on which they act, but hopping operators require knowledge of \textit{five} links---the three links on which their path $\gamma$ resides, and the two diagonal links adjacent to the endpoints of $\gamma$.
In other words, our qutrit Hamiltonians must implement plaquette interactions as eight-body operators, while hopping terms are five-body operators.
The latter is also why we adopted the notation $(\ell^*_{\bullet}, \ell_{\bullet}, \ell, \ell_{\circ}, \ell^*_{\circ})$ for the five links surrounding $\ell$ that participate in hopping transitions.
In particular, when writing down truncated Hamiltonians, we will use this notation along with the plaquette notation $p \equiv (p_1, \dots, p_8)$ to directly label qutrit operators by the links at which those qutrits reside.

\subsubsection{Truncation \TTone}

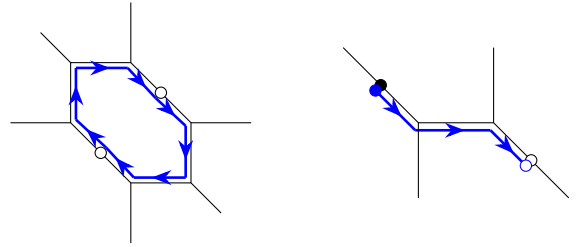
\begin{figure}[htp]
    \centering
    \begin{minipage}{0.5\linewidth}
    \begin{tikzpicture}
    \pgfmathsetmacro{\side}{0.8}
    \pgfmathsetmacro{\norm}{0.707}
    \pgfmathsetmacro{\offset}{0.1}
    \pgfmathsetmacro{\x}{3*\side+\offset*\norm}
    \pgfmathsetmacro{\y}{-3*\side-\offset*\norm}
    \pgfmathsetmacro{\xx}{4*\side-\offset*(2*\norm-1)}
    \pgfmathsetmacro{\yy}{-3*\side-\offset*\norm}
    \pgfmathsetmacro{\xxx}{4.5*\side-\offset*\norm}
    \pgfmathsetmacro{\yyy}{-3.5*\side-\offset*\norm}
    \pgfmathsetmacro{\xxxx}{5*\side-\offset*\norm}
    \pgfmathsetmacro{\yyyy}{-4*\side-\offset*(2*\norm-1)}
    \pgfmathsetmacro{\xxxxx}{5*\side-\offset*\norm}
    \pgfmathsetmacro{\yyyyy}{-5*\side+\offset*\norm}
    \pgfmathsetmacro{\xxxxxx}{4*\side+\offset*(2*\norm-1)}
    \pgfmathsetmacro{\yyyyyy}{-5*\side+\offset*\norm}
    \pgfmathsetmacro{\xxxxxxx}{3.5*\side+\offset*\norm}
    \pgfmathsetmacro{\yyyyyyy}{-4.5*\side+\offset*\norm}
    \pgfmathsetmacro{\xxxxxxxx}{3*\side+\offset*\norm}
    \pgfmathsetmacro{\yyyyyyyy}{-4*\side+\offset*(2*\norm-1)}

    \draw (2.5*\side,-2.5*\side) -- (3*\side, -3*\side);
    \draw (4*\side, -2*\side) -- (4*\side, -3*\side);
    \draw (5*\side, -4*\side) -- (6*\side, -4*\side);
    \draw (5*\side, -5*\side) -- (5.5*\side, -5.5*\side);
    \draw (4*\side, -5*\side) -- (4*\side, -6*\side);
    \draw (3*\side, -4*\side) -- (2*\side, -4*\side);
    
    \OddHexPlaquetteReduced{3*\side}{-3*\side}{\side}{black}
    
    \OrientedEdge{\x}{\y}{\xx}{\yy}{blue}
    \OrientedEdge{\xx}{\yy}{\xxx}{\yyy}{blue}
    \OrientedEdge{\xxx}{\yyy}{\xxxx}{\yyyy}{blue}
    \OrientedEdge{\xxxx}{\yyyy}{\xxxxx}{\yyyyy}{blue}
    \OrientedEdge{\xxxxx}{\yyyyy}{\xxxxxx}{\yyyyyy}{blue}
    \OrientedEdge{\xxxxxx}{\yyyyyy}{\xxxxxxx}{\yyyyyyy}{blue}
    \OrientedEdge{\xxxxxxx}{\yyyyyyy}{\xxxxxxxx}{\yyyyyyyy}{blue}
    \OrientedEdge{\xxxxxxxx}{\yyyyyyyy}{\x}{\y}{blue}
    \end{tikzpicture}
    \end{minipage}%
    \begin{minipage}{0.5\linewidth}
    \begin{tikzpicture}
    \pgfmathsetmacro{\side}{1.0}
    \pgfmathsetmacro{\norm}{0.707}
    \draw (0,0) -- (\side, -\side) -- (2*\side, -\side) -- (3*\side, -2*\side);
    \SmallCircle{0.5*\side}{-0.5*\side}{1}{0.075}
    \SmallCircle{2.5*\side}{-1.5*\side}{0}{0.075}
    \pgfmathsetmacro{\offset}{0.1}
    \pgfmathsetmacro{\x}{0.5*\side-\offset*\norm}
    \pgfmathsetmacro{\y}{-0.5*\side-\offset*\norm}
    \pgfmathsetmacro{\xx}{\side-\offset*(2*\norm-1)}
    \pgfmathsetmacro{\yy}{-\side-\offset}
    \pgfmathsetmacro{\xxx}{2*\side-(2*\norm-1)*\offset}
    \pgfmathsetmacro{\yyy}{-\side-\offset}
    \pgfmathsetmacro{\xxxx}{2.5*\side-\offset*\norm}
    \pgfmathsetmacro{\yyyy}{-1.5*\side-\offset*\norm}
    \OrientedEdge{\x}{\y}{\xx}{\yy}{blue}
    \OrientedEdge{\xx}{\yy}{\xxx}{\yyy}{blue}
    \OrientedEdge{\xxx}{\yyy}{\xxxx}{\yyyy}{blue}
    \SmallCircle[draw=blue, fill=blue]{\x}{\y}{1}{0.075}
    \SmallCircle[draw=blue]{\xxxx}{\yyyy}{0}{0.075}
    \draw (\side, -\side) -- (\side, -2*\side);
    \draw (2*\side, -\side) -- (2*\side, 0);
    \end{tikzpicture}
    \end{minipage}
    
    \caption{Interaction operators (blue) showing an isolated plaquette transition (left) and isolated hopping transition (right).}\label{fig:TT1new}
\end{figure}

At truncation \TTone, there are three types of excitations supported: (i) clockwise plaquettes; (ii) counterclockwise plaquettes; and (iii) quark-antiquark pairs between neighboring pairs of lattice sites.
These excitations must remain isolated from each other to prevent two excitations on the same link or lattice site.
\Cref{fig:TT1new} shows an example of these transitions.

Every excited link contributes the same amount of electric energy, so we can straightforwardly write down the electric term of the qutrit Hamiltonian as
\begin{equation}
    H^{\mathrm{elec}}_{\text{\TTone}} = -\frac{g^2}{4} \frac{N_c^2 - 1}{2N_c} \sum_{\ell\in\lk^{\mathrm{hex}}} \left(\mathcal{Z}^{\mathbf{N}}_{\ell} + \mathcal{Z}^{\overline{\mathbf{N}}}_{\ell}\right).
\end{equation}

For the mass term, we need to include a mass contribution $2m$ for every quark-antiquark pair.
This means we should use \eqref{eq:flow}, which leads to
\begin{equation}\label{eq:mass1}
    H^{\mathrm{mass}}_{\text{\TTone}} = -m \sum_{\ell \in \lk^{\mathrm{hex}}} \left(\mathcal{Z}^{\mathbf{N}}_j - \mathcal{Z}^{\overline{\mathbf{N}}}_j\right).
\end{equation}
There is one catch, however: this only works if we choose an irrep labeling convention on links where any contiguous field line in a state alternates between $\ket{\mathbf{N}}$ and $\ket{\overline{\mathbf{N}}}$ on its links.

In that case, this can actually be simplified by observing that the cardinal links must altogether carry the same number of $\ket{\mathbf{N}}$ and $\ket{\overline{\mathbf{N}}}$ excitations, which follows from gauge invariance (and the fact that charges do not live at the endpoints of any cardinal links).
By removing the cardinal links from this sum, we can write the equivalent (and slightly simpler) mass term
\begin{equation}\label{eq:mass2}
    H^{\mathrm{mass}}_{\text{\TTone}} \cong -m \sum_{\ell \in \lk^{\mathrm{hex}}_d} \left(\mathcal{Z}^{\mathbf{N}}_j - \mathcal{Z}^{\overline{\mathbf{N}}}_j\right),
\end{equation}
where we use the $\cong$ symbol to indicate equivalence on the gauge-invariant Hilbert space.
Even though \eqref{eq:mass2} contains less terms, an actual quantum simulation might turn out to be easier to implement using \eqref{eq:mass1}, just because it leads to the same coefficients on the Pauli-$Z$'s for every qutrit after combining with the electric term.

An alternative convention (and one that makes calculations slightly simpler) is to assume that all cardinal links take the same irrep label that they would on the corresponding contracted square lattice, and then extend their convention (i.e. which direction of electric flux is labeled as $\mathbf{N}$ vs. $\overline{\mathbf{N}}$) to the two diagonal links adjacent to each cardinal link.
This is consistent because conventions on the square lattice are based on the orientation of the link relative to an odd endpoint of the link, and every cardinal link adjacent to a given diagonal link shares the same orientation relative to the lattice site that the diagonal link is incident on.

In this case, the mass term should exactly parallel the mass term for the contracted square lattice, i.e. by summing over only cardinal links:
\begin{equation}
    H^{\mathrm{mass}}_{\text{\TTone}} \cong -m \sum_{\ell \in \lk^{\mathrm{hex}}_c} \left(\mathcal{Z}^{\mathbf{N}}_j - \mathcal{Z}^{\overline{\mathbf{N}}}_j\right).
\end{equation}
Due to the familiarity between this convention and the square lattice, as well as the simplification it provides in specific circumstances, we will use this convention throughout this section.

To write down the interaction terms, we can use \eqref{eq:Omaster} to see that the matrix element for creating or annihilating an isolated plaquette excitation is $1$, regardless of orientation.
Similarly, the matrix element for creating or annihilating an isolated quark-antiquark pair is $\sqrt{N_c}$.

A plaquette $p = (p_1, \dots, p_8)$ can only be excited if all eight participating links are in the singlet representation.
Similarly, $p$ can only be de-excited if its links are in one of the two orientations consistent with a clockwise or counterclockwise loop.
Therefore, if we define the operators
\begin{align}
    B^{\mathrm{CW}}_p &= b^{\mathbf{N},\mathbf{1}}_{p_1} b^{\mathbf{N},\mathbf{1}}_{p_2} b^{\overline{\mathbf{N}},\mathbf{1}}_{p_3} b^{\overline{\mathbf{N}},\mathbf{1}}_{p_4} b^{\mathbf{N},\mathbf{1}}_{p_5} b^{\mathbf{N},\mathbf{1}}_{p_6} b^{\overline{\mathbf{N}},\mathbf{1}}_{p_7} b^{\overline{\mathbf{N}},\mathbf{1}}_{p_8},\nonumber \\
    B^{\mathrm{CCW}}_p &= b^{\overline{\mathbf{N}},\mathbf{1}}_{p_1} b^{\overline{\mathbf{N}},\mathbf{1}}_{p_2} b^{\mathbf{N},\mathbf{1}}_{p_3} b^{\mathbf{N},\mathbf{1}}_{p_4} b^{\overline{\mathbf{N}},\mathbf{1}}_{p_5} b^{\overline{\mathbf{N}},\mathbf{1}}_{p_6} b^{\mathbf{N},\mathbf{1}}_{p_7} b^{\mathbf{N},\mathbf{1}}_{p_8},
\end{align}
where $b^{\mathbf{u},\mathbf{v}}_{l} = \ket{\mathbf{u}}_l\bra{\mathbf{v}}_l$, then the truncated plaquette term becomes simply
\begin{align}
    H^{\mathrm{plaq}}_{\text{\TTone}} = &-\frac{1}{2g^2} \sum_{p\in\pq^{\mathrm{hex}}} B^{\mathrm{CW}}_p +B^{\mathrm{CW} \dagger}_p \nonumber\\
    & -\frac1{2g^2}\sum_{p\in\pq^{\mathrm{hex}}} B^{\mathrm{CCW}}_p +B^{\mathrm{CCW} \dagger}_p.
\end{align}

At any cardinal link $\ell = (\ell^*_{\bullet}, \ell_{\bullet}, \ell, \ell_{\circ}, \ell^*_{\circ})$, the hopping transitions require the links $\ell^*_{\bullet}$ and $\ell^*_{\circ}$ to be in the singlet representation.
Moreover, a hopping operation can only transform the set of links $\ell_{\bullet}$, $\ell$, and $\ell_{\circ}$ from the all-singlet to the all-fundamental configuration, or vice versa.
Therefore, if we define
\begin{align}
    B^{\mathbf{N},\mathbf{1}}_{\ell} &\equiv b^{\mathbf{N},\mathbf{1}}_{\ell_{\bullet}} b^{\mathbf{N},\mathbf{1}}_{\ell} b^{\mathbf{N},\mathbf{1}}_{\ell_{\circ}},
\end{align}
then the hopping term takes the form
\begin{equation}
    H^{\hop}_{\text{\TTone}} = \frac{\sqrt{N_c}}2 \sum_{\ell \in \lk_c^{\mathrm{hex}}} P^{\mathbf{1}}_{\ell^*_{\bullet}}\left(B^{\mathbf{N},\mathbf{1}}_{\ell} + B^{\mathbf{N},\mathbf{1} \dagger}_{\ell} \right)P^{\mathbf{1}}_{\ell^*_{\circ}}.
\end{equation}

Note that in the strict large $N_c$ limit, plaquette interactions are actually sub-leading by a factor $1/\sqrt{N_c}$ relative to hopping interactions; however, the electric energy tends to infinity in this limit.
The large $N_c$ limit that controls the electric energy is the 't Hooft limit, where $N_c\to \infty$ while holding $\lambda \equiv g^2 N_c$ fixed.
In the 't Hooft limit, it is the hopping interactions that are sub-leading by a factor $1/\sqrt{N_c}$ relative to the plaquette interactions.

\subsubsection{Truncation \TTtwo}

The modification of the Hamiltonian from truncation \TTone\ to \TTtwo\ is a straightforward application of \eqref{eq:Omaster}.
This will require computing the $N_c$ scaling of all matrix elements for all interaction terms, since \TTtwo\ needs to contain the $\mathcal O(1/N_c)$ corrections to the Hamiltonian at $(n_e, n_f) = (1, 1)$.

To begin, we can write down the electric and mass terms by reusing the exact same expressions from truncation \TTone.
\begin{align}
    H^{\mathrm{elec}}_{\text{\TTtwo}} &= -\frac{g^2}{4} \frac{N_c^2 - 1}{2N_c} \sum_{\ell\in\lk^{\mathrm{hex}}} \left(\mathcal{Z}^{\mathbf{N}}_{\ell} + \mathcal{Z}^{\overline{\mathbf{N}}}_{\ell}\right), \nonumber \\
    H^{\mathrm{mass}}_{\text{\TTtwo}} &= -m \sum_{\ell \in \lk^{\mathrm{hex}}} \left(\mathcal{Z}^{\mathbf{N}}_j - \mathcal{Z}^{\overline{\mathbf{N}}}_j\right).
\end{align}
As before, the mass term can be slightly simplified by replacing the sum over all links by a sum over only diagonal links, if desired.

To understand the structure of plaquette matrix elements, begin by considering an arbitrary Wilson loop operator $O_{\gamma}$ where $\gamma$ is a non-intersecting loop whose circumference contains $r$ links.
It is straightforward to see by using \eqref{eq:Omaster} that the matrix elements for $O_{\gamma}$ have unit magnitude for transitions that create or annihilate a single isolated loop of electric flux along $\gamma$.
In the most general case, $O_{\gamma}$ can act on an initial state $\ket{i}$ which contains $k$ isolated, contiguous stretches of an electric field line on the links of $\gamma$ (oriented opposite to $\gamma$).
In that case, the resulting normalized state $\ket{f} \propto O_{\gamma} \ket{i}$ will also contain $k$ isolated, contiguous electric field line segments on the links of $\gamma$ (this time, oriented in the same direction as $\gamma$).
Label these contiguous segments by index $j = 1, \dots, k$, and denote the lengths of the $j^{\mathrm{th}}$ segments in $\ket{i}$ and $\ket{f}$ by $\alpha_j$ and $\beta_j$, respectively.
The number of vertex singlets with two excited legs present in the initial and final states are respectively given by
\begin{align}
    v_i &= v + \sum_{j=1}^k \left(\alpha_j + 1\right), \nonumber\\
    v_f &= v + \sum_{j=1}^k \left(\beta_j + 1\right),
\end{align}
where $v$ denotes the number of vertex singlets with two excited legs that \textit{do not} contain a leg used in the path $\gamma$.
Additionally, note that the number of initial vertex singlets where $O_{\gamma}$ acts on two excited links is simply given by
\begin{equation}
    s = \sum_{j=1}^k\left(\alpha_j - 1\right)\,.
\end{equation}
We thus find
\begin{align}
    v_f - v_i - r + 2s &= \left(\sum_{j=1}^k \alpha_j + \sum_{j=1}^k \beta_j\right) - r - 2k \nonumber\\
    & = -2k\,,
\end{align}
where we used the fact that
\begin{equation}
    \sum_{j=1}^k \alpha_j + \sum_{j=1}^k \beta_j = r\,,
\end{equation}
because every link on $\gamma$ must either be excited in $\ket{i}$ or in $\ket{f}$, but not both.
By using \eqref{eq:Omaster}, we arrive at the interesting conclusion that
\begin{equation}
    \bra{f}O_{\gamma}\ket{i} = \frac1{N_c^k}\,.
\end{equation}
In other words, the $1/N_c$ scaling of a Wilson loop depends entirely on the number of contiguous field lines it intersects.

\begin{figure}[htp]
    \centering
    \begin{minipage}{0.5\linewidth}
    \begin{tikzpicture}
    \pgfmathsetmacro{\side}{0.8}
    \pgfmathsetmacro{\norm}{0.707}
    \pgfmathsetmacro{\offset}{0.1}
    \pgfmathsetmacro{\x}{3*\side+\offset*\norm}
    \pgfmathsetmacro{\y}{-3*\side-\offset*\norm}
    \pgfmathsetmacro{\xx}{4*\side-\offset*(2*\norm-1)}
    \pgfmathsetmacro{\yy}{-3*\side-\offset*\norm}
    \pgfmathsetmacro{\xxx}{4.5*\side-\offset*\norm}
    \pgfmathsetmacro{\yyy}{-3.5*\side-\offset*\norm}
    \pgfmathsetmacro{\xxxx}{5*\side-\offset*\norm}
    \pgfmathsetmacro{\yyyy}{-4*\side-\offset*(2*\norm-1)}
    \pgfmathsetmacro{\xxxxx}{5*\side-\offset*\norm}
    \pgfmathsetmacro{\yyyyy}{-5*\side+\offset*\norm}
    \pgfmathsetmacro{\xxxxxx}{4*\side+\offset*(2*\norm-1)}
    \pgfmathsetmacro{\yyyyyy}{-5*\side+\offset*\norm}
    \pgfmathsetmacro{\xxxxxxx}{3.5*\side+\offset*\norm}
    \pgfmathsetmacro{\yyyyyyy}{-4.5*\side+\offset*\norm}
    \pgfmathsetmacro{\xxxxxxxx}{3*\side+\offset*\norm}
    \pgfmathsetmacro{\yyyyyyyy}{-4*\side+\offset*(2*\norm-1)}

    \draw (2.5*\side,-2.5*\side) -- (3*\side, -3*\side);
    \draw (4*\side, -2*\side) -- (4*\side, -3*\side);
    \draw (5*\side, -4*\side) -- (6*\side, -4*\side);
    \draw (5*\side, -5*\side) -- (5.5*\side, -5.5*\side);
    \draw (4*\side, -5*\side) -- (4*\side, -6*\side);
    \draw (3*\side, -4*\side) -- (2*\side, -4*\side);

    \OrientedEdge{6*\side}{-4*\side}{5*\side}{-4*\side}{red}
    \OrientedEdge{5*\side}{-4*\side}{4.5*\side+0.08*\norm}{-3.5*\side-0.08*\norm}{red}
    \OrientedEdge{4.5*\side-0.08*\norm}{-3.5*\side+0.08*\norm}{4*\side}{-3*\side}{red}
    \OrientedEdge{4*\side}{-3*\side}{3*\side}{-3*\side}{red}
    \OrientedEdge{3*\side}{-3*\side}{2.5*\side}{-2.5*\side}{red}
    
    \OddHexPlaquetteReduced{3*\side}{-3*\side}{\side}{black}
    
    \OrientedEdge{\x}{\y}{\xx}{\yy}{blue}
    \OrientedEdge{\xx}{\yy}{\xxx}{\yyy}{blue}
    \OrientedEdge{\xxx}{\yyy}{\xxxx}{\yyyy}{blue}
    \OrientedEdge{\xxxx}{\yyyy}{\xxxxx}{\yyyyy}{blue}
    \OrientedEdge{\xxxxx}{\yyyyy}{\xxxxxx}{\yyyyyy}{blue}
    \OrientedEdge{\xxxxxx}{\yyyyyy}{\xxxxxxx}{\yyyyyyy}{blue}
    \OrientedEdge{\xxxxxxx}{\yyyyyyy}{\xxxxxxxx}{\yyyyyyyy}{blue}
    \OrientedEdge{\xxxxxxxx}{\yyyyyyyy}{\x}{\y}{blue}
    \end{tikzpicture}
    \end{minipage}%
    \begin{minipage}{0.5\linewidth}
    \begin{tikzpicture}
    \pgfmathsetmacro{\side}{0.8}
    \pgfmathsetmacro{\norm}{0.707}
    \pgfmathsetmacro{\offset}{0.1}
    \pgfmathsetmacro{\x}{3*\side+\offset*\norm}
    \pgfmathsetmacro{\y}{-3*\side-\offset*\norm}
    \pgfmathsetmacro{\xx}{4*\side-\offset*(2*\norm-1)}
    \pgfmathsetmacro{\yy}{-3*\side-\offset*\norm}
    \pgfmathsetmacro{\xxx}{4.5*\side-\offset*\norm}
    \pgfmathsetmacro{\yyy}{-3.5*\side-\offset*\norm}
    \pgfmathsetmacro{\xxxx}{5*\side-\offset*\norm}
    \pgfmathsetmacro{\yyyy}{-4*\side-\offset*(2*\norm-1)}
    \pgfmathsetmacro{\xxxxx}{5*\side-\offset*\norm}
    \pgfmathsetmacro{\yyyyy}{-5*\side+\offset*\norm}
    \pgfmathsetmacro{\xxxxxx}{4*\side+\offset*(2*\norm-1)}
    \pgfmathsetmacro{\yyyyyy}{-5*\side+\offset*\norm}
    \pgfmathsetmacro{\xxxxxxx}{3.5*\side+\offset*\norm}
    \pgfmathsetmacro{\yyyyyyy}{-4.5*\side+\offset*\norm}
    \pgfmathsetmacro{\xxxxxxxx}{3*\side+\offset*\norm}
    \pgfmathsetmacro{\yyyyyyyy}{-4*\side+\offset*(2*\norm-1)}

    \draw (2.5*\side,-2.5*\side) -- (3*\side, -3*\side);
    \draw (4*\side, -2*\side) -- (4*\side, -3*\side);
    \draw (5*\side, -4*\side) -- (6*\side, -4*\side);
    \draw (5*\side, -5*\side) -- (5.5*\side, -5.5*\side);
    \draw (4*\side, -5*\side) -- (4*\side, -6*\side);
    \draw (3*\side, -4*\side) -- (2*\side, -4*\side);

    \OddHexPlaquetteReduced{3*\side}{-3*\side}{\side}{black}

    \OrientedEdge{2.5*\side}{-2.5*\side}{3*\side}{-3*\side}{red}
    \OrientedEdge{3*\side}{-3*\side}{3*\side}{-4*\side}{red}
    \OrientedEdge{3*\side}{-4*\side}{3.5*\side}{-4.5*\side}{red}
    \SmallCircle[draw=red]{3.5*\side}{-4.5*\side}{0}{0.075}
    
    \OrientedEdge{\x}{\y}{\xx}{\yy}{blue}
    \OrientedEdge{\xx}{\yy}{\xxx}{\yyy}{blue}
    \OrientedEdge{\xxx}{\yyy}{\xxxx}{\yyyy}{blue}
    \OrientedEdge{\xxxx}{\yyyy}{\xxxxx}{\yyyyy}{blue}
    \OrientedEdge{\xxxxx}{\yyyyy}{\xxxxxx}{\yyyyyy}{blue}
    \OrientedEdge{\xxxxxx}{\yyyyyy}{\xxxxxxx}{\yyyyyyy}{blue}
    \OrientedEdge{\xxxxxxx}{\yyyyyyy}{\xxxxxxxx}{\yyyyyyyy}{blue}
    \OrientedEdge{\xxxxxxxx}{\yyyyyyyy}{\x}{\y}{blue}
    \end{tikzpicture}
    \end{minipage}
    
    \caption{Plaquette interaction (blue) on a state with two excited external links (left), and a pair of an excited external link and a charge excitation (right). Although these illustrations show all $8+6=14$ links involved, the external link irreps (and charge excitations, if any) can be deduced from Gauss's law applied to the $8$ central links.}\label{fig:TT2new1}
\end{figure}
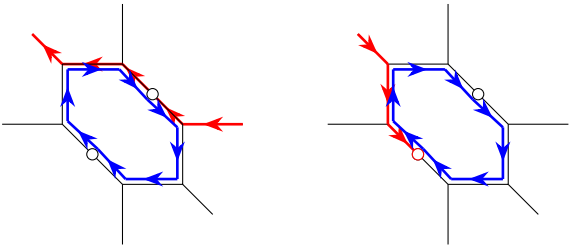

Order $1/N_c$ contributions to the plaquette operator only occur at this truncation when the plaquette intersects exactly one line of electric flux. This means that for these transitions, either two neighboring links of the plaquette are excited and the matter sites on the plaquette are unoccupied, or one external link is excited, and one of the matter sites is occupied.
These two cases are illustrated in \Cref{fig:TT2new1}.

For a given set of external fluxes flowing into the plaquette and matter content on the plaquette, there are two possible gauge-invariant field configurations, and the plaquette operator will flip between them. Denoting a state on a plaquette with a single contiguous line of electric flux by $\ket{c}$, we can define $\ket{F_p(c)}$, which is the flipped field configuration on the plaquette that keeps the external flux and matter content on the plaquette $p$ unchanged.
The plaquette operator for truncation \TTtwo\ is then
\begin{align}
    H^{\mathrm{plaq}}_{\text{\TTtwo}} &= H^{\mathrm{plaq}}_{\text{\TTone}}-\frac1{2g^2 N_c} \sum_{p\in\pq^{\mathrm{hex}}} \sum_{c \in \mathcal{C}_p} \ket{F_p(c)} \bra{c},
\end{align}
where $\mathcal{C}_p$ is the set of states where plaquette $p$ has a single contiguous field line on it.

The logic for hopping interactions is similar.
Begin by considering an arbitrary meson-like string operator $O_{\gamma}$ where $\gamma$ is a non-intersecting path comprised of $r$ links ($r$ must be odd for open strings, which also implies that $\gamma$ does not close into a loop).
The amplitude of $O_{\gamma}$ to create or annihilate an entire string in isolation is $\sqrt{N_c}$, according to \eqref{eq:Omaster}.
Now suppose the path $\gamma$ is not totally isolated in the initial state $\ket{i}$.
Without loss of generality, we can assume that neither of the endpoints of $\gamma$ are initially excited (which corresponds to $O_{\gamma}$ being a \textit{raising operation}).
Then the initial state contains $k$ isolated, contiguous field line segments along the links of $\gamma$.
Label these segments with indices $j = 1, \dots, k$, and denote the length of the $j^{\mathrm{th}}$ segment by $\alpha_j$.
In this case, the final state $\ket{f}$ contains $k-1$ isolated, contiguous field line segments along $\gamma$ that do not touch the endpoints of $\gamma$.
Additionally, $\ket{f}$ contains exactly two isolated, contiguous field line segments that terminate in a fermion or anti-fermion excitation at the endpoints of $\gamma$.
Label the full collection of segments with indices $j = 0, \dots, k$, where the segments that terminate are indexed $j=0$ and $j=k$, and denote the length of the $j^{\mathrm{th}}$ segment by $\beta_j$ (it is possible for $\beta_0$ or $\beta_k$ to be zero, if the final state contains excitations at the endpoints of $\gamma$ that do not extend into $\gamma$).
Then by the same logic as for the loop operators, we have
\begin{align}
    v_i &= v + \sum_{j=1}^k (\alpha_j + 1), \nonumber\\
    v_f &= v + \sum_{j=0}^k (\beta_j + 1),\nonumber\\
    r &= \sum_{j=1}^k \alpha_j + \sum_{j=0}^k \beta_j,\nonumber\\
    s &= \sum_{j=1}^k (\alpha_j - 1),
\end{align}
where $v$ again denotes the number of vertices that have two excited legs and do not lie on $\gamma$ or its endpoints.
This implies that
\begin{equation}
    v_f - v_i - r + 2s = -2k+1,
\end{equation}
so that
\begin{equation}
    \bra{f} O_{\gamma} \ket{i} = \frac1{N_c^k} \cdot \sqrt{N_c}.
\end{equation}
As for the case with closed strings, we see that the matrix elements for open strings are also entirely determined by the number of contiguous field line segments that the path $\gamma$ overlaps with, but there is an offset in matrix elements by a factor of $\sqrt{N_c}$. 

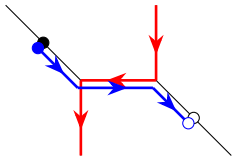
\begin{figure}[htp]
    \centering
    
    \begin{minipage}{0.5\linewidth}
    \begin{tikzpicture}
    \pgfmathsetmacro{\side}{1.0}
    \pgfmathsetmacro{\norm}{0.707}
    \draw (0,0) -- (\side, -\side) -- (2*\side, -\side) -- (3*\side, -2*\side);
    \SmallCircle{0.5*\side}{-0.5*\side}{1}{0.075}
    \SmallCircle{2.5*\side}{-1.5*\side}{0}{0.075}
    \pgfmathsetmacro{\offset}{0.1}
    \pgfmathsetmacro{\x}{0.5*\side-\offset*\norm}
    \pgfmathsetmacro{\y}{-0.5*\side-\offset*\norm}
    \pgfmathsetmacro{\xx}{\side-\offset*(2*\norm-1)}
    \pgfmathsetmacro{\yy}{-\side-\offset}
    \pgfmathsetmacro{\xxx}{2*\side-(2*\norm-1)*\offset}
    \pgfmathsetmacro{\yyy}{-\side-\offset}
    \pgfmathsetmacro{\xxxx}{2.5*\side-\offset*\norm}
    \pgfmathsetmacro{\yyyy}{-1.5*\side-\offset*\norm}
    
    \draw (\side, -\side) -- (\side, -2*\side);
    \draw (2*\side, -\side) -- (2*\side, 0);

    \OrientedEdge{2*\side}{0}{2*\side}{-\side}{red}
    \OrientedEdge{2*\side}{-\side}{\side}{-\side}{red}
    \OrientedEdge{\side}{-\side}{\side}{-2*\side}{red}
    
    \OrientedEdge{\x}{\y}{\xx}{\yy}{blue}
    \OrientedEdge{\xx}{\yy}{\xxx}{\yyy}{blue}
    \OrientedEdge{\xxx}{\yyy}{\xxxx}{\yyyy}{blue}
    \SmallCircle[draw=blue, fill=blue]{\x}{\y}{1}{0.075}
    \SmallCircle[draw=blue]{\xxxx}{\yyyy}{0}{0.075}
    \end{tikzpicture}
    \end{minipage}
    
    \caption{Hopping operator (blue) acting on a state with one intermediate line of flux (with no charge excitations on the matter sites).}\label{fig:TT2new2}
\end{figure}

For each cardinal link $\ell \in \lk_c^{\mathrm{hex}}$ and a given set of external fluxes flowing into the link, there are at most two possible allowed gauge invariant states depending on whether or not the sites are occupied. The hopping term has $1/\sqrt{N_c}$ contributions at this truncation for states that have a single contiguous line of electric flux flowing through the path $(\ell_{\bullet},\ell,\ell_{\circ})$.
See \Cref{fig:TT2new2} for an example illustrating a transition of this type.

Analogous to the plaquette, we can define $\mathcal{C}_\ell$, the set of basis states on the cardinal link $\ell$ with a single contiguous line of electric flux flowing through the path $(\ell_{\bullet},\ell,\ell_{\circ})$ (in either direction)---and with either both matter sites occupied or both unoccupied.
For each $c\in \mathcal{C}_\ell$ where both matter sites are unoccupied, we can define $F_\ell(c)$ which takes $c$ to the gauge invariant state with the same external flux and both matter sites occupied. The hopping term then takes the form
\begin{align}\nonumber
    H^{\hop}_{\text{\TTtwo}} =&\frac{1}{2\sqrt{N_c}} \sum_{\ell \in \lk_c^{\mathrm{hex}}} \sum_{c \in \mathcal{C}_\ell} \big(\ket{F_\ell(c)} \bra{c} + \ket{c}\bra{F_\ell(c)}\big) \\
    &+H^{\hop}_{\text{\TTone}}, \,
\end{align}
for truncation \TTtwo.
As with the plaquette term, note that truncating to $\mathcal O(1)$ in the large $N_c$ expansion produces exactly the same result as in truncation \TTone.

\subsection{External Charges}\label{sec:external}
Adding external charges to a lattice gauge theory changes the sector of the Hilbert space, and a different initial state will be needed to construct a Hamiltonian.
In order to simulate the dynamics of a single string with endpoints pinned by external charges on the lattice, we therefore need to derive \textit{modified} Hamiltonians that can parallel the truncations discussed previously.
To keep the discussion concise, we will focus on modifying truncations \Ttwo (in $1+1$D) and \TTtwo (in $2+1$D), for very specific lattice sizes and initial string configurations, so that the modified theories we simulate in \Cref{sec:simuls} are well-defined.

The key aspect of our approach here is that we will no longer use the free vacuum $\ket{0}$ of the Kogut-Susskind Hamiltonian to generate Krylov states, because $\ket{0}$ does not exist as a state in the modified Hilbert space.
Instead, our Krylov states will be generated by an initial state $\ket{\psi_0}$ that represents a string spanning a total of $r$ links, pinned by two external charges (one at each endpoint).
Since we ultimately want to simulate the time evolution of $\ket{\psi_0}$, this plays nicely with the intuition that a Krylov-based truncation can be used to approximate the time evolution of the same initial state that generates the Krylov subspace in the first place.

The standard method to add external charges to a Hamiltonian lattice gauge theory involves modifying Gauss's law at the specific lattice sites where the charges are added, without interfering with the rest of the dynamics.
In our setting, if the endpoints of the initial string are two lattice sites $v_1$ and $v_2$, then we can implement this change by appending a new Hilbert space to both $v_1$ and $v_2$, with the new spaces furnishing appropriate irreps of $\SU(N_c)$.
The aim is to allow links in the fundamental representation to be contracted into a gauge-invariant state against the new states supported at $v_1$ and $v_2$.
In doing so, we effectively modify the total gauge-invariant Hilbert space for the lattice \textit{without} modifying the Kogut-Susskind Hamiltonian itself, thus enabling time evolution of a state that is not gauge-invariant from the perspective of Kogut-Susskind dynamics.

To organize our discussion, we will begin by modifying truncation \Ttwo\ on a periodic 1D spatial lattice with $L$ lattice sites, assuming an initial string of odd length $r$.
We will follow this by showing how the logic changes for modifying truncation \TTtwo\ on a 2D chain of four hexagonal plaquettes with open boundaries, assuming an initial string that lives on nine links.
Larger lattices in $2+1$D and higher dimensions follow completely analogous logic, but their description is more cumbersome, and therefore beyond the scope of our present discussion.

\subsubsection{$1+1$D Case: Modifying \Ttwo}

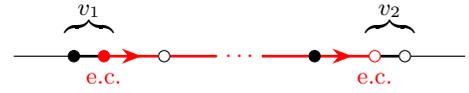
\begin{figure}[htp]
    \centering

    \begin{minipage}[][2cm]{\linewidth}
    \begin{tikzpicture}
    \draw (0,0) -- (2.7,0);
    \PlaceEq{3}{0}{\textcolor{red}{\cdots}}
    \draw (3.3,0) -- (6,0);
    \SmallCircle{0.8}{0}{1}{0.075}
    \SmallCircle[draw=red, fill=red]{1.2}{0}{1}{0.075}
    \SmallCircle{2}{0}{0}{0.075}
    \SmallCircle{4}{0}{1}{0.075}
    \SmallCircle[draw=red]{4.8}{0}{0}{0.075}
    \SmallCircle{5.2}{0}{0}{0.075}
    \PlaceEq{1.2}{-0.3}{\text{\textcolor{red}{e.c.}}}
    \PlaceEq{4.8}{-0.3}{\text{\textcolor{red}{e.c.}}}
    \PlaceEq{1}{0.5}{\overbrace{\quad\quad}^{\displaystyle v_1}}
    \PlaceEq{5}{0.5}{\overbrace{\quad\quad}^{\displaystyle v_2}}

    \draw[line width=1pt] (0.88,0) -- (1.12,0);
    \draw[line width=1pt] (4.88,0) -- (5.12,0);
    \OrientedEdge{1.28}{0}{1.92}{0}{red}
    \draw[draw=red, line width=1pt] (2.08,0)--(2.7,0);
    \OrientedEdge{4.08}{0}{4.72}{0}{red}
    \draw[draw=red, line width=1pt] (3.3,0)--(3.92,0);
    \end{tikzpicture}
    \end{minipage}
    
    \caption{Graphical illustration of point-splitting the external charges pinning a long string (red). The original vertex singlets at the endpoints $v_1$ and $v_2$ have each been split into two lattice sites by a virtual link (bolded).}\label{fig:extcharge}
\end{figure}

Let us assume the initial state $\ket{\psi_0}$ is a string of length $r$ between an odd site $v_1$ and an even site $v_2$, and furthermore suppose that the $r$ links comprising the string are indexed $\ell_0, \dots, \ell_{r-1}$, in order from $v_1$ to $v_2$, where $\ell_j = \ell_0 + j$.
For simplicity, we will discard the special cases $r = 1$ and $r = L-1$ (where $L$ is the size of the lattice), which have a different character compared to the general setting.
Then we can express the local data of the initial state in the form
\begin{widetext}
\begin{equation}
    \ket{\psi_0} = \frac1{\sqrt{N_c^{r+1}}}\ket{\overline{\mathbf{N}},a_0}^{\mathrm{e.c.}}_{v_1} \ket{\mathbf{N},a_1,a_0}_{\ell_0} \ket{\mathbf{1}} \ket{\overline{\mathbf{N}},a_2,a_1}_{\ell_1}\ket{\mathbf{1}}\cdots\ket{\mathbf{N},a_r,a_{r-1}}_{\ell_{r-1}} \ket{\mathbf{N},a_r}^{\mathrm{e.c.}}_{v_1}\,,
\end{equation}
\end{widetext}
where the superscript $\mathrm{e.c.}$ denotes that the state is part of the ``external charge Hilbert space" and therefore does not participate in hopping transitions (see \Cref{fig:extcharge} for a graphical depiction of this initial state).
This is the reason one can think of the external charges as an entirely separate species of fermions that do not contribute anything to the Hamiltonian.
We also omit the indices for lattice sites in the singlet representation, for brevity.

The aim is now to determine which states can be reached from $\ket{\psi_0}$ under the Krylov constraints $(n_h, n_\chi) = (1, 2)$.
Away from the initial string, the constraints allow for precisely the same operations as in truncation \Ttwo.
On the string itself, but away from the external charges, the operations from truncation \Ttwo\ are only permitted as long as they don't reverse the orientation of any link, because breaking the string and deleting isolated quark-antiquark pairs requires at most two fermion operations per site; but further exciting a quark-antiquark pair in the opposite orientation requires an additional link and fermion operation that is not allowed at this truncation.
The creation or annihilation of a quark-antiquark pair is permitted if and only if the excitation does not overlap with already occupied sites. 
Note that sites $v_1$ and $v_2$ cannot participate in any string-joining or string-breaking operations, given the energy cutoff $(n_e, n_f) = (1, 1)$.

The modifications we need to make to the truncated Hamiltonian $H_{\text{\Ttwo}}$ are therefore two-fold: (i) we must modify the hopping terms adjacent to the external charges; and (ii) we must remove terms that can create excitations with the opposite orientation inside the string.
The latter is achieved by simply omitting terms in the Hamiltonian that do not have the desired character.

Before we can write down the modified hopping terms, we must first understand the truncated Hilbert space after modifying Gauss's law.
The truncated singlet subspaces at $v_1$ and $v_2$ should each have dimension $3$, because an orthogonal basis of allowed states is obtained from the following actions (applied sequentially): (i) doing nothing (keeping the original vertex tied to the string); (ii) creating a quark-antiquark pair on the \textit{opposite side} from the initial string; (iii) deleting a quark-antiquark pair on the \textit{same side} as the initial string.
Any further operations will exceed the Krylov constraints near the external charges.

The fact that the Krylov constraints generate an orthogonal basis of states is a special feature of the low-lying truncation with $(n_h, n_\chi) = (1, 2)$.
At higher truncations, Krylov states can be generated that are linearly independent but \textit{not orthgonal} to each other.
In such cases, the representation of the system on a quantum computer would require a qudit assignment based on an orthonormal basis, such as the point-split basis.
\Cref{fig:extcharge} shows one possible virtual point-splitting in graphical form , which is useful to visualize in the forthcoming discussion.
At the low-lying truncation we are dealing with, the key point is that there is only one Krylov state $\ket{\psi}$ that requires activating the virtual link in the point-split basis, and this state can be uniquely identified from specifying the irreps only on physical links.
This implies that the original qutrit mapping still suffices to uniquely label all Krylov states on the unsplit lattice.

There are now two hopping transitions that we must consider to fully specify the truncated hopping terms in the Krylov basis.
The first transition creates a quark-antiquark pair sharing a lattice site with the external charge; and can only occur on the side of the external charge that lies outside the original string.
This transition can be understood via point-splitting, which separates the external charge Hilbert space from the physical charge Hilbert space by a virtual link in the singlet representation, thereby creating a quark-antiquark pair that does not intertwine with the external charge.
According to \Cref{tab:hops}, such a transition has matrix element $\sqrt{N_c}$.

The second transition annihilates the outermost link of the initial string, by hopping a charge from inside the string through the endpoint where the external charge is located, and annihilating a massive (not external) charge, if present there.
This transition does not already appear in \Cref{tab:hops}, but can be computed explicitly by working with local data as
\begin{widetext}
\begin{align}
    H_{\hop}(\ell_0)^{\dagger} \left(\frac1{N_c} \ket{\mathbf{N},a} \ket{\mathbf{N},a,b}_{\ell_0-1} \ket{\overline{\mathbf{N}},b}_{v_1}\right) \otimes \left(\frac1{N_c} \ket{\overline{\mathbf{N}},c}_{v_1}^{\mathrm{e.c.}} \ket{\mathbf{N},d,c}_{\ell_0} \ket{\mathbf{N},d}\right) &= \frac1{\sqrt{N_c}} \ket{\psi},
\end{align}
\end{widetext}
where $\ket{\psi}$ is a normalized quantum state on the local data---in particular, it is the state referenced earlier that requires an activated virtual link in the point-split basis.
In terms of local data on the physical lattice, the state can simply be expressed as
\begin{equation}
    \ket{\psi} = \frac 1{N_c} \ket{\mathbf{N},a'} \ket{\mathbf{N},a',b'}_{\ell_0-1} \ket{\mathbf{1}}_{v_1}\ket{\overline{\mathbf{N}},b'}_{v_1}^{\mathrm{e.c.}} \ket{\mathbf{1}}_{\ell_0} \ket{\mathbf{1}},
\end{equation}
where we omit indices for links and sites in the singlet representation.
The case of $v_2$ is similar, giving rise to the same hopping matrix elements for the analogous transitions.

The modified hopping piece of the Hamiltonian needs to include most of the PXP terms from truncation \Ttwo, but we must replace the PXP terms around the external charges with the modified transitions as discussed above.
This leads to
\begin{widetext}
\begin{align}
    H_{\text{\Ttwo, }\mathrm{e.c.}}^{\hop} &= \frac{\sqrt{N_c}}2\sum\limits_{\substack{j=1 \\ j \notin A\cup B}}^L P^{\mathbf{1}}_{j-1} \mathcal{X}^{\mathbf{1}\mathbf{N}}_j P^{\mathbf{1}}_{j+1} + \frac1{2\sqrt{N_c}} \sum\limits_{\substack{j=1 \\ j \notin A}}^L P^{\mathbf{N}}_{j-1} \mathcal{X}^{\mathbf{1}\overline{\mathbf{N}}}_j P^{\mathbf{N}}_{j+1} \nonumber
    + \frac{\sqrt{N_c}}2 \left(P^{\mathbf{1}}_{\ell_0-2} \mathcal{X}^{\mathbf{1}\mathbf{N}}_{\ell_0-1} P^{\mathbf{N}}_{\ell_0} + P^{\mathbf{N}}_{\ell_{r-1}} \mathcal{X}^{\mathbf{1}\mathbf{N}}_{\ell_{r-1}+1} P^{\mathbf{1}}_{\ell_{r-1}+2}\right) \nonumber\\
    &\qquad + \frac1{2\sqrt{N_c}} \left(P^{\mathbf{N}}_{\ell_0-1} \mathcal{X}^{\mathbf{1}\mathbf{N}}_{\ell_0} P^{\mathbf{1}}_{\ell_0+1} + P^{\mathbf{1}}_{\ell_{r-1}-1} \mathcal{X}^{\mathbf{1}\mathbf{N}}_{\ell_{r-1}} P^{\mathbf{N}}_{\ell_{r-1}+1}\right),
\end{align}
\end{widetext}
where $A \equiv \{\ell_0-1,\ell_0,\ell_{r-1},\ell_{r-1}+1\}$ is the set of links adjacent to any external charge, and
\begin{equation}
    B \equiv \left\{\ell \in [\ell_0, \ell_{r-1}] \mid \ell - \ell_0 \equiv 1 \pmod 2\right\}
\end{equation}
is the set of links in the string that cannot flip their initial orientation.
The full Hamiltonian with external charges is given by adding these terms to the otherwise unchanged electric and mass terms from truncation \Ttwo:
\begin{equation}
    H_{\text{\Ttwo, }\mathrm{e.c.}} = H_{\text{\Ttwo}}^{\mathrm{elec}} + H_{\text{\Ttwo}}^{\mathrm{mass}} + H_{\text{\Ttwo, }\mathrm{e.c.}}^{\hop}.
\end{equation}

\subsubsection{$2+1$D Case: Modifying \TTtwo}

\begin{figure}[htp]
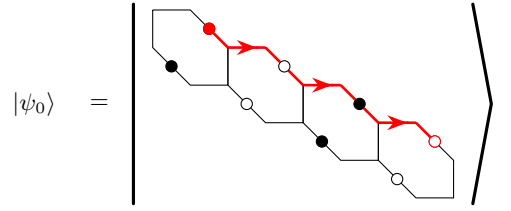

    \centering
    \begin{minipage}{0.1\linewidth}
    $\ket{\psi_0}$
    \end{minipage}%
    \begin{minipage}{0.1\linewidth}
    $=$
    \end{minipage}%
    \begin{minipage}{0.5\linewidth}
    \KetPic{
    \pgfmathsetmacro{\side}{0.5}
    \pgfmathsetmacro{\norm}{0.707}
    \EvenHexPlaquetteReduced{0}{0}{\side}{black}
    \OddHexPlaquetteReduced{2*\side}{-1*\side}{\side}{black}
    \EvenHexPlaquetteReduced{4*\side}{-2*\side}{\side}{black}
    \OddHexPlaquetteReduced{6*\side}{-3*\side}{\side}{black}
    \draw[draw=red, line width=1pt] (1.5*\side+0.08*\norm,-0.5*\side-0.08*\norm) -- (2*\side,-1*\side);
    \draw[draw=red, line width=1pt] (2*\side,-1*\side) -- (3*\side,-1*\side);
    \draw[draw=red, line width=1pt] (3*\side,-1*\side) -- (3.5*\side-0.08*\norm,-1.5*\side+0.08*\norm);
    \draw[draw=red, line width=1pt] (3.5*\side+0.08*\norm,-1.5*\side-0.08*\norm) -- (4*\side,-2*\side);
    \draw[draw=red, line width=1pt] (4*\side,-2*\side) -- (5*\side,-2*\side);
    \draw[draw=red, line width=1pt] (5*\side,-2*\side) -- (5.5*\side-0.08*\norm,-2.5*\side+0.08*\norm);
    \draw[draw=red, line width=1pt] (5.5*\side+0.08*\norm,-2.5*\side-0.08*\norm) -- (6*\side,-3*\side);
    \draw[draw=red, line width=1pt] (6*\side,-3*\side) -- (7*\side,-3*\side);
    \draw[draw=red, line width=1pt] (7*\side,-3*\side) -- (7.5*\side-0.08*\norm,-3.5*\side+0.08*\norm);
    \SmallCircle[draw=red, fill=red]{1.5*\side}{-0.5*\side}{1}{0.075}
    \SmallCircle[draw=red]{7.5*\side}{-3.5*\side}{0}{0.075}
    \OrientedEdge{2.7*\side}{-1*\side}{2.8*\side}{-1*\side}{red}
    \OrientedEdge{4.7*\side}{-2*\side}{4.8*\side}{-2*\side}{red}
    \OrientedEdge{6.7*\side}{-3*\side}{6.8*\side}{-3*\side}{red}
    }
    \end{minipage}
    \caption{Hexagonal plaquette chain with four plaquettes. The initial string is shown in red.}\label{fig:hexstring}
\end{figure}

Let us now assume that we are dealing with an initial string pinned by external charges in the $2+1$D hexagonal plaquette chain of size four, as shown in \Cref{fig:hexstring}.
One key point here is that the Kogut-Susskind Hamiltonian for this lattice can only allow fermion hops between nearest-neighbor pairs of lattice sites.
Since we assume open boundary conditions, this implies that there are only $9$ distinct pairs of neighboring sites, substantially restricting the number of hopping terms in the Hamiltonian (as compared to the periodic boundary case).
This simplification allows for a more concise description of the truncated model.
The strategy for modifying the Hamiltonian to include external charges follows logic that is precisely analogous to the $1+1$D case.

We must begin by understanding the truncated singlet subspaces around $v_1$ and $v_2$, with Krylov parameters $(n_p, n_h, n_\ell, n_\chi) = (1, 1, 2, 2)$.
Let us focus on $v_1$, because $v_2$ is similar.
Without breaking $(n_e, n_f) = (1, 1)$, the only interaction operator that can be applied to $\ket{\psi_0}$ in the vicinity of $v_1$ is a counterclockwise plaquette.
After that interaction has been applied, we have consumed our only available plaquette operation usage through $v_1$, so the only reachable states are obtained by applying a single hopping operation (any of the two available raising operations in the most general case).
This again leads to a singlet subspace of dimension $3$ around $v_1$ (and hence also around $v_2$), with orthogonal Krylov states. Note that on a larger 2D lattice, the Krylov states would \textit{not} be orthogonal, thus requiring a new point-splitting step, or the introduction of qudits in general.

Therefore, by the same logic as in the $1+1$D case, we can keep our qutrit notation and adjust the hopping terms adjacent to the external charges to reflect the matrix elements for the new allowed transitions.
Note in particular that the plaquette terms do not need to be modified, because the matrix elements are identical when $v_1$ (or $v_2$ as appropriate) contains a single charge; and the previous plaquette terms already annihilate the new vertex state with two charges, as necessary for this truncation.

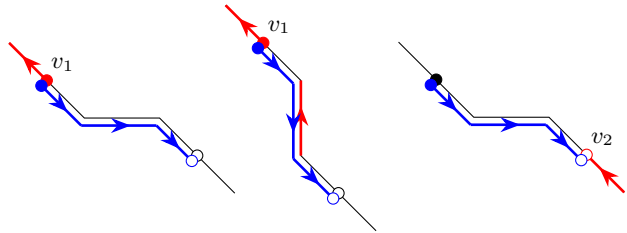
\begin{figure}[htp]
    \centering

    \begin{minipage}{0.3\linewidth}
    \begin{tikzpicture}
    \pgfmathsetmacro{\side}{1.0}
    \pgfmathsetmacro{\norm}{0.707}
    \draw (0,0) -- (\side, -\side) -- (2*\side, -\side) -- (3*\side, -2*\side);
    \SmallCircle[draw=red, fill=red]{0.5*\side}{-0.5*\side}{1}{0.075}
    \SmallCircle{2.5*\side}{-1.5*\side}{0}{0.075}
    \OrientedEdge{0.5*\side}{-0.5*\side}{0}{0}{red}
    \pgfmathsetmacro{\offset}{0.1}
    \pgfmathsetmacro{\x}{0.5*\side-\offset*\norm}
    \pgfmathsetmacro{\y}{-0.5*\side-\offset*\norm}
    \pgfmathsetmacro{\xx}{\side-\offset*(2*\norm-1)}
    \pgfmathsetmacro{\yy}{-\side-\offset}
    \pgfmathsetmacro{\xxx}{2*\side-(2*\norm-1)*\offset}
    \pgfmathsetmacro{\yyy}{-\side-\offset}
    \pgfmathsetmacro{\xxxx}{2.5*\side-\offset*\norm}
    \pgfmathsetmacro{\yyyy}{-1.5*\side-\offset*\norm}
    \OrientedEdge{\x}{\y}{\xx}{\yy}{blue}
    \OrientedEdge{\xx}{\yy}{\xxx}{\yyy}{blue}
    \OrientedEdge{\xxx}{\yyy}{\xxxx}{\yyyy}{blue}
    \SmallCircle[draw=blue, fill=blue]{\x}{\y}{1}{0.075}
    \SmallCircle[draw=blue]{\xxxx}{\yyyy}{0}{0.075}
    \PlaceEq{0.5*\side+0.3*\norm}{-0.5*\side+0.3*\norm}{v_1}
    \end{tikzpicture}
    \end{minipage}%
    \begin{minipage}{0.3\linewidth}
    \begin{tikzpicture}
    \pgfmathsetmacro{\side}{1.0}
    \pgfmathsetmacro{\norm}{0.707}
    \draw (0,0) -- (\side, -\side) -- (\side, -2*\side) -- (2*\side, -3*\side);
    \SmallCircle[draw=red, fill=red]{0.5*\side}{-0.5*\side}{1}{0.075}
    \SmallCircle{1.5*\side}{-2.5*\side}{0}{0.075}
    \OrientedEdge{0.5*\side}{-0.5*\side}{0}{0}{red}
    \OrientedEdge{\side}{-2*\side}{\side}{-\side}{red}
    \pgfmathsetmacro{\offset}{0.1}
    \pgfmathsetmacro{\x}{0.5*\side-\offset*\norm}
    \pgfmathsetmacro{\y}{-0.5*\side-\offset*\norm}
    \pgfmathsetmacro{\xx}{\side-\offset}
    \pgfmathsetmacro{\yy}{-\side-\offset*(2*\norm-1)}
    \pgfmathsetmacro{\xxx}{\side-\offset}
    \pgfmathsetmacro{\yyy}{-2*\side-(2*\norm-1)*\offset}
    \pgfmathsetmacro{\xxxx}{1.5*\side-\offset*\norm}
    \pgfmathsetmacro{\yyyy}{-2.5*\side-\offset*\norm}
    \OrientedEdge{\x}{\y}{\xx}{\yy}{blue}
    \OrientedEdge{\xx}{\yy}{\xxx}{\yyy}{blue}
    \OrientedEdge{\xxx}{\yyy}{\xxxx}{\yyyy}{blue}
    \SmallCircle[draw=blue, fill=blue]{\x}{\y}{1}{0.075}
    \SmallCircle[draw=blue]{\xxxx}{\yyyy}{0}{0.075}
    \PlaceEq{0.5*\side+0.3*\norm}{-0.5*\side+0.3*\norm}{v_1}
    \end{tikzpicture}
    \end{minipage}%
    \begin{minipage}{0.3\linewidth}
    \begin{tikzpicture}
    \pgfmathsetmacro{\side}{1.0}
    \pgfmathsetmacro{\norm}{0.707}
    \draw (0,0) -- (\side, -\side) -- (2*\side, -\side) -- (3*\side, -2*\side);
    \OrientedEdge{3*\side}{-2*\side}{2.5*\side}{-1.5*\side}{red}
    \SmallCircle{0.5*\side}{-0.5*\side}{1}{0.075}
    \SmallCircle[draw=red]{2.5*\side}{-1.5*\side}{0}{0.075}
    \pgfmathsetmacro{\offset}{0.1}
    \pgfmathsetmacro{\x}{0.5*\side-\offset*\norm}
    \pgfmathsetmacro{\y}{-0.5*\side-\offset*\norm}
    \pgfmathsetmacro{\xx}{\side-\offset*(2*\norm-1)}
    \pgfmathsetmacro{\yy}{-\side-\offset}
    \pgfmathsetmacro{\xxx}{2*\side-(2*\norm-1)*\offset}
    \pgfmathsetmacro{\yyy}{-\side-\offset}
    \pgfmathsetmacro{\xxxx}{2.5*\side-\offset*\norm}
    \pgfmathsetmacro{\yyyy}{-1.5*\side-\offset*\norm}
    \OrientedEdge{\x}{\y}{\xx}{\yy}{blue}
    \OrientedEdge{\xx}{\yy}{\xxx}{\yyy}{blue}
    \OrientedEdge{\xxx}{\yyy}{\xxxx}{\yyyy}{blue}
    \SmallCircle[draw=blue, fill=blue]{\x}{\y}{1}{0.075}
    \SmallCircle[draw=blue]{\xxxx}{\yyyy}{0}{0.075}
    \PlaceEq{2.5*\side+0.3*\norm}{-1.5*\side+0.3*\norm}{v_2}
    \end{tikzpicture}
    \end{minipage}
    
    \caption{The new hopping transitions allowed on a four-plaquette hexagonal chain with external charges at $v_1$ and $v_2$. Red indicates electric field lines and (external) charges in the initial state. Blue indicates the Wilson line used to transition to the final state. To avoid clutter, only two legs of both intermediate trivalent vertices are drawn, but the remaining irreps can be inferred from Gauss's law. The matrix element are denoted $M_{\ell,c} = \sqrt{N_c}$ (left), $M_{\ell,c} = 1/\sqrt{N_c}$ (middle), and $M_{\ell,c} = \sqrt{N_c}$ (right).}\label{fig:exthops}
\end{figure}

Moreover, there are very few newly allowed hopping transitions involving $v_1$ and $v_2$, which can easily be listed out as shown in \Cref{fig:exthops}, and their matrix elements follow directly from \eqref{eq:Omaster}.
These transitions are not already part of the Hamiltonian $H_{\text{\TTtwo}}$ because they result from hops that leave two charges on the same lattice site.
To incorporate these transitions into the truncated hopping term, $H_{\text{\TTtwo, }\mathrm{e.c.}}^{\hop}$, we can redefine $F_\ell$ to take into account the presence of the external charges in the gauge constraints.

Specifically, we need to add in the three transitions discussed above, and subtract out the transitions that correspond to annihilating any external charge.
If $A$ is the set of cardinal links that reside only a single diagonal link away from any external charge, then
\begin{align}
H_{\text{\TTtwo, }\mathrm{e.c.}}^{\hop} &= H_{\text{\TTtwo}} - \frac{\sqrt{N_c}}2 \sum_{\ell \in A} P^{\mathbf{1}}_{\ell^*_{\bullet}}\left(B^{\mathbf{N},\mathbf{1}}_{\ell} + B^{\mathbf{N},\mathbf{1} \dagger}_{\ell} \right)P^{\mathbf{1}}_{\ell^*_{\circ}}\nonumber\\
& + \sum_{\ell\in A} \sum_{c\in \mathcal C_{\ell}^{\mathrm{e.c.}}} \frac{M_{\ell,c}}{2} \big(\ket{c}\bra{F_{\ell}^{\mathrm{e.c.}}(c)} + \ket{F_{\ell}^{\mathrm{e.c.}}(c)}\bra{c}\big),
\end{align}
where $\mathcal C_{\ell}^{\mathrm{e.c.}}$ denotes the set of basis states around a cardinal link that contain a single external charge together with link or matter excitations in any of the configurations shown in \Cref{fig:exthops} along the path $(\ell_{\bullet},\ell,\ell_{\circ})$; and $F_{\ell}^{\mathrm{e.c.}}$ modifies every $c \in \mathcal C_{\ell}^{\mathrm{e.c.}}$ by shifting the excitations at the links $(\ell_{\bullet},\ell,\ell_{\circ})$ by one unit of flux in the direction of the hopping Wilson line, while also adding dynamical charge excitations at the endpoints.
We used here the notation $M_{\ell, c}$ to denote the value of the matrix elements for each transition specified by cardinal link $\ell$ and state $c$, with the exact values listed in the caption of \Cref{fig:exthops}.
Given this modification, the truncated Hamiltonian with the external charges is given by
\begin{equation}
    H_{\text{\TTtwo, }\mathrm{e.c.}} = H_{\text{\TTtwo}}^{\mathrm{elec}} + H_{\text{\TTtwo}}^{\mathrm{mass}} + H_{\text{\TTtwo}}^{\mathrm{plaq}} + H_{\text{\TTtwo, }\mathrm{e.c.}}^{\hop}.
\end{equation}

\section{Simulating Real-Time Lattice QCD}\label{sec:simuls}

Real-time simulations of lattice gauge theories (LGTs) have recently seen a resurgence of interest, many decades since the Kogut-Susskind formulation was originally developed.
One reason is that quantum computers offer the most practical solution for scaling up any Hamiltonian simulation when Hilbert spaces become too large for classical hardware.
Another key insight has been that many-body systems of interest in condensed matter physics have recently been realized as discrete truncations of lattice gauge theories \cite{Halimeh2023ColdatomQuantumSimulators,Halimeh2025QuantumSimulationOutofequilibrium}.
Conversely, it is now possible to understand lattice gauge theories by re-expressing their truncations in the many-body framework.
Prior work in this direction has investigated domain walls, string breaking, and glueball formation, in both abelian LGTs \cite{Borla2025StringBreaking$2+1$D,Xu2025StringBreakingDynamics,Tian2025RolePlaquetteTerm,Barata2025RealTimeSimulationJetEnergyLoss,Marcantonio2025RougheningDynamicsElectric,Xu2025TensorNetworkStudyRoughening,artiaco2025outofequilibriumdynamicsu1lattice,cao2026stringbreakingglueballdynamics} and non-abelian LGTs \cite{Kuhn2015NonAbelianStringBreaking,cataldi2025realtimestringdynamics21d}.
In this section, we provide simulation results for truncated non-abelian LGTs with fermions, taking $N_c \ge 3$.
This bridges the gap between prior work and lattice QCD, for the case of a single staggered quark flavor.

Our main focus is the real-time evolution of an initial state (typically the free vacuum or an open string), according to the truncated Hamiltonians derived in \Cref{sec:hams}.
We performed these simulations by exact diagonalization on classical hardware, allowing us to emulate a high-performance quantum processor.
The discussion is organized as follows: in \Cref{sec:freevac}, we showcase the time evolution of the free vacuum and the smallest possible meson-like excitation; in \Cref{sec:openstring}, we compare the time evolution of a plain open string vs. an open string pinned by external charges, and discuss the resonances for string breaking.
Our simulations will be performed in both $1+1$D (periodic spatial lattice) and $2+1$D (open boundary plaquette chain).

In order to gain an understanding of the physics underlying the initial state's time evolution, we will concern ourselves with several observables, which can be measured by compiling their matrix elements in the basis of gauge-invariant lattice states.
These observables include, but are not limited to:
\begin{itemize}
    \item Fidelity against an arbitrary state $\ket{\phi}$, which can be computed as $\langle \mathcal F_\phi \rangle$, where $\mathcal F_{\phi}$ denotes the projector $\mathcal F_{\phi} = \ket{\phi}\bra{\phi}$.
    \item Total number of quarks and antiquarks on the lattice (which is also twice the number of open strings), denoted $\mathcal N_q$.
    \item Average length among all distinct strings on the lattice, denoted $\mathcal L_s$.
    \item Number of closed strings that surround only one plaquette, denoted $\mathcal G$.\footnote{In the limit of strong coupling, this counts glueballs~\cite{Kogut1975HamiltonianFormulationWilsons}.} (Only in $2+1$D.)
\end{itemize}

\subsection{Free Vacuum and Meson-Like Excitation}\label{sec:freevac}

For low-lying states such as the free vacuum $\ket{0}$ and a meson-like excitation $\ket{\phi}$, we perform simulations using the truncated Hamiltonians from \Cref{sec:1+1d} and \Cref{sec:2+1d}, i.e., without external charges.
Our discussion is organized by the spatial dimension of the lattice.

\begin{figure}[htp]
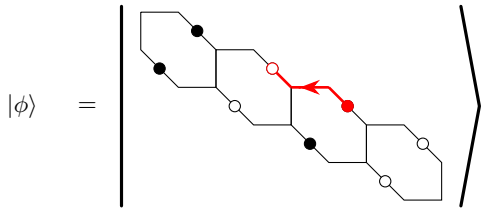

    \centering
    \begin{minipage}{0.1\linewidth}
    $\ket{\phi}$
    \end{minipage}%
    \begin{minipage}{0.1\linewidth}
    $=$
    \end{minipage}%
    \begin{minipage}{0.5\linewidth}
    \KetPic{
    \pgfmathsetmacro{\side}{0.5}
    \pgfmathsetmacro{\norm}{0.707}
    \EvenHexPlaquetteReduced{0}{0}{\side}{black}
    \OddHexPlaquetteReduced{2*\side}{-1*\side}{\side}{black}
    \EvenHexPlaquetteReduced{4*\side}{-2*\side}{\side}{black}
    \OddHexPlaquetteReduced{6*\side}{-3*\side}{\side}{black}
    \draw[draw=red, line width=1pt] (3.5*\side+0.08*\norm,-1.5*\side-0.08*\norm) -- (4*\side,-2*\side);
    \draw[draw=red, line width=1pt] (4*\side,-2*\side) -- (5*\side,-2*\side);
    \draw[draw=red, line width=1pt] (5*\side,-2*\side) -- (5.5*\side-0.08*\norm,-2.5*\side+0.08*\norm);
    \SmallCircle[draw=red, fill=red]{5.5*\side}{-2.5*\side}{1}{0.075}
    \SmallCircle[draw=red]{3.5*\side}{-1.5*\side}{0}{0.075}
    \OrientedEdge{4.3*\side}{-2*\side}{4.2*\side}{-2*\side}{red}
    }
    \end{minipage}
    \caption{Initial meson-like excitation (red) that we simulate on the $2+1$D hexagonal plaquette chain.}\label{fig:hexmeson}
\end{figure}

In both $1+1$D and $2+1$D, the free vacuum $\ket{0}$ can be defined in the irrep basis as simply the state where all links are in the singlet representation.
The meson-like excitation $\ket{\phi}$ is obtained by exciting a single quark-antiquark pair on top of the free vacuum.
In the $1+1$D case, it doesn't matter which link holds the quark-antiquark pair, since the entire system is translation-invariant due to periodic boundary conditions; but it matters in $2+1$D, because we use open boundary conditions on our plaquette chain.
In the latter case, the initial state we use is shown in~\Cref{fig:hexmeson}.

In accordance with our notation above, we use $\langle \mathcal F_0 \rangle$ and $\langle \mathcal F_{\phi}\rangle$ to denote the fidelities against $\ket{0}$ and $\ket{\phi}$, respectively, which will be plotted alongside the observables $\mathcal N_q$ and $\mathcal G$, as applicable.

\subsubsection{$1+1$D Case}

Our simulations are performed for a 1D spatial lattice with periodic boundaries containing $L = 10$ links (equal to the number of lattice sites).
Roughly speaking, there are two regimes of parameter values that can be explored in the truncated theories: (i) the strongly-coupled regime, where $g$ and $m$ are both large; and (ii) the weakly-coupled regime, where $g$ and $m$ are both small.
By the nature of our truncation scheme, the highest accuracy is achieved in the strongly-coupled regime, but we find that even the weakly-coupled regime presents interesting qualitative features that persist through all truncations \Tone, \Ttwo, \Tthreep, \Tthree, \Tfour, and \Tfive.

\begin{figure}[!th]
    \centering

    \begin{minipage}{\twocolumnwidth}
        \centering
        \includegraphics[width=\twocolumnwidth]{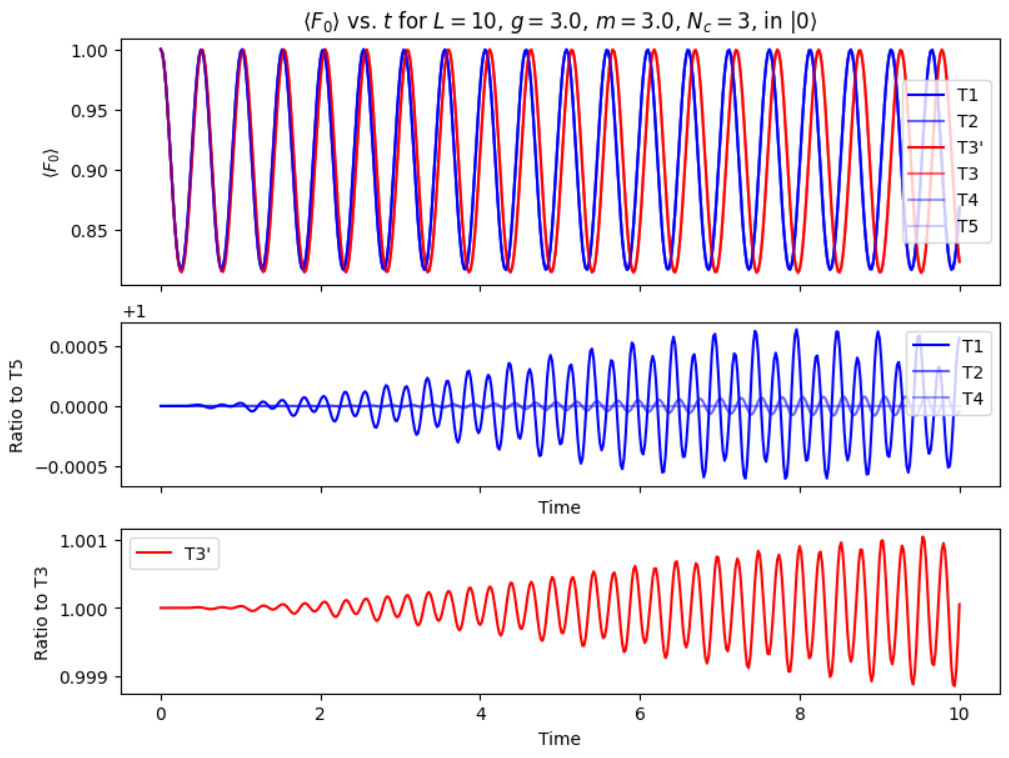}
    \end{minipage}%
    \hfill
    \begin{minipage}{\twocolumnwidth}
        \centering
        \includegraphics[width=\twocolumnwidth]{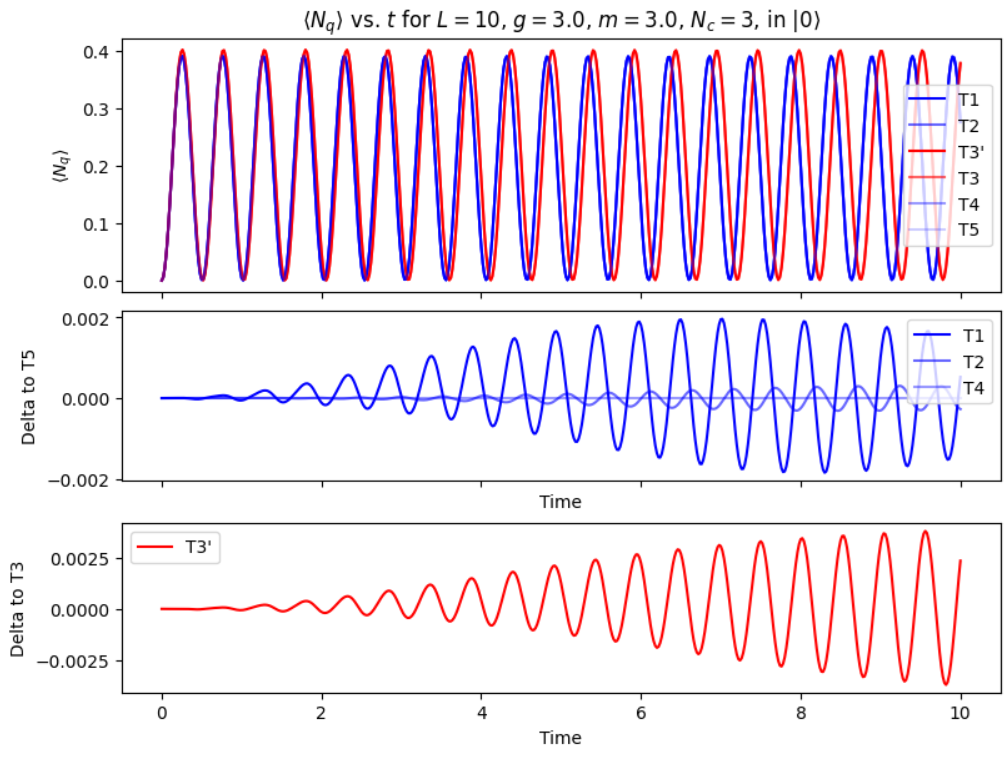}
    \end{minipage}

    \caption{Real-time evolution of $\mathcal F_0$ and $\mathcal N_q$ for the free vacuum state $\ket{0}$ at large coupling and mass (top). The six truncations give rise to only two distinguishable curves, with slightly different frequencies. Truncations \Tone, \Ttwo, \Tfour, and \Tfive\ all fall on the slightly faster-oscillating curve (blue); whereas \Tthree' and \Tthree\ fall on the slower-oscillating curve (red). Plotting ratios to the highest truncation of a given color helps to zoom in on the differences. For observables like $\mathcal N_q$ that nearly reach zero, the difference to the highest truncation is given instead.}\label{fig:strongvac1D}
\end{figure}

The time evolution of the free vacuum at large parameter values (namely, $g=3.0$ and $m=3.0$) is illustrated by the plots in~\Cref{fig:strongvac1D}.
Qualitatively, we see that the initial free vacuum state is highly preferred by the dynamics, since the fidelity tends to oscillate near $\langle \mathcal F_0 \rangle \sim 0.9$.
This makes sense, because at large $g$ and $m$, it requires a lot of energy to excite quark-antiquark pairs on top of the free vacuum.
Additionally, we can see that truncations \Tone, \Ttwo, \Tfour, and \Tfive\ (i.e. those with $n_f=1$) have slightly distinguished themselves from truncations \Tthreep\ and \Tthree\ (i.e. those with $n_f=2$).
To highlight these two categories of truncations, we color them differently (blue for $n_f = 1$ and red for $n_f = 2$).

All truncations with a given fermion cutoff $n_f$ have visibly identical oscillation frequency; but the frequencies between the two groups are slightly different.
This slight shift in frequency can be understood from the perspective of perturbation theory, for large $g$ and $m$.
At zeroth order (infinite $g$ and $m$), no virtual transitions are allowed, and the state remains in the free vacuum for all times.
At finite $g$ and $m$, quark-antiquark pairs can spawn in the vacuum---but the character of their appearance depends on the value of $n_f$ for a given truncation.
If $n_f=1$, then quark-antiquark pairs can never spawn on adjacent links, due to Krylov constraints; but if $n_f=2$, then there is no such restriction.
However, if either of $g$ or $m$ is large, then spawning multiple quark-antiquark pairs in a virtual process is suppressed compared to spawning just one pair, and the distinction posed by the Krylov constraints becomes irrelevant.
That is the effect we see in~\Cref{fig:strongvac1D}, leading to similar oscillation frequencies based on the value of $n_f$.
In fact, perturbation theory also implies that the oscillation frequencies should be further split within each $n_f$ group, but this splitting is a higher order effect (and thus only visible when plotting ratios to the highest truncation), since it would require creating multiple quark-antiquark pairs, and then virtually joining them into a longer string.

A similar effect also leads to the slight difference we see in the matter production, $\langle \mathcal N_q\rangle$.
Indeed, if the matter truncation cutoff $n_f$ is relaxed, then more matter can be produced.
But for the large parameter values we have chosen, the virtual transitions to states containing two charges at a single lattice site are a higher-order effect compared to the production of a single quark-antiquark pair.
This leads to similar levels of real-time matter production across all truncations.

One way to see a more pronounced difference between these truncations, even at large parameter values, is by taking an initial state with the following property: the lowest-order nonzero correction to time evolution resides at the same order in perturbation theory for both the $n_f=1$ and $n_f=2$ cases, \textit{and} that correction is numerically different for $n_f=1$ vs. $n_f=2$.
This is completely satisfied by taking an initial meson-like excitation, $\ket{\phi}$.

\begin{figure}[htp]
    \centering

    \begin{minipage}{\twocolumnwidth}
        \centering
        \includegraphics[width=\twocolumnwidth]{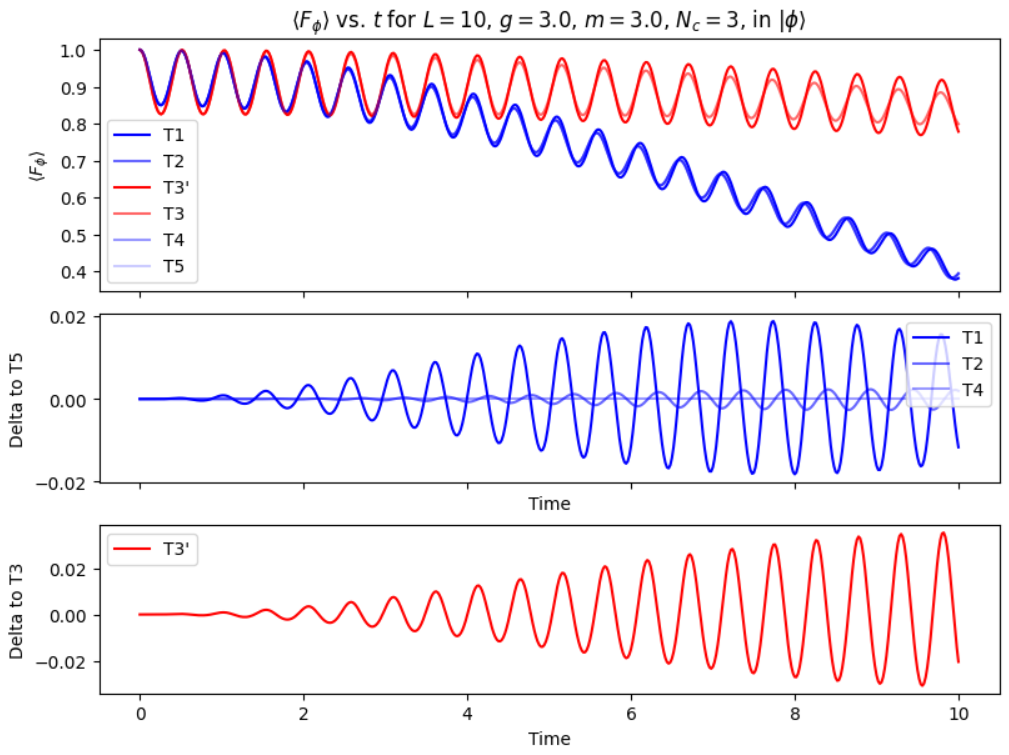}
    \end{minipage}%
    \hfill
    \begin{minipage}{\twocolumnwidth}
        \centering
        \includegraphics[width=\twocolumnwidth]{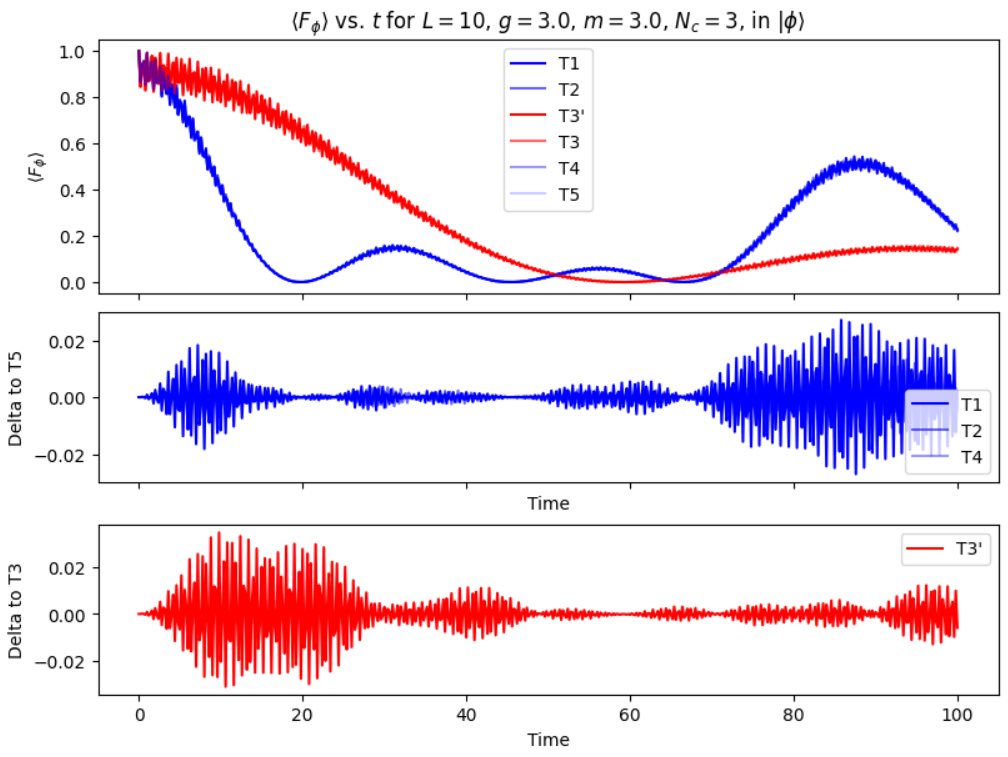}
    \end{minipage}

    \caption{Real-time evolution of $\mathcal F_{\phi}$ for the meson-like excitation $\ket{\phi}$ at large coupling and mass, plotted up to duration $t=10$ (top) and $t=100$ (bottom). At early times, all truncations coincide, but they drift apart at much longer times. The large revival at $t \sim 85$ suggests that the meson has traveled around the lattice.}\label{fig:strongmeson1D}
\end{figure}

The real-time evolution of $\ket{\phi}$ is shown in~\Cref{fig:strongmeson1D}, by plotting the fidelity $\langle \mathcal F_{\phi}\rangle$ at both short and long time intervals.
This time, the fidelity diverges between the blue and red truncations after around $t \sim 2$.
At strong coupling,~\Cref{fig:strongmeson1D} can essentially be explained by imagining the initial meson state as being comprised of left-moving and right-moving modes, which take equal times to travel around the lattice and return to their starting points (with a growing dispersion through each round trip).
The main difference between the blue and red curves is the speed of the meson, largely dictated by the mass truncation $n_f$.

From the perspective of perturbation theory, this is explainable as a consequence of the fact that all single-link meson-like states (there are $L$ of them) are physical at large $g$ and $m$, and the leading order virtual transitions between them are different depending on $n_f$: if $n_f = 1$, then the only leading-order channel enabling the excitation to transport to a neighboring link is by annihilating the excitation at the original link, and re-spawning it at the next link; if $n_f=2$, there is an additional option to spawn the excitation at the next link \textit{first}, and then annihilate the original excitation.

\begin{figure}[htp]
    \centering

    \begin{minipage}{\twocolumnwidth}
        \centering
        \includegraphics[width=\twocolumnwidth]{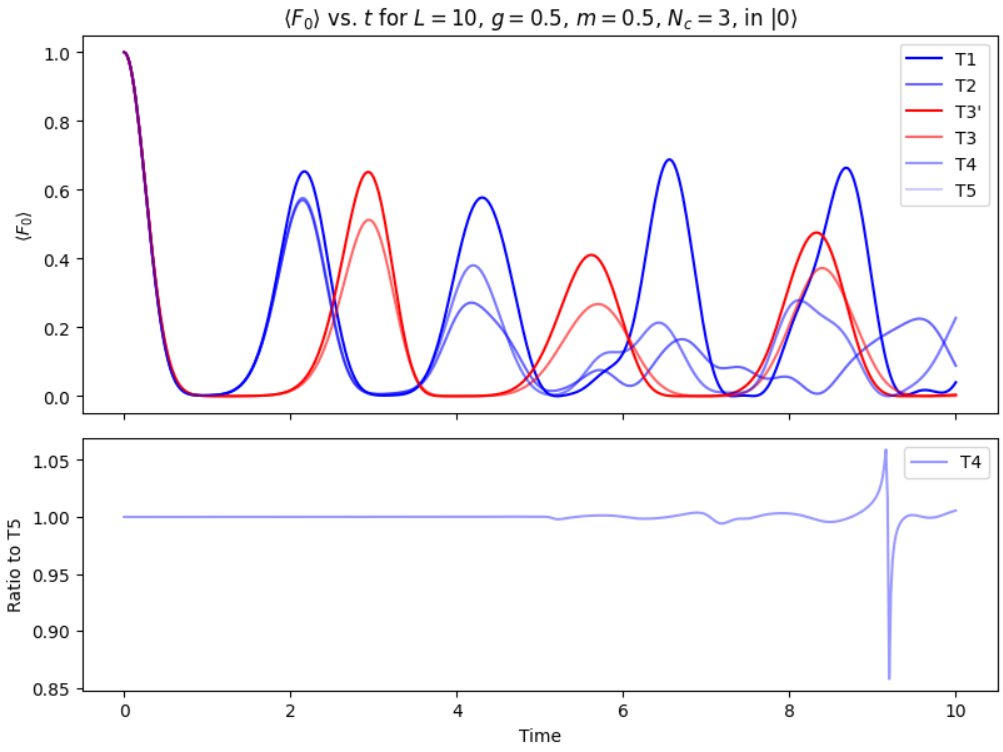}
    \end{minipage}%
    \hfill
    \begin{minipage}{\twocolumnwidth}
        \centering
        \includegraphics[width=\twocolumnwidth]{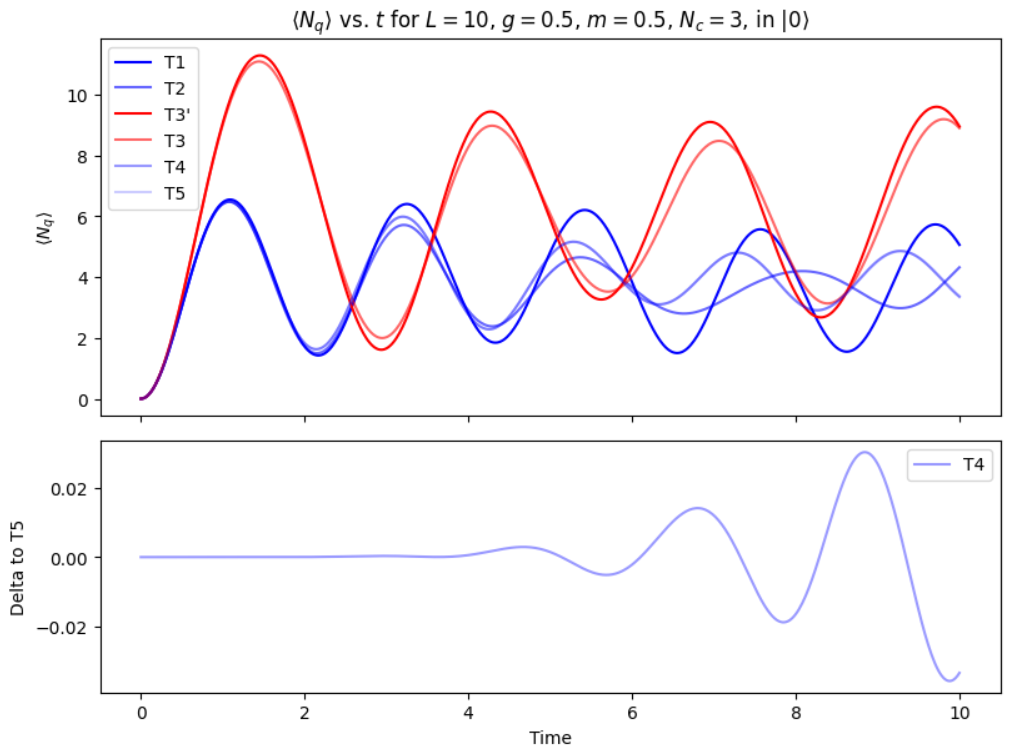}
    \end{minipage}

    \caption{Real-time evolution of $\mathcal F_0$ and $\mathcal N_q$ for the free vacuum state $\ket{0}$ at small coupling and mass. Most truncations are distinguishable, and we don't include them in the ratio or delta plots; however, truncations \Tfour\ and \Tfive\ remain very close together, even up to time $t = 10$.}\label{fig:weakvac1D}
\end{figure}

Now let us come to the weakly-coupled regime, which is no longer perturbatively accessible.
In~\Cref{fig:weakvac1D}, we set $g=0.5$ and $m=0.5$, and repeat the plots for time-evolving the free vacuum $\ket{0}$.
The interesting observation from this simulation is that there is a brief period of time near the beginning when the free vacuum contribution to the overall quantum state is entirely lost---only to re-emerge at a later time.
These revivals resemble quantum many-body scarring \cite{Turner2018WeakErgodicityBreaking,Moudgalya2018EntanglementExactExcitedStates,Moudgalya2018ExactExcitedStates,Zhao2020QuantumManyBodyScars,Serbyn2021QuantumManyBodyScars,Moudgalya2022QuantumManyBodyScarsHilbertSpaceFragmentation,Chandran2023QuantumManyBodyScars}, originally discovered in the PXP model \cite{Bernien2017ProbingManyBodyDynamics} and the subject of many experiments since \cite{Bluvstein2021ControllingQuantumManyBodyDynamics,Bluvstein2022QuantumProcessor,Su2023ObservationManybodyScarring,Zhang2023ManyBodyHilbertSpaceScarring,Dong2023DisorderTunableEntanglement}. The PXP model is equivalent to truncation \Tone, but our simulation results show that a similar phenomenon occurs even at higher truncations, with perhaps a shifted revival time depending on how many charged excitations are allowed per lattice site. This finding is particularly interesting, especially since quantum many-body scarring has been shown to occur in various toy model LGTs with (non-)Abelian gauge groups in one and two spatial dimensions \cite{Halimeh2023robustquantummany,Desaules2023WeakErgodicityBreaking,Desaules2023ProminentQuantumManyBodyScars,Iadecola2020QuantumManyBodyScar,Aramthottil2022ScarStates,Banerjee2021QuantumScarsZeroModes,Biswas2022ScarsFromProtectedZeroModes,Daniel2023BridgingQuantumCriticality,Ebner2024EntanglementEntropy,Sau2024sublatticescarsbeyond,Osborne2024QuantumManyBodyScarring,Budde2024QuantumManyBodyScars,Calajo2025QuantumManyBodyScarringNonAbelian,Hartse2025StabilizerScars}, but their fate in $d+1$D SU$(3)$ LGTs with higher truncations is an open question that our formulation can directly address.

\begin{figure}[htp]
    \centering

    \begin{minipage}{\twocolumnwidth}
        \centering
        \includegraphics[width=\twocolumnwidth]{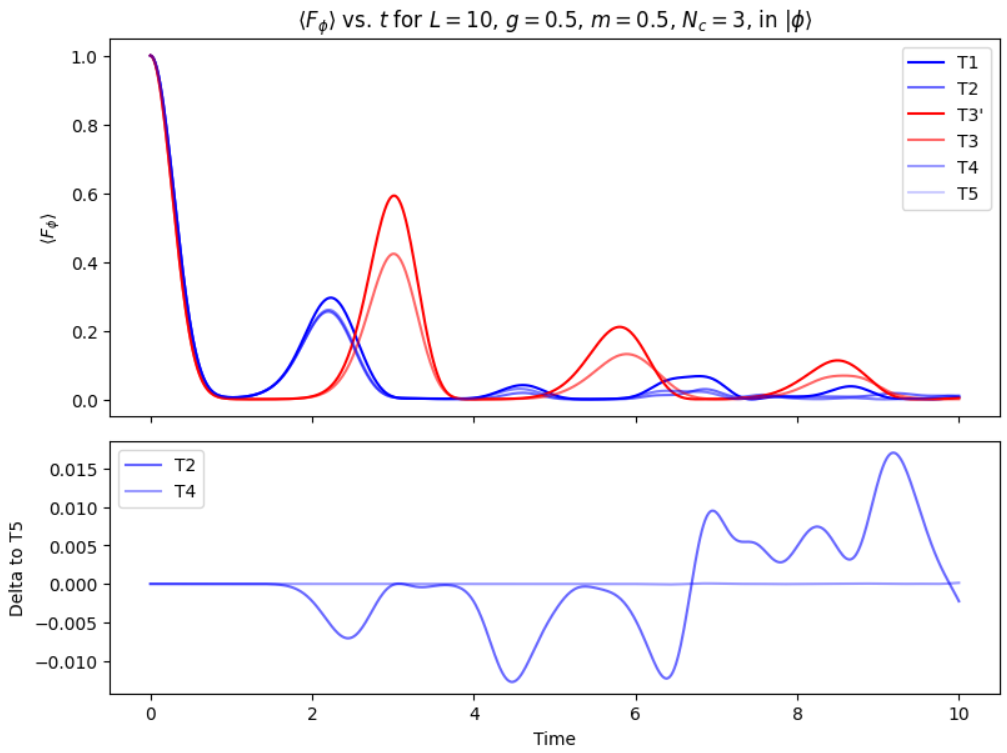}
    \end{minipage}%
    \hfill
    \begin{minipage}{\twocolumnwidth}
        \centering
        \includegraphics[width=\twocolumnwidth]{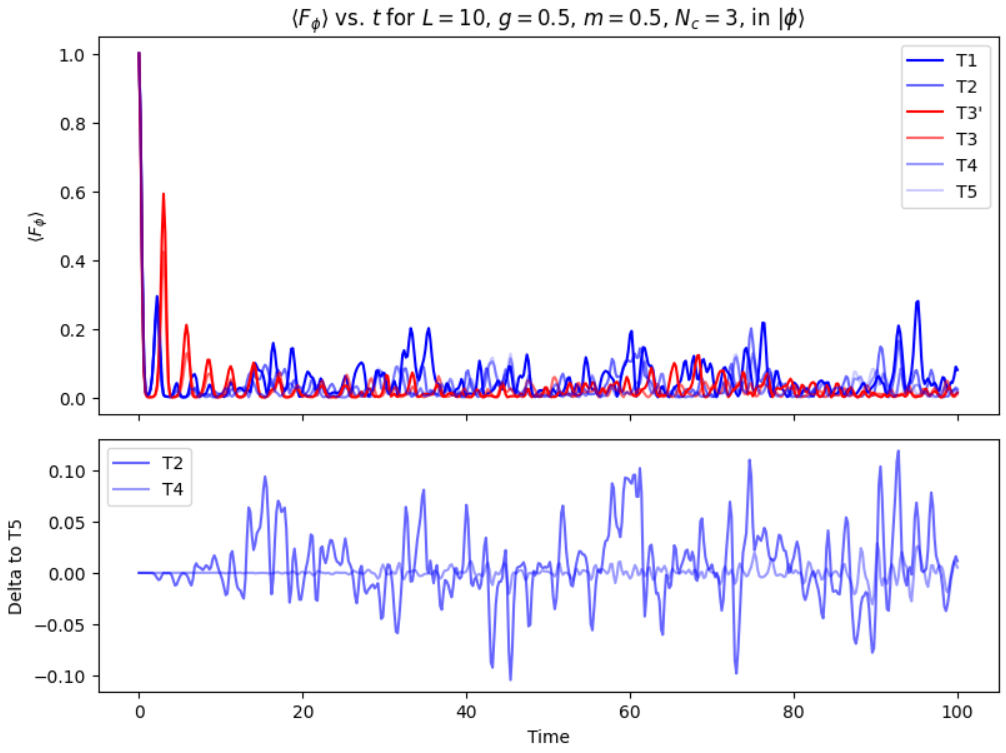}
    \end{minipage}

    \caption{Real-time evolution of $\mathcal F_{\phi}$ for the meson-like excitation $\ket{\phi}$ at small coupling and mass, plotted up to duration $t=10$ (top) and $t=100$ (bottom). At early times, the meson loses its entire fidelity to the physical Hilbert space, as if thermalizing; but revivals occur, even at long times.}\label{fig:weakmeson1D}
\end{figure}

Keeping $g=0.5$ and $m=0.5$, we also simulate the initial meson-like excitation $\ket{\phi}$ drawn in ~\Cref{fig:hexmeson}.
The results are shown in~\Cref{fig:weakmeson1D}.
There are some qualitative similarities in this simulation with the case of the free vacuum, such as a revival of the initial state after some time.
That being said, the revivals shrink much quicker in amplitude than for the case of the free vacuum, and this suggests that the meson disperses too quickly to coherently return to its original position on the lattice.

One final important point to make for the results we obtained is that, even at weak coupling, the dynamics at truncations \Tfour\ and \Tfive\ remain very close for substantially long times.
For instance, in~\Cref{fig:weakvac1D} and ~\Cref{fig:weakmeson1D}, the divergence between these truncations is hardly visible, even up to $t = 10$, and appears to remain controlled at much longer times.
Since the only difference between \Tfour\ and \Tfive\ is the order in $1/N_c$ at which interactions are kept, this suggests that $N_c = 3$ is already ``large" in the sense that the higher order adjustment made by \Tfive\ is very weakly explored (at least until quite late times).

\subsubsection{$2+1$D Case}

For our $2+1$D simulations, we choose the lattice to be a hexagonal plaquette chain with $P=4$ plaquettes, equipped with the truncated Hamiltonians defined in \Cref{sec:2+1d}.
We again find that there is a significant qualitative distinction between the large parameter and small parameter regimes.
The precise regions in parameter space that correspond to these regimes are also slightly shifted compared to the $1+1$D case---in particular, we generally need smaller values of $g$ and $m$ to draw qualitative similarities between our plaquette chain results and  the results from $1+1$D shown previously.

\begin{figure*}[htp]
    \centering
    
    \begin{minipage}{0.5\linewidth}
        \centering
        \includegraphics[width=0.9\linewidth]{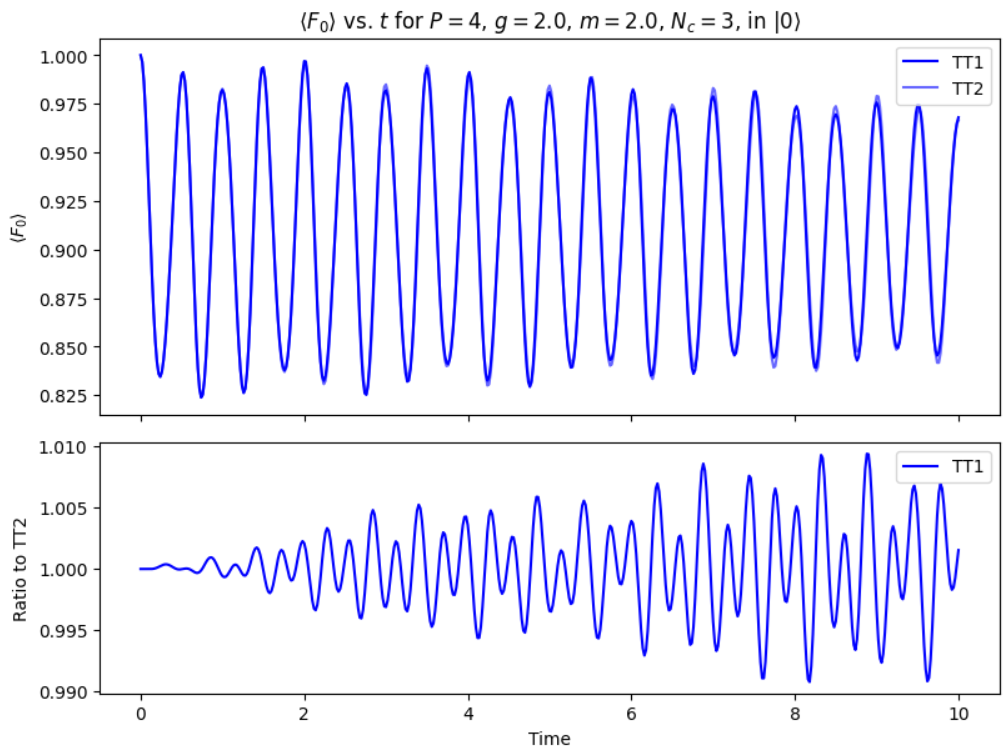}
    \end{minipage}%
    \begin{minipage}{0.5\linewidth}
        \centering
        \includegraphics[width=0.9\linewidth]{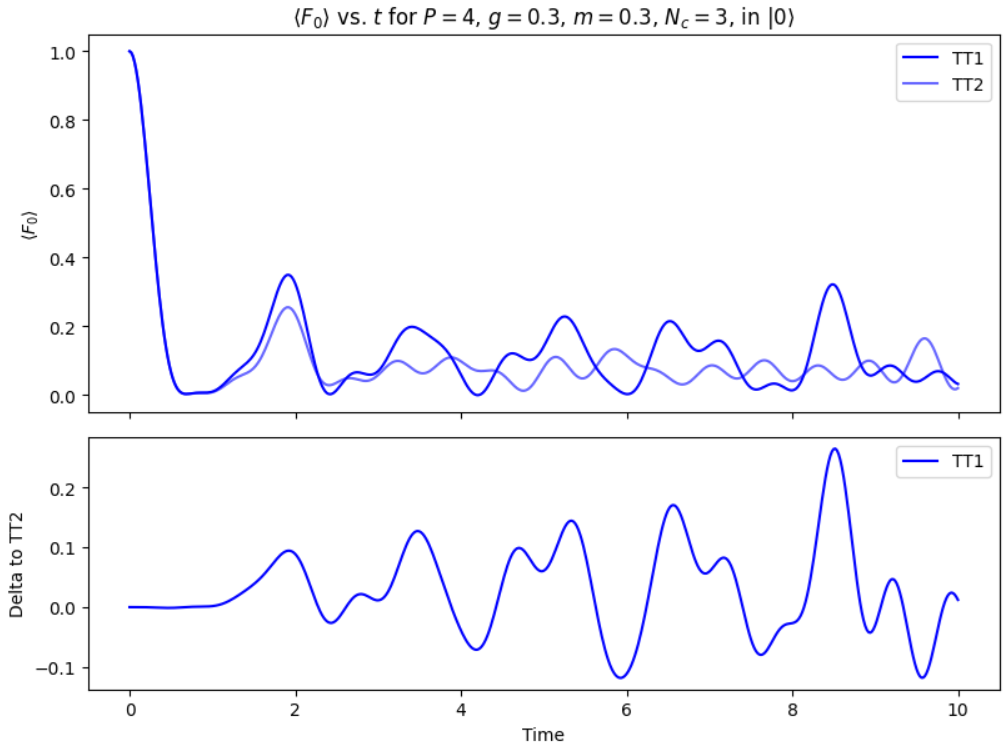}
    \end{minipage}
    \hfill
    \begin{minipage}{0.5\linewidth}
        \centering
        \includegraphics[width=0.9\linewidth]{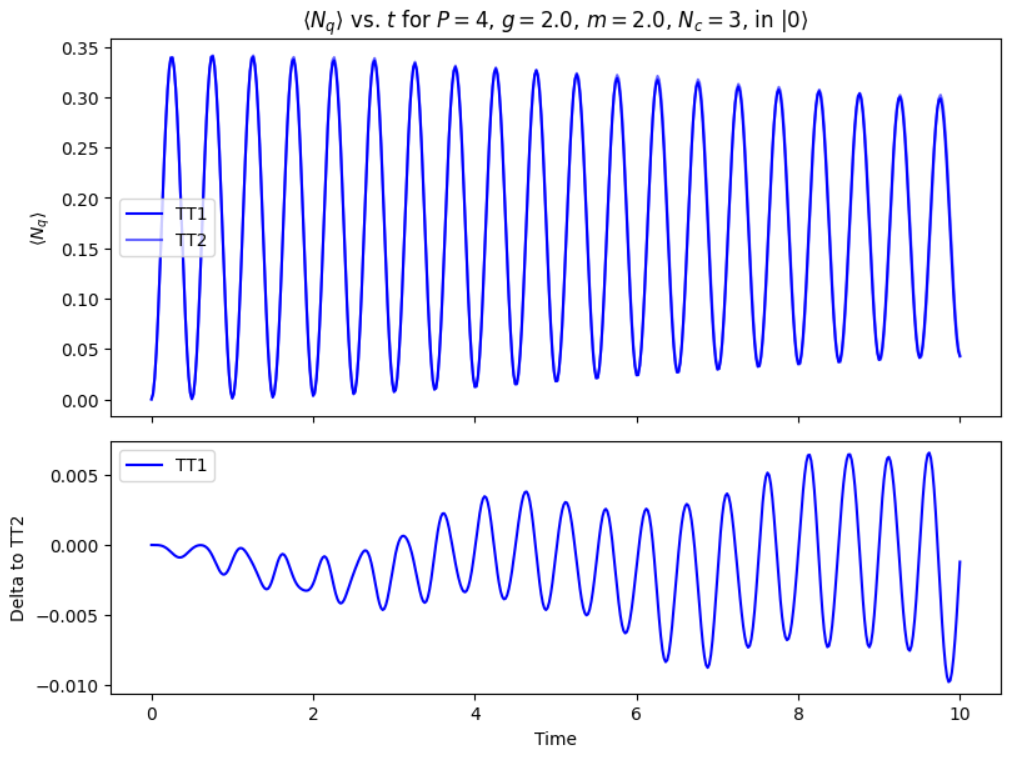}
    \end{minipage}%
    \begin{minipage}{0.5\linewidth}
        \centering
        \includegraphics[width=0.9\linewidth]{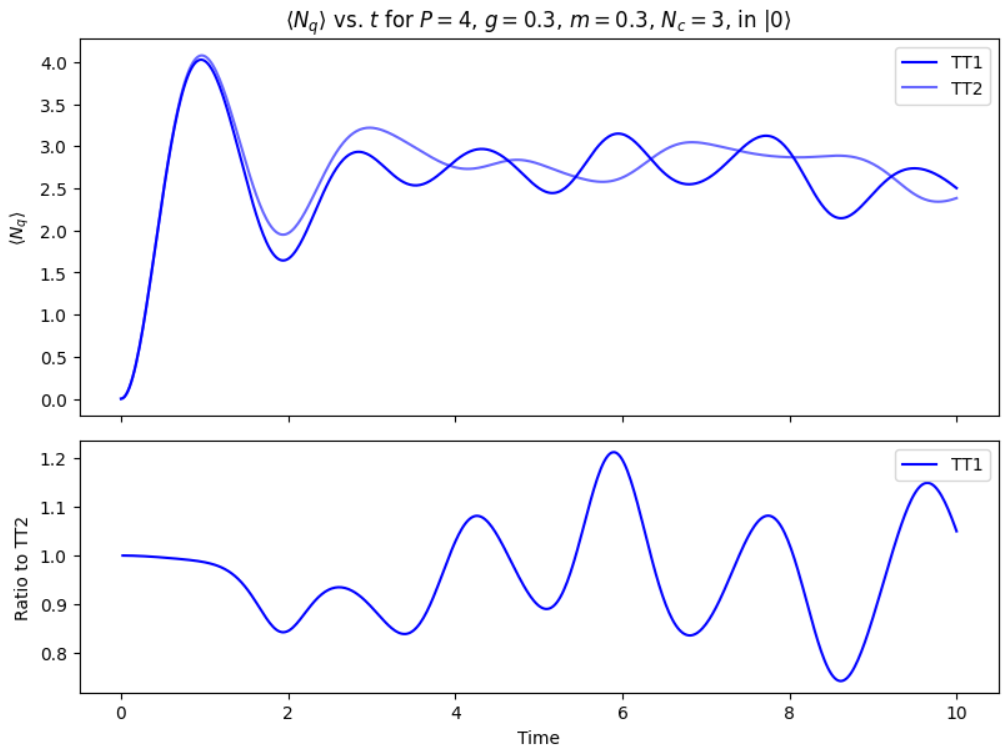}
    \end{minipage}
    \hfill
    \begin{minipage}{0.5\linewidth}
        \centering
        \includegraphics[width=0.9\linewidth]{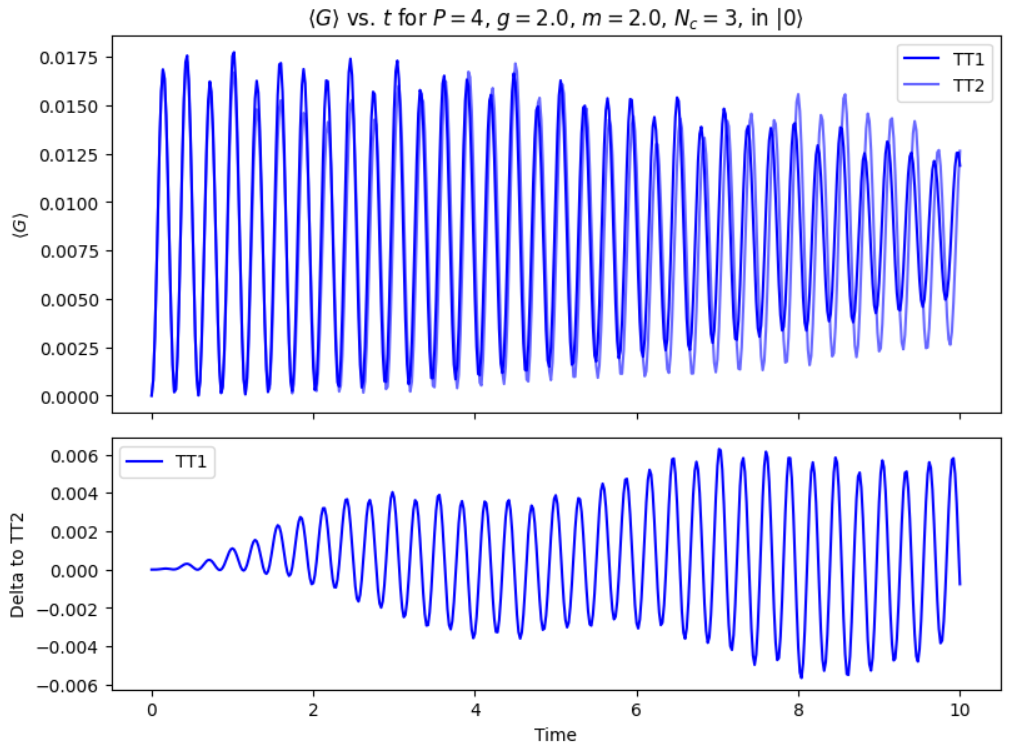}
    \end{minipage}%
    \begin{minipage}{0.5\linewidth}
        \centering
        \includegraphics[width=0.9\linewidth]{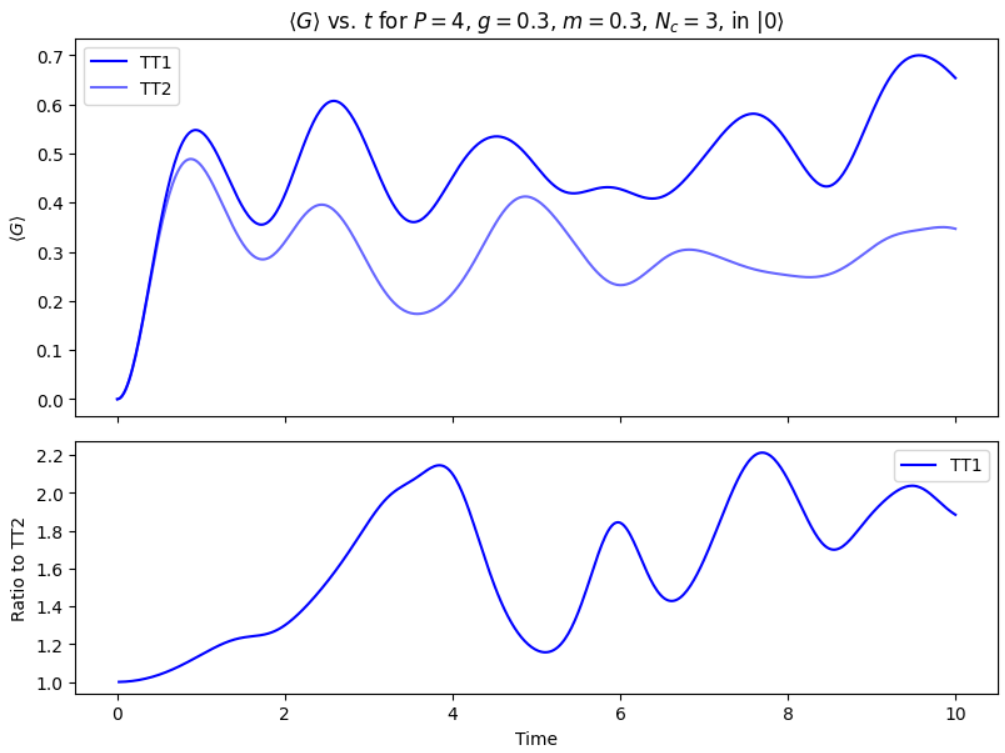}
    \end{minipage}
    \caption{Real-time evolution of $\mathcal F_0$, $\mathcal N_q$, and $\mathcal G$ for the free vacuum state $\ket{0}$ at large parameters (left) and small parameters (right) on the plaquette chain. All $2+1$D truncations computed in this paper have $n_f = 1$, and are shown with different shades of blue, in analogy with the $1+1$D plots. Note that the choice of ratio or delta plotting is dependent on whether the observables get very close to zero. The observables that reach zero are swapped between the strongly-coupled and weakly-coupled cases.}\label{fig:strongweakvac2D}
\end{figure*}

The time evolution of the free vacuum $\ket{0}$ is shown in \Cref{fig:strongweakvac2D}.
This time, we plot both the strongly-coupled ($g = 2.0$ and $m = 2.0$) and weakly-coupled ($g=0.5$ and $m=0.5$) results side by side, which highlights the qualitative differences between them.
As in the $1+1$D case, the plots for each truncation lie very close together in the strongly-coupled regime, and the splitting of the oscillation frequency between different truncations is barely visible.
In the weakly-coupled regime, the plots continue to remain qualitatively very similar, but are only strictly indistinguishable at very early times.
Additionally, it is apparent from the plots that reducing the coupling constant and fermion mass promotes the production of quarks and gluons, as one would expect when the energy cost of doing so is reduced.
If the fermion cutoff $n_f$ is raised, one can reasonably expect that these results will pick up modifications that directly parallel the differences found in the $1+1$D case.

\begin{figure*}[htp]
    \centering

    \begin{minipage}{0.5\linewidth}
        \centering
        \includegraphics[width=0.9\linewidth]{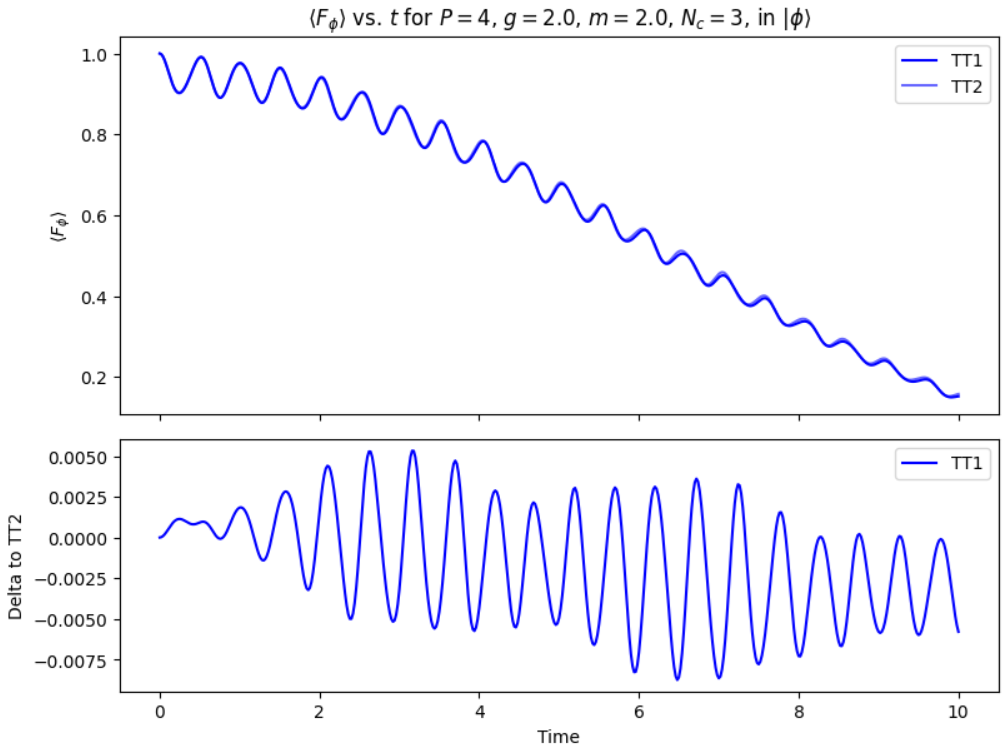}
    \end{minipage}%
    \begin{minipage}{0.5\linewidth}
        \centering
        \includegraphics[width=0.9\linewidth]{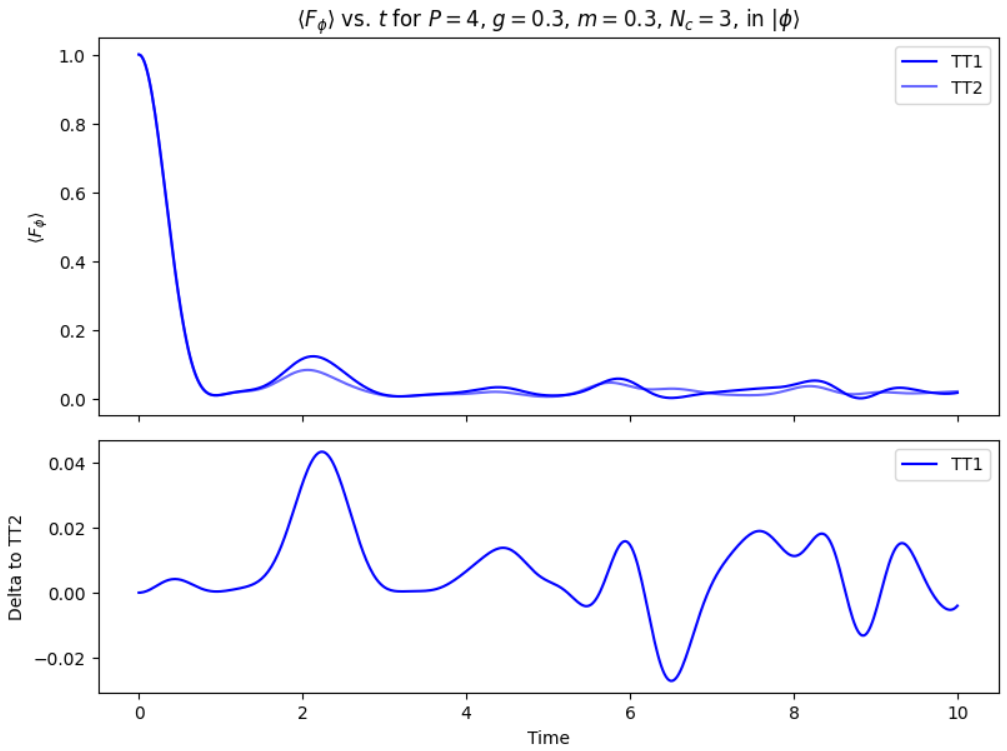}
    \end{minipage}
    \hfill
    \begin{minipage}{0.5\linewidth}
        \centering
        \includegraphics[width=0.9\linewidth]{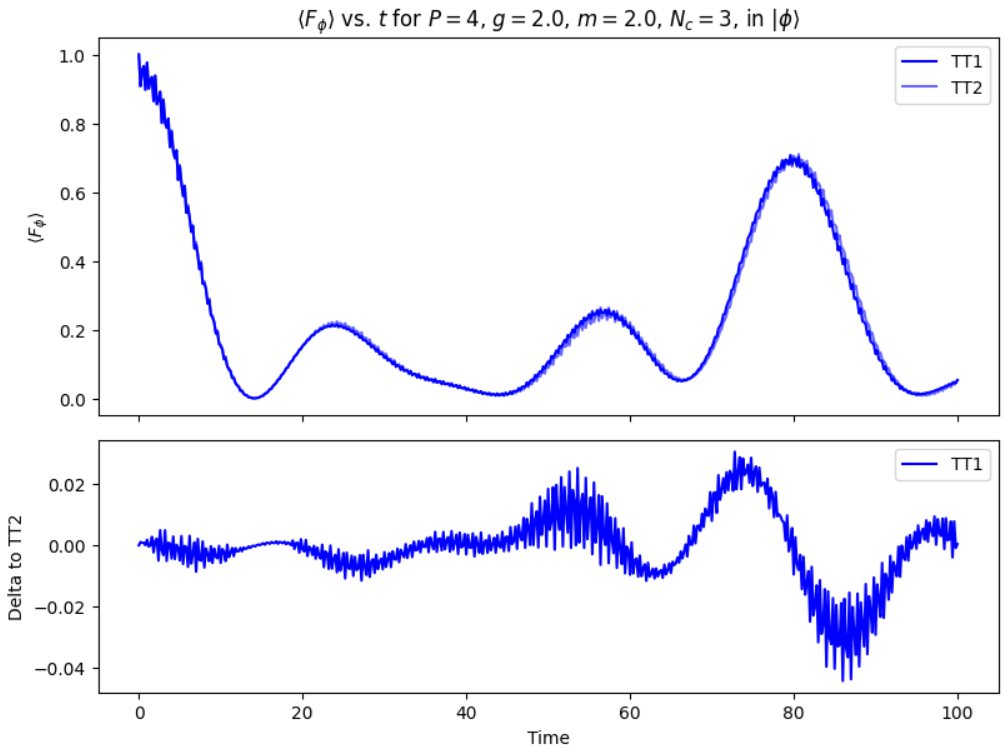}
    \end{minipage}%
    \begin{minipage}{0.5\linewidth}
        \centering
        \includegraphics[width=0.9\linewidth]{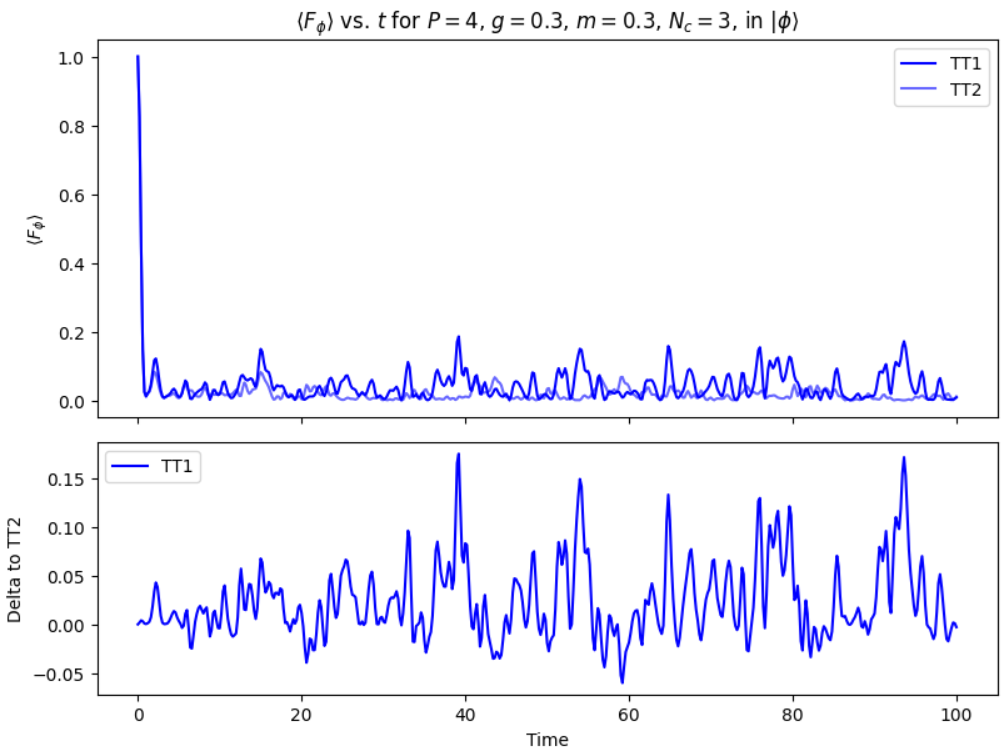}
    \end{minipage}

    \caption{Real-time evolution of $\mathcal F_{\phi}$ for the meson-like excitation $\ket{\phi}$ at large parameters (left) and small parameters (right), plotted up to duration $t=10$ (top) and $t=100$ (bottom).}\label{fig:strongweakmeson2D}
\end{figure*}

For the same sets of parameter values, we again simulate the meson-like excitation $\ket{\phi}$ (drawn in~\Cref{fig:hexmeson}), and plot the results in \Cref{fig:strongweakmeson2D}, again showing the strongly-coupled case alongside the weakly-coupled case.
All plots we obtain bear a qualitative resemblance to the $1+1$D case, with the main differences being the revival times, amplitudes, and oscillation frequencies.
As discussed for the free vacuum simulations, we can expect that the exact numerics of the dynamical trajectory will change substantially if the fermion truncation level is increased---while keeping similar qualitative features to those found on the $1+1$D lattice.

\subsection{Open Strings and String Breaking}\label{sec:openstring}

We now present the real-time evolution results for an open string on the lattice.
For any open string, its time evolution can either be performed by directly using the Kogut-Susskind Hamiltonian, or by pinning the endpoints of the string to external charges and evolving according to the modified Hamiltonian.
We will showcase a comparison between the string breaking mechanism in both types of simulations.
This means our $1+1$D simulations will be based on both $H_{\text{\Ttwo}}$ and $H_{\text{\Ttwo, }\mathrm{e.c.}}$, and our $2+1$D simulations will be based on both $H_{\text{\TTtwo}}$ and $H_{\text{\TTtwo, }\mathrm{e.c.}}$ (the Hamiltonians with external charges are denoted with the `e.c.' subscript, and were derived in \Cref{sec:external}).
String breaking at higher truncations of QCD is beyond the scope of our present work.

In addition to computing the real-time evolution of all observables, we will also utilize the \textit{diagonal ensemble average} of observables as a method to diagnose the occurrence of string breaking.
For any operator $A$, its diagonal ensemble average under Hamiltonian $H$ (and relative to initial state $\ket{\psi_0}$) is given by
\begin{equation}
    \overline A \equiv \sum_{j=1}^{n} \left|\braket{\psi_0}{E_j}\right|^2 \cdot \bra{E_j} A \ket{E_j},
\end{equation}
where $n$ is the dimension of the physical Hilbert space, and $\ket{E_1},\dots,\ket{E_n}$ are the eigenstates of $H$.
Generically speaking, this represents the long-time average of the observable $A$, under the time evolution of the initial state from $\ket{\psi_0}$ to $\ket{\psi(t)}$, i.e.
\begin{equation}
    \overline A = \lim\limits_{T\to\infty} \frac 1{T} \int_0^T \dd t\, \bra{\psi(t)} A \ket{\psi(t)}.
\end{equation}

For any conjectured feature (e.g. confinement, etc.) of real-time dynamics, the diagonal ensemble average can tell us whether the feature is expected to survive in the long term.
Since our focus is on string breaking, we will primarily use the diagonal ensemble average of the average string length to assess whether the initial string genuinely breaks or not.

\subsubsection{String Breaking in \texorpdfstring{$1+1$}{1+1}D}

Let us take the $1+1$D periodic spatial lattice with $L = 10$ lattice sites, and set $m = 1.5$ and $N_c = 3$, but allow $g$ to vary.
This choice of $m$ is small enough to show deviations from the strongly-coupled perturbative regime, but still large enough that we can qualitatively see remnants of the perturbative effects.
Our simulations here begin with an initial string state $\ket{s_r}$ of length $r$, and we directly investigate the question ``Does this string break dynamically?" at truncation \Ttwo\ in both the cases with and without external charges pinning the endpoints of the string.

As a matter of fact, there are only very specific values of $g$ for which string breaking can dynamically occur in the $1+1$D theory.
Ultimately, these ``resonant" values of the coupling must be determined by performing several simulations, each at a different value of $g$.

The intuition behind this point is easiest to understand in the strongly-coupled regime.
In this regime, the real-time dynamics can only break conservation of energy \textit{virtually} (i.e., in the perturbative sense), and this becomes harder to permit at longer times and larger energy deviations.
The initial string state contains electric energy on the links it spans; if it breaks, then some of that electric energy is converted into producing a quark-antiquark pair.
If $g$ is much larger than $m$, while $m$ itself is large, then the energy gained in the production of matter/antimatter can never make up for the energy lost by de-exciting the electric field, and string breaking will not occur dynamically.
If $g$ and $m$ are not too small, the same intuition can be used, with the main difference being that the resonance will be wider, because virtual violations of conservation of energy don't deviate too far from the physical expectation value, allowing for dynamical string breaking at a larger range of couplings.

\begin{figure}[htp]
    \centering
    \begin{minipage}{\twocolumnwidth}
        \centering
        \includegraphics[width=\twocolumnwidth]{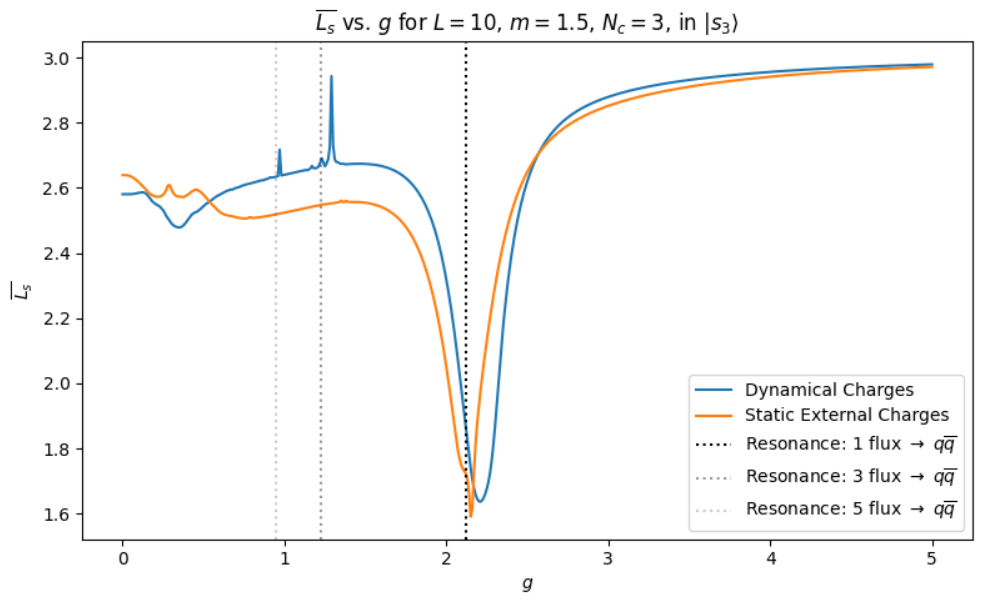}
    \end{minipage}%
    \hfill
    \begin{minipage}{\twocolumnwidth}
        \centering
        \includegraphics[width=\twocolumnwidth]{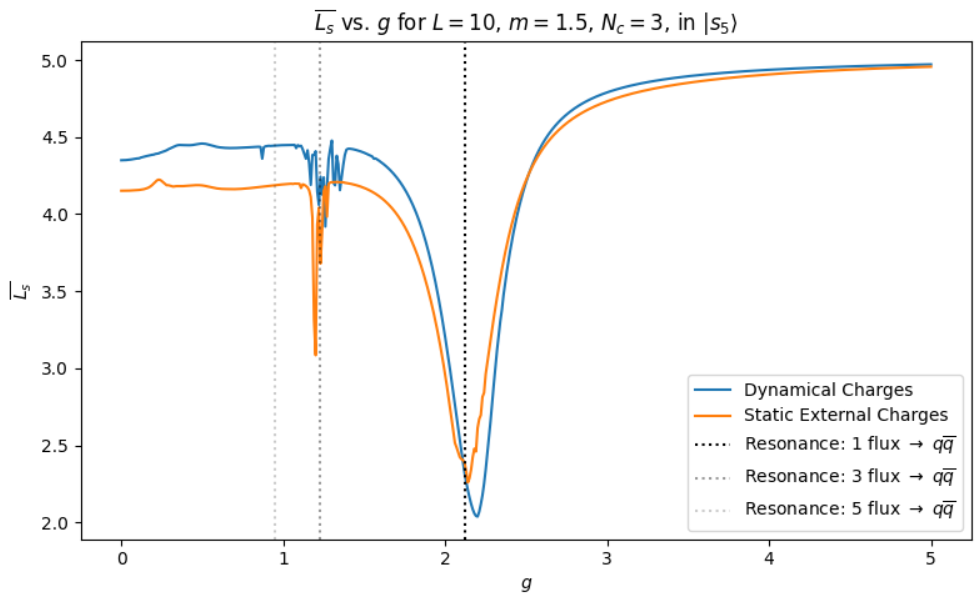}
    \end{minipage}
    \caption{Diagonal ensemble average of the average string length operator $\mathcal L_s$ vs. $g$ in truncation \Ttwo, both with and without static external charges pinning the endpoints of the string. A comparison to the expected resonances is provided by the vertical lines. String-breaking occurs at the resonances shown for a length-$3$ string (top) and a length-$5$ string (bottom).}\label{fig:res1D}
\end{figure}

This well-known aspect of lattice QCD can be observed in our truncated theory by plotting the diagonal ensemble average of the average string length as a function of $g$, as shown in \Cref{fig:res1D}.
The locations of the resonances can be estimated from our intuition above by recognizing that quark-antiquark pair production costs an energy $2m$, and de-exciting $k$ links gives energy $k\cdot g^2 C_2(\mathbf{N})/2$.
Therefore, there should exist one resonance per each allowed value of $k$, with coupling
\begin{equation}\label{eq:resk}
    g(k) \sim \sqrt{\frac{8mN_c}{k\cdot(N_c^2 - 1)}}.
\end{equation}
These expected resonances are plotted for a length-$3$ string and a length-$5$ string in \Cref{fig:res1D}.
In most cases, the resonance allows the string to break, thus substantially reducing the value of $\overline{\mathcal L_s}$.
An notable exception is the $k=3$ resonance for a length-$3$ string that has \textit{not} been pinned by external charges---this resonance actually allows the string to grow, which is impossible for the case when the endpoints have been pinned in place.

\begin{figure}[htp]
    \centering
    \begin{minipage}{\twocolumnwidth}
        \centering
        \includegraphics[width=\twocolumnwidth]{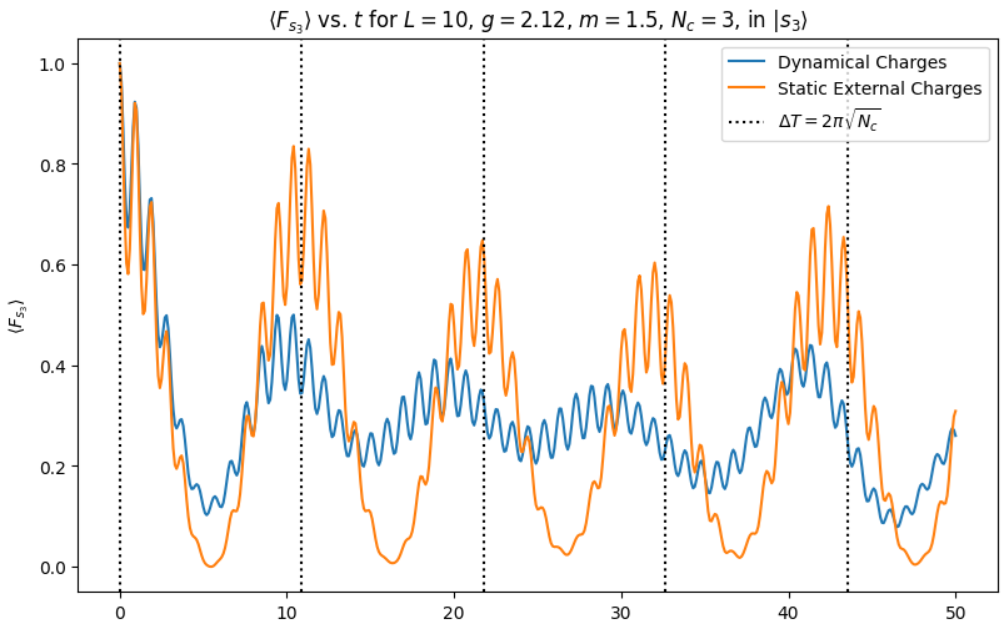}
    \end{minipage}%
    \hfill
    \begin{minipage}{\twocolumnwidth}
        \centering
        \includegraphics[width=\twocolumnwidth]{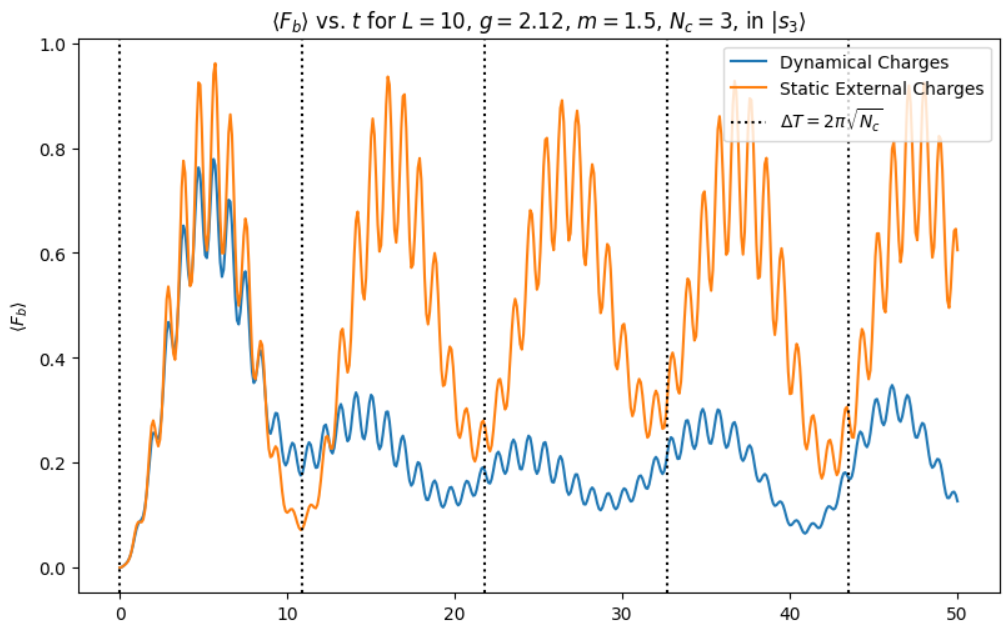}
    \end{minipage}
    \caption{Time-evolution of fidelities for an initial string state $\ket{s_3}$ with length $3$. We plot the fidelity against the initial state $\mathcal F_{s_3}$ (top) and the expectation value of the projector $\mathcal F_b$ to the broken string subspace (bottom). The overall oscillation is compared against the period cycle for the Rabi frequency $\omega_R = 1/\sqrt{N_c}$. The inner oscillations are due to vacuum fluctuations outside the string.}\label{fig:string1D}
\end{figure}

To understand string breaking more concretely in the real-time setting, we can directly plot the dynamics on resonance for the length-$3$ string state $\ket{s_3}$, as shown in \Cref{fig:string1D} for the $k=1$ resonance.
In addition to the projection $\mathcal F_{s_3}$ to the initial string state, we also plot the expectation value of the projector $\mathcal F_b \equiv \sum_b \ket{b}\bra{b}$, where each $\ket{b}$ represents a state containing only the broken string fragments, with minimum possible length (and no other excitations on the lattice).
There are four such states at this truncation, since the broken string is split into two fragments (one fragment attached to each external charge), and each fragment has two allowed configurations, corresponding to whether it resides on the left or right side of its external charge.
Thus $\mathcal F_{s_3}$ in some sense measures whether the string is intact (with no other matter surrounding it), and $\mathcal F_b$ measures whether the string is broken (with no additional excitations).

The results in \Cref{fig:string1D} clearly show that the trajectory of the quantum state slowly oscillates between the intact and broken states, while there are faster, smaller oscillations that simply correspond to vacuum fluctuations surrounding the transition.
The approximate frequency of the slow oscillations can be estimated from a standard first-order perturbative argument as follows: write the Dyson series for the resolvent of the Hamiltonian $H \equiv H_{\text{\Ttwo, }\mathrm{e.c.}}$ as
\begin{equation}
    R_H(E) \equiv \frac1{E-H} = \frac1{E-T} \sum_{j=0}^\infty \left(V \frac{1}{E-T}\right)^j,
\end{equation}
where $T$ and $V$ are the diagonal and off-diagonal parts of $H$, respectively.
In a regime where $g$ and $m$ are sufficiently large, $T$ dominates $V$, and the series converges.
Now let $P$ be the projector to a two-dimensional subspace consisting of the intact string state $\ket{s_3}$, and a single other normalized state defined by $\ket{b_1} \propto H_{\hop}(\ell) \ket{s_3}$.
The state $\ket{b_1}$ contains a broken string (because the hopping operation purely de-excites the middle link at this truncation), and at strong coupling, it is the preferred intermediate state in any sequence of transitions from $\ket{s_3}$ to any broken string state (because other transitions require exciting virtual quark-antiquark pairs outside the string).
Then the effective Hamiltonian on the two dimensional subspace spanned by $\ket{s_3}$ and $\ket{b_1}$ is given by solving
\begin{equation}
    P R_H(E) P = P R_{H_{\mathrm{eff}}}(E) P
\end{equation}
to first order in the ratio of spectral norms $\|V\| / \|T\|$ (which is small since $T$ dominates $V$).
This leads to the effective Hamiltonian
\begin{equation}
    H_{\mathrm{eff}} \approx PTP + PVP,
\end{equation}
which is a $2\times 2$ matrix of the form
\begin{equation}
    H_{\mathrm{eff}} \approx \begin{pmatrix} 3\cdot E_{\ell}(g) && \frac 1{2\sqrt{N_c}}\\ \frac 1{2\sqrt{N_c}} && 2m + 2\cdot E_{\ell}(g) \end{pmatrix},
\end{equation}
where $E_{\ell}(g) \equiv g^2 C_2(\mathbf{N})/2$ is the electric energy contained in a single link.
At the resonance plotted in \Cref{fig:string1D}, $2m \approx E_{\ell}(g)$, so that the diagonal terms are nearly equal.
This implies that the effective dynamics are Rabi oscillations with frequency $\omega_R = 1/\sqrt{N_c}$, which indeed shows agreement with \Cref{fig:string1D} to first order.

\subsubsection{String Breaking in $2+1$D}
For a hexagonal plaquette chain consisting of $P = 4$ plaquettes, we can now repeat much of the same analysis as in the $1+1$D case above.
We perform simulations for an initial string state $\ket{s_9}$ spanning 9 links (drawn in~\Cref{fig:hexstring}) at $m = 1.5$ and $N_c = 3$, leaving $g$ arbitrary for now.

\begin{figure}[htp]
    \centering
    \includegraphics[width=\twocolumnwidth]{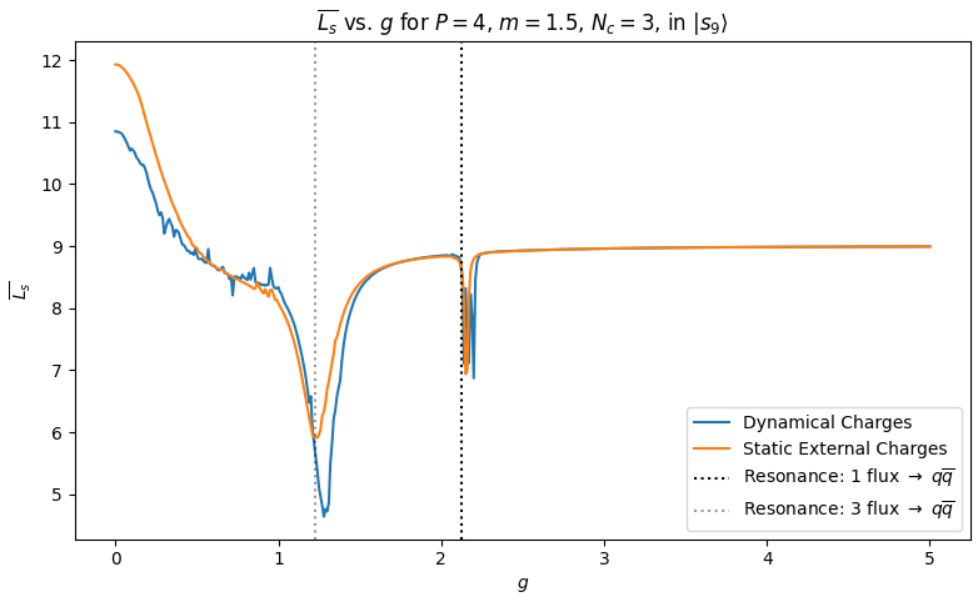}
    \caption{Diagonal ensemble average of the average string length operator $L_s$ vs. $g$ in truncation \TTtwo, both with and without static external charges pinning the endpoints of the string. A comparison to the expected resonances is provided by the vertical lines.}\label{fig:res2D}
\end{figure}

On the hexagonal lattice, the simplest way to break this string is by de-exciting three links on the string, in exchange for exciting a quark-antiquark pair.
A slightly more complicated string-breaking transition involves effectively de-exciting only a single link, by deforming the shape of the string through an appropriate sequence of virtual hopping and plaquette transitions, in exchange for exciting a quark-antiquark pair.
These two resonances correspond to the $k = 3$ and $k = 1$ cases dictated by \eqref{eq:resk}.
The plot of the diagonal ensemble average of the average string length $\overline{\mathcal L_s}$ is shown in \Cref{fig:res2D}, which highlights these two resonances.

\begin{figure}[htp]
    \centering
    \begin{minipage}{\twocolumnwidth}
        \centering
        \includegraphics[width=\twocolumnwidth]{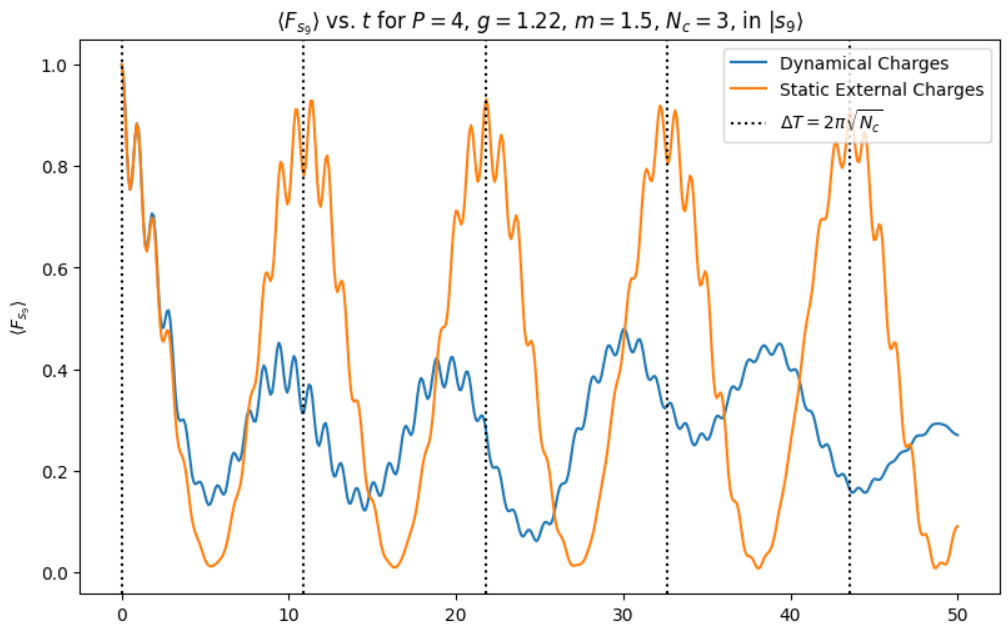}
    \end{minipage}%
    \hfill
    \begin{minipage}{\twocolumnwidth}
        \centering
        \includegraphics[width=\twocolumnwidth]{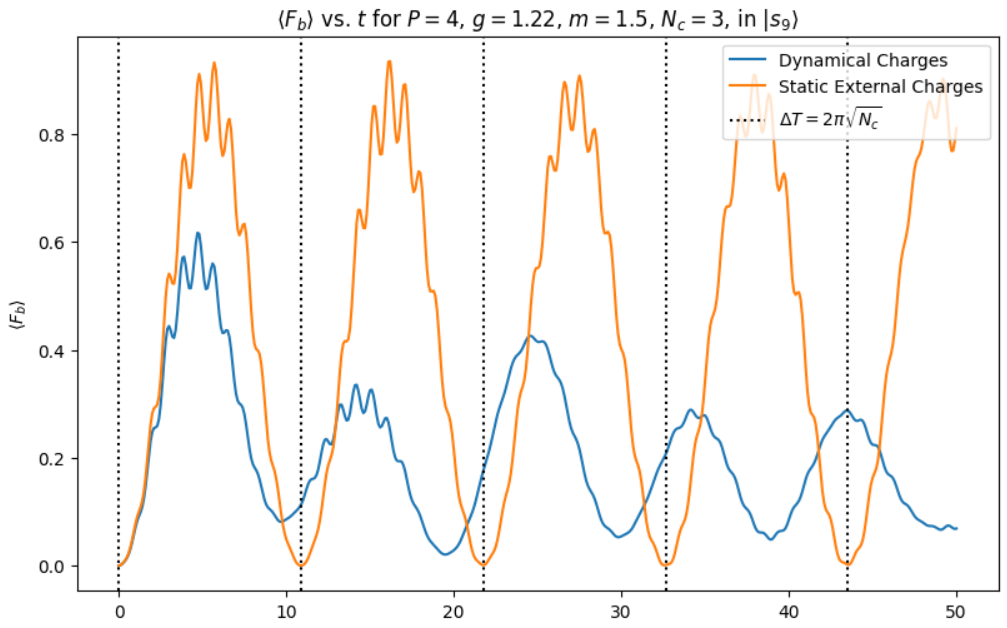}
    \end{minipage}
    \caption{Time-evolution of fidelities for an initial string state $\ket{s_9}$ with length $9$. We plot the fidelity against the initial state $\mathcal F_{s_9}$ (top) and the expectation value of the projector $\mathcal F_b$ to the broken string subspace (bottom). The overall oscillation is compared against the period cycle for the Rabi frequency $\omega_R = 1/\sqrt{N_c}$. The inner oscillations are due to vacuum fluctuations outside the string.}\label{fig:string2D}
\end{figure}

To keep the discussion concise, we simulate time evolution only for the $k = 3$ resonance, because this leads to the same Rabi frequency from first order perturbation theory as we found for $k = 1$ in the $1+1$D case.
The results are shown in \Cref{fig:string2D}, where we plot the periods corresponding to the Rabi frequency $\omega_R = 1/\sqrt{N_c}$, finding a high degree of agreement as before.

One might notice that the agreement is even sharper in the $2+1$D case, and that is partly because of the fact that the boundary between the strongly-coupled and weakly-coupled regimes is at higher parameter values than in our $1+1$D theory.
In other words, while $m = 1.5$ was small enough in the $1+1$D case to see increased deviations from perturbation theory, this parameter value falls well within the range where perturbation theory is accurate for our $2+1$D simulations.

\subsubsection{String Breaking at Large $N_c$}

To conclude our discussion about string breaking, we investigate the connection between our large $N_c$ truncations and the well-known fact that string breaking becomes suppressed in the large $N_c$ limit of $\SU(N_c)$ lattice gauge theory.
We work in a theory where $m = 3.0$ (which is the strongly-coupled regime for both our $1+1$D and $2+1$D simulations), and we take the `t Hooft limit $N_c\to\infty$, which holds the `t Hooft coupling $\lambda \equiv g^2 N_c$ fixed.

We would like to understand whether string breaking is suppressed dynamically, and if so, by which mechanism.
Our truncation scheme already implies that string breaking should strictly not exist at very large $N_c$, because the $N_c \to \infty$ limit at energy cutoff $(n_e, n_f) = (1,1)$ collapses to \Tone\ and \TTone, in $1+1$D and $2+1$D, respectively, and neither of these truncations permits string breaking, even kinematically.

But how exactly is string breaking dynamically suppressed when $N_c$ is large, from the perspective of higher truncations such as \Ttwo\ and \TTtwo?
To understand this, we can choose a trajectory where $N_c \to \infty$ by setting a fixed value of $\lambda$ that is greedily chosen to coerce string breaking, and then we can measure the diagonal ensemble average of an operator to diagnose the trend in the large $N_c$ limit.

To this end, we consider two candidate trajectories, denoted $\lambda = \lambda_{\mathrm{res}}(N_c = 3)$ and $\lambda = \lambda_{\mathrm{res}}(N_c = \infty)$.
The first trajectory $\lambda_{\mathrm{res}}(N_c = 3)$ is chosen by fixing the `t Hooft coupling based on the string-breaking resonances $g(k)$ discussed above, specifically for the $N_c = 3$ theory.
This means we take
\begin{equation}
    \lambda_{\mathrm{res}}(N_c = 3) \equiv \frac{8 m N_c^2}{k\cdot(N_c^2-1)} = \frac{9m}{k},
\end{equation}
where $k$ is chosen depending on which resonance we want to match (we'll use the same resonances simulated above, so $k=1$ for the $1+1$D case and $k=3$ for the $2+1$D case).

The second trajectory $\lambda_{\mathrm{res}}(N_c = \infty)$ is chosen by fixing the `t Hooft coupling to be on resonance at $N_c = \infty$.
This is possible because \eqref{eq:resk} implies that $g(k) \propto N_c^{-1/2}$ as $N_c$ gets large,
which is exactly the correct scaling behavior for a consistent `t Hooft coupling at $N_c = \infty$.
More specifically, we choose
\begin{equation}
    \lambda_{\mathrm{res}}(N_c = \infty) \equiv \lim\limits_{N_c\to\infty} g(k)^2 N_c = \frac{8m}{k},
\end{equation}
where again $k$ is chosen depending on which resonance we want to aim towards as $N_c \to \infty$.

The observable we simulate is denoted $\mathcal F_s \equiv \sum_s \ket{s} \bra{s}$, which is defined to be the projector to the subspace of the lattice Hilbert space that contains any string connecting the two external charges.
In other words, the sum runs over all states $\ket{s}$ including the initial string state $\ket{s_r}$, as well as any surrounding matter or deformation of the string, provided it remains intact.
Therefore, measurements of $\mathcal F_s$ directly tell us whether there is an intact or broken string in a given lattice state.

\begin{figure}[htp]
    \centering
    \begin{minipage}{\twocolumnwidth}
        \centering
        \includegraphics[width=\twocolumnwidth]{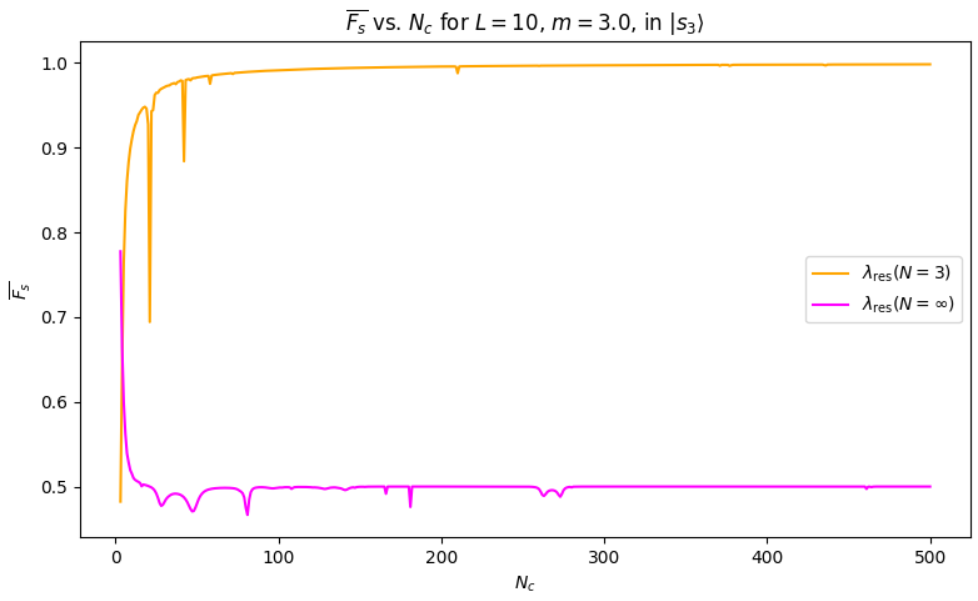}
    \end{minipage}%
    \hfill
    \begin{minipage}{\twocolumnwidth}
        \centering
        \includegraphics[width=\twocolumnwidth]{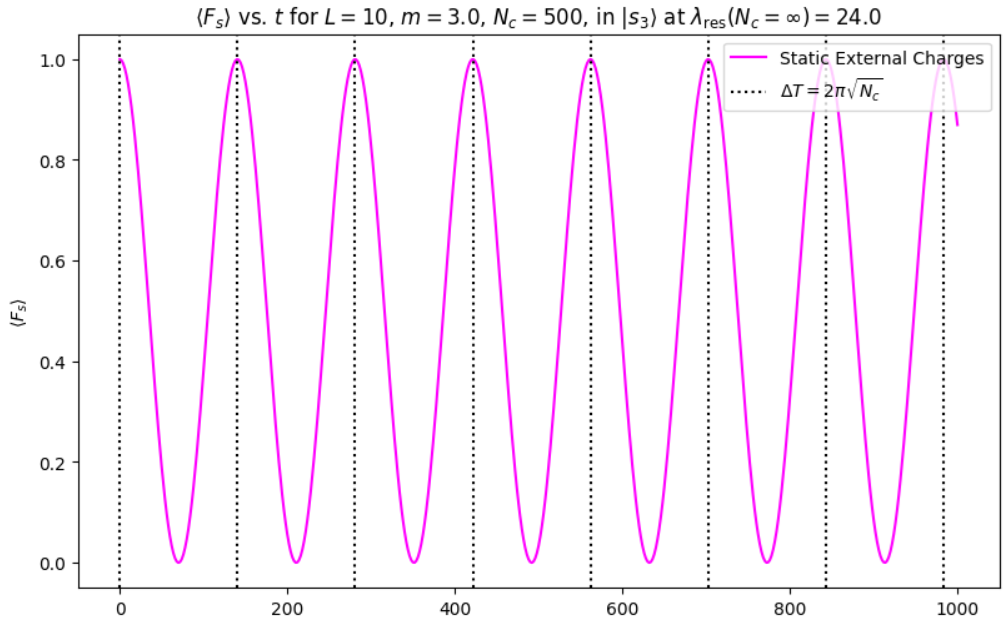}
    \end{minipage}
    \caption{String-breaking of a length-$3$ string in a $1+1$D lattice as a function of $N_c$, explored by plotting the diagonal ensemble average $\overline{\mathcal{F_s}}$ (top), and dynamics of $\mathcal F_s$ on resonance at fixed large $N_c$ (bottom). The dynamics in the `t Hooft limit perfectly match the result of perturbation theory, with Rabi frequency $\omega_R = 1/\sqrt{N_c}$.}\label{fig:largeN1D}
\end{figure}

The results for a $1+1$D periodic lattice with $L = 10$ lattice sites are shown in \Cref{fig:largeN1D}, with `t Hooft couplings tuned to the $k=1$ resonance .
For these simulations, we pinned an initial string state $\ket{s_3}$ spanning $3$ links at truncation \Ttwo\ with external charges.
When $\lambda = \lambda_{\mathrm{res}}(N_c = 3)$ is chosen, the large $N_c$ limit converges to high fidelity on the subspace where the string is intact on the lattice.
This is essentially because although this $\lambda$ is on-resonance at $N_c = 3$, it quickly goes off-resonance when $N_c$ is increased, and string breaking is no longer dynamically preferred.
From the plots, one can also see $\overline{\mathcal F_s}$ drop sharply at specific values of $N_c$, corresponding to instances where the `t Hooft coupling just happens to pass through another resonance at finite $N_c$.

Perhaps more interesting is the case when $\lambda = \lambda_{\mathrm{res}}(N_c=\infty)$.
In this case, the large $N_c$ limit approaches an equal superposition between the intact-string subspace and the broken-string subspace.
This is the only scenario where we can possibly expect to find the string breaking dynamically.
Upon plotting the time evolution of $\mathcal F_s$ at this coupling for large $N_c$, as shown in~\Cref{fig:largeN1D}, we find that the string exhibits almost perfectly oscillatory behavior between the intact subspace and the broken subspace.
In fact, the oscillation frequency is precisely the Rabi frequency $\omega_R = 1/\sqrt{N_c}$ derived previously from perturbation theory.
Therefore, the sense in which string breaking is dynamically suppressed in the large $N_c$ limit, even on resonance, is the sense in which the Rabi oscillations become drawn out over time, with $\omega_R \to 0$ as $N_c \to \infty$.

\begin{figure}[htp]
    \centering
    \begin{minipage}{\twocolumnwidth}
        \centering
        \includegraphics[width=\twocolumnwidth]{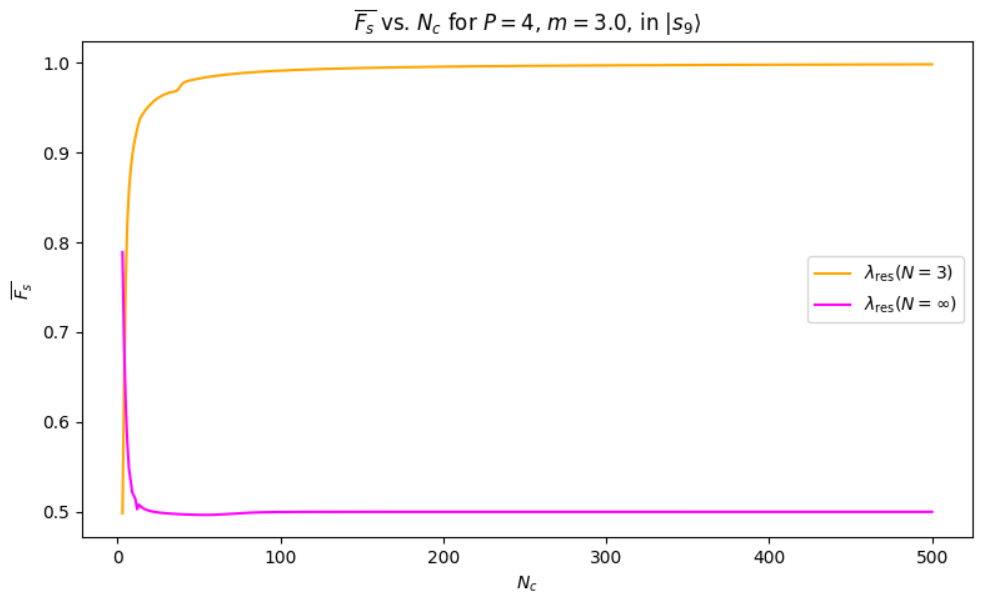}
    \end{minipage}%
    \hfill
    \begin{minipage}{\twocolumnwidth}
        \centering
        \includegraphics[width=\twocolumnwidth]{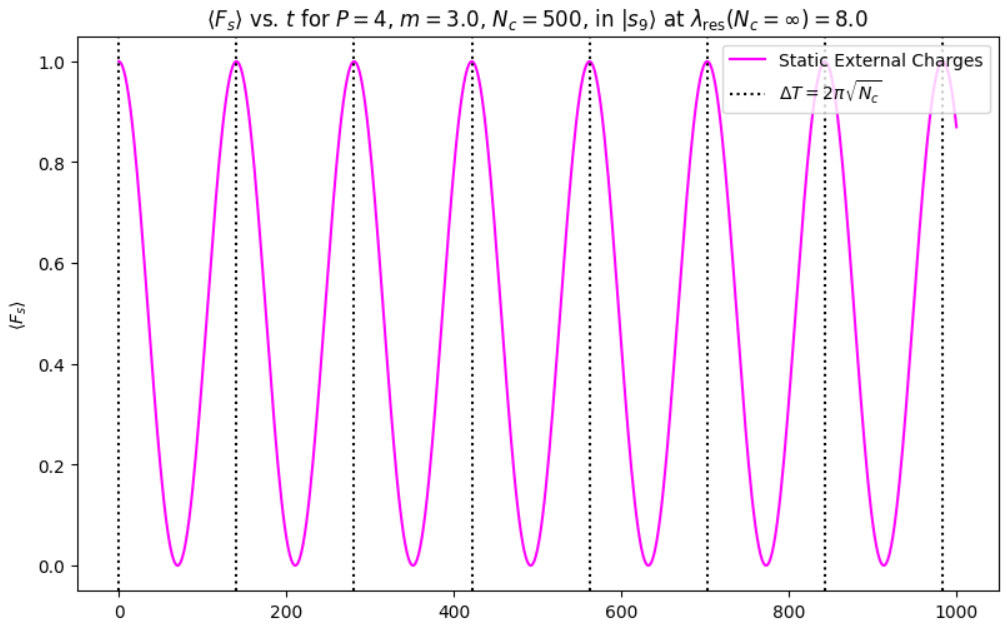}
    \end{minipage}
    \caption{String-breaking of a length-$9$ string in a $2+1$D lattice as a function of $N_c$, explored by plotting the diagonal ensemble average $\overline{\mathcal{F_s}}$ (top), and dynamics of $\mathcal F_s$ on resonance at fixed large $N_c$ (bottom). The dynamics in the `t Hooft limit perfectly match the result of perturbation theory, with Rabi frequency $\omega_R = 1/\sqrt{N_c}$.}\label{fig:largeN2D}
\end{figure}

Similar results are obtained for the $2+1$D hexagonal chain with $P=4$ plaquettes, as plotted in~\Cref{fig:largeN2D}.
We follow our previous $2+1$D simulations by choosing the $k=3$ resonance, and pinning an initial string state $\ket{s_9}$ spanning $9$ links (see~\Cref{fig:hexstring} as before) at truncation \TTtwo\ with external charges.
This leads to precisely the same type of behavior as in the $1+1$D case.

\section{Summary and Conclusion}\label{sec:summary}

This work has introduced a new scheme for truncating lattice QCD, aiming to improve the feasibility of real-time quantum simulation, both on classical and quantum hardware.
Our truncation scheme leads to verifiable results in the strongly-coupled, heavy-quark, and large $N_c$ limits, but also shows that the regime at smaller parameter values (relevant for observables in the continuum limit) still seems to be accessible to quantum simulation.

In particular, the physical Hilbert space, while growing exponentially with lattice size and truncation level, is massively redundant.
This was to the benefit of the simulations we performed on classical hardware, simply because the truncated Hilbert space is always substantially smaller than the qubit or qutrit counts would suggest.

This also indicates an opportunity to simulate our truncated theories on near-term hardware by merging this redundancy with quantum error correction (QEC).
Prior work \cite{Rajput:2021trn, Stryker:2018efp,Spagnoli:2024mib, Yao:2025cxs} has already laid a foundation for simulating lattice gauge theories on quantum hardware by taking advantage of the fact that gauge-invariance enables a reduction in the cost for error correction. 

Future work will extend previously developed techniques for quantum simulation of pure lattice gauge theories to the truncations considered in this work, which will enable the first scalable quantum simulation of QCD physics to explore real-time observables and their non-perturbative behaviors.

\begin{acknowledgments}
    This work was supported by the US DOE and other Agencies. C.W.B, A.N.C, and N.S.M were supported by the US Department of Energy, Office of Science, National Quantum Information Science Research Centers, Quantum Systems Accelerator (Award No. DE-SCL0000121). C.W.B also acknowledges support from the U.S. Department of Energy (DOE), Office of Science under contract DE-AC02-05CH11231, partially through Quantum Information Science Enabled Discovery (QuantISED) for High Energy Physics (KA2401032).
    J.C.H.~acknowledges funding by the Max Planck Society, the Deutsche Forschungsgemeinschaft (DFG, German Research Foundation) under Germany’s Excellence Strategy – EXC-2111 – 390814868, and the European Research Council (ERC) under the European Union’s Horizon Europe research and innovation program (Grant Agreement No.~101165667)—ERC Starting Grant QuSiGauge. Views and opinions expressed are, however, those of the author(s) only and do not necessarily reflect those of the European Union or the European Research Council Executive Agency. Neither the European Union nor the granting authority can be held responsible for them. This work is part of the Quantum Computing for High-Energy Physics (QC4HEP) working group.
\end{acknowledgments}
\appendix

\section{Electric Basis}\label{app:electric}

Our truncation scheme is most clearly specified using an eigenbasis of the electric field known as the \textit{Peter-Weyl basis}.
We briefly review the definition of this basis here, by relating it to the wavefunction picture of Wilson lines in a pure gauge theory.

Every normalized state in a pure lattice gauge theory with gauge group $G$ is a global wavefunction $\psi(\sigma)$ over all possible assignments of a group element to each link, denoted $\sigma : \lk \to G$.
The Hilbert space of all such wavefunctions is denoted $\mathcal H$, and it factorizes into a space of Haar-square-integrable wavefunctions at each link:
\begin{equation}\label{eq:hilb}
    \mathcal H = \bigotimes_{\ell \in \lk} \mathrm{L}^2(G)_{\ell},
\end{equation}
where we use the $\ell$ subscript to emphasize that one copy of $\mathrm{L}^2(G)$ is allocated to each link.
In other words, $\mathcal H$ is spanned by products of \textit{link wavefunctions} of the form $\psi_\ell : G \to \mathbb C$ for $\ell \in \lk$.

For each $g \in G$, we assume the convention that the link wavefunction $\psi_\ell(g)$ computes the probability amplitude that $W(\ell^-,\ell^+) = g$, where $W(x_0,x_1)$ denotes the parallel transporter
\begin{equation}
    W(x_0,x_1) \equiv \mathcal P \exp\left(i\int_{x_0}^{x_1} A_\mu \,\dd x^\mu\right)
\end{equation}
along the straight contour from $x_0$ to $x_1$.
This amplitude can also be represented in the magnetic basis by using the notation
\begin{equation}
    \braket{g}{\psi_\ell} \equiv \psi_\ell(g), \quad \forall g \in G,
\end{equation}
where the states $\{\ket{g}\}$ satisfy the Dirac delta normalization $\braket{g}{h} = \delta_h(g)$, and under the unit normalized Haar measure,
\begin{equation}
    \int_G \dd g\, \ket{g}\bra{g} = \mathrm{Id}_{\mathrm{L}^2(G)},
\end{equation}
where $\mathrm{Id}_{\mathrm{L}^2(G)}$ is the identity operator on $\mathrm{L}^2(G)$.

Under a finite gauge transformation $\Omega : \vx \to G$, the Wilson line $W(\ell^-, \ell^+)$ transforms as
\begin{equation}
    W(\ell^-,\ell^+) \mapsto W'(\ell^-,\ell^+) \equiv \Omega(\ell^+) W(\ell^-,\ell^+) \Omega(\ell^-)^{-1}.
\end{equation}
Equivalently, link wavefunctions transform as
\begin{equation}
    \psi_\ell(g) \mapsto \psi'_\ell(g) \equiv \psi_\ell\left(\Omega(\ell^+)^{-1} g \Omega(\ell^-)\right).
\end{equation}
This gauge transformation law can be interpreted locally as a simultaneous left/right $G$-action on link wavefunctions.
The left action $L_h \psi_\ell(g) \equiv \psi_\ell(h^{-1} g)$ and right action $R_h \psi_\ell(g) \equiv \psi_\ell(g h)$ both preserve wavefunction norm, so they furnish unitary representations of $G$. Additionally, note that $[L_{h_1}, R_{h_2}] = 0$ for every $h_1,h_2\in G$.
The entire gauge transformation $\Omega$ can thus be represented by a unitary operator $U(\Omega)$ that satisfies
\begin{align}
    \bra{g} U(\Omega) \ket{\psi_\ell}
    & = \bra{g} L_{\Omega(\ell^+)} R_{\Omega(\ell^-)} \ket{\psi_\ell}\nonumber\\
    & = \psi_\ell\left(\Omega(\ell^+)^{-1} g \Omega(\ell^-)\right)
\end{align}
for every link wavefunction $\psi_\ell$.

The Peter-Weyl theorem states that the link Hilbert space $\mathrm{L}^2(G)$, equipped with these left/right $G$-actions, canonically decomposes into orthogonal sectors as
\begin{equation}\label{eq:pw}
    \mathrm{L}^2(G) \cong \bigoplus_{\textbf{R}} \left[V_{\overline{\textbf{R}}} \otimes V_{\textbf{R}}\right],
\end{equation}
where the sum runs over all finite-dimensional irreducible unitary representations (irreps) of $G$, and we use the notation $V_{\textbf{R}}$ and $V_{\overline{\textbf{R}}}$ to denote vector spaces carrying representation $\textbf{R}$ and $\overline{\textbf{R}}$, respectively where $\overline{\textbf{R}}$ denotes the dual or conjugate representation to $\textbf{R}$.

Equation \eqref{eq:pw} is really an equivalence between representations, so its practical interpretation is as follows: for any vector space $V_{\textbf{R}}$ associated to some representation $\textbf{R} : G \to \mathrm{GL}(V_{\textbf{R}})$, denote an orthonormal basis for $V_{\textbf{R}}$ with the notation
\begin{equation}
    \left\{\ket{\textbf{R},1}, \ket{\textbf{R},2}, \dots, \ket{\textbf{R}, \dim \textbf{R}}\right\} \subset V_{\textbf{R}},
\end{equation}
and define the representation matrix entry functions
\begin{equation}
    \textbf{R}_{ab}(g) \equiv \bra{\textbf{R},a} \textbf{R}(g) \ket{\textbf{R},b},\quad \forall g \in G.
\end{equation}
In other words, we declare that a group element $g \in G$ acts on the basis of $V_{\textbf{R}}$ as
\begin{equation}
    g : \ket{\textbf{R},a} \mapsto \textbf{R}(g) \ket{\textbf{R},a} = \textbf{R}_{ba}(g) \ket{\textbf{R},b}.
\end{equation}
Analogous definitions can be made for irrep $\overline{\textbf{R}}$, culminating in the observation that $\overline{\textbf{R}}_{ab}(g) = \textbf{R}_{ab}(g)^*$.
We can thus write down an action of $G \times G$ in the direct product representation. We use the term ``direct product representation" in the sense of the direct product group, e.g. $G \times G$, in contrast with the more common ``product representation" which carries a simultaneous action on the tensor factors by $G$ alone. For direct sums or wedge products, we will always form representations of the original group $G$ and never its direct products. $V_{\overline{\textbf{R}}}\otimes V_{\textbf{R}}$ as
\begin{align}
    (h_1, h_2)
    & : \ket{\overline{\textbf{R}},a} \otimes \ket{\textbf{R},b}\nonumber\\
    & \mapsto \left[\overline{\textbf{R}}(h_1) \ket{\overline{\textbf{R}},a}\right] \otimes \left[\textbf{R}(h_2) \ket{\textbf{R},b}\right]\nonumber\\
    & = \left[\textbf{R}_{ca}(h_1)^* \cdot \textbf{R}_{db}(h_2)\right] \ket{\overline{\textbf{R}},c} \otimes \ket{\textbf{R},d}.\label{eq:rep}
\end{align}
Now, note that $\textbf{R}_{ab}(g)$ is a Haar-square-integrable function in its own right, and it thus transforms under simultaneous left/right $G$-actions as
\begin{align}
    L_{h_1} R_{h_2} \textbf{R}_{ab}(g)
    & = \textbf{R}_{ab}(h_1^{-1} g h_2)\nonumber\\
    & = \textbf{R}_{ac}(h_1^{-1}) \textbf{R}_{cd}(g) \textbf{R}_{db}(h_2)\nonumber\\
    & = \left[\textbf{R}_{ca}(h_1)^* \cdot \textbf{R}_{db}(h_2)\right] \textbf{R}_{cd}(g),
\end{align}
coinciding exactly with the transformation law for the representation $V_{\overline{\textbf{R}}}\otimes V_{\textbf{R}}$ as in \eqref{eq:rep}.
Therefore, for any collection of complex coefficients
\begin{equation}
    \left\{c_{ab} \mid a,b \in \{1,\dots,\dim\textbf{R}\}\right\},
\end{equation}
the link wavefunction
\begin{equation}
    \psi_{\ell}(g) \equiv c_{ab} \textbf{R}_{ab}(g),\quad \forall g \in G
\end{equation}
can be viewed as living in the direct product representation $V_{\overline{\textbf{R}}} \otimes V_{\textbf{R}}$, in the sense that the left/right group actions (and thus all finite gauge transformations) correspond exactly to the actions carried by the representations on the left/right tensor factors, respectively.

The Peter-Weyl theorem goes one step further and states that if we define the correspondence
\begin{gather}
    \ket{\textbf{R},a,b} \equiv \ket{\overline{\textbf{R}},a}\otimes\ket{\textbf{R},b}\label{eq:factor}\nonumber\\
    \big\updownarrow\nonumber\nonumber\\
    \braket{g}{\textbf{R},a,b} \equiv \sqrt{\dim \textbf{R}}\cdot \textbf{R}_{ab}(g),
\end{gather}
then the set of states $\{\ket{\textbf{R},a,b}\}$, running over all irreps $\textbf{R}$ of $G$ and all representation indices $1 \le a,b\le \dim \textbf{R}$, forms an orthonormal basis for $\mathrm{L}^2(G)$.
This is the electric (or Peter-Weyl) basis for each individual link Hilbert space, and the thus-defined product basis on the global Hilbert space $\mathcal H$ is the global electric basis.
Note that in diagrams where we draw an odd site on the left and even site on the right (so that the Wilson line is oriented from left-to-right, e.g., \Crefpanel{fig:cut}{b}), the left-to-right order of tensor factors should be $\ket{\mathbf{R},a,b} = \ket{\mathbf{R},b}\otimes\ket{\overline{\mathbf{R}},a}$ to match the diagram.

We can now say that a link state carries ``definite representation $\textbf{R}$" if its left-transforming tensor factor is always in the irrep $\overline{\textbf{R}}$ and its right-transforming tensor factor is always in the irrep $\textbf{R}$, as in \eqref{eq:factor}. In this sense, electric basis states carry definite representations on their links.

The ``electric basis" terminology is motivated by the fact that, at any link $\ell \in \lk$, it can be shown that each state $\ket{\textbf{R},a,b}_{\ell}$ is an eigenstate of the squared electric operator $E^2(\ell)$, satisfying
\begin{equation}
    E^2(\ell) \ket{\textbf{R},a,b}_{\ell} = C_2(\textbf{R}) \ket{\textbf{R},a,b}_{\ell},
\end{equation}
where $C_2(\textbf{R})$ denotes the quadratic Casimir of irrep $\textbf{R}$. Note that the $\ell$ subscript on quantum states emphasizes that they are link states at link $\ell$.

\section{Gauge Invariance}\label{app:gauge}

The Hilbert space $\mathcal H$ defined in \Cref{app:electric} is not the physical Hilbert space $\mathcal H_{\phys}$, because it contains states which are not invariant under arbitrary gauge transformations. In this section, we review the conditions satisfied by global lattice states satisfying gauge invariance in
\begin{enumerate}[label=(\roman*)]
    \item the case of pure gauge theory; and
    \item the case of gauge theory coupled to fermions.
\end{enumerate}

\subsection{Pure Gauge Theory}

A gauge-invariant state $\ket{\psi}$ on the entire lattice must remain invariant under every gauge transformation that acts only on a single vertex, i.e., transformations $\Omega_{\textbf{n}_0, g}: \vx \to G$ of the form
\begin{equation}
    \Omega_{\textbf{n}_0, g}(\textbf{n}) \equiv
    \begin{cases}
        g & \text{if } \textbf{n} = \textbf{n}_0,\\
        \mathrm{Id}_G & \text{if } \textbf{n} \neq \textbf{n}_0,
    \end{cases}
\end{equation}
where $\mathrm{Id}_G$ denotes the identity element in $G$.
In fact, any finite gauge transformation on the lattice is generated by the set $\{\Omega_{\textbf{n}_0,g} \mid \textbf{n}_0 \in \vx, g \in G\}$, so invariance under these specific transformations is also sufficient for full gauge-invariance.

To write down the constraint obeyed by a lattice state $\ket{\psi}$ that remains invariant under some $\Omega_{\textbf{n}_0,g}$, it is useful to define some intermediate constructions. We define the vertex Hilbert space $\mathcal H_{\textbf{n}}$ at each lattice site $\textbf{n} \in \vx$ as the product representation (i.e., using the group action by the same group element)
\begin{equation}
    \mathcal H_{\textbf{n}} \equiv
    \begin{dcases}
        \bigotimes_{\ell \in \lk(\textbf{n})} \left(\bigoplus_{\textbf{R}} V_{\textbf{R}}\right)_{\ell} & \text{if } \textbf{n} \text{ is odd},\\
        \bigotimes_{\ell \in \lk(\textbf{n})} \left(\bigoplus_{\textbf{R}} V_{\overline{\textbf{R}}}\right)_{\ell} & \text{if } \textbf{n} \text{ is even},
    \end{dcases}
\end{equation}
where $\lk(\textbf{n})$ denotes the set of links incident to \textbf{n}, the sums run over all irreps $\textbf{R}$ of $G$, and the $\ell$ subscripts emphasize that one factor is allocated to each incident link.
We subsequently construct the enlarged Hilbert space as the direct product representation (i.e., using an independent copy of the group action on each lattice site)
\begin{equation}
    \mathcal H_{\vx} \equiv \bigotimes_{\textbf{n}\in\vx} \mathcal H_{\textbf{n}}.
\end{equation}
Note that each link is incident to exactly one even site and one odd site, so the tensor factors in $\mathcal H_{\vx}$ can be rearranged into
\begin{equation}
    \mathcal H_{\vx} \cong \bigotimes_{\ell \in \lk} \mathcal H_{\ell},
\end{equation}
where the direct product representation
\begin{equation}\label{eq:prod}
    \mathcal H_{\ell} \equiv \left(\bigoplus_{\textbf{R}} V_{\overline{\textbf{R}}}\right)_{\ell}\otimes\left(\bigoplus_{\textbf{R}'} V_{\textbf{R}'}\right)_{\ell}
\end{equation}
is referred to as the enlarged link Hilbert space at $\ell \in \lk$.

The ``enlarged" terminology alludes to the fact that we can define projectors $P_{\ell} : \mathcal H_{\ell} \to \mathcal H_{\ell}$ by linear continuation of
\begin{equation}
    P_{\ell}\Big\vert_{V_{\overline{\textbf{R}}}\otimes V_{\textbf{R}'}} =
    \begin{cases}
        \mathrm{Id}_{V_{\overline{\textbf{R}}}\otimes V_{\textbf{R}'}} & \text{if } \textbf{R} \cong \textbf{R}',\\
        0 & \text{if } \textbf{R} \ncong \textbf{R}',
    \end{cases}
\end{equation}
which satisfy the relation $P_{\ell}(\mathcal H_{\ell}) \cong \mathrm{L}^2(G)_\ell$, according to the Peter-Weyl theorem \eqref{eq:pw}.
It also follows that the global projector $P \equiv \prod_{\ell\in\lk} P_{\ell}$ satisfies $P(\mathcal H_{\vx}) \cong \mathcal H$.

The benefit of this construction becomes apparent after recognizing that $P$ is equivariant under the group actions on $\mathcal H$ and $\mathcal H_{\vx}$.
By distinguishing the ``product representations'' from ``direct product representations'' throughout our above definitions, it is ensured that the group action on the representation $\mathcal H_{\vx}$ is naturally identified with the group of gauge transformations on $\mathcal H$. In particular, this implies that the gauge-invariant sector of $\mathcal H$ is precisely the image under $P$ of the singlet sector of $\mathcal H_{\vx}$, where the latter is much simpler to directly characterize.

To this end, consider an arbitrary state $\ket{\psi} \in \mathcal H_{\vx}$ that remains invariant under some $\Omega_{\textbf{n}_0,g}$.
For starters, let us assume $\ket{\psi}$ factorizes over each $\mathcal H_{\textbf{n}}$, so that we can write
\begin{equation}
    \ket{\psi} \equiv \ket{\psi_{\textbf{n}_0}} \otimes \ket{\phi},
\end{equation}
where $\ket{\psi_{\textbf{n}_0}}$ lives in the vertex Hilbert space $\mathcal H_{\textbf{n}_0}$, and $\ket{\phi}$ resides on the remainder of the vertex Hilbert spaces.
The important observation here is that $\Omega_{\textbf{n}_0,g}$ acts non-trivially only on $\mathcal H_{\textbf{n}_0}$, so that $\ket{\psi}$ is kept invariant if and only if $\ket{\psi_{\textbf{n}_0}}$ lives in the singlet sector of $\mathcal H_{\textbf{n}_0}$.
Therefore, in the general case, a global state on $\mathcal H_{\vx}$ transforms as a singlet if and only if it can be expressed as a linear combination of product states whose factors each transform as singlets on their respective vertex Hilbert spaces.

By projecting down to $\mathcal H$, this establishes that we can form an orthonormal basis $\mathcal B$ for the gauge-invariant sector of $\mathcal H$, where every $\ket{\psi} \in \mathcal B$ is a suitable linear combination of electric basis states such that:
\begin{enumerate}[label=(\roman*)]
    \item $\ket{\psi}$ carries a definite representation on each link;
    \item $\ket{\psi}$ factorizes over every vertex Hilbert space; and
    \item the Gauss law constraint, which requires each factor to live in the singlet sector of the product of all irreps on its incident links.
\end{enumerate}
Lattice states can easily be expressed in the electric basis to satisfy the first two conditions by simply labeling each link with an arbitrary irrep and allowing any representation indices within each irrep.
The third condition takes significantly more effort to implement, as it requires determining which linear combinations of product basis states in a product of irreps can form a singlet.
An additional complication is that, although an orthonormal basis exists satisfying (i)-(iii), it is not the case that every basis satisfying (i)-(iii) is orthonormal.
These issues are discussed in \Cref{sec:2+1d}, but for our truncations it is possible to sidestep any further complications.

To conclude this section, we introduce some terminology that is useful when discussing our constructions.
First, it is common to use the term half-link for the concept associated with either just the left-transforming or just the right-transforming tensor factor in any link Hilbert space or enlarged link Hilbert space.
Second, we use the term vertex singlet to describe a state living in the singlet sector within some vertex Hilbert space $\mathcal H_{\textbf{n}}$.

For instance, any vertex singlet at an odd site lives on the product of the right-half of the links incident to that site, and any vertex singlet at an even site lives on the product of the left-half of the links incident to that site.
This will be modified slightly as we incorporate fermions into the discussion.

\subsection{Gauge Theory with Fermions}

Given the constructions in the pure gauge case, we now need to take into account the extra fermion degrees of freedom to upgrade our previous definitions.

The primary object that enters the picture is a copy of the fermionic Fock space at each lattice site
\begin{equation}
    \pq_{\textbf{n}} \equiv \bigoplus_{p=0}^{N_c} \left(\bigwedge_{j=1}^p V_{\textbf{N}}\right),
\end{equation}
where $V_{\textbf{N}}\cong \mathbb C^{N_c}$ denotes the target space of the fundamental representation (denoted \textbf{N}) of $\SU(N_c)$, and we use the convention that the $p=0$ term contributes a singlet (denoted \textbf{1}) to the sum.
The anti-symmetrized product ensures consistency with the Pauli exclusion principle.
Note that $\pq_{\textbf{n}}$ is a representation of $\SU(N_c)$ in its own right, and has dimension $\dim \pq_{\textbf{n}} = 2^{N_c}$ corresponding to all possible subsets of the color indices $\{1,\dots, N_c\}$ to be represented among the excited fermions at lattice site $\textbf{n}$.

Under a gauge transformation $\Omega : G \to \vx$, a fermion state in $\pq_{\textbf{n}}$ should transform only by the action of the group element $\Omega(\textbf{n})$.
We can thus define the vertex Hilbert space with fermions as the product representation
\begin{equation}
    \mathcal H_{\textbf{n}}^{\mathrm F} \equiv \pq_{\textbf{n}} \otimes \mathcal H_{\textbf{n}},
\end{equation}
where $\mathcal H_{\textbf{n}}$ is the vertex Hilbert space without fermions, as defined previously in the case of pure gauge theory.
The enlarged Hilbert space with fermions is then defined as the direct product representation
\begin{equation}
    \mathcal H_{\vx}^{\mathrm F} \equiv \bigotimes_{\textbf{n} \in \vx} \mathcal H_{\textbf{n}}^{\mathrm F},
\end{equation}
which ensures that its group action coincides with the group of lattice gauge transformations.

Note that $\mathcal H_{\vx}^{\mathrm F} \cong \mathcal H_{\vx} \otimes \left(\bigotimes_{\textbf{n} \in \vx} \pq_{\textbf{n}}\right)$, where $\mathcal H_{\vx}$ is the enlarged Hilbert space from the pure gauge case.
Therefore, the projector $P$ from the previous section extends naturally to this setting, by defining it to act as the identity on all fermionic Fock space factors, and keeping its action as before on the $\mathcal H_{\vx}$ factor.
We can thus define the global Hilbert space with fermions as
\begin{equation}
    \mathcal H^{\mathrm F} \equiv P(\mathcal H_{\vx}^{\mathrm F}),
\end{equation}
which naturally carries a representation of the group of lattice gauge transformations.
The projector $P$ is equivariant as before, so gauge-invariant states on $\mathcal H^{\mathrm F}$ (which span the physical Hilbert space with fermions, $\mathcal H_{\phys}^{\mathrm F}$) correspond to projections of the singlet sector of $\mathcal H_{\vx}^{\mathrm F}$.

In complete analogy to the pure gauge case, any state $\ket{\psi} \in \mathcal H_{\vx}^{\mathrm F}$ that transforms as a singlet must be expressible as a linear combination of states that factorize over each $\mathcal H_{\textbf{n}}^{\mathrm F}$, such that each factor separately transforms as a singlet on its respective vertex space.
This implies the existence of an orthonormal basis $\mathcal B$ for $\mathcal H_{\phys}^{\mathrm F}$, where every $\ket{\psi} \in \mathcal B$ factorizes over each $\mathcal H_{\textbf{n}}^{\mathrm F}$ and satisfies the properties:
\begin{enumerate}[label=(\roman*)]
    \item $\ket{\psi}$ carries a definite representation on each lattice site and link;
    \item $\ket{\psi}$ factorizes over each vertex space $\mathcal H_{\textbf{n}}^{\mathrm F}$; and
    \item each factor lives in the singlet sector of the product of all associated irrep data (i.e., the irreps on half-links incident to the associated site, and the irrep on the site itself).
\end{enumerate}
As before, the first two conditions are easily satisfied by assigning irrep labels to all lattice sites and links, but the third condition requires intertwining the local irrep data into singlets that we call vertex singlets.

\section{Hopping Master Equation}\label{app:calc}

To derive the equation \eqref{eq:master} which we ultimately use for all hopping transitions, we need to make some important observations regarding the vertex tensors used to define any vertex singlet state on the lattice.
We will assume the same notation used in \Cref{sec:trunc} in the main text.

First, by gauge-invariance, we must have
\begin{align}
    V^1_{a'b'c'} &= V^1_{abc} \mathbf{R^i}_{a'a}(g) \overline{\mathbf{A^p}}_{b'b}(g) \mathbf R_{c'c}(g),\nonumber\\
    V^2_{d'e'f'} &= V^2_{def} \overline{\mathbf R}_{d'd}(g) \mathbf{A^q}_{e'e}(g) \overline{\mathbf{R^j}}_{f'f}(g),
\end{align}
for every $g \in \SU(N_c)$.
Additionally, if the singlet state is \textit{normalized}, then we must also have
\begin{align}
    \left(V^1_{abc}\right)^* V^1_{abc} &= 1,\nonumber\\
    \left(V^2_{def}\right)^* V^2_{def} &= 1,
\end{align}
where the complex conjugation is necessary because Clebsch-Gordan coefficients for $\SU(N_c)$ cannot always be taken as real-valued.
Similar properties exist for the vertex tensors $W^1$ and $W^2$ in the final state $\ket{W}$, with the main differences being that $\mathbf{R} \to \mathbf{R'}$, $p \to p+1$, and $q\to q+1$.

An important corollary of \eqref{eq:T} is that
\begin{equation}\label{eq:T2}
    \chi_b \ket{\overline{\mathbf{A^r}},c} = T^{N_c-r-1}_{cba} \ket{\overline{\mathbf{A^{r+1}}},a},
\end{equation}
by identifying $\overline{\mathbf{A^r}} = \mathbf{A^{N_c-r}}$ and $\overline{\mathbf{A^{r+1}}} = \mathbf{A^{N_c-r-1}}$.
The conjugation symmetry between fermions and anti-fermions allows us to assume without loss of generality that the phases of the irrep basis states in the Fock space are chosen such that in fact
\begin{equation}
    \chi_b \ket{\overline{\mathbf{A^r}},c} = T^r_{abc} \ket{\overline{\mathbf{A^{r+1}}},a},
\end{equation}
implying that $T^{N_c-r-1}_{cba} = T^r_{abc}$.

By plugging these expressions into \eqref{eq:hop}, we find
\begin{align}\label{eq:contractedintegral}
    \bra{W} H_\hop(\ell) \ket{V}
    & = V^1_{abc} \left(W^1_{a\beta\gamma}\right)^* T^{p}_{\beta\sigma b}\times\nonumber\\
    & \times \bra{\mathbf{R'},\delta,\gamma} U_{\rho\sigma} \ket{\mathbf{R},d,c}\times\nonumber\\
    & \times V^2_{def} \left(W^2_{\delta\eps f}\right)^* T^{q}_{\eps\rho e}.
\end{align}
The only matrix element left to compute is the middle piece
\begin{widetext}
\begin{align}\label{eq:Uintegral}
    \bra{\mathbf{R'},\delta,\gamma} U_{\rho\sigma} \ket{\mathbf{R},d,c}
    & = \int_{\SU(N_c)} \dd g\, \braket{\mathbf{R'},\delta,\gamma}{g} \bra{g} U_{\rho\sigma} \ket{\mathbf{R},d,c}\nonumber\\
    & = \sqrt{\dim\mathbf R' \cdot \dim\mathbf R} \cdot \int_{\SU(N_c)} \dd g\, \overline{\mathbf{R'}}_{\delta\gamma}(g) \cdot \mathbf{N}_{\rho\sigma}(g) \cdot \mathbf{R}_{dc}(g).
\end{align}
\end{widetext}
This is proportional to an outer product of Clebsch-Gordan coefficients for combining $\mathbf R \otimes \mathbf N \to \mathbf{R'}$, and in principle we could replace those coefficients in our previous expression to obtain a closed-form hopping matrix element.
But it turns out that a substantial simplification exists, if we take into account the contractions already present in \eqref{eq:contractedintegral}.

To see this, we must first make use of the transformation law of the fermion fields under an arbitrary gauge transformation $\Omega : \vx \to \SU(N_c)$.
Specifically, if $U(\Omega)$ is the unitary representation of the gauge transformation on quantum states, and if $\Omega(\mathbf{n}) = g$, then we have
\begin{equation}
    U(\Omega) \chi^{\dagger}(\mathbf{n})_a U(\Omega)^{\dagger} = \mathbf{N}_{ba}(g) \chi^{\dagger}(\mathbf{n})_b,
\end{equation}
for any $\mathbf{n} \in \vx$.
This follows from the requirement that gauge transformations must act as the irreps that quantum states live in:
\begin{equation}
    U(\Omega) \ket{\mathbf{N},a}_{\mathbf{n}} = \mathbf{N}_{ba}(g) \ket{\mathbf{N},b}_{\mathbf{n}}.
\end{equation}
Keeping the lattice site $\mathbf{n}$ implicit, this implies that for $0 \le r \le N_c - 1$,
\begin{align}
    T^r_{abc} &= \bra{\mathbf{A^{r+1}},a} \chi^\dagger_b \ket{\mathbf{A^r},c}\nonumber\\
    & = \bra{\mathbf{A^{r+1}},a} U(\Omega)^{\dagger} U(\Omega) \chi^\dagger_b U(\Omega)^\dagger U(\Omega) \ket{\mathbf{A^r},c}\nonumber\\
    & = \bra{\mathbf{A^{r+1}},a'} \overline{\mathbf{A^{r+1}}}_{a' a}(g) \mathbf{N}_{b' b}(g) \chi^{\dagger}_{b'} \mathbf{A^r}_{c' c}(g)\ket{\mathbf{A^r},c'}\nonumber\\
    & = T^r_{a'b'c'} \overline{\mathbf{A^{r+1}}}_{a' a}(g) \mathbf{N}_{b' b}(g)\mathbf{A^r}_{c' c}(g),
\end{align}
where $g \in \SU(N_c)$ can be taken as arbitrary.

By transforming indices according to the rules discussed above, we can rewrite the first factor of \eqref{eq:contractedintegral} in the form
\begin{widetext}
\begin{align}
    V^1_{abc} \left(W^1_{a\beta\gamma}\right)^* T^{p}_{\beta\sigma b} \cdot  = &\,V^1_{a'b'c'} \mathbf{R^i}_{a'a}(h) \overline{\mathbf{A^p}}_{b'b}(h) \mathbf R_{c'c}(h)
\left(W^1_{a''\beta'\gamma'}\right)^* \mathbf{R^i}_{a''a}(h)^* \overline{\mathbf{A^{p+1}}}_{\beta'\beta}(h)^* \mathbf{R'}_{\gamma'\gamma}(h)^*\nonumber\\
    & \times T^p_{\beta''\sigma'b''} \overline{\mathbf{A^{p+1}}}_{\beta'' \beta}(h) \mathbf{N}_{\sigma' \sigma}(h)\mathbf{A^p}_{b'' b}(h)\nonumber\\
    = &\, V^1_{a'b'c'} \left(W^1_{a'\beta'\gamma'}\right)^* T^p_{\beta'\sigma'b'} \cdot \mathbf R_{c'c}(h) \mathbf{R'}_{\gamma'\gamma}(h)^* \mathbf{N}_{\sigma' \sigma}(h),
\end{align}
after simplifying the products of conjugate representation matrices due to unitarity.
Note that this is true for any $h \in \SU(N_c)$

Contracting this against the second factor in \eqref{eq:contractedintegral} requires plugging in the integrand from \eqref{eq:Uintegral}.
We will first discuss the contraction against just the integrand.
To this end, let $g^*$ denote the complex conjugate of $g$ for any $g \in \SU(N_c)$.
By using the above transformation laws with $h = g^*$, we find
\begin{align}
    V^1_{abc} \left(W^1_{a\beta\gamma}\right)^* T^{p}_{\beta\sigma b} \cdot \overline{\mathbf{R'}}_{\delta\gamma}(g) \mathbf{N}_{\rho\sigma}(g) \mathbf{R}_{dc}(g) &= V^1_{a'b'c'} \left(W^1_{a'\beta'\gamma'}\right)^* T^p_{\beta'\sigma'b'} \cdot \mathbf R_{c'c}(g^*) \mathbf{R'}_{\gamma'\gamma}(g^*)^* \mathbf{N}_{\sigma' \sigma}(g^*) \cdot \overline{\mathbf{R'}}_{\delta\gamma}(g) \mathbf{N}_{\rho\sigma}(g) \mathbf{R}_{dc}(g)\nonumber \\
    & = V^1_{abd} \left(W^1_{a\beta\delta}\right)^* T^{p}_{\beta\rho b}.
\end{align}

This contracted integrand can be directly plugged into the integral \eqref{eq:Uintegral}, which yields
\begin{align}
V^1_{abc} \left(W^1_{a\beta\gamma}\right)^* T^{p}_{\beta\sigma b}\cdot \bra{\mathbf{R'},\delta,\gamma} U_{\rho\sigma} \ket{\mathbf{R},d,c} & = \sqrt{\dim\mathbf R' \cdot \dim\mathbf R} \cdot \int_{\SU(N_c)} \dd g\, \overline{\mathbf{R'}}_{\delta\gamma}(g) \cdot \mathbf{N}_{\rho\sigma}(g) \cdot \mathbf{R}_{dc}(g) \nonumber\\ 
& = \sqrt{\dim\mathbf{R'} \cdot \dim\mathbf{R}} \cdot V^1_{abd} \left(W^1_{a\beta\delta}\right)^* T^{p}_{\beta\rho b},
\end{align}
due to the unit-normalized Haar measure.

Finally, plugging this into \eqref{eq:contractedintegral}, we obtain the simplified form
\begin{equation}
    \bra{W} H_\hop(\ell) \ket{V} = \sqrt{\dim\mathbf{R'} \cdot \dim\mathbf{R}} \cdot V^1_{abd} \left(W^1_{a\beta\delta}\right)^* T^{p}_{\beta\rho b} \cdot V^2_{def} \left(W^2_{\delta\eps f}\right)^* T^{q}_{\eps\rho e},
\end{equation}
\end{widetext}
which matches \eqref{eq:master} up to a re-labeling of indices, as desired.

The fact that this simplification exists is a realization of the fact that the gauge field in $1+1$D Yang-Mills theory (with staggered fermions) can be fully integrated out of the theory, leading to interactions that directly couple between fermions.
In particular, this simplification highlights that in \textit{any dimension}, we can re-frame hopping terms as local representations of a one-dimensional interaction, which must then be glued together through linear combinations dictated by the representation theory of $\SU(N_c)$ and graph structure of the lattice, as used in the main text.

\section{Tabulating Singlet States}\label{app:tab}

Here, we perform a systematic tabulation strategy to enumerate all gauge-invariant double-vertex cut states, as presented in \Cref{tab:cuts}.
We work at energy cutoffs dictated by
\begin{equation}
    (n_e, n_f) \in \left\{(1,1), (1,2), (2,1)\right\}
\end{equation}
as used in the main text.
This means the irreps we have access to are
\begin{equation}
    \left\{\mathbf{1},\mathbf{N},\overline{\mathbf{N}},\mathbf{Ad},\mathbf{A^2},\overline{\mathbf{A^2}},\mathbf{S^2},\overline{\mathbf{S^2}}\right\}.
\end{equation}

To begin, let us tabulate all possible gauge-invariant states that can reside at a single vertex (i.e. the vertex singlets).
We are allowed to use the irreps listed above with two constraints: (i) fermions / anti-fermions must live in anti-symmetrized irreps with $n_f \le 2$; and (ii) we cannot simultaneously have $n_e = 2$ and $n_f = 2$ in a vertex state.
To make the tabulation slightly easier, note that specifying the link irreps on the two half-links adjacent to a given lattice site uniquely determines the irrep on the lattice site, if it can exist at the level of fusion rules.
(This follows from the discussion in \Cref{sec:trunc}.)

Additionally, we will only tabulate the vertex singlets that are allowed around an odd (anti-fermion-like) lattice site, because charge conjugation implies that these states are in one-to-one correspondence with the vertex singlets around an even lattice site.
This is also convenient because, in our notation convention, the irrep used to label a link is the irrep on the half-link adjacent to its odd endpoint.

Most pairs of allowed irreps from the above list will define an allowed odd vertex singlet if the sum of their net fluxes (defined in \Cref{tab:flow}) is between $0$ and $2$ (inclusive), which means the implied charge occupancy at the lattice site is not outright forbidden.
The only exceptions are for the pairs $(\mathbf{Ad},\mathbf{1})$, $(\mathbf{A^2},\mathbf{1})$, and $(\mathbf{S^2},\mathbf{1})$ (along with permutations).
Interestingly, the exact reason for why each pair is disallowed are all different: in the case of $(\mathbf{Ad},\mathbf{1})$, the implied charge occupancy from the net flux is $0$, but $\mathbf{Ad}\otimes\mathbf{1}\otimes\mathbf{1}$ does not contain a singlet, so gauge-invariance cannot be satisfied; in the case of $(\mathbf{A^2},\mathbf{1})$, gauge-invariance can be satisfied, but the implied charge occupancy is $2$, which cannot co-exist with $\mathbf{A^2}$ on a link unless the energy cutoff is at least $(n_e,n_f)=(2,2)$; and in the case of $(\mathbf{S^2},\mathbf{1})$, both the energy cutoff and existence of a gauge-invariant singlet are not satisfied.

\begin{table*}[htp]
\centering

\begin{minipage}[t][][]{0.33\linewidth}
\begin{tabular}{|c|c|}

\hline
\textbf{Odd Singlet} & \textbf{Anti-Fermions}\\
\hline\hline

\begin{minipage}[c][1cm][c]{0.15\linewidth}
$\big\langle\mathbf{1},\mathbf{1}$
\end{minipage}
&
\begin{minipage}[c][1cm][c]{0.15\linewidth}
0
\end{minipage}
\\
\hline

\begin{minipage}[c][1cm][c]{0.15\linewidth}
$\big\langle\mathbf{N},\mathbf{1}$
\end{minipage}
&
\begin{minipage}[c][1cm][c]{0.15\linewidth}
1
\end{minipage}
\\
\hline

\begin{minipage}[c][1cm][c]{0.15\linewidth}
$\big\langle\mathbf{1},\mathbf{N}$
\end{minipage}
&
\begin{minipage}[c][1cm][c]{0.15\linewidth}
1
\end{minipage}
\\
\hline

\begin{minipage}[c][1cm][c]{0.15\linewidth}
$\big\langle\mathbf{N},\mathbf{N}$
\end{minipage}
&
\begin{minipage}[c][1cm][c]{0.15\linewidth}
2
\end{minipage}
\\
\hline

\begin{minipage}[c][1cm][c]{0.15\linewidth}
$\big\langle\mathbf{N},\overline{\mathbf{N}}$
\end{minipage}
&
\begin{minipage}[c][1cm][c]{0.15\linewidth}
0
\end{minipage}
\\
\hline

\begin{minipage}[c][1cm][c]{0.15\linewidth}
$\big\langle\overline{\mathbf{N}},\mathbf{N}$
\end{minipage}
&
\begin{minipage}[c][1cm][c]{0.15\linewidth}
0
\end{minipage}
\\
\hline
\end{tabular}
\end{minipage}%
\begin{minipage}[t][][]{0.33\linewidth}
\begin{tabular}{|c|c|}

\hline
\textbf{Odd Singlet} & \textbf{Anti-Fermions}\\
\hline\hline

\begin{minipage}[c][1cm][c]{0.15\linewidth}
$\big\langle\mathbf{Ad},\mathbf{N}$
\end{minipage}
&
\begin{minipage}[c][1cm][c]{0.15\linewidth}
1
\end{minipage}
\\
\hline

\begin{minipage}[c][1cm][c]{0.15\linewidth}
$\big\langle\mathbf{N},\mathbf{Ad}$
\end{minipage}
&
\begin{minipage}[c][1cm][c]{0.15\linewidth}
1
\end{minipage}
\\
\hline

\begin{minipage}[c][1cm][c]{0.15\linewidth}
$\big\langle\mathbf{A^2},\overline{\mathbf{N}}$
\end{minipage}
&
\begin{minipage}[c][1cm][c]{0.15\linewidth}
1
\end{minipage}
\\
\hline

\begin{minipage}[c][1cm][c]{0.15\linewidth}
$\big\langle\overline{\mathbf{N}},\mathbf{A^2}$
\end{minipage}
&
\begin{minipage}[c][1cm][c]{0.15\linewidth}
1
\end{minipage}
\\
\hline

\begin{minipage}[c][1cm][c]{0.15\linewidth}
$\big\langle\mathbf{S^2},\overline{\mathbf{N}}$
\end{minipage}
&
\begin{minipage}[c][1cm][c]{0.15\linewidth}
1
\end{minipage}
\\
\hline

\begin{minipage}[c][1cm][c]{0.15\linewidth}
$\big\langle\overline{\mathbf{N}},\mathbf{S^2}$
\end{minipage}
&
\begin{minipage}[c][1cm][c]{0.15\linewidth}
1
\end{minipage}
\\
\hline
\end{tabular}
\end{minipage}%
\begin{minipage}[t][][]{0.33\linewidth}
\begin{tabular}{|c|c|}

\hline
\textbf{Odd Singlet} & \textbf{Anti-Fermions}\\
\hline\hline

\begin{minipage}[c][1cm][c]{0.15\linewidth}
$\big\langle\mathbf{Ad},\mathbf{Ad}$
\end{minipage}
&
\begin{minipage}[c][1cm][c]{0.15\linewidth}
0
\end{minipage}
\\
\hline

\begin{minipage}[c][1cm][c]{0.15\linewidth}
$\big\langle\mathbf{A^2},\overline{\mathbf{A^2}}$
\end{minipage}
&
\begin{minipage}[c][1cm][c]{0.15\linewidth}
0
\end{minipage}
\\
\hline

\begin{minipage}[c][1cm][c]{0.15\linewidth}
$\big\langle\overline{\mathbf{A^2}},\mathbf{A^2}$
\end{minipage}
&
\begin{minipage}[c][1cm][c]{0.15\linewidth}
0
\end{minipage}
\\
\hline

\begin{minipage}[c][1cm][c]{0.15\linewidth}
$\big\langle\mathbf{S^2},\overline{\mathbf{S^2}}$
\end{minipage}
&
\begin{minipage}[c][1cm][c]{0.15\linewidth}
0
\end{minipage}
\\
\hline

\begin{minipage}[c][1cm][c]{0.15\linewidth}
$\big\langle\overline{\mathbf{S^2}},\mathbf{S^2}$
\end{minipage}
&
\begin{minipage}[c][1cm][c]{0.15\linewidth}
0
\end{minipage}
\\
\hline
\end{tabular}
\end{minipage}%

\caption{All possible odd singlet states that can be written down consistently, and continue to obey our energy cutoffs. There are $17$ in total, where we have kept states related by discrete lattice symmetries as \textit{distinct}.}\label{tab:oddsinglets}
\end{table*}

We will use the unclosed-bracket notation $\big\langle\mathbf{R},\mathbf{R'}$ for odd lattice sites and $\mathbf{R},\mathbf{R'}\big\rangle$ for even lattice sites.
Additionally, the irreps used in the even site notation will always be the \textit{conjugates} of the irreps that actually contract with the fermion state at the even site.
This is convenient, because it makes the bijection between odd singlet states and even singlet states as simple as
\begin{equation}
\big\langle\mathbf{R},\mathbf{R'} \to \mathbf{R},\mathbf{R'}\big\rangle.
\end{equation}
This particular bijection is charge conjugation, but other bijections can also be defined, for instance corresponding to the (discrete) group of local lattice symmetries.
For our purposes, we will disregard all equivalences besides strictly charge conjugation.
Every allowed odd vertex singlet in this notation is tabulated in \Cref{tab:oddsinglets}.

Another convenience this notation implies is that a consistent double-vertex singlet $\ket{\mathbf{R^1},\mathbf{R^2},\mathbf{R^3}}$ (in the triple-irrep notation used throughout the main text) exists if and only if we can overlap the odd singlet notation $\big\langle\mathbf{R^1},\mathbf{R^2}$ with the even singlet notation $\mathbf{R^2},\mathbf{R^3}\big\rangle$ at the middle irrep, to write down $\big\langle\mathbf{R^1},\mathbf{R^2},\mathbf{R^3}\big\rangle$ as a consistent expression of three irreps.
The conjugation embedded in the even site notation simultaneously ensures compatibility with our irrep convention, and also enforces the fact that on any link, the two half-link states must transform under irreps that are conjugate to each other.

\begin{table*}[htp]
\centering

\begin{minipage}{\linewidth}
\begin{tabular}{|c|c|c|c|c|c|c|}
\hline
\multicolumn{7}{|c|}{\textbf{Allowed Gauge-Invariant Double-Vertex Cut States}} \\
\hline
\begin{minipage}[c][1cm][c]{0.13\linewidth}
$\big\langle\mathbf{1},\mathbf{1},\mathbf{1}\big\rangle$
\end{minipage}
&
\begin{minipage}[c][1cm][c]{0.13\linewidth}
$\big\langle\mathbf{1},\mathbf{1},\mathbf{N}\big\rangle$
\end{minipage}
&
\begin{minipage}[c][1cm][c]{0.13\linewidth}
$\big\langle\mathbf{N},\mathbf{1},\mathbf{1}\big\rangle$
\end{minipage}
&
\begin{minipage}[c][1cm][c]{0.13\linewidth}
$\big\langle\mathbf{N},\mathbf{1},\mathbf{N}\big\rangle$
\end{minipage}
&
\begin{minipage}[c][1cm][c]{0.13\linewidth}
$\big\langle\mathbf{1},\mathbf{N},\mathbf{1}\big\rangle$
\end{minipage}
&
\begin{minipage}[c][1cm][c]{0.13\linewidth}
$\big\langle\mathbf{1},\mathbf{N},\mathbf{N}\big\rangle$
\end{minipage}
&
\begin{minipage}[c][1cm][c]{0.13\linewidth}
$\big\langle\mathbf{1},\mathbf{N},\overline{\mathbf{N}}\big\rangle$
\end{minipage}
\\
\hline
\begin{minipage}[c][1cm][c]{0.13\linewidth}
$\big\langle\mathbf{1},\mathbf{N},\mathbf{Ad}\big\rangle$
\end{minipage}
&
\begin{minipage}[c][1cm][c]{0.13\linewidth}
$\big\langle\mathbf{N},\mathbf{N},\mathbf{1}\big\rangle$
\end{minipage}
&
\begin{minipage}[c][1cm][c]{0.13\linewidth}
$\big\langle\mathbf{N},\mathbf{N},\mathbf{N}\big\rangle$
\end{minipage}
&
\begin{minipage}[c][1cm][c]{0.13\linewidth}
$\big\langle\mathbf{N},\mathbf{N},\overline{\mathbf{N}}\big\rangle$
\end{minipage}
&
\begin{minipage}[c][1cm][c]{0.13\linewidth}
$\big\langle\mathbf{N},\overline{\mathbf{N}},\mathbf{N}\big\rangle$
\end{minipage}
&
\begin{minipage}[c][1cm][c]{0.13\linewidth}
$\big\langle\mathbf{N},\overline{\mathbf{N}},\mathbf{A^2}\big\rangle$
\end{minipage}
&
\begin{minipage}[c][1cm][c]{0.13\linewidth}
$\big\langle\mathbf{N},\overline{\mathbf{N}},\mathbf{S^2}\big\rangle$
\end{minipage}
\\
\hline
\begin{minipage}[c][1cm][c]{0.13\linewidth}
$\big\langle\overline{\mathbf{N}},\mathbf{N},\mathbf{1}\big\rangle$
\end{minipage}
&
\begin{minipage}[c][1cm][c]{0.13\linewidth}
$\big\langle\overline{\mathbf{N}},\mathbf{N},\mathbf{N}\big\rangle$
\end{minipage}
&
\begin{minipage}[c][1cm][c]{0.13\linewidth}
$\big\langle\overline{\mathbf{N}},\mathbf{N},\overline{\mathbf{N}}\big\rangle$
\end{minipage}
&
\begin{minipage}[c][1cm][c]{0.13\linewidth}
$\big\langle\overline{\mathbf{N}},\mathbf{N},\mathbf{Ad}\big\rangle$
\end{minipage}
&
\begin{minipage}[c][1cm][c]{0.13\linewidth}
$\big\langle\mathbf{Ad},\mathbf{N},\mathbf{1}\big\rangle$
\end{minipage}
&
\begin{minipage}[c][1cm][c]{0.13\linewidth}
$\big\langle\mathbf{Ad},\mathbf{N},\overline{\mathbf{N}}\big\rangle$
\end{minipage}
&
\begin{minipage}[c][1cm][c]{0.13\linewidth}
$\big\langle\mathbf{Ad},\mathbf{N},\mathbf{Ad}\big\rangle$
\end{minipage}
\\
\hline
\begin{minipage}[c][1cm][c]{0.13\linewidth}
$\big\langle\mathbf{N},\mathbf{Ad},\mathbf{N}\big\rangle$
\end{minipage}
&
\begin{minipage}[c][1cm][c]{0.13\linewidth}
$\big\langle\mathbf{N},\mathbf{Ad},\mathbf{Ad}\big\rangle$
\end{minipage}
&
\begin{minipage}[c][1cm][c]{0.13\linewidth}
$\big\langle\mathbf{A^2},\overline{\mathbf{N}},\mathbf{N}\big\rangle$
\end{minipage}
&
\begin{minipage}[c][1cm][c]{0.13\linewidth}
$\big\langle\mathbf{A^2},\overline{\mathbf{N}},\mathbf{A^2}\big\rangle$
\end{minipage}
&
\begin{minipage}[c][1cm][c]{0.13\linewidth}
$\big\langle\mathbf{A^2},\overline{\mathbf{N}},\mathbf{S^2}\big\rangle$
\end{minipage}
&
\begin{minipage}[c][1cm][c]{0.13\linewidth}
$\big\langle\overline{\mathbf{N}},\mathbf{A^2},\overline{\mathbf{N}}\big\rangle$
\end{minipage}
&
\begin{minipage}[c][1cm][c]{0.13\linewidth}
$\big\langle\overline{\mathbf{N}},\mathbf{A^2},\overline{\mathbf{A^2}}\big\rangle$
\end{minipage}
\\
\hline
\begin{minipage}[c][1cm][c]{0.13\linewidth}
$\big\langle\mathbf{S^2},\overline{\mathbf{N}},\mathbf{N}\big\rangle$
\end{minipage}
&
\begin{minipage}[c][1cm][c]{0.13\linewidth}
$\big\langle\mathbf{S^2},\overline{\mathbf{N}},\mathbf{A^2}\big\rangle$
\end{minipage}
&
\begin{minipage}[c][1cm][c]{0.13\linewidth}
$\big\langle\mathbf{S^2},\overline{\mathbf{N}},\mathbf{S^2}\big\rangle$
\end{minipage}
&
\begin{minipage}[c][1cm][c]{0.13\linewidth}
$\big\langle\overline{\mathbf{N}},\mathbf{S^2},\overline{\mathbf{N}}\big\rangle$
\end{minipage}
&
\begin{minipage}[c][1cm][c]{0.13\linewidth}
$\big\langle\overline{\mathbf{N}},\mathbf{S^2},\overline{\mathbf{S^2}}\big\rangle$
\end{minipage}
&
\begin{minipage}[c][1cm][c]{0.13\linewidth}
$\big\langle\mathbf{Ad},\mathbf{Ad},\mathbf{N}\big\rangle$
\end{minipage}
&
\begin{minipage}[c][1cm][c]{0.13\linewidth}
$\big\langle\mathbf{Ad},\mathbf{Ad},\mathbf{Ad}\big\rangle$
\end{minipage}
\\
\hline
\begin{minipage}[c][1cm][c]{0.13\linewidth}
$\big\langle\mathbf{A^2},\overline{\mathbf{A^2}},\mathbf{A^2}\big\rangle$
\end{minipage}
&
\begin{minipage}[c][1cm][c]{0.13\linewidth}
$\big\langle\overline{\mathbf{A^2}},\mathbf{A^2},\overline{\mathbf{N}}\big\rangle$
\end{minipage}
&
\begin{minipage}[c][1cm][c]{0.13\linewidth}
$\big\langle\overline{\mathbf{A^2}},\mathbf{A^2},\overline{\mathbf{A^2}}\big\rangle$
\end{minipage}
&
\begin{minipage}[c][1cm][c]{0.13\linewidth}
$\big\langle\mathbf{S^2},\overline{\mathbf{S^2}},\mathbf{S^2}\big\rangle$
\end{minipage}
&
\begin{minipage}[c][1cm][c]{0.13\linewidth}
$\big\langle\overline{\mathbf{S^2}},\mathbf{S^2},\overline{\mathbf{N}}\big\rangle$
\end{minipage}
&
\begin{minipage}[c][1cm][c]{0.13\linewidth}
$\big\langle\overline{\mathbf{S^2}},\mathbf{S^2},\overline{\mathbf{S^2}}\big\rangle$
\end{minipage}
&
\\
\hline
\end{tabular}
\end{minipage}

\caption{All possible gauge-invariant double-vertex cut states that can be written down consistently, and continue to obey our energy cutoffs. There are $41$ in total, where we have kept states related by discrete lattice symmetries as \textit{distinct}.}\label{tab:doublesinglets}
\end{table*}

To construct all possible singlet states that can reside at a double-vertex cut, we must combine every compatible pair of an odd site singlet with an even site singlet.
This is necessary to ensure a consistent wavefunction for the Wilson line on the whole link built from the tensor product.
Doing so yields $43$ gauge-invariant double-vertex cut states with a consistent Wilson line wavefunction---but two of them, namely $\big\langle \mathbf{Ad}, \mathbf{N}, \mathbf{N}\big\rangle$ and $\big\langle\mathbf{N},\mathbf{N},\mathbf{Ad}\big\rangle$, can only exist at energy cutoff $(n_e, n_f) = (2,2)$ or higher.
\Cref{tab:doublesinglets} shows the list of all allowed gauge-invariant double-vertex cut states, after these states are removed.

\begin{table*}[htp]
\centering

\begin{tabular}{|c|c|c|}

\hline
\textbf{Raising Operation} & $(n_e, n_f)$ \textbf{Transition} & \textbf{Minimum} $(n_e,n_f)$\\
\hline\hline

\begin{minipage}[c][1cm][c]{0.3\linewidth}
$\big\langle\mathbf{1},\mathbf{1},\mathbf{1}\big\rangle \to \big\langle\mathbf{1},\mathbf{N},\mathbf{1}\big\rangle$
\end{minipage}
&
\begin{minipage}[c][1cm][c]{0.2\linewidth}
$(1,1)\to(1,1)$
\end{minipage}
&
\begin{minipage}[c][1cm][c]{0.2\linewidth}
$(1,1)$
\end{minipage}
\\
\hline

\begin{minipage}[c][1cm][c]{0.3\linewidth}
$\big\langle\mathbf{1},\mathbf{1},\mathbf{N}\big\rangle \to \big\langle\mathbf{1},\mathbf{N},\mathbf{N}\big\rangle$
\end{minipage}
&
\begin{minipage}[c][1cm][c]{0.2\linewidth}
$(1,1)\to(1,2)$
\end{minipage}
&
\begin{minipage}[c][1cm][c]{0.2\linewidth}
$(1,2)$
\end{minipage}
\\
\hline

\begin{minipage}[c][1cm][c]{0.3\linewidth}
$\big\langle\mathbf{N},\mathbf{1},\mathbf{1}\big\rangle \to \big\langle\mathbf{N},\mathbf{N},\mathbf{1}\big\rangle$
\end{minipage}
&
\begin{minipage}[c][1cm][c]{0.2\linewidth}
$(1,1)\to(1,2)$
\end{minipage}
&
\begin{minipage}[c][1cm][c]{0.2\linewidth}
$(1,2)$
\end{minipage}
\\
\hline

\begin{minipage}[c][1cm][c]{0.3\linewidth}
$\big\langle\mathbf{N},\mathbf{1},\mathbf{N}\big\rangle \to \big\langle\mathbf{N},\mathbf{N},\mathbf{N}\big\rangle$
\end{minipage}
&
\begin{minipage}[c][1cm][c]{0.2\linewidth}
$(1,1) \to (1,2)$
\end{minipage}
&
\begin{minipage}[c][1cm][c]{0.2\linewidth}
$(1,2)$
\end{minipage}
\\
\hline

\begin{minipage}[c][1cm][c]{0.3\linewidth}
$\big\langle\mathbf{N},\overline{\mathbf{N}},\mathbf{N}\big\rangle \to \big\langle\mathbf{N},\mathbf{1},\mathbf{N}\big\rangle$
\end{minipage}
&
\begin{minipage}[c][1cm][c]{0.2\linewidth}
$(1,1) \to (1,1)$
\end{minipage}
&
\begin{minipage}[c][1cm][c]{0.2\linewidth}
$(1,1)$
\end{minipage}
\\
\hline

\begin{minipage}[c][1cm][c]{0.3\linewidth}
$\big\langle\mathbf{N},\overline{\mathbf{N}},\mathbf{N}\big\rangle \to \big\langle\mathbf{N},\mathbf{Ad},\mathbf{N}\big\rangle$
\end{minipage}
&
\begin{minipage}[c][1cm][c]{0.2\linewidth}
$(1,1) \to (2,1)$
\end{minipage}
&
\begin{minipage}[c][1cm][c]{0.2\linewidth}
$(2,1)$
\end{minipage}
\\
\hline

\begin{minipage}[c][1cm][c]{0.3\linewidth}
$\big\langle\overline{\mathbf{N}},\mathbf{N},\overline{\mathbf{N}}\big\rangle \to \big\langle\overline{\mathbf{N}},\mathbf{A^2},\overline{\mathbf{N}}\big\rangle$
\end{minipage}
&
\begin{minipage}[c][1cm][c]{0.2\linewidth}
$(1,1) \to (2,1)$
\end{minipage}
&
\begin{minipage}[c][1cm][c]{0.2\linewidth}
$(2,1)$
\end{minipage}
\\
\hline

\begin{minipage}[c][1cm][c]{0.3\linewidth}
$\big\langle\overline{\mathbf{N}},\mathbf{N},\overline{\mathbf{N}}\big\rangle \to \big\langle\overline{\mathbf{N}},\mathbf{S^2},\overline{\mathbf{N}}\big\rangle$
\end{minipage}
&
\begin{minipage}[c][1cm][c]{0.2\linewidth}
$(1,1) \to (2,1)$
\end{minipage}
&
\begin{minipage}[c][1cm][c]{0.2\linewidth}
$(2,1)$
\end{minipage}
\\
\hline
\rowcolor{gray!25}
\begin{minipage}[c][1cm][c]{0.3\linewidth}
$\big\langle\mathbf{N},\mathbf{Ad},\mathbf{N}\big\rangle \to \big\langle\mathbf{N},\mathbf{N},\mathbf{N}\big\rangle$
\end{minipage}
&
\begin{minipage}[c][1cm][c]{0.2\linewidth}
$(2,1) \to (1,2)$
\end{minipage}
&
\begin{minipage}[c][1cm][c]{0.2\linewidth}
$(2,2)$
\end{minipage}
\\
\hline

\begin{minipage}[c][1cm][c]{0.3\linewidth}
$\big\langle\mathbf{Ad},\mathbf{Ad},\mathbf{Ad}\big\rangle \to \big\langle\mathbf{Ad},\mathbf{N},\mathbf{Ad}\big\rangle$
\end{minipage}
&
\begin{minipage}[c][1cm][c]{0.2\linewidth}
$(2,1) \to (2,1)$
\end{minipage}
&
\begin{minipage}[c][1cm][c]{0.2\linewidth}
$(2,1)$
\end{minipage}
\\
\hline

\begin{minipage}[c][1cm][c]{0.3\linewidth}
$\big\langle\mathbf{A^2},\overline{\mathbf{A^2}},\mathbf{A^2}\big\rangle \to \big\langle\mathbf{A^2},\overline{\mathbf{N}},\mathbf{A^2}\big\rangle$
\end{minipage}
&
\begin{minipage}[c][1cm][c]{0.2\linewidth}
$(2,1) \to (2,1)$
\end{minipage}
&
\begin{minipage}[c][1cm][c]{0.2\linewidth}
$(2,1)$
\end{minipage}
\\
\hline

\begin{minipage}[c][1cm][c]{0.3\linewidth}
$\big\langle\mathbf{S^2},\overline{\mathbf{S^2}},\mathbf{S^2}\big\rangle \to \big\langle\mathbf{S^2},\overline{\mathbf{N}},\mathbf{S^2}\big\rangle$
\end{minipage}
&
\begin{minipage}[c][1cm][c]{0.2\linewidth}
$(2,1) \to (2,1)$
\end{minipage}
&
\begin{minipage}[c][1cm][c]{0.2\linewidth}
$(2,1)$
\end{minipage}
\\
\hline

\end{tabular}

\caption{All possible raising operations between the previously tabulated double-vertex cut states from \Cref{tab:doublesinglets}. Lowering operations are simply in the reverse direction. The second column shows the minimum energy cutoffs needed for the initial and final states in the transition, and the third column deduces the minimum overall energy cutoff required. The disallowed rows are grayed out, requiring $(n_e, n_f) = (2, 2)$.}\label{tab:allhops}
\end{table*}

Finally, to complete our systematic tabulation process, we must use selection rules established in \Cref{sec:largeN} to write down all possible hopping transitions between these double-vertex cut states.
It suffices to write down only the raising transitions, as in \Cref{sec:largeN}, because the lowering transitions are perfectly dual.
All such operations are provided in \Cref{tab:allhops}, which elaborates on the logic behind \Cref{tab:cuts} and \Cref{tab:hops}.

\section{Hopping Matrix Elements}\label{app:calc2}

Here, we calculate explicitly all hopping matrix elements shown in \Cref{tab:hops}.
In doing so, we will explicitly write down the vertex tensors that define the vertex singlets, and therefore this section also serves to set the phase conventions for the gauge-invariant basis states.

There are some symmetries apparent in \Cref{tab:hops}, but we will still write out all $11$ calculations explicitly, for the sake of clarity.
Additionally, we rewrite the ``hopping master equation" \eqref{eq:master} here for convenience, since it will be reused extensively:
\begin{widetext}
\begin{equation}\label{eq:master2}
    \bra{W} H_\hop(\ell) \ket{V}
    = \sqrt{\dim \mathbf R'\cdot\dim\mathbf R} \cdot V^1_{abc} \left(W^1_{a\beta\gamma}\right)^* T^{p}_{\beta \rho b} \cdot V^2_{cef} \left(W^2_{\gamma\eps f}\right)^* T^{q}_{\eps\rho e}.
\end{equation}
\end{widetext}
We will always refer to $\ket{V}$ as the ``initial state" and $\ket{W}$ as the ``final state," using the same symbols $V$ and $W$ throughout this section for each separate calculation.
When writing out vertex tensors for any symbol like $V$ or $W$ in component form, sometimes there will be indices that range over only a single value (if the corresponding irrep is a singlet)---in those cases, we will omit the index by using the $\cdot$ symbol, rather than writing something in its place.

Additionally, we need to define special symbols that will aid us in writing vertex singlets for states involving higher Casimir irreps.
For $1\le \alpha \le N_c^2-1$, we use $\lambda_{\alpha}$ to denote a set of linearly independent, traceless $N_c \times N_c$ Hermitian matrices that satisfy the trace orthonormality condition
\begin{equation}
    \tr\left(\lambda_{\alpha}\lambda_{\beta}\right) = \delta_{\alpha\beta}.
\end{equation}
For instance, one can choose a basis for the Lie algebra $\mathfrak{su}(N_c)$, e.g., proportional to the Gell-Mann matrices for $\SU(N_c)$.
In our case, the import conclusion will be
\begin{equation}
    \left(\lambda_{\alpha}\right)_{ab}^* \left(\lambda_{\alpha}\right)_{ab} = N_c^2 - 1,
\end{equation}
where $\left(\lambda_{\alpha}\right)_{ab}$ is the matrix entry of $\lambda_{\alpha}$ for $1\le a,b\le N_c$.
These symbols can intertwine $\mathbf{N}\otimes\overline{\mathbf{N}} \to \mathbf{Ad}$.

To intertwine $\mathbf{N}\otimes\mathbf{N} \to \mathbf{A^2}$ and $\mathbf{N}\otimes\mathbf{N}\to\mathbf{S^2}$ (and the conjugate cases), we will need the following symbols that generalize the Kronecker delta for indices $a$, $b$, and $c$, with $1 \le a, b \le N_c$ and $1 \le c \le \frac12 N_c(N_c+1)$:
\begin{subequations}
\begin{align}
    \Delta^A_{(ab)c} &\equiv \begin{cases}0 & a = b,\\ \delta_{\phi_{ab} c} & a < b,\\ -\delta_{\phi_{ba} c} & a > b.\end{cases}\\
    \Delta^S_{(ab)c} &\equiv \begin{cases}\sqrt{2} & a = b = c,\\ \delta_{\phi_{ab} c} & a < b, \\ \delta_{\phi_{ba} c} & a > b.\end{cases}
\end{align}
\end{subequations}
We use here the notation $\phi_{ij}$ to mean any bijection between the set of index pairs $(i,j)$ with $1\le i < j\le N_c$ and the set of indices $N_c+1 \le k \le \frac 12 N_c(N_c + 1)$.\footnote{This works for any integer $N_c \ge 2$.}
The symbol $\delta$ continues to denote the plain Kronecker delta between indices coming from equivalent index sets.

For future reference, it will be useful to note that
\begin{subequations}
\begin{align}
    \Delta^A_{(ab)c} \Delta^A_{(ab)c} &= N_c(N_c - 1),\\
    \Delta^S_{(ab)c} \Delta^S_{(ab)c} &= N_c(N_c + 1).
\end{align}
\end{subequations}
The top line follows from index chasing: there are $N_c$ values that $a$ runs over, leaving $N_c-1$ values of $b$ that can positively contribute; the set of such pairs $(a,b)$ is in a $2$-to-$1$ correspondence with the $\frac 12 N_c(N_c-1)$ values of $c$ that are allowed.
For the bottom line, the exact same argument holds, but we have to add in the $N_c$ cases where $a=b=c$, each with a weight $(\sqrt{2})^2 = 2$.

Before we present the calculations, we will first need to write down the fermion operator matrix elements (the $T$-symbols).
In doing so, we also implicitly define the phase conventions for the Fock space basis states (our convention is equivalent to a wedge product that preserves the norm of the state).
We choose
\begin{subequations}
\begin{align}
    T^0_{ij\cdot} &= \delta_{ij},\\
    T^1_{ijk} &= \Delta^A_{(jk)i}.
\end{align}
\end{subequations}
Higher $T$-symbols are not needed for the transitions considered in this work.

The first transition has initial state $\ket{V} = \ket{\mathbf{1},\mathbf{1},\mathbf{1}}$ and final state $\ket{W} = \ket{\mathbf{1},\mathbf{N},\mathbf{1}}$.
The vertex tensors for these states are given by
\begin{subequations}
\begin{align}
    V^1_{\cdot\cdot\cdot} &= 1,\\
    V^2_{\cdot\cdot\cdot} &= 1,\\
    W^1_{\cdot\beta\gamma} &= \delta_{\beta\gamma} \times N_c^{-1/2},\\
    W^2_{\gamma\eps\cdot} &= \delta_{\gamma\eps} \times N_c^{-1/2}.
\end{align}
\end{subequations}

Then \eqref{eq:master2} is easily evaluated as
\begin{equation}
    \sqrt{N_c} \cdot \delta_{\beta\gamma} N_c^{-1/2} \delta_{\beta\rho} \cdot \delta_{\gamma\eps}N_c^{-1/2} \delta_{\eps\rho} = \boxed{\sqrt{N_c}}.
\end{equation}

The second transition has initial state $\ket{V} = \ket{\mathbf{N},\overline{\mathbf{N}},\mathbf{N}}$ and final state $\ket{W} = \ket{\mathbf{N},\mathbf{1},\mathbf{N}}$.
The vertex tensors for these states are given by
\begin{subequations}
\begin{align}
    V^1_{a\cdot c} &= \delta_{ac}\times N_c^{-1/2},\\
    V^2_{d\cdot f} &= \delta_{df}\times N_c^{-1/2},\\
    W^1_{\alpha\beta\cdot} &= \delta_{\alpha\beta} \times N_c^{-1/2},\\
    W^2_{\cdot\eps\zeta} &= \delta_{\eps\zeta} \times N_c^{-1/2}.
\end{align}
\end{subequations}

Then \eqref{eq:master2} is similarly evaluated as
\begin{widetext}
\begin{equation}
    \sqrt{N_c} \cdot \delta_{ac}N_c^{-1/2} \delta_{a\beta}N_c^{-1/2} \delta_{\beta\rho} \cdot \delta_{cf}N_c^{-1/2} \delta_{\eps f}N_c^{-1/2} \delta_{\eps\rho} = \boxed{\frac1{\sqrt{N_c}}}.
\end{equation}

The next transition has initial state $\ket{V} = \ket{\mathbf{N},\mathbf{1},\mathbf{1}}$ and final state $\ket{W} = \ket{\mathbf{N},\mathbf{N},\mathbf{1}}$.
The vertex tensors for these states are given by
\begin{subequations}
\begin{align}
    V^1_{ab\cdot} &= \delta_{ab}\times N_c^{-1/2},\\
    V^2_{\cdot\cdot\cdot} &= 1,\\
    W^1_{\alpha\beta\gamma} &= -\Delta^A_{(\alpha\gamma)\beta}\times(N_c^2-N_c)^{-1/2},\\
    W^2_{\delta\eps\cdot} &= \delta_{\delta\eps}\times N_c^{-1/2}.
\end{align}
\end{subequations}

Then \eqref{eq:master2} is evaluated as
\begin{align}
    -\sqrt{N_c} \cdot \delta_{ab}N_c^{-1/2} \Delta^A_{(a\gamma)\beta}(N_c^2-N_c)^{-1/2} \Delta^A_{(\rho b)\beta} \cdot  \delta_{\gamma\eps}N_c^{-1/2} \delta_{\eps\rho} &= (N_c^3-N_c^2)^{-1/2} \Delta^A_{(\eps b)\beta}\Delta^A_{(\eps b)\beta}\nonumber\\
    & = \boxed{\sqrt{N_c-1}}.
\end{align}
Note that we introduced a minus sign in the coefficients for the odd vertex singlet $\big\langle \mathbf{N}, \mathbf{N}$ (see \Cref{app:tab} for this notation) in order to absorb the phase of the matrix element.
This means we will have to reuse that minus sign (keeping the same index order) anytime the odd vertex singlet $\big\langle \mathbf{N}, \mathbf{N}$ reappears in future calculations.\footnote{Technically, we merely have to preserve the phase for \textit{global} gauge-invariant \textit{global} lattice states, which is much weaker than this local condition, but the local approach certainly suffices.}

The next transition has initial state $\ket{V} = \ket{\mathbf{1},\mathbf{1},\mathbf{N}}$ and final state $\ket{W} = \ket{\mathbf{1},\mathbf{N},\mathbf{N}}$.
The vertex tensors for these states are given by
\begin{subequations}
\begin{align}
    V^1_{\cdot\cdot\cdot} &= 1,\\
    V^2_{\cdot ef} &= \delta_{ef}\times N_c^{-1/2},\\
    W^1_{\cdot\beta\gamma} &= \delta_{\beta\gamma}\times N_c^{-1/2},\\
    W^2_{\delta\eps\zeta} &= \Delta^A_{(\delta\zeta)\eps}\times (N_c^2-N_c)^{-1/2}.
\end{align}
\end{subequations}

Then \eqref{eq:master2} is similarly evaluated as
\begin{align}
    \sqrt{N_c} \cdot \delta_{\beta\gamma}N_c^{-1/2} \delta_{\beta\rho} \cdot \delta_{ef}N_c^{-1/2} \Delta^A_{(\gamma f)\eps}(N_c^2-N_c)^{-1/2} \Delta^A_{(\rho e)\eps} &= (N_c^3-N_c^2)^{-1/2} \Delta^A_{(\rho e)\eps} \Delta^A_{(\rho e)\eps}\nonumber\\
    & = \boxed{\sqrt{N_c-1}}.
\end{align}
Note that despite adding a phase factor to the odd vertex singlet $\big\langle \mathbf{N},\mathbf{N}$, we \textit{do not} need to add one to the even vertex singlet $\mathbf{N},\mathbf{N}\big\rangle$.
This is because odd and even vertex singlets are entirely different states on the lattice, and therefore can have their phase conventions chosen independently from each other.
Another way to set the phase conventions more geometrically is by encoding them directly in the discrete symmetry group.
For instance, one can view the even vertex singlet state as a charge-conjugated, parity-reversed odd vertex singlet, and accordingly switch the index order in the definition of $W^2$, while matching the phase convention already chosen for the corresponding odd vertex singlet---this would have the same effect as choosing the phase ad hoc as we have done here.

The next transition has initial state $\ket{V} = \ket{\mathbf{N},\mathbf{1},\mathbf{N}}$ and final state $\ket{W} = \ket{\mathbf{N},\mathbf{N},\mathbf{N}}$.
The vertex tensors for these states are given by
\begin{subequations}
\begin{align}
    V^1_{ab\cdot} &= \delta_{ab}\times N_c^{-1/2},\\
    V^2_{\cdot ef} &= \delta_{ef}\times N_c^{-1/2},\\
    W^1_{\alpha\beta\gamma} &= -\Delta^A_{(\alpha\gamma)\beta}\times(N_c^2-N_c)^{-1/2},\\
    W^2_{\delta\eps\zeta} &= \Delta^A_{(\delta\zeta)\eps}\times(N_c^2-N_c)^{-1/2}.
\end{align}
\end{subequations}
Note that the minus sign on $W^1$ is needed, as previously discussed for the odd vertex singlet $\big\langle\mathbf{N},\mathbf{N}$.

Then \eqref{eq:master2} is evaluated as
\begin{align}\nonumber
&-\sqrt{N_c} \cdot \delta_{ab}N_c^{-1/2} \Delta^A_{(a\gamma)\beta}(N_c^2-N_c)^{-1/2} \Delta^A_{(\rho b)\beta} \cdot \delta_{ef}N_c^{-1/2} \Delta^A_{(\gamma f)\eps}(N_c^2-N_c)^{-1/2} \Delta^A_{(\rho e)\eps}\\
&= N_c^{-1/2}(N_c^2-N_c)^{-1}\Delta^A_{(\gamma b)\beta} \Delta^A_{(\rho b)\beta}\Delta^A_{(\gamma e)\eps}\Delta^A_{(\rho e)\eps}.
\end{align}
To evaluate the final contraction, we perform index chasing.
Index $\beta$ enforces that indices $\gamma$ and $\rho$ are equal; index $\eps$ always agrees with that enforcement.
Let us say $\gamma = \rho \equiv z$.
Then there are $N_c$ choices for index $z$, which leaves $N_c-1$ independent choices for index $b$, and $N_c-1$ independent choices for index $e$.
Every single one of these independent choices uniquely determines corresponding values of $\beta$ and $\eps$, and contributes with weight $1$.
Therefore, the contraction of the $\Delta^A$ symbols above evaluates to $N_c(N_c-1)^2$.
Plugging this in, we find the matrix element
\begin{equation}
    N_c^{-1/2}(N_c^2-N_c)^{-1} \cdot N_c(N_c-1)^2 = \boxed{\frac{N_c-1}{\sqrt{N_c}}}.
\end{equation}

The next transition has initial state $\ket{V} = \ket{\mathbf{N},\overline{\mathbf{N}},\mathbf{N}}$ and final state $\ket{W} = \ket{\mathbf{N},\mathbf{Ad},\mathbf{N}}$.
The vertex tensors for these states are given by
\begin{subequations}
\begin{align}
    V^1_{a\cdot c} &= \delta_{ac}\times N_c^{-1/2},\\
    V^2_{d\cdot f} &= \delta_{df}\times N_c^{-1/2},\\
    W^1_{\alpha\beta\gamma} &= \left(\lambda_{\gamma}\right)_{\alpha\beta}\times(N_c^2-1)^{-1/2},\\
    W^2_{\delta\eps\zeta} &= \left(\lambda_{\delta}\right)_{\eps\zeta}\times(N_c^2-1)^{-1/2}.
\end{align}
\end{subequations}

Then \eqref{eq:master2} is evaluated as
\begin{align}
    &\sqrt{N_c(N_c^2-1)} \cdot \delta_{ac}N_c^{-1/2} \left(\lambda_{\gamma}\right)_{a\beta}^*(N_c^2-1)^{-1/2} \delta_{\beta\rho} \cdot \delta_{cf}N_c^{-1/2} \left(\lambda_{\gamma}\right)_{\eps f}^*(N_c^2-1) \delta_{\eps\rho}\nonumber\\
    =\,& (N_c^3-N_c)^{-1/2}\left(\lambda_{\gamma}\right)_{\beta c} \left(\lambda_{\gamma}\right)_{\beta c}^*\nonumber\\
    =\,& \boxed{\sqrt{\frac{N_c^2-1}{N_c}}}.
\end{align}

The next transition has initial state $\ket{V} = \ket{\mathbf{Ad},\mathbf{Ad},\mathbf{Ad}}$ and final state $\ket{W} = \ket{\mathbf{Ad},\mathbf{N},\mathbf{Ad}}$.
The vertex tensors for these states are given by
\begin{subequations}
\begin{align}
    V^1_{a\cdot c} &= \delta_{ac}\times(N_c^2-1)^{-1/2},\\
    V^2_{d\cdot f} &= \delta_{df}\times(N_c^2-1)^{-1/2},\\
    W^1_{\alpha\beta\gamma} &= \left(\lambda_{\alpha}\right)_{\beta\gamma}\times(N_c^2-1)^{-1/2},\\
    W^2_{\delta\eps\zeta} &= \left(\lambda_{\zeta}\right)_{\delta\eps}\times(N_c^2-1)^{-1/2}.
\end{align}
\end{subequations}

Then \eqref{eq:master2} is evaluated as
\begin{align}
&\sqrt{N_c(N_c^2-1)} \cdot \delta_{ac}(N_c^2-1)^{-1/2} \left(\lambda_a\right)_{\beta\gamma}^*(N_c^2-1)^{-1/2} \delta_{\beta\rho} \cdot \delta_{cf}(N_c^2-1)^{-1/2} \left(\lambda_f\right)_{\gamma\eps}^*(N_c^2-1)^{-1/2} \delta_{\eps\rho}\nonumber\\
=\,& N_c^{1/2}(N_c^2-1)^{-3/2}\left(\lambda_c\right)_{\gamma\beta}\left(\lambda_c\right)_{\gamma\beta}^*\nonumber\\
=\,& \boxed{\sqrt{\frac{N_c}{N_c^2-1}}}.
\end{align}

The next transition has initial state $\ket{V} = \ket{\overline{\mathbf{N}},\mathbf{N},\overline{\mathbf{N}}}$ and final state $\ket{W} = \ket{\overline{\mathbf{N}},\mathbf{A^2},\overline{\mathbf{N}}}$.
The vertex tensors for these states are given by
\begin{subequations}
\begin{align}
    V^1_{a\cdot c} &= \delta_{ac}\times N_c^{-1/2},\\
    V^2_{d\cdot f} &= \delta_{df}\times N_c^{-1/2},\\
    W^1_{\alpha\beta\gamma} &= \Delta^A_{(\alpha\beta)\gamma}\times(N_c^2-N_c)^{-1/2},\\
    W^2_{\delta\eps\zeta} &= -\Delta^A_{(\eps\zeta)\delta}\times(N_c^2-N_c)^{-1/2}.
\end{align}
\end{subequations}

Then \eqref{eq:master2} is evaluated as
\begin{align}
    &-\sqrt{\frac 12 N_c (N_c^2-N_c)} \cdot \delta_{ac}N_c^{-1/2} \Delta^A_{(a\beta)\gamma}(N_c^2-N_c)^{-1/2} \delta_{\beta\rho} \cdot \delta_{cf}N_c^{-1/2} \Delta^A_{(\eps f)\gamma}(N_c^2-N_c)^{-1/2} \delta_{\eps\rho}\nonumber\\
    &= (2N_c^3-2N_c^2)^{-1/2}\Delta^A_{(\beta c)\gamma}\Delta^A_{(\beta c)\gamma}\nonumber\\
    & = \boxed{\sqrt{\frac{N_c-1}{2}}}.
\end{align}
Note the minus sign we added into the convention for the even vertex singlet $\mathbf{A^2},\overline{\mathbf{N}}\big\rangle$ in order to ensure the positive sign on the matrix element.

The next transition has initial state $\ket{V} = \ket{\mathbf{A^2},\overline{\mathbf{A^2}},\mathbf{A^2}}$ and final state $\ket{W} = \ket{\mathbf{A^2},\overline{\mathbf{N}},\mathbf{A^2}}$.
The vertex tensors for these states are given by
\begin{subequations}
\begin{align}
    V^1_{a\cdot c} &= \delta_{ac}\times \sqrt{2/(N_c^2-N_c)},\\
    V^2_{d\cdot f} &= \delta_{df}\times \sqrt{2/(N_c^2-N_c)},\\
    W^1_{\alpha\beta\gamma} &= -\Delta^A_{(\beta\gamma)\alpha}\times(N_c^2-N_c)^{-1/2},\\
    W^2_{\delta\eps\zeta} &= \Delta^A_{(\delta\eps)\zeta}\times(N_c^2-N_c)^{-1/2}.
\end{align}
\end{subequations}

Then \eqref{eq:master2} is evaluated as
\begin{align}
    &-\sqrt{\frac12 N_c(N_c^2-N_c)} \cdot \delta_{ac}\sqrt{2/(N_c^2-N_c)} \Delta^A_{(\beta\gamma)a}(N_c^2-N_c)^{-1/2} \delta_{\beta\rho} \cdot \delta_{cf}\sqrt{2/(N_c^2-N_c)} \Delta^A_{(\gamma\eps)f}(N_c^2-N_c)^{-1/2} \delta_{\eps\rho}\nonumber\\
    &= \sqrt{2} (N_c-1)^{-3/2}N_c^{-1} \Delta^A_{(\gamma\beta)c} \Delta^A_{(\gamma\beta)c}\nonumber\\
    &=\boxed{\sqrt{\frac{2}{N_c-1}}}.
\end{align}
This time, we added the phase to the vertex coefficients for the odd singlet $\big\langle\mathbf{A^2},\overline{\mathbf{N}}$.

The next transition has initial state $\ket{V} = \ket{\overline{\mathbf{N}},\mathbf{N},\overline{\mathbf{N}}}$ and final state $\ket{W} = \ket{\overline{\mathbf{N}},\mathbf{S^2},\overline{\mathbf{N}}}$.
The vertex tensors for these states are given by
\begin{subequations}
\begin{align}
    V^1_{a\cdot c} &= \delta_{ac}\times N_c^{-1/2},\\
    V^2_{d\cdot f} &= \delta_{df}\times N_c^{-1/2},\\
    W^1_{\alpha\beta\gamma} &= \Delta^S_{(\alpha\beta)\gamma}\times(N_c^2+N_c)^{-1/2},\\
    W^2_{\delta\eps\zeta} &= \Delta^S_{(\eps\zeta)\delta}\times(N_c^2+N_c)^{-1/2}.
\end{align}
\end{subequations}

Then \eqref{eq:master2} is evaluated as
\begin{align}
    &\sqrt{\frac 12 N_c(N_c^2+N_c)} \cdot \delta_{ac}N_c^{-1/2} \Delta^S_{(a\beta)\gamma}(N_c^2+N_c)^{-1/2} \delta_{\beta\rho} \cdot \delta_{cf}N_c^{-1/2} \Delta^S_{(\eps f)\gamma}(N_c^2+N_c)^{-1/2} \delta_{\eps\rho}\nonumber\\
    =\,& (2N_c^3+2N_c^2)^{-1/2} \Delta^S_{(c\beta)\gamma} \Delta^S_{(\beta c)\gamma}\nonumber\\
    =\,& \boxed{\sqrt{\frac{N_c+1}{2}}}.
\end{align}

The next transition has initial state $\ket{V} = \ket{\mathbf{S^2},\overline{\mathbf{S^2}},\mathbf{S^2}}$ and final state $\ket{W} = \ket{\mathbf{S^2},\overline{\mathbf{N}},\mathbf{S^2}}$.
The vertex tensors for these states are given by
\begin{subequations}
\begin{align}
    V^1_{a\cdot c} &= \delta_{ac}\times\sqrt{2/(N_c^2+N_c)},\\
    V^2_{d\cdot f} &= \delta_{df}\times\sqrt{2/(N_c^2+N_c)},\\
    W^1_{\alpha\beta\gamma} &= \Delta^S_{(\beta\gamma)\alpha}\times(N_c^2+N_c)^{-1/2},\\
    W^2_{\delta\eps\zeta} &= \Delta^S_{(\delta\eps)\zeta}\times(N_c^2+N_c)^{-1/2}.
\end{align}
\end{subequations}

Then \eqref{eq:master2} is evaluated as
\begin{align}
    &\sqrt{\frac12 N_c(N_c^2+N_c)} \cdot \delta_{ac}\sqrt{2/(N_c^2+N_c)} \Delta^S_{(\beta\gamma)a}(N_c^2+N_c)^{-1/2} \delta_{\beta\rho} \cdot \delta_{cf}\sqrt{2/(N_c^2+N_c)} \Delta^S_{(\gamma\eps)f}(N_c^2+N_c)^{-1/2} \delta_{\eps\rho}\nonumber\\
    =\,& \sqrt{2}(N_c+1)^{-3/2}N_c^{-1} \Delta^S_{(\beta\gamma)c} \Delta^S_{(\gamma\beta)c}\nonumber\\
    =\,& \boxed{\sqrt{\frac{2}{N_c+1}}}.
\end{align}
This completes the full derivation of \Cref{tab:hops}.
\end{widetext}

\bibliographystyle{apsrev4-1}
\bibliography{formalism_bib}

\end{document}